%% file: ms.tex
\newcommand{\mydriver}{pdftex} 

\documentclass[12pt,\mydriver,openright]{thesis}

\usepackage{titlesec}
   \titleformat{\chapter}
      {\normalfont\large}{Chapter \thechapter:}{1em}{}

\usepackage{pdfpages} 
\usepackage{graphicx}
\usepackage{epstopdf}
\usepackage{cite}
\usepackage{lscape}
\usepackage{indentfirst}
\usepackage{latexsym}
\usepackage{amsmath}
\usepackage{multirow}
\usepackage{tabls}
\usepackage{wrapfig}
\usepackage{slashbox}
\usepackage{longtable}
\usepackage{threeparttablex}
\usepackage{tabularx}
\newcolumntype{R}{>{\raggedright\arraybackslash}m}  
\newcolumntype{L}{>{\raggedright\arraybackslash}m}  
\usepackage{supertabular}
\usepackage{subfigure}
\usepackage{csquotes}
\usepackage[pdftex, bookmarks, colorlinks=true, plainpages=false, citecolor=blue, urlcolor=blue, filecolor=blue, linkcolor=blue]{hyperref} 

\renewcommand{\baselinestretch}{2}
\begin{document}

\let\oldcaption\caption
\def\splitter #1. #2@@{\oldcaption[#1]{#1. #2}}
\def\caption#1{\splitter #1@@}

\include{Abstract} 
\include{Titlepage} 
\include{Copyright} 

\pagestyle{plain}
\pagenumbering{roman}
\setcounter{page}{2}

\addcontentsline{toc}{chapter}{Dedication}
\include{Dedication} 
\addcontentsline{toc}{chapter}{Acknowledgments}
\include{Acknowledgments} 

\renewcommand{\baselinestretch}{1}
\small\normalsize
\tableofcontents 
\newpage
\addcontentsline{toc}{chapter}{List of Tables}
\listoftables 
\newpage
\addcontentsline{toc}{chapter}{List of Figures}
\listoffigures 
\newpage
\addcontentsline{toc}{chapter}{List of Abbreviations}
\include{Abbreviations}

\newpage
\setlength{\parskip}{0em}
\renewcommand{\baselinestretch}{2}
\small\normalsize

\makeatletter
\def\env@matrix{\hskip -\arraycolsep
  \let\@ifnextchar\new@ifnextchar
  \linespread{1}\selectfont
  \renewcommand{\arraystretch}{0.5}%
  \array{*\c@MaxMatrixCols c}}
\makeatother


\setcounter{page}{1}
\pagenumbering{arabic}
\newpage \thispagestyle{empty}\addtocounter{page}{-1} \mbox{} \pagebreak
\include{1.Chapter}

\include{2.Chapter}

\include{3.Chapter}

\include{4.Chapter}
\newpage \thispagestyle{empty}\addtocounter{page}{-1} \mbox{} \pagebreak
\include{5.Chapter}
\include{6.Chapter}

\newpage \thispagestyle{empty}\addtocounter{page}{-1} \mbox{} \pagebreak
\include{7.Chapter}

\include{8.Chapter}
\newpage \thispagestyle{empty}\addtocounter{page}{-1} \mbox{} \pagebreak
\include{9.Chapter}

\newpage \thispagestyle{empty}\addtocounter{page}{-1} \mbox{} \pagebreak

\include{10.Chapter}

\newpage \thispagestyle{empty}\addtocounter{page}{-1} \mbox{} \pagebreak
\include{11.Chapter}

\include{12.Chapter}

\titleformat{\chapter}
      {\normalfont\large}{Appendix \thechapter:}{1em}{}
\appendix
\newpage \thispagestyle{empty}\addtocounter{page}{-1} \mbox{} \pagebreak
\include{A.Appendix}

\include{B.Appendix}

\newpage \thispagestyle{empty}\addtocounter{page}{-1} \mbox{} \pagebreak
\include{C.Appendix}

\newpage \thispagestyle{empty}\addtocounter{page}{-1} \mbox{} \pagebreak
\include{D.Appendix}

\include{E.Appendix}
\include{F.Appendix}
\newpage \thispagestyle{empty}\addtocounter{page}{-1} \mbox{} \pagebreak
\include{G.Appendix}

\newpage \thispagestyle{empty}\addtocounter{page}{-1} \mbox{} \pagebreak
\include{H.Appendix} 
\newpage \thispagestyle{empty}\addtocounter{page}{-1} \mbox{} \pagebreak
\include{I.Appendix}

\include{J.Appendix}

\include{K.Appendix}

\newpage \thispagestyle{empty}\addtocounter{page}{-1} \mbox{} \pagebreak

\let\caption\oldcaption 

\include{L.Appendix}

\renewcommand{\baselinestretch}{1}
\small\normalsize
\newpage \thispagestyle{empty}\addtocounter{page}{-1} \mbox{} \pagebreak
\newpage
\addcontentsline{toc}{chapter}{Bibliography}
\bibliographystyle{unsrt} 
\bibliography{mainthesis} 
\end{document}

%% file: Abstract.tex

\hbox{\ }

\renewcommand{\baselinestretch}{1}
\small \normalsize

\begin{center}
\large{{ABSTRACT}} 

\vspace{3em} 

\end{center}
\hspace{-.15in}
\begin{tabular}{ll}
Title of dissertation:  & {\large DESIGN OF A NONLINEAR QUASI-}\\
						& {\large INTEGRABLE LATTICE FOR RESONANCE } \\
						& {\large SUPPRESSION AT THE UNIVERSITY OF } \\
						& {\large MARYLAND ELECTRON RING } \\
\ \\
&                          {\large Kiersten Ruisard, Doctor of Philosophy, 2018} \\
\ \\
Dissertation directed by: & {\large Doctor Timothy Koeth} \\
&  							{\large	Department of Physics } \\
\end{tabular}

\vspace{3em}

\renewcommand{\baselinestretch}{2}
\large \normalsize

Conventional particle accelerators use linear focusing forces for transverse confinement. As a consequence of linearity, accelerating rings are sensitive to myriad resonances and instabilities. 
At high beam intensity, uncontrolled resonance-driven losses can deteriorate beam quality and cause damage or radio-activation in beam line components and surrounding areas. 
This is currently a major limitation of achievable current densities in state-of-the-art accelerators.
Incorporating nonlinear focusing forces into machine design should provide immunity to resonances through nonlinear detuning of particle orbits from driving terms. 
A theory of nonlinear integrable beam optics is currently being investigated for use in accelerator rings.
Such a system has potential to overcome the limits on achievable beam intensity.

This dissertation presents a plan for implementing a proof-of-principle quasi-integrable octupole lattice at the University of Maryland Electron Ring (UMER). 
UMER is an accelerator platform that supports the study of high-intensity beam dynamics. 
In this dissertation, two designs are presented that differ in both complexity and strength of predicted effects. 
A configuration with a single, relatively long octupole magnet is expected to be more stabilizing than an arrangement of many short, distributed octupoles. 

Preparation for this experiment required the development and characterization of a low-intensity regime previously not operated at UMER. 
Additionally, required tolerances for the control of first and second order beam moments in the proposed experiments have been determined on the basis of simulated beam dynamics. 
In order to achieve these tolerances, a new method for improved orbit correction is developed. 
Finally, a study of resonance-driven losses in the linear UMER lattice is discussed.

%% file: Titlepage.tex

\thispagestyle{empty}
\hbox{\ }
\vspace{1in}
\renewcommand{\baselinestretch}{1}
\small\normalsize
\begin{center}

\large{{DESIGN OF A NONLINEAR QUASI-INTEGRABLE LATTICE FOR DEMONSTRATION OF RESONANCE SUPPRESSION AT THE UNIVERSITY OF MARYLAND ELECTRON RING}}\\
\ \\
\ \\
\large{by} \\
\ \\
\large{Kiersten Ruisard}
\ \\
\ \\
\ \\
\ \\
\normalsize
Dissertation submitted to the Faculty of the Graduate School of the \\
University of Maryland, College Park in partial fulfillment \\
of the requirements for the degree of \\
Doctor of Philosophy \\
2018
\end{center}

\vspace{7.5em}

\noindent Advisory Committee: \\
Professor Thomas Antonsen,Co-Chair \\
Dr. Timothy Koeth, Co-Chair/Advisor \\
Professor Andrew Baden \\
Dr. Brian Beaudoin \\
Dr. Irving Haber	\\
Professor Patrick O'Shea

%% file: Copyright.tex

\thispagestyle{empty}
\hbox{\ }

\vfill
\renewcommand{\baselinestretch}{1}
\small\normalsize

\vspace{-.65in}

\begin{center}
\large{\copyright \hbox{ }Copyright by\\
Kiersten Ruisard  
\\
2018}
\end{center}

\vfill

%% file: Dedication.tex

\renewcommand{\baselinestretch}{2}
\small\normalsize
\hbox{\ }
 
\vspace{-.65in}

\begin{center}
\large{Dedication}

For George.

\end{center} 

%% file: Acknowledgments.tex

\renewcommand{\baselinestretch}{2}
\small\normalsize
\hbox{\ }
 
\vspace{-.65in}

\begin{center}
\large{Acknowledgments} 
\end{center} 

\vspace{1ex}
The biggest acknowledgment is for Tim Koeth, who not only advised my dissertation research but is responsible for introducing me to accelerator physics, bringing me to Maryland, guiding me through the project and introducing me to the wonderful accelerator community. Tim, your passion for research and appreciation of the beauty in science inspires me every day. 

Thank you to my committee for your time and attention. Especially thanks to Brian Beaudoin and Irv Haber, who painstakingly sifted through every word in this document and offered very useful comments. Even more, your guidance and support at every step of this journey has been essential and very appreciated. 

Thanks to everyone else in the UMER group. Your help, engagement, discussions and arguments have made this an enriching experience. In no particular order, thank you Rami Kishek, Dave Sutter, Eric Montgomery, Santiago Bernal and Max Cornacchia. To the UMER graduate students (Will Stem, Yichao Mo, Hao Zhang, Kamal PorRezai, Heidi Baumgartner, Levon Dovlatyan, Moiz Siddiqi, Dave Matthew) thanks for being friends as well as co-workers. Special thanks to Dave for creating the UMER octupoles and Heidi for picking up the torch and building components for this experiment. Thanks also to the many undergraduate students who contributed to this project.

Much of what I understand about this project is due to interactions and exchange with the IOTA collaboration. Thank you Sergei Nagaitsev for supporting the UMER project and teaching a course on the theory behind my thesis project. Thanks also to Sasha Valishev and Sergei Antipov, who hosted my visit to Fermilab during the start of this project. Also thanks to the folks at Radiasoft, for many thoughtful discussions and good advice.

Much of what I understand about accelerators is thanks to the US Particle Accelerator School. Much thanks to Sue Winchester and Bill Barletta for organizing the USPAS meetings and to all my course instructors, TA's and fellow students, who are too numerous to name. 
Thanks are also due to my many physics teachers, who sparked the initial interest and led me down this path. Especially Sean Chappe and Andrew Baker; this whole venture starts with your good mentorship when I was still deciding my path in life.

Thank you to my family for supporting my graduate studies and celebrating my academic milestones. Your enthusiasm for education held me up when I was tired and set my path from a young age. I hope I've made you proud.

Finally, thank you to my College Park family: Joe, Ginny, Kevin, Nat, Meredith, Madi, Alireza and Rachel. The College Park Cooking Collective has kept me going and nourished me, in both mind and through copious amounts of food. Your friendship, not Town Hall, is truely the gem of College Park. Also, I owe \textit{most} of my sanity to the people and horses in the Maryland Equestrian Club. I treasure the hours spent doing barn work, riding and socializing with this wonderful group.

Thanks to everybody, named and unnamed, who has supported me in this journey. You have filled my graduate school years with love, good times and fond memories. Especially thanks to George, I couldn't have done it without you. 

%% file: Abbreviations.tex

\renewcommand{\baselinestretch}{1}
\small\normalsize
\hbox{\ }

\vspace{-3em}

\begin{center}
\large{List of Symbols and Abbreviations}
\end{center} 

\vspace{3pt}

\begin{tabular}{ll}
AG & Alternating gradient \\
BD & Bending dipole \\
$\beta$& Betatron lattice function \\
$\beta_*$ & Betatron function at beam waist \\
BGRD & Gridded field element in WARP \\
BPM & Beam position monitor \\
DA & Dynamic aperture \\
$\epsilon$& emittance\\
FMA & Frequency map analysis \\
FODO & Periodic lattice of alternating focusing and de-focusing quads\\
FOFO & Lattice with periodic focusing elements \\
$G_3$ & octupole gradient\\
$H_N$ & Normalized single-particle Hamiltonian \\
I & Beam current \\
IOTA & Integrable Optics Test Accelerator \\
K & beam perveance \\
$K_x$, $K_y$ & geometric focusing strength in $x$, $y$ planes\\
KAM & Kolmogorov Arnold Moser theorem\\
KV & Kapchinskij-Vladimirskij distribution\\
MMLT & Multipole element in WARP \\
NAFF & Numerical analysis of fundamental frequency\\
NLIO & Nonlinear integrable optics \\
$\nu$ & tune\\
PCB & Printed circuit board \\
PD & Pulsed injection dipoles \\
QIO & Quasi-integrable optics \\
QR & Ring quadrupole \\
RC & Ring chamber \\
RSV & Ring steerer vertical \\
SD & Steering dipole \\
SSV & Short steerer vertical\\
UMER & University of Maryland Electron Ring\\
WCM & Wall current monitor \\
YQ & Pulsed ``Y-section" injection quadrupole\\
\end{tabular}

%% file: 1.Chapter.tex

\renewcommand{\thechapter}{1}

\chapter{Introduction}

\section{Motivation}

Modern day accelerators have far exceeded expectations of the early accelerator pioneers. 
Since Rutherford first used naturally accelerated alpha particles to probe atomic structure, advancements in accelerator capabilities continue to access previously incomprehensible regimes in both beam energy and intensity. 
On the energy frontier, research in super-conducting magnet and radio-frequency (RF) cavities as well as plasma-based acceleration reach towards higher total energy. The current state of the art is the Large Hadron Collider, at 14 TeV collision energy, with future plans aimed at a 100 TeV ``Future Circular Collider."  

An increasing number of applications require comparable advances in beam intensity. 
Intensity scales with the density of particles in a beam and can be measured in terms of beam power, luminosity or brightness depending on the application. 
Some high-intensity applications include high-luminosity colliders, high-brightness light sources and medical radioisotope production.
Many research fields rely on secondary beams generated by energetic hadrons colliding with targets. Examples include spallation neutron sources for neutron spectroscopy, accelerator-driven systems for nuclear waste treatment and neutrino factories for high-energy physics research. As beam energy is set by desired target interaction, increased power for next-generation machines requires increased beam current on target.

At high beam current, Coulomb interactions between charged particles become significant and induce a \textit{space charge} force on the beam that complicates dynamics. This force is highly dependent on the evolving beam distribution and is typically nonlinear. The dynamics of high-intensity accelerators extend beyond the scope of conventional accelerator theory, which assumes linearity and negligible space charge.
Space charge induced nonlinearities may drive beam loss but also decrease the loss threshold. To avoid excessive radio-activation of the surrounding environment, loss rates must be less than one Watt per meter, effectively capping the maximum beam intensity that can be safely transported in a beam line. On-going research seeks to raise the intensity ceiling by better understanding of intensity-driven losses and development of mitigation strategies. This dissertation describes implementation of a novel theory of nonlinear focusing proposed to mitigate space-charge driven resonant losses in future high-intensity rings.

\section{Historical perspective}

Modern accelerator design is based on Courant and Snyder`s theory of the alternating-gradient (AG) synchrotron, developed in 1952.\cite{Courant1958}
In an alternating-gradient accelerator, quadrupole magnets with linear restoring forces provide transverse confinement of the beam. The breakthrough of alternating gradients (where the beam is alternatively focused and defocused as it propagates) allowed a higher net focusing effect compared to the conventional approach using only continuously-focusing gradients. With smaller beam sizes, the magnets (the main cost of any accelerator) could be made smaller and larger radius rings became feasible. As Courant reminisced in 1980 \cite{Courant1980},

\begin{displayquote}
We have succeeded in building the Cosmotron, the world's first accelerator above one billion volts. ... Stan [Livingston] suggested one particular improvement: In the Cosmotron, the magnets all faced outward. ... Why not have some magnets face inward so that the positive secondaries have a clear path to experimental apparatus inside the ring? ... 

I did the calculation and found to my surprise that the focusing would be strengthened simultaneously for both vertical and horizontal motion. ... Thus it seemed that aperture could be made as small as one or two inches ...

With these slimmer magnets, it seemed one could now afford to string them out over a much bigger circle and thus go to 30 or even 100 billion volts.
\end{displayquote}

In the linear focusing of an AG accelerator, particle orbits are regular and bounded.  
All particles oscillate transversely with a characteristic frequency called the beam \textit{tune}. Such a system is sensitive to resonances and instabilities, as small magnetic field errors can resonantly couple to the beam. 
The need to avoid resonances imposes strict limitations on magnetic field precision and machine design. Ring tunes are chosen as far as possible from known resonance conditions.
Sophisticated feedback systems are built to control instabilities and damp resonances and state-of-the-art accelerator modeling codes compute millions of turns to predict which small nonlinearities lead to beam loss. 
At the scale of long confinement times, even perturbative nonlinearities can introduce and drive resonant losses. 

From the great lengths taken to maximize linearity, and the accompanying loss in performance when any nonlinearity is introduced, it is not surprising that even weak space charge forces can drive losses. In general, space charge acts to complicate the resonance landscape, shifting both the beam tune and resonant conditions. The nonlinearities that arise due to space charge are at odds with the underlying assumptions of a linear AG accelerator. For a leap in the intensity frontier comparable to beam energy after the invention of AG focusing, a new approach must be considered.

\section{Nonlinear transverse focusing for accelerators}

There is no fundamental reason why accelerator focusing must be linear, other than the linear system has well-understood equations of motion that are known to be bounded. Boundedness is related to the concept of integrability in dynamics. An integrable trajectory has conserved invariants that are a function of the phase space coordinates. With invariants of motion, there is certainty that an orbit which appears to be bounded over short time scales will continue to bounded for all times. This is a necessary condition for accelerator focusing, as the beam may be confined for many millions of turns. 

An integrable system that includes nonlinear fields will be relatively insensitive to the resonant phenomena that drives losses in AG systems. In the presence of strong (as opposed to perturbative) nonlinearities, regular driving forces cannot resonantly couple energy into a single trajectory. If a particle gains energy from an external force or field error, its oscillation frequency will shift away from the resonant condition. 

Theoretical research into nonlinear focusing seeks to identify integrable (or near-integrable) systems that include nonlinear focusing forces. Early work found integrable solutions for round colliding beams with nonlinear beam-beam interactions.\cite{Danilov1997} 
A numerical approach using Lie algebra methods showed a system with arbitrarily strong sextupole and octupole magnets can be optimized for ``near-integrability."\cite{Sonnad2004} More recently, Danilov and Nagaitsev propose an accelerator design for which two invariants of transverse motion exist for an arbitrarily strong nonlinear potential of a particular form.\cite{Danilov2010} Assuming linear focusing such that the transported beam is round, there is a family of nonlinear potentials for which transverse particle motion is fully integrable.

\section{Experimental tests of nonlinear integrable systems}

The Integrable Optics Test Accelerator (IOTA) is currently under construction at Fermilab to test the implementation of the Danilov-Nagaitsev theory.\cite{Antipov2017} This includes design of a custom nonlinear magnet that satisfies the condition for integrability.  
The University of Maryland Electron Ring (UMER) is identified as another testbed for the integrable optics concept. 
UMER is a 10 keV, 11.52 meter diameter ring that serves as a scaled, low-cost experiment with space charge physics relevant to higher-energy hadron machines. 
As UMER can access variable space charge regimes, it is attractive for an initial demonstration of integrable optics under different space charge conditions.

While IOTA will test the fully integrable system, design of the custom nonlinear element to high tolerance is a relatively complex task that exceeds the scale of the UMER experiment. 
In \cite{Danilov2010}, Danilov and Nagaitsev also discuss the case in which the nonlinear potential is purely octupolar. While motion in this case is only \textit{quasi-integrable} (having only one invariant of motion), orbits are predicted to remain bounded. The quasi-integrable lattice is identical to the fully integrable case except for the form of the nonlinear potential. Experimentally, the observed effects should be similar. 
The goal of this dissertation is to re-design UMER to include nonlinear quasi-integrable optics and outline a program to experimentally demonstrate stable transport and resonance suppression in this novel type of lattice.

\section{Key terms and definitions}

At this stage it seems necessary to introduce and clarify a few key terms. In the context of transverse accelerator focusing, \textit{linear} and \textit{nonlinear} refer to the dependence of the transverse restoring forces on beam distance from magnetic center. Restoring force goes as $x^{n-1}$ for multipole component $n$ (Dipole has order 1, quadrupole 2, and so on). \textit{Integrability}, as mentioned above, is the property of dynamical systems which possess sufficient invariants of motion. For a fully integrable system, there must be as many invariants as degrees of freedom. For transverse focusing, we restrict discussion to two-dimensional motion and require two invariants. 
\textit{Lattice} is used to describe the arrangement of magnetic focusing elements in an accelerator. \textit{Optics} is used synonymously with lattice due to the similarity between linear transverse focusing and conventional ray optics. The nonlinear magnet in the integrable and quasi-integrable lattices is referred to as an \textit{insert}, as it is a specialized element that is incorporated into the accelerator lattice.

\section{Organization of the dissertation}

Chapter \ref{ch:theory} covers relevant background theory for the linear-focusing AG accelerator, while Ch. \ref{ch:theory-nlio} describes the Hamiltonian approach to finding nonlinear integrable lattice with analytic invariants. 
Chapter \ref{ch:numeric} briefly describes the accelerator modeling codes used in this dissertation, as well as analysis technique applied to the nonlinear lattice. In Ch. \ref{ch:qio} a simplified model of the nonlinear system is used to study dynamics, quantifying ``best case" performance and examining the dependence of nonlinear damping on space charge intensity. This study motivates the need for a mode of UMER operation at low space charge density for initial nonlinear optics experiments. 

Chapter \ref{ch:apparatus} describes the UMER apparatus, including available diagnostics and techniques for data collection developed in this dissertation. 
Chapter \ref{ch:design} describes the changes needed to transform UMER from a linear-focusing to quasi-integrable octupole lattice, including octupole magnet design and generation and detection of an ultra-low-current ``DC beam." 
Also in Ch. \ref{ch:design}, the simplified model is used to study dynamics in the presence of lattice errors and set error tolerances. 

Chapter \ref{ch:lattice} describes the design of the linear ring optics to accommodate the octupole experiments and meet requirements for quasi-integrability. Chapter \ref{ch:lattice} also includes initial results from full ring simulations of the proposed experiments.
Finally, a variation of the quasi-integrable lattice proposed by \cite{Danilov2010} is modeled and preliminary measurements made. The key results are summarized in Chapter \ref{ch:distr}.

Chapter \ref{ch:steering} shows the development and application of an orbit-correction algorithm to improve beam steering. This is crucial for the nonlinear experiments, as lattice performance suffers when the beam centroid is allowed to deviate from the octupole magnetic center.
Experimental measurements of resonant structure and beam transmission in linear UMER are discussed in Ch. \ref{ch:res}, including results for the low-current test beam. Characterizing the linear lattice resonance landscape is  preparation for the nonlinear experiments, as the most irrefutable positive result is demonstrating resonance suppression that depends on octupole strength.

%% file: 2.Chapter.tex

\renewcommand{\thechapter}{2}

\chapter{Theory of Transverse Focusing in Accelerators}
\label{ch:theory}

This chapter provides a brief introduction to transverse particle dynamics in a linear focusing accelerator.
Section \ref{sec:theory:linear} introduces the Hamiltonian approach for single particle equations of motion. This includes identification of the normalized frame in which the particle motion is reduced to simple harmonic motion.
Section \ref{sec:theory:integrability} discusses integrability in dynamical systems and identifies the invariants of motion in the linear focusing accelerator.
Section \ref{sec:theory:resonance} describes the condition for resonant particle orbits and qualitatively describes particle motion near resonances for perturbative nonlinearities.
Finally, Section \ref{sec:theory:collective} covers the formalism for describing collective motion in a distribution of particles. This includes the equations of motion for the beam edge as well as treatment of space charge effects in high-intensity beams.

\section{Particle dynamics in a linear accelerator} \label{sec:theory:linear}

Most modern accelerating rings use quadrupole magnets to provide transverse confinement. This approach is based on the theory of alternating gradient focusing, first introduced by Courant and Snyder \cite{Courant1958}, in which linear restoring forces of alternating polarity (focus-defocus-focus) provide net focusing.\footnote{Also independently discovered by Nicholas Christofilos.\cite{christofilos}} Vertical dipole fields provide steering, keeping the beam inside the vacuum pipe.
This section describes single particle dynamics in an alternating gradient accelerator containing only linear (dipole and quadrupole) fields. Further notes on the derivations can be found in Appendix \ref{ap:Hamiltonian}, while a much more thorough treatment can be found in \cite{SYLee} and \cite{uspasEM}.

\subsection{Single particle equations of motion}

\begin{figure}[tb]
\centering
\includegraphics[width=0.5 \textwidth]{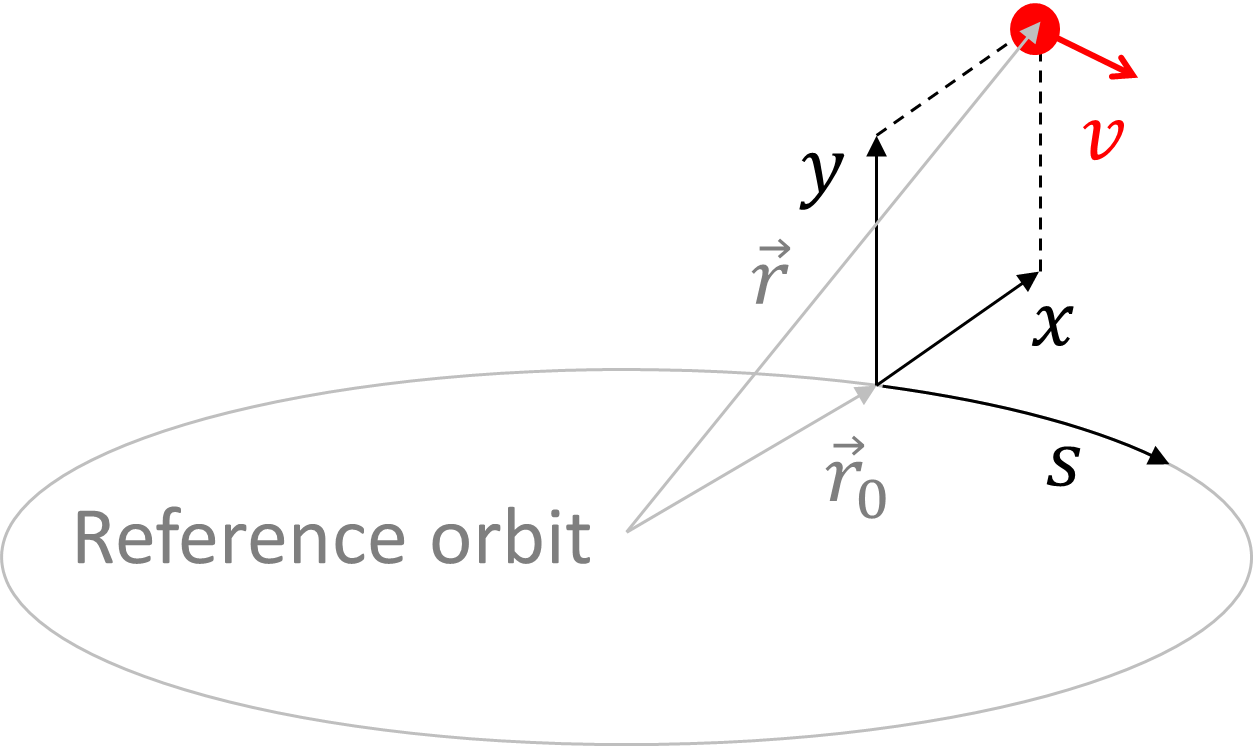}
\caption{Frenet Serret coordinate system for accelerator rings. The coordinates $(x,y,s)$ follow the beam frame along a reference orbit $\vec{r_0}$. The position of a particle (red) is $x \hat{x} + y\hat{y}$ in the beam frame and $\vec{r} = \vec{r_0} + x \hat{x} + y\hat{y}$ in the lab frame.}
\label{fig:frenet-serret}
\end{figure}

The natural frame for describing particle dynamics in an accelerator is the curvilinear Frenet-Serret coordinate system.\cite{frenet,serret} In this frame, the coordinate axes $\left(x,y,s\right)$ follow the beam along a reference orbit as shown in Fig. \ref{fig:frenet-serret}. Coordinate $s$ is the propagation distance along this reference orbit. In the Hamiltonian approach, we choose to use $s$ as the independent variable, where $s=v_0 t$ for beam velocity $v_0$.

The conventional accelerator contains only dipole and quadrupole fields, for bending and focusing respectively. In the beam frame, all restoring forces are linear in transverse displacement and the Hamiltonian describing single particle motion is
\begin{equation}
H =  \frac{1}{2} \left(p_{x}^2 + p_{y}^2\right) + \frac{1}{2} \left(K_x(s)x^2 +K_y(s)y^2 \right).
\label{eq:H}
\end{equation}

\noindent Canonical momenta $p_x$ and $p_y$ are dimensionless variables. They are related to physical momenta as $p_x = P_x/P_0$ and $p_y = P_y/P_0$ where $P_0$ is the nominal or ``design" momentum of the beam.\footnote{$x'$ is often used in place of $p_x$, as $x' = dx/ds = P_x/P_s \approx P_x/P_0 = p_x$.} In the derivation of $H$, a small angle approximation is applied, assuming transverse momenta $P_x$ and $P_y$ are much smaller than total momentum $P$. 

External focusing is expressed in the terms $K_x(s)$ and $K_y(s)$. The focusing functions are related to quadrupole strength as $K_x(s) = eG_1(s)/P_0$ and  $K_y(s) = -eG_1(s)/P_0$ where $G_1 = dB_x/dy = dB_y/dx$ is the quadrupole gradient.\footnote{If dipole fields are included, $K_x(s) = \rho(s)^{-2} + eG_1(s)/P_0$ for bending radius $\rho(s)$.}
The sign difference in $G_1$ indicates the alternating gradient nature of quadrupole focusing: when $G_1 > 0$, the horizontal force is focusing while the vertical force is defocusing, and vice versa. 
From Eq. \ref{eq:H} we see that $x$ and $y$ motion are decoupled, such that $H = H_x + H_y$. For this reason, transverse dynamics are often studied as a 1D Hamiltonian system.


Applying Hamilton's equations of motion,

\begin{align}
\dot{z} &= \frac{\partial H}{\partial p_z} \\ 
\dot{p_z} &= -\frac{\partial H}{\partial z},
\label{eq:H-eqns}
\end{align}

\noindent we arrive at Hill's equation:
\begin{equation}
z''(s) + K(s)z = 0.
\label{eq:diff-eqn-motion}
\end{equation}

\noindent Here $z$ indicates either transverse plane, $z \in \{x,y\}$. 
In general, $K(s)$ is a piecewise constant function that has value zero between focusing elements.
In an accelerator ring, $K(s)$ is periodic in s, $K(s+C) = K(s)$ for ring circumference $C$. 
With this periodicity, the solution to Hill's equation has the form of a Floquet transformation,

\begin{equation}
z(s) = \sqrt{\epsilon \beta(s) }e^{\pm i \psi(s)},
\label{eq:eqn-motion}
\end{equation}

\noindent for amplitude constant $\sqrt{\epsilon}$, amplitude function $\sqrt{\beta(s)}$ and phase function $\psi(s)$. 
The oscillatory motion of $z(s)$ is referred to as \textit{betatron motion}.
The amplitude $\epsilon$ is known as the single-particle \textit{emittance}.
 
$\beta(s)$ is also called the betatron or envelope function, as the beam edge follows the trajectory $\sigma(s) = \sqrt{\epsilon \beta(s)}$ when $\epsilon$ is the emittance of the highest amplitude particle in the distribution. $\beta(s)$ is only dependent on the linear focusing function $K(s)$.

\subsection{Phase advance and tune}

At this point it becomes useful to define two quantities used to characterize particle motion in a ring. The phase advance is found by integrating over the inverse of the betatron function:\footnote{Origin of this relationship is shown in Appendix \ref{ap:Hamiltonian}.}

\begin{equation}
\Delta \psi_{s_1 \to s_2} \equiv \int_{s_1}^{s_2} \frac{ds}{\beta(s)}.
\label{eq:phase-adv}
\end{equation}
 
Phase advance per cell (the minimum length over which $K(s)$ and $\beta(s)$ are periodic) is a useful quantity for quantifying lattice focusing strength. Another common metric is tune $\nu$, defined as the number of betatron oscillations per revolution. In terms of ring phase advance,

\begin{equation}
\nu = \frac{\Delta \psi_{0 \to C}}{2\pi} = \frac{1}{2\pi} \int_0^C \frac{ds}{\beta(s)}.
\label{eq:tune}
\end{equation}

\subsection{FODO lattice}
\begin{figure}[tb]
\centering
\includegraphics[width=\textwidth]{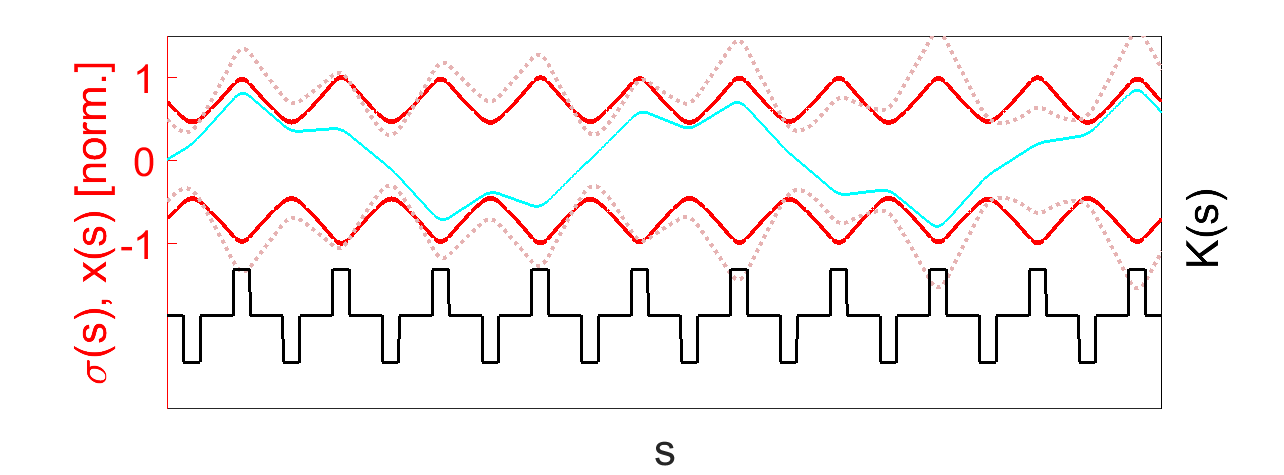}
\caption{Beam evolution in FODO lattice, including single particle trajectory (light blue) and beam edge $\sigma_x$ (solid red). An injection error in the initial beam distribution creates envelope mismatch oscillations (dotted light red).}
\label{fig:FODO-cartoon}
\end{figure}

In a ring, focusing function $K(s)$ is chosen such that there exists a periodic solution for $\beta(s)$. 
The most ubiquitous choice is the FODO lattice, consisting of quadrupoles of alternating polarity (F-D-F-D-F-D) separated by field-free drift spaces. A plot of $K(s)$, $x(s)$ and $\sigma(s)$ is shown for an example lattice in Fig. \ref{fig:FODO-cartoon}. UMER is designed and operated as a FODO lattice, although the focusing $K(s)$ can easily be adjusted.

The periodic solution $\beta(s)$ is a property of the linear focusing lattice. For the most efficient transport of beam, the size and divergence of an injected beam should be \textit{matched} to the envelope function at the injection point. An initially \textit{mismatched} beam will oscillate about the equilibrium solution $\sqrt{\epsilon \beta(s)}$, as illustrated by the dotted curve in Fig. \ref{fig:FODO-cartoon}.

\subsection{Smooth-focusing lattice} \label{sec:theory:smooth}

A common theoretical approach is to simplify dynamics by considering a continuous or smooth-focusing lattice with constant focusing coefficient $k = \left<K(s)\right>$.\cite{Reiser} In this lattice, $\beta(s) = k^{-1}$ constant and the beam edge is constant during transport. Motion is purely sinusoidal, with $z(s) = \sqrt{\frac{\epsilon}{k}}e^{\pm i \psi(s)}$. The smooth focusing model is generally not a good approximation except in the case of low phase advance per cell. In this thesis, a smooth focusing approximation of the UMER FODO lattice is used to estimate tune in Chapter \ref{ch:res}.

\section{Integrability in accelerators} \label{sec:theory:integrability}

Integrability is a crucial concept for ring design, as the beam may be stored or accelerated over millions of turns. A Hamiltonian system H with $N$ degrees of freedom is integrable if there exist $N$ invariant quantities $J_i$ of the motion which are in involution (Poisson bracket $\left[J_i,J_j\right] = 0$). $J_i$ are known as the isolating integrals or constants of motion.\cite{NagaitsevUSPAS}, \cite{Lichtenberg} 

A time-independent system with one degree of freedom is always integrable. For $N$ degrees of freedom, existence of $N$ isolating integrals is not guaranteed (there may be any number from zero to $N$). There is no universal procedure to find all integrals for a general system or even to identify the number of independent integrals that exist. Even if $N$ integrals exist, they may not be easily recognizable. For a given Hamiltonian, the system can be shown to be integrable (by identifying closed form expressions for the conserved invariants), but cannot conclusively be shown to be non-integrable. 

Accelerator systems should be both bounded and (ideally) integrable. However, integrability is violated once we consider realistic perturbations (including magnetic field errors and space charge forces, not to mention higher order terms that were neglected in the derivation of Eq. \ref{eq:H}). Thankfully, according to the KAM theorem, for small perturbations invariant surfaces continue to exist for most initial conditions.\cite{KAM} Therefore, in a real system it is only necessary to operate near integrability for long term stability.

\subsection{Action-angle variables}
For any integrable system, there exists a canonical transformation into action-angle variables, where Hamiltonian depends only on the ``action" (or integrals of motion) $J_i$.\cite{Lichtenberg} For each degree of freedom in the transverse accelerator Hamiltonian Eq. \ref{eq:H}, 

\begin{equation}
H_z =  \frac{1}{2} p_{z}^2  + \frac{1}{2} K_z(s)z^2,
\label{eq:Hz}
\end{equation}

\noindent transformation from phase space coordinates $\left( z,p_z\right)$ to action-angle coordinates $\left( J_z,\psi_z \right)$ is made with the generating function 

\begin{equation}
F_1(z,\psi_z) = -\frac{z^2}{2\beta_z}\left( \tan{\psi_z} - \frac{\beta'_z}{2}\right)
\end{equation}

\noindent for $z\in{x,y}$. The resulting Hamiltonian is 

\begin{equation}
\tilde{H}_z =  H_z + \frac{\partial F_1}{\partial s} = \frac{J_z}{\beta_z}.
\label{eq:Hz-AA}
\end{equation}

As $\tilde{H}_z$ is independent of angle $\psi_z$ (betatron phase), the action $J_z$ is constant; $J_z$ is the invariant of motion in the linear focusing accelerator. As $xy$ motion is uncoupled, $H=H_x+H_y$, there are two invariants of 2D motion, $J_x$ and $J_y$, and the system is fully integrable. 
The action $J_z$ is a measure of single-particle orbit amplitude. It is equivalent to the volume of phase space enclosed by the particle orbit:

\begin{equation}
J_z = \frac{1}{2\pi} \oint z' dz.
\end{equation}

This invariant quantity is essentially a special case of Liouville's theorem, which states that phase space volume occupied by a distribution of non-interacting particles in a Hamiltonian system is conserved. In the case of the time-independent linear-focusing Hamiltonian, the phase space volume is a conserved quantity of single particle orbits.

\subsection{Courant-Snyder parameters}

The accelerator literature typically describes the orbit amplitude in terms of the Courant-Snyder invariant and parameterizes the phase space volume in terms of Courant-Snyder parameters.
The Courant-Snyder invariant is the same emittance $\epsilon$ introduced earlier. Orbits are confined to invariant surfaces in phase space defined by the ellipse 

\begin{equation}
\epsilon_z = \gamma z^2 + 2\alpha z z' + \beta z'^2
\label{eq:CSinvar}
\end{equation}

\noindent where $\beta$ is the betatron function previously introduced and $\alpha$ and $\gamma$ are lattice functions defined in terms of $\beta$ as:

\begin{align}
\alpha &= -\frac{1}{2}\frac{d\beta}{ds} \\
\gamma &= \frac{1+\alpha^2}{\beta}.
\label{eq:abg}
\end{align}

\noindent $\pi \epsilon_z$ is the phase space area of the ellipse, therefore the Courant-Snyder invariant $\epsilon_z$ is related to the action $J_z$ as $\epsilon_z = 2J_z$.
The total phase space area inhabited by a distribution of particles is defined by the maximum single-particle emittance.
Beam spot size and divergence are related to the Courant-Snyder parameters as 
\begin{subequations}
\begin{equation}X(s) = \sqrt{\epsilon \beta(s)} \end{equation}
\begin{equation}X'(s) = \frac{dX(s)}{ds} = \sqrt{\epsilon \gamma(s)}\end{equation}
\end{subequations}

\subsection{Hamiltonian in normalized coordinates} \label{sec:theory:norm-coord}
As mentioned above, in the appropriate frame the single-particle motion reduces to simple harmonic oscillation. This is done through canonical transformation to the normalized coordinates: 
\begin{align}
z_N &\equiv \frac{z}{\sqrt{\beta(s)}}, \\
p_{z,N} &\equiv p_z\sqrt{\beta(s)}+\frac{\alpha z}{\sqrt{\beta(s)}}. 
\label{eq:norm-coord}
\end{align}

In this frame, the normalized Hamiltonian is

\begin{equation}
H_N = \frac{1}{2} \left( p_{x,N}^2 + p_{y,N}^2 +x_N^2 + y_N^2 \right)
\end{equation}

which is instantly recognizable as a simple harmonic oscillator. The particle orbits $x(s)$ and $y(s)$ trace a circle in phase space coordinates of radius equal to particle amplitude $\sqrt{\epsilon \beta}$.

\section{Transverse resonances} \label{sec:theory:resonance}

\begin{figure}[tb]
\centering
\includegraphics[width=0.7\textwidth]{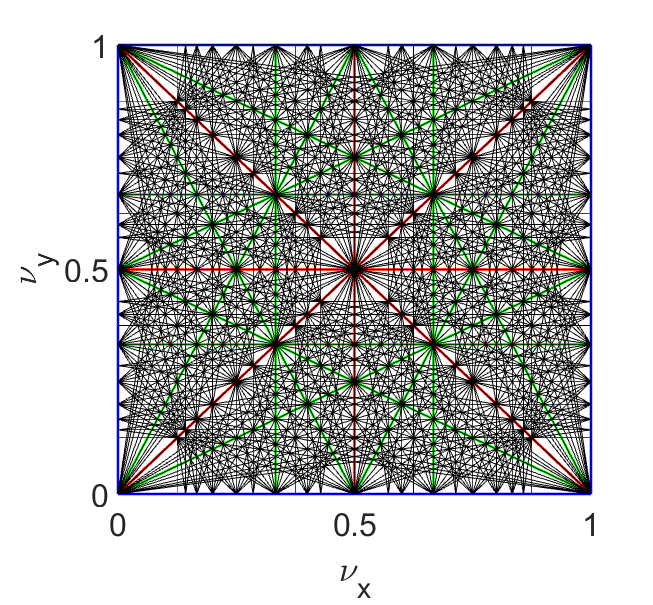}
\caption{Tune resonance diagram with all relationships $m \nu_x + n \nu_y = p$ up to order eight. Orders 1-3 are shown with blue, red and green highlights respectively.}
\label{fig:res-diag}
\end{figure}

In the single particle equations of motion, resonant orbits occur for all rational tune relationships $m \nu_x + n \nu_y = p$ where $m,n,p$ are integers.\cite{Reiser} 
Due to the fact that tranverse focusing in a ring is periodic, resonant orbits can be excited by field errors in the magnetic lattice. These field errors act as a driving term in the equations of motion. A particle on a periodic orbit will experience resonant growth if the appropriate driving term is present. When designing an accelerator we assume all resonances are driven and choose a off-resonant operating point.  Figure \ref{fig:res-diag} shows the resonant relationships up to order eight. When considering very high order resonances, the ``safe" regions shrink dramatically.

Not all resonances are equally damaging.
Generally speaking, resonant growth is slower the higher the resonance order, defined as $m+n$.
Difference resonances, where sign($m$) = sign($n$), are generally stable as energy is transferred between planes but does not grow without bound. Sum resonances, sign($m$) $\neq$ sign($n$), allow energy to couple into the particle oscillation.\cite{Ruth1986}

\begin{figure}[tb]
\centering
\includegraphics[width=0.7\textwidth]{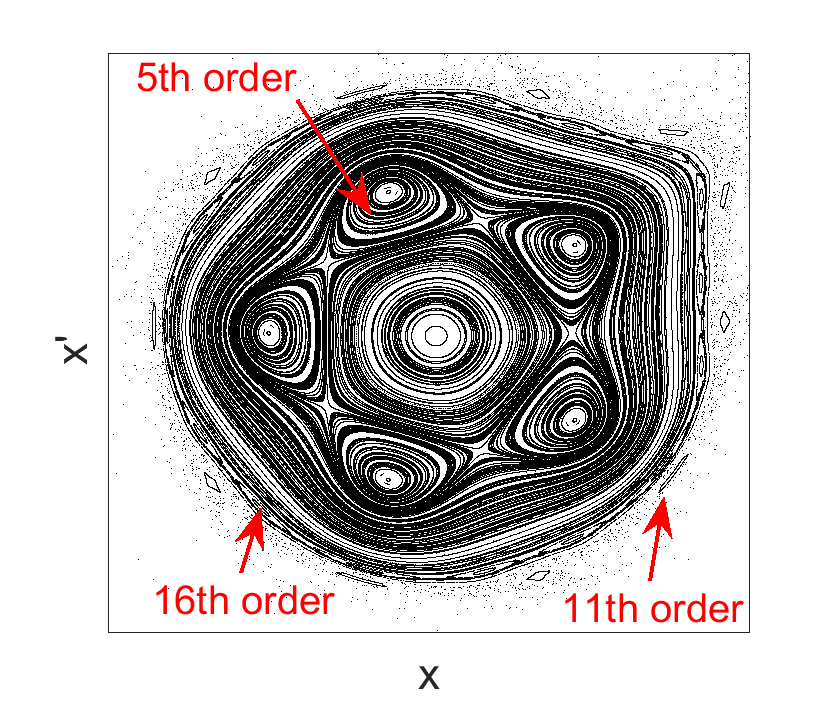}
\caption{Surface of section for a 1D linear accelerator with nonlinear perturbation. Noticeable periodic orbits are indicated by arrows.}
\label{fig:poincare-plot}
\end{figure}

In the perfectly linear case, non-interacting particles all occupy an infinitesimal point in tune space. In reality, many effects act to increase the range of tunes occupied by the beam (the ``tune footprint"). Space charge forces (discussed below, Section \ref{sec:theory:collective}) and chromatic effects (see Appendix \ref{ap:Hamiltonian}) both introduce tune spreads that depend on beam distribution in configuration and momentum space, respectively.
An additional source of tune spread is nonlinearities in the lattice elements. While nonlinear components are minimized in accelerator magnet design, some amount of unwanted harmonic content is unavoidable.   
In this way, magnetic field errors act two-fold: they are both the driving term and the source of tune spread that overlaps resonant conditions.

As nonlinear field errors grow with distance from magnet center, the induced tune shift increases with particle amplitude.
The bare tune $\nu_0$ (without errors and higher order effects) is chosen to be irrational, but the tune footprint encompasses infinitely many rational relationships. For 1D motion, periodic orbits appear as ``dot chains" in phase space, between layers of irrational orbits that trace a circle. This effect is demonstrated in Fig. \ref{fig:poincare-plot} for a surface-of-section in a 1D accelerator map. Despite the perturbation, these orbits still follow invariant surfaces as predicted by the KAM theorem. There are finite ``islands" around the periodic orbits where nearby orbits are distorted but motion is still regular. Stochastic motion emerges in regions where islands overlap. At higher amplitude and stronger nonlinearity, the island overlap leads to a stochastic continuum beyond which particle motion is unstable. 
In Fig. \ref{fig:poincare-plot}, stochastic orbits surround the 11th order islands. 
The largest phase space amplitude with stable motion is the \textit{dynamic aperture}. \cite{Chao1999}
The balance between driving terms (reduced as much as possible through careful magnet design) and resonance order (higher orders have slower growth rates) ultimately determines the dynamic aperture of a ring.

\section{Collective motion and space charge effects} \label{sec:theory:collective}

The prior discussion on linear lattices was mostly limited to single particle dynamics. This section describes the treatment for an interacting distribution of particles. Intra-beam space charge forces are considered as a collective self-force due to the Coulomb mean-field, neglecting particle-particle interactions and collisions.

\subsection{Collective emittance}
Just as emittance $\epsilon$ is a single particle invariant in the linear focusing lattice, the RMS emittance of a distribution is also a conserved quantity.
The beam moments are calculated as integrals over the normalized beam distribution $\rho(z,z')$. For example, the RMS width $\sigma_z$ is calculated as 

\begin{equation}
\sigma_z^2 = \int \left(z - \left<z\right>\right)^2 \rho(z,z')dz dz'
\end{equation} 

\noindent The RMS emittance is defined in terms of the second order beam moments as 

\begin{equation}
\epsilon_{z,RMS} \equiv \sqrt{\sigma_z^2\sigma_{z'}^2-\sigma^2_{zz'}}
\end{equation}

\subsection{Space charge effects} 

The self-force of a beam on itself introduces additional terms in the equations of motion that are dependent on the beam intensity. In general, the effect of space charge is to introduce a defocusing force. This reduces the frequency of particle oscillation, an effect called ``tune depression." Tune depression is often expressed in terms $\nu / \nu_0$ where $\nu_0$ is the ``bare tune" in the zero-charge limit. 

Betatron motion is divisible into coherent and incoherent motion. Coherent motion occurs in the beam distribution as a whole. If the beam is displaced from quadrupole centers, it will oscillate coherently at the betatron frequency. Incoherent motion describes the motions of individual particles under the combined effect of space charge and external forces. 
Space charge causes both a coherent and incoherent tune shift. The incoherent shift is the amplitude-dependent shift in tune $\delta \nu = \nu - \nu_0$ for each particle. A coherent shift manifests as a decrease in the centroid oscillation frequency due to image charge forces from the pipe wall. Typically the coherent shift is much smaller than the incoherent shift.
The space charge force generated by an arbitrary beam distribution is generally  nonlinear.
As any nonlinearity creates amplitude-dependent tune spreads, the effect of space charge is to increase the range of the tune distribution (the \textit{tune footprint}) through incoherent tune spread.

\subsubsection{Kapchinskij-Vladimirskij distribution}
 
The Kapchinskij-Vladimirskij (KV) distribution is a special case for which the space charge force is linear.\cite{KV} While this distribution is not physical, it is a useful tool for the theoretical treatment of beams, as analytic solutions are possible.
Particles are distributed on a constant emittance surface in phase space:

\begin{equation}
\rho(x,p_x,y,p_y) = \frac{\lambda e}{\pi^2 \epsilon_x \beta_x \epsilon_y \beta_y} \delta\left(\frac{1}{\epsilon_x \beta_x} (x^2 + p_x^2) + \frac{1}{\epsilon_y \beta_y} (y^2 + p_y^2) -1\right)
\end{equation}

\noindent for line density $\lambda$ and edge emittances $\epsilon_x$, $\epsilon_y$.
Every phase space projection of the KV distribution has a uniform distribution out to the beam edge. 
The RMS emittance is related to edge emittance (the emittance which contains 100\% of the beam) as

\begin{equation}
\epsilon_{RMS} = \frac{\epsilon_{edge}}{4}.
\end{equation}

Similarly, the RMS size is related to beam edge as 

\begin{equation}
\sigma_{RMS} = \frac{\sigma_{edge}}{2}.
\end{equation}

A unique (and unphysical) property of the KV distribution is that the incoherent tune shift is the same for every particle (therefore, there is no incoherent tune spread).
For a long bunch with a round KV distribution, \cite{Schindl1999}

\begin{equation}
\Delta \nu = \nu - \nu_0 = -\frac{r_e C I}{2 \pi \epsilon c e \beta^3 \gamma^3}
\label{eq:tune-spread}
\end{equation}

for beam current $I$, transverse edge emittance $\epsilon$ (containing $100\%$ of particles) and velocity $\beta c$ in a ring of circumference C. $r_e$ is the classical electron radius and $\gamma$ is the relativistic factor.
This tune shift is proportional to $I/\epsilon$ and scales with beam energy as $\left(\beta \gamma\right)^{-3}$. For this reason, space charge is typically a concern near injectors during low energy transport as well as in high intensity rings.

\subsection{RMS beam evolution}

The evolution of the RMS beam edge can also be described as a Hamiltonian system. Here the resulting equations of motion are defined. These equations are valid for any system subject to linear focusing forces.\cite{Lapostolle1971,Sacherer1971} Full detail can be found in many references, including \cite{LundTCE} and \cite{Reiser}. The equations of motion for RMS beam sizes $X \equiv \sigma_{x,RMS}$ and $Y \equiv \sigma_{y,RMS}$ are:

\begin{subequations}
\begin{equation}
X" +K_x(s)X - \frac{2K}{X+Y} - \frac{\epsilon_x^2}{X^3} = 0
\end{equation}
\begin{equation}
Y" +K_y(s)Y - \frac{2K}{X+Y} - \frac{\epsilon_y^2}{Y^3} = 0.
\end{equation}
\label{eq:env-eq}
\end{subequations}

The space charge force is included as the beam perveance K, defined as

\begin{equation}
K = \frac{q \lambda}{2 \pi \epsilon_0 m \gamma_b^3 \beta_b^2 c^2}
\label{eq:perv1}
\end{equation}

for line charge density $\lambda$. An alternative expression is 

\begin{equation}
K = \frac{I}{I_0}\cdot \frac{2}{\gamma_b^3 \beta_b^3}.
\label{eq:perv2}
\end{equation}
Here $I$ is the beam current and $I_0$ is the Alfven current, $I_0 = 4\pi \epsilon_0 \frac{mc^3}{e} \approx 17$ kA for electrons.

\section{Chapter summary}

This chapter reviewed the basic principles of particle dynamics in a linear focusing lattice. Linear orbits conserve the invariant of motion $\epsilon$, which corresponds to orbit phase space area. Variations of linear focusing form the backbone of modern accelerator design. While the linear motion is well understood, these systems are not robust to perturbations which tend to couple energy into resonant orbits. One of the main challenges when designing an accelerator is ensuring there will be sufficient dynamic aperture to transport the injected beam for the desired number of turns with minimal losses. Many complex feedback systems are implemented to damp destructive resonances. Transportable beam intensity is limited due to the need to limit losses for machine protection, while space charge tune spreads also excite further resonances.

%% file: 3.Chapter.tex
\renewcommand{\thechapter}{3}

\chapter{Theory of Nonlinear Integrable Optics}
\label{ch:theory-nlio}

Nonlinear integrable optics is currently being investigated as a strategy for mitigating resonance-driven beam loss.\cite{Danilov2010,Antipov2017} 
Identification of nonlinear integrable systems reconciles two sometimes contradictory goals: long term stability of dynamics and strongly nonlinear forces for resonance suppression. 
As reviewed in the previous chapter, integrability (the existence of conserved invariants in particle motion) guarantees that the system is dynamically stable and confined for arbitrary time scales. All accelerating rings are designed to operate near integrability with minimal perturbation.

Linear lattice focusing as presented above in Chapter \ref{ch:theory} is attractive due to the existence of the Courant-Snyder invariant for any linear focusing function $K(s)$. 
The FODO arrangement was provided as an example, but many other options are implemented according to the optics requirements of the ring. In all cases, particle orbits follow invariant surfaces defined by $\epsilon_x$ and $\epsilon_y$. The weakness of the linear lattice is the sensitivity to resonant excitation, as discussed in Section \ref{sec:theory:resonance}. For perturbative nonlinearities and small tune spreads, field errors drive resonant losses that limit the dynamic aperture. 

While small nonlinearities limit dynamic aperture via resonant losses, external nonlinear tune spreads can also damp resonances. 
Two effects work to suppress resonant growth and instability. 
Landau damping with octupole fields is a collective effect in which coherent instabilities are damped through the introduction of incoherent motion.\cite{Herr2013} For sufficiently strong nonlinear fields with large amplitude-dependent tune spreads, single-particle detuning is also possible. A resonantly excited particle that gains transverse energy will decohere from the driving term, which will limit particle amplitude growth.\cite{Webb2012} 

The difficulty with designing intentionally nonlinear lattices is maintaining integrability. As mentioned above, there is no general test for integrability, and the form of the invariant may not be easily recognizable. This chapter follows the approach in \cite{Danilov2010} to find integrable solutions with arbitrarily strong nonlinear elements. 
Their approach uses a Hamiltonian formalism to identify a system that (a) includes arbitrarily strong nonlinear potentials, (b) has conserved invariants of motion and (c) is realizable in an accelerator lattice. 

\section{Approach to identifying a nonlinear integrable system}

A generic nonlinear potential $V(x,y,s)$ is added to the single-particle linear-focusing Hamiltonian in Eq. \ref{eq:H}. The new Hamiltonian is:

\begin{equation}
H = \frac{1}{2} \left( p_{x}^2 + p_{y}^2 \right) + \frac{1}{2} K(s) \left(x^2 + y^2 \right)  + V(x,y,s).
\end{equation}

\noindent There are no requirements for the transverse fields $V(x,y)$ except that it has nonlinear components $x^{n+1}$ for $n>1$. 

The first assumption in the search for integrable solutions is that the horizontal and vertical linear focusing is equal. This sets the condition $\beta_x(s) = \beta_y(s) = \beta(s) $. 
Through canonical transformation into normalized coordinates (Eq. \ref{eq:norm-coord}), the Hamiltonian becomes 

\begin{equation}
H_N = \frac{1}{2} \left( p_{x,N}^2 + p_{y,N}^2 +x_N^2 + y_N^2 \right) + \beta(s) V(x_N \sqrt{\beta(s)},y_N \sqrt{\beta(s)},s)
\end{equation}

\noindent $H_N$ is an invariant of motion if the $V$ term is $s$-independent. This can be done for any multipole component $n$ of the potential $V$ by choosing longitudinal profile $K_n(s)$ such that the $s$-dependence is canceled. For ease of notation, and also following with the convention in \cite{Danilov2010}, I define $U$ as the potential in the normalized frame,

\begin{equation}
U(x_N,y_N,s) \equiv \beta(s) V(x_N \sqrt{\beta(s)},y_N \sqrt{\beta(s)},s).
\end{equation}

\noindent For $ U(x_N,y_N,s)$ to be independent of $s$, the $s$-dependence in $\beta(s)$ and $V(x,y,s)$ must cancel.
Arbitrary field $V(x,y)$ can be constructed as a multipole expansion, where the fields of order $n$ depend on position as $V \propto z^{n+1}$.
From here it is clear that for each order $n$, the $s$-dependence can be removed by appropriate scaling of $V_s(s) \propto \beta^{-\frac{n+1}{2}}(s)$.

In the case of a pure octupole field, where $V \propto z^4$, if

\begin{equation}
V_{xy}(x,y,s) = \frac{\kappa}{\beta^3(s)} \frac{1}{4}\left(x^4 + y^4 - 6y^2x^2\right)
\label{eq:octu-scaling}
\end{equation}

\noindent then

\begin{equation}
U(x_N,y_N,s) = \frac{\kappa}{4} \left(x_N^4 + y_N^4 - 6y_N^2x_N^2\right)
\end{equation}

\noindent and the normalized Hamiltonian is,

\begin{equation}
H_N = \frac{1}{2} \left( p_{x,N}^2 + p_{y,N}^2 +x_N^2 + y_N^2 \right) + \frac{\kappa}{4} \left(x_N^4 + y_N^4 - 6 y_N^2 x_N^2\right).
\label{eq:H-norm}
\end{equation}

\noindent Note the addition of the scaling factor $\kappa$ to parametrize the strength of the potential $U$. The invariant is conserved for arbitrary $\kappa$, therefore this type of lattice should remain invariant for arbitrarily strong nonlinearities.

\subsection{Quasi-integrable lattice}

As shown above, the Hamiltonian $H_N$ can be made time-independent for appropriate scaling of $V$ in $s$. The case of invariant $H_N$ is referred to the ``quasi-integrable" case, as there is a single integral of 2D transverse motion. As the invariant is the orbit ``energy," particle motion is bounded (but chaotic). The quasi-integrable case with octupole potential is the focus of this dissertation, as the magnet configuration is significantly simpler than in the fully integrable case while still providing nonlinear detuning.

\subsection{Fully integrable solution}

A fully integrable solution is beyond the scope of this thesis, which covers design of a quasi-integrable lattice. However, the integrable solution found in \cite{Danilov2010} is summarized for the purpose of completeness. The existence of nonlinear integrable systems with physically realizable potentials (obeying Laplace's equation) have huge implications for accelerator design, as most modern accelerators are a variation on the linear system described by Eq. \ref{eq:H}. If this concept is successful, future accelerators may have very different dynamics and lead to a different formalism than has been developed for the existing type.

Danilov and Nagaitsev \cite{Danilov2010} identified a family of physically-realizable potentials $U(x,y)$ such that the motion described by $H_N$ has two known invariants.
As discussed, the longitudinal scaling of $V_s(s)$ to remove s-dependence guarantees that $H_N$ is the first invariant. The search for a second invariant quadratic in position/momenta assumes a form 

\begin{equation}
I = (ay^2+c^2)p_x^2 - 2axyp_xp_y + ax^2p_y^2 + D
\end{equation}

\noindent for constants $a$, $c$ and $D$. For $a=1$, $c\neq0$, the potential $U(x,y)$ with invariant $I$ is given by the solution to the Bertrand-Darboux equation \cite{Darboux1901}:  

\begin{equation}
xy(\frac{d^2U}{dx^2}-\frac{d^2U}{dy^2}) + (y^2 - x^2 + c^2)\frac{d}{dy}\frac{dU}{dx}+2y\frac{dU}{dx}-3x\frac{dU}{dy} = 0
\label{eq:darboux}
\end{equation}

\noindent The general solution to this partial differential equation has the form

\begin{equation}
U(x,y) = \frac{f(\xi) + g(\eta)}{\xi^2 - \eta^2}
\end{equation}

\noindent for arbitrary functions $f$ and $g$ and elliptic coordinates

\begin{align}
\xi &= \frac{\sqrt{(x+c)^2 + y^2} + \sqrt{(x-c)^2 + y^2}}{2c} \\
\eta &= \frac{\sqrt{(x+c)^2 + y^2} - \sqrt{(x-c)^2 + y^2}}{2c}. 
\end{align}

\noindent The fully integrable Hamiltonian is therefore

\begin{equation}
H = \frac{1}{2} \left( p_{x,N}^2 + p_{y,N}^2 + x_N^2 + y_N^2 \right)  + \frac{f(\xi) + g(\eta)}{\xi^2 - \eta^2}.
\end{equation}

\noindent The condition that a potential be physically realizable in an accelerator requires that the fields be generated by magnets external to the beam pipe, therefore $U$ must satisfy Laplace's equation, $\Delta U = 0$.
The freedom allowed through the definition of functions $f$ and $g$ allows for  this. For Laplacian fields, $f$ and $g$ must have the form

\begin{align}
f(\xi) &= \xi \sqrt{\xi^2-1}\left(d + ta \cosh{\xi} \right) \\
g(\eta) &= \eta \sqrt{1-\eta^2}\left(b + ta \cosh{\eta} \right)
\end{align}

\noindent for arbitrary constants $a$, $b$, $d$ and $t$.
The case of $a=1$, $b=\frac{\pi}{2} t$, $c=1$ and $d=0$ (with $t$ left as a strength scaling factor) is being tested in the IOTA lattice.\cite{Antipov} Additional IOTA experiments include the exploration of integrable systems with electron lenses as focusing elements, which avoids the requirement that $U$ satisfy Laplace's equation.

\section{Implementation of a nonlinear integrable lattice}

Implicit in the Hamiltonian Eq. \ref{eq:H-norm} is the necessity of external linear focusing $K(s)$. This term is present in the definition of $\beta(s)$, which is used to define the normalized coordinates Eq. \ref{eq:norm-coord}. Linear focusing must be present in order for $\beta(s)$ to have a solution. The implemented integrable or quasi-integrable nonlinear lattice therefore consists of two sections: a nonlinear \textit{insertion} element with potential $V(x,y,s)$ paired with a region of linear (quadrupole) fields for transverse confinement.

The requirement that $\beta_x = \beta_y$ in the nonlinear insertion demands that the insertion region be free of quadrupole fields. Inside a quadrupole element, $K_x=-K_y$, and a horizontally focusing quadrupole is vertically defocusing. The effect of quadrupole gradients is to cause ellipticity, but the insertion region must contain a round beam to meet the integrable condition.

As the envelope function $\beta(s)$ depends only on linear forces and the planned octupole insertion contains only third order focusing terms, it is natural to consider the insertion as a field-free \textit{drift} space when solving for second order beam moments. The most natural design, when considering a relatively long insertion element, is to allow the beam to come to a symmetric waist inside the element as pictured in Fig. \ref{fig:toy-model}. This design is reminiscent of the low beta insertions in a collider ring, where the transverse spot size must be reduced as much as possible for highest luminosity at the collision point. However, in the collider ring, $\beta_x(s)$ and $\beta_y(s)$ are in general not required to be equal.

The discussion of integrable lattice design is continued in Chapter \ref{ch:qio}. A simple (but not physically realizable) representation of an integrable lattice uses FOFO transverse focusing between beam waists/insertion regions. A description of the FOFO lattice is given in Appendix \ref{ap:fofo}.

\section{Chapter summary}

Nonlinear detuning is proposed as a method to decrease sensitivity to resonances and, in doing so, radically increase dynamic aperture in circular accelerators. This chapter followed the derivation of nonlinear integrable lattices proposed by Danilov and Nagaitsev. The quasi-integrable lattice is identified as the focus of the UMER nonlinear optics program.

%% file: 4.Chapter.tex
\renewcommand{\thechapter}{4}

\chapter{Numerical Tools}
\label{ch:numeric}

An important part of the apparatus are the simulation codes used for numerical studies of the proposed lattice. This chapter covers the four accelerator codes used to model beam dynamics in the nonlinear lattices. The technique of frequency map analysis for identifying regions of chaotic orbits and diffusion in accelerator lattices is also discussed. 

\section{Accelerator modeling codes}
\label{sec:numeric:codes}

A large portion of the experimental planning has employed simulation studies to examine dynamics and predict lattice behavior both with and without the nonlinear octupole elements. This section reviews the simulation codes used to study features of the quasi-integrable experiments.

\subsection{VRUMER}
\label{sec:numeric:vrumer}

VRUMER (Virtual UMER) is a simple orbit integrator written in Matlab.\cite{vrumer} 
VRUMER integrates the linearized single-particle equations of motion subject to hard-edged magnetic field elements, including ring quadrupoles and dipoles as well as all steering corrector magnets.
The model also includes the background Earth field, applied as a continuously acting, position-dependent force based on linear interpolation between measurement points at the 36 dipoles.

In this thesis, VRUMER is applied to test the performance of orbit correction algorithms, described in Chapter \ref{ch:steering}. VRUMER is also used to calibrate beam position in the quad-as-BPM method (Section \ref{sec:apparatus:quad-as-bpm}). As a result of this work, VRUMER has been integrated into the Matlab-based UMER interface, and can be implemented as a "virtual machine" with the same routines used for machine control and data collection. Additional information on the VRUMER model can be found in Appendix \ref{ap:model}.

\subsection{MENV}
\label{sec:numeric:menv}

MENV (Matlab ENVelope Integrator) is an in-house tool for integrating RMS envelope equations.\cite{menv} The envelope equations (Eq. \ref{eq:env-eq}) consider only linear terms in external and internal forces. Therefore, MENV (and any envelope integrator) includes only the linear portion of the space charge force on the beam envelope, $\frac{2K}{X+Y}$ for perveance $K$. 
Although fully customizable, at this point MENV only includes hard-edged representation of the magnetic elements. 
In this thesis, MENV is used for optimization of ring lattice solutions for quasi-integrable experiments, as described in Chapter \ref{ch:lattice}.

\subsection{WARP}
\label{sec:numeric:WARP}

WARP is an open source, particle-in-cell (PIC) code used to model self-consistent beam evolution in accelerator lattices.\cite{warp} In PIC simulation, the beam distribution is approximated by a distribution of charge-weighted macro particles. The self-force of the beam is
calculated by interpolating the macro-particle distribution onto a uniform grid and solving the field equations on that grid. The gridded beam self-force is applied while integrating particle equations of motion over each time step. 
The name WARP refers to the use of warped coordinate system, where geometric \verb|BEND| transformations are used to follow the co-moving beam frame. 
The WARP PIC mode has three geometries: 3D, transverse slice, and cylindrically symmetric RZ. All PIC simulations in this thesis are run with the WARP 2D transverse slice package, which has minimal transverse-longitudinal coupling. This is a suitable approximation for UMER, which transports a long coasting bunch without synchrotron motion. 

WARP is written in Fortran and C with a Python wrapper. The Python user interface means WARP is easily customizable and is compatible with many existing Python modules for data analysis and visualization. 
WARP includes many analytic element definitions but also easily allows the inclusion of arbitrary magnetic fields through an interpolated, gridded \verb|BGRD| field element. The WARP model of the UMER lattice includes gridded field elements for ring quadrupoles and dipoles, calculated using the in-house Biot-Savart integrator MAGLI. 

In this thesis, WARP is used for simulations where accurate resolution of the space charge forces is necessary. In addition, the MAGLI-generated gridded field elements are more representative of the UMER magnets than the hard-edged field models used in other codes. When benchmarking different codes, WARP results with gridded fields are assumed to be the most accurate.

\subsection{Elegant} \label{sec:numeric:elegant}

Elegant (ELEctron Generation ANd Tracking) is a 6D accelerator tracking code.\cite{elegant} Elegant also supports many different analytic field elements, including matrix-based tracking and canonical kick (symplectic) elements.
Elegant includes space charge effects, with transverse space charge implemented as kick elements based on a frozen-in model of beam charge distribution. 

Elegant includes many powerful built-in modules for accelerator modeling, including frequency map analysis (below, Section \ref{sec:numeric:fma}) and lattice matching/optimization routines, and is widely used to model a variety of rings. However, the space charge model is not self-consistent. 
Additionally, the limited element models available in Elegant are not the most accurate representations of the fringe-dominated UMER magnets. 
For work in this thesis, Elegant is only used for calculations in a zero-charge limit. Elegant optimization routines are applied to lattice matching problems using the built-in envelope integrator (Chapter \ref{ch:lattice}), and a reduced model of the quasi-integrable system is examined in Elegant to study the effects of errors in external focusing (Chapter \ref{ch:qio}).

\section{Frequency Map Analysis} 
\label{sec:numeric:fma}

A standard approach to understanding long path length particle dynamics in an accelerator, particularly the effects of nonlinearities and resonances on dynamic aperture, is frequency map analysis (FMA). Originally applied to the study of celestial mechanics, this technique has been applied to accelerator dynamics \cite{Laskar2003} and is incorporated into the general cookbook of accelerator tools. This is a powerful technique for simulation studies, but has also been applied to experimental data as well. 

A frequency map is a plot of particle initial position in frequency (tune) and configuration space. The color axis $d\nu$ corresponds with the nonlinearity of a particle orbit. 
In this application of FMA, the orbit is divided in time ($s_o \to s_{mid}$ and $s_{mid} \to s_{final}$) to calculate two frequency values, $\nu_1$ and $\nu_2$ and $d \nu \equiv \nu_1-\nu_2$. High $d\nu$ indicates an orbit with shifting frequency (due to diffusion) or chaotic behavior (due to orbit irregularity), while $d \nu \to 0$ for regular orbits. 
Lines of high $d \nu$ indicate resonance structures, which may contribute to aperture limitation through particle diffusion.

FMA is a built-in feature of the Elegant code. While not included in standard WARP packages, I wrote an FMA module accessed at the user-interface (Python) level. The general approach is to define an initial particle distribution on a transverse (xy) grid. These are zero-current simulations: particles are not weighted macro-particles, but non-interacting test particles that sample the lattice dynamics. In this way, we isolate nonlinearities in the lattice from nonlinear behavior driven by space charge.  The distribution is initiated at a chosen s-location, and propagated for a number of turns. The tune and $d \nu$ are calculated as described above, using Numerical Analysis of Fundamental Frequency (NAFF) to determine the lowest frequency component of transverse motion. 

\subsection{Numerical Analysis of Fundamental Frequency} 
\label{sec:numeric:naff}

In accelerators, position data used for frequency calculation is often limited in number of turns. Frequency resolution using Fourier transformation scales as $\frac{1}{N}$ for number of sample points $N$. The NAFF algorithm allows frequency calculation to a higher resolution $\sim N^{-4}$. The trade-off is the increase in computational time. The fundamental frequency $\omega_1$ of the signal $f(t)$ is the maximum of the overlap integral 

\begin{equation}
\phi(\omega) = \frac{1}{2T}\int_{-T}^T e^{-i \omega t} f(t) dt.
\end{equation}

$\omega_1$ is found by applying a minimum-seeking routine to $-\phi(\omega)$. Pseudo-code for the implementation of NAFF for particle data in this thesis is shown in Appendix \ref{ap:naff}.

\subsection{Calculating dynamic aperture and tune spread from FMA}

\begin{figure}[]
\centering
\subfigure[Test particles plotted in configuration space (corresponding with particle initial positions)]{
\includegraphics[width=0.7\textwidth]{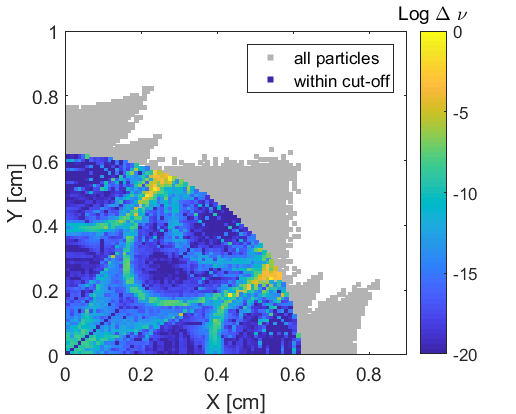}
\label{fig:fma-xy-label}
}
\hspace{.5in}
\subfigure[Tune footprint of test particles.]{
\includegraphics[width=0.7\textwidth]{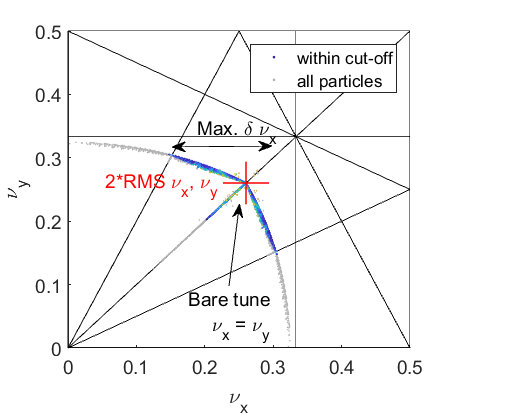}
\label{fig:fma-tune-label}
}
\caption{Example of frequency map analysis for a simple model of quasi-integrable octupole lattice. A radial cut is made as indicated in (a), this corresponds with colored region in (b) tune footprint.}
\label{fig:fma-label}
\end{figure}

FMA is applied to predict dynamic aperture and tune spread for a given nonlinear lattice configuration. Fig. \ref{fig:fma-label} shows typical results for a simple model of a single-channel quasi-integrable octupole lattice, discussed in further detail in Chapter \ref{ch:qio}. Fig. \ref{fig:fma-xy-label} shows particle $d \nu$ (color-axis) versus initial particle position in X and Y. Particles that are lost during the simulation are outside the dynamic aperture and are not plotted. As the shape of the dynamic aperture is not round but the matched beam is initially round through the octupole channel, effective dynamic aperture is defined to be the maximum radius circle that contains only stable initial conditions. Particles outside this boundary in Fig. \ref{fig:fma-xy-label} are masked in gray. 

Octupole-induced tune spread is calculated based only on the particle distribution within this radial cut-off. This is important, as considering the tune spread across all stable particles will result in over-estimation since the largest tune shifts occur near the edge of the dynamic aperture. 
To distinguish tune spread from the ``chaos metric" $d \nu$, notation $\delta \nu$ is used to indicate the half-width of the tune footprint (equivalently shift from the bare linear lattice tune $\nu_0$). The spread of tunes around the linear lattice tune can be thought of as the tune spread that is imprinted on the particle distribution due to external nonlinearity. Two figures of merit are used for this tune spread: the maximum tune spread, $\max{|\nu - \nu_0|}$, and the RMS tune spread, RMS$(\nu - \nu_{0})$. 
As seen in Fig. \ref{fig:fma-tune-label}, for the error-free, simple model of the single-channel lattice the tune footprint is symmetric around the bare linear lattice tune, $\nu_x = \nu_y = \nu_{0}$. The RMS and maximum spreads are calculated relative to this bare tune.

\section{Chapter summary}

The numerical tools applied to the design of the quasi-integrable lattices are discussed. WARP and Elegant are the main tools for exploring nonlinear particle dynamics, while VRUMER and MENV are primarily used to explore tuning and optimization of the UMER lattice for the proposed experiments. As the WARP model includes gridded field elements that are a more faithful representation of UMER magnetic fields than the hard-edged approximation in other codes, this is used as the baseline for accuracy when benchmarking between codes.

%% file: 5.Chapter.tex

\renewcommand{\thechapter}{5}

\chapter{Particle Dynamics in a Quasi-Integrable Octupole Lattice}
\label{ch:qio}

As introduced in Chapter 1, a quasi-integrable octupole (QIO) lattice is proposed as a way to mitigate resonant beam loss in accelerating rings. Large amplitude-dependent tune spreads, driven by external nonlinear potentials, detune incoherent single-particle resonances and damp collective oscillations that otherwise lead to instability, deteriorate beam quality and limit dynamic aperture. While strong nonlinearity can reduce intensity-driven beam loss, quasi-integrability ensures that stable trajectories exist. In this chapter I use simulation to study properties of the QIO octupole lattice as described in \cite{Danilov2010} using a simple reduced model.


\begin{figure}[t]
\centering
\includegraphics[width=.5\textwidth]{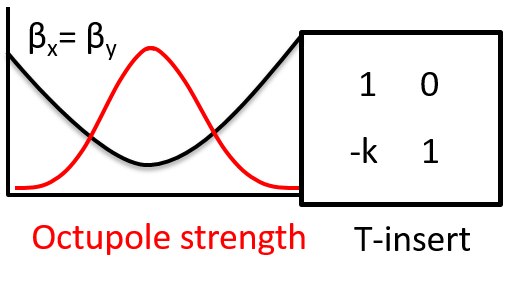}
\caption{Simple quasi-integrable system consisting of focusing lens and octupole insert. }
\label{fig:toy-model}
\end{figure}

The basic recipe for nonlinear integrable optics presented in \cite{Danilov2010} consists of non-interacting particles propagating through a constant nonlinear potential in the normalized frame (Eq. \ref{eq:norm-coord}).
For the case of an octupole potential, transverse particle motion has one conserved invariant (originally shown in Eq. \ref{eq:H-norm} and rewritten here):

\begin{equation}
H_N = \frac{1}{2} \left( p_{x,N}^2 + p_{y,N}^2 +x_N^2 + y_N^2 \right) + \frac{\kappa}{4} \left(x_N^4 + y_N^4 - 6y_N^2x_N^2\right).
\label{eq:H-norm-2}
\end{equation}

\noindent This simple model assumes external linear focusing, which is implicitly included through the $x_N^2 + y_N^2$ term. The nonlinear potential is contained in the second term.
Eq. \ref{eq:H-norm-2} places constraints on both the external focusing and nonlinear potential.
For the case of a ``round" (or XY symmetric) beam, $\beta_x = \beta_y$, we preserve this invariant if the lab-frame nonlinear potential scales properly with the betatron function (this maintains the constant potential in the normalized frame). Linear focusing elements external to the nonlinear element must provide a round beam through the nonlinear insert.  

To achieve this in a ring, we design a linear lattice in which the beam comes to a symmetric waist over a long drift (field-free) section. A nonlinear element with appropriate longitudinal field profile is then inserted in this drift space.
The simplest model contains two alternating elements: the nonlinear insertion device, and the linear focusing lattice between insertions.  
Particle evolution through linear focusing elements is reduced to a transfer function applied as  a matrix operation (see Appendix \ref{ap:matrix}). The transfer function must be equivalent to an XY symmetric focusing lens for the desired beta function. This is shown schematically in Fig. \ref{fig:toy-model}, where the linear lattice transfer function is labeled ``T-insert." 

Without the nonlinear insert, a periodic arrangement of focusing elements and drift spaces is referred to as FOFO lattice. The periodic, matched solution for $\beta(s)$ comes to a \textit{waist} between focusing elements. Appendix \ref{ap:fofo} contains analytic expressions for $\beta(s)$. In the following discussion, $\beta_*$ is used to indicate the beam size at the waist. 

\begin{figure}[t]
   \centering
    \includegraphics[width=\textwidth]{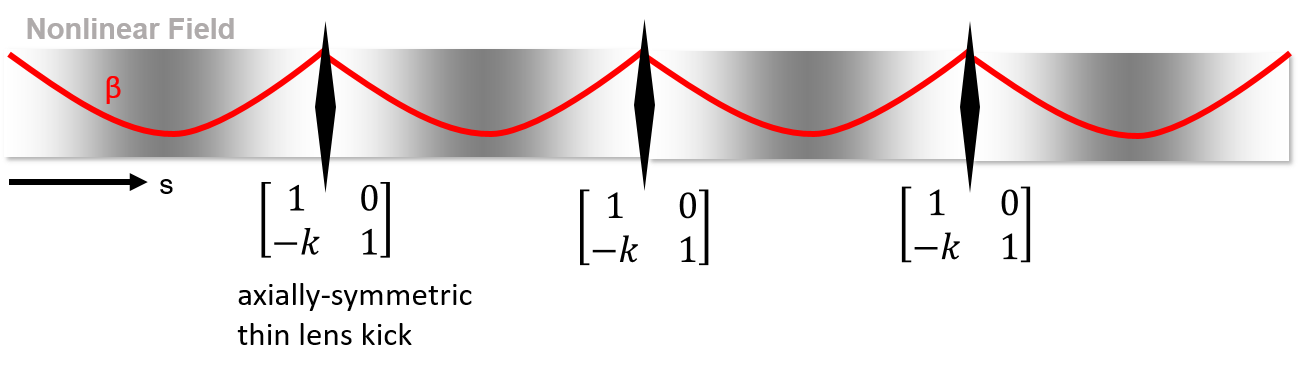}
 	\caption{Diagram of reduced model for QIO lattice, including octupole potential (gray), betatron lattice function (red) and thin-lens kicks (black). }
   \label{fig:toy-model-diagram}
\end{figure}

The reduced model consists of only the nonlinear insert and a thin lens ``T-insert."
This model is visualized in Fig. \ref{fig:toy-model-diagram}.
Particle propagate while immersed in the nonlinear fields and experience periodic focusing impulses. 
With this simple model, I investigate the fundamental properties of a quasi-integrable octupole lattice of this type, including dynamic aperture, tune spread and invariant conservation. I study the effect of errors in the nonlinear insertion, but assume that the linear lattice sections external to the insertion are error-free.

\section{Parametrization of the octupole lattice}

There are only three lattice parameters that define the simple model of a quasi-integrable lattice: insertion length, amplitude of nonlinear field and strength of external focusing. Space charge density \textit{could} be considered a fourth, but initially dynamics are considered in the zero-current limit. 
The insertion length is set as $L=64$ cm, which is the length of a UMER $20^o$ section. 
The remaining two parameters are tuned for maximum dynamic aperture and largest possible tune spread. In this case, dynamic aperture is determined by the stability of particles after 1024 ``turns" (passes through one period of simple model) and defined in terms of particle initial position in configuration space. 

Octupole strength is defined in terms of the peak octupole gradient in the insertion element. Geometric strength $K_3$ is defined in terms of gradient $G_3$, 

\begin{equation}
K_3 \equiv \frac{G_3}{B\rho}
\end{equation}

\noindent for beam rigidity $B\rho$.\footnote{$B\rho = p / q$ for momentum $p$ and charge $q$} and octupole gradient

\begin{equation}
G_3(s) \equiv \frac{1}{6} \frac{\partial^3 B_x}{\partial y^3}.
\end{equation}

\noindent To meet the integrable condition (that the octupole potential is constant in the normalized frame), the longitudinal octupole gradient profile should be equal to $K_3(s) = \kappa \beta(s)^{-3}$. 
Here $\kappa$ is a free scaling parameter for octupole strength that appears in the normalized Hamiltonian, Eq. \ref{eq:H-norm-2}. While $\kappa$ is like a natural choice to parameterize octupole strength from a dynamics perspective, I choose to use the gradient $G_{3,max}$ instead for easier conversion to required octupole excitation. With this choice, the ideal longitudinal profile of the octupole fields is

\begin{equation}
G_3(s) = G_{3,max} \left( \frac{\beta_*}{\beta(s)} \right) ^3 
\label{eq:K3s}
\end{equation}

\noindent For 1 A peak current in the octupoles, $G_{3,max} \sim 50 T/m^3$. For a waist size of $\beta_* = 0.3 $ m and 10 keV electrons, this corresponds to a strength factor of $\kappa = 3980$ m$^{-1}$.

There are several convenient parameters that may be used to define external focusing strength. A natural choice is $k$, which appears in linear lattice transfer function for a thin focusing impulse,\footnote{See Appendix \ref{ap:matrix} for matrix representation of focusing optics.}

\begin{equation}
T = \begin{bmatrix} 1 & 0 \\ -k & 1 \\ \end{bmatrix}.
\end{equation}

\noindent $k$ has units $m^{-1}$ and is related to the thin-lens focal length as $k=f^{-1}$ (also $k=K_1 l$ for geometric quadrupole strength $K_1$). However, this is not the most ideal choice to define external focusing strength, as it has no intuitive relation to observable quantities (such as beam size or lattice tune). The periodic envelope solution $\beta(s)$ depends only on $k$ and $L$ (see Eq. \ref{eq:betas}).
In this dissertation, focusing strength is parameterized in terms of beam waist size $\beta_*$. An expression for $\beta_*$ in terms of $k$ and $L$ is given in Eq. \ref{eq:betastar2}. 

Phase advance of a particle orbit through the insertion is another useful quantity, as it directly correlates with maximum possible octupole-induced tune spread. An expression for $\psi_{dr}$ in terms of $k$ and $L$ is given in Eq. \ref{eq:phi-analytic}. To better connect focusing strength with maximum tune spread, it is easier to define a ``tune advance" for particle motion in the octupole insert, where $\nu_{dr} \equiv \psi_{dr} / 2 \pi$. As the desired phase advance in the linear lattice is $n\pi$, $\nu_{dr}$ is the fractional lattice tune. 
In the following sections, focusing strength is defined in terms of both tune advance $\nu_{dr}$ and $\beta_*$.

\section{Simulations with reduced model for QIO octupole lattice}
\label{sec:qio:toy-model}

Simulations were run in both Elegant \cite{elegant} and WARP \cite{warp}. In general, both codes gave nearly identical results. Choice of code was dictated by the physics being addressed. As described in Section \ref{sec:numeric:WARP}, WARP is a very customizable code that is capable of accurately resolving space charge effects. Existing WARP models of UMER include realistic gridded field measurements based on PCB configuration. Elegant, while less customizable, includes a suite of powerful, built-in tools and is more widely used in the accelerator community.
Further details on the simulation models used here are discussed in Appendix \ref{ap:simple-model}.

\subsection{Invariant tracking in the simple octupole lattice.}

The appeal of the quasi-integrable lattice is that the nonlinear particle motion has a conserved invariant. For a well-behaved (non-diverging) invariant, this guarantees that orbits are bounded for arbitrarily long times. 
In the lab, particle orbits are subject to non-ideal forces, including magnetic field errors and nonlinear space charge forces. Thankfully, invariant surfaces are theorized to still exist for perturbations from integrability.\cite{KAM}
For real systems, minimizing perturbations has been sufficient for million-turn storage of beam in rings. This should also be sufficient for the nonlinear lattice - if we are ``close-enough" to the quasi-integrable solution particle orbits will be stable.

\begin{figure}[tb]
   \centering
    \includegraphics[width=\textwidth]{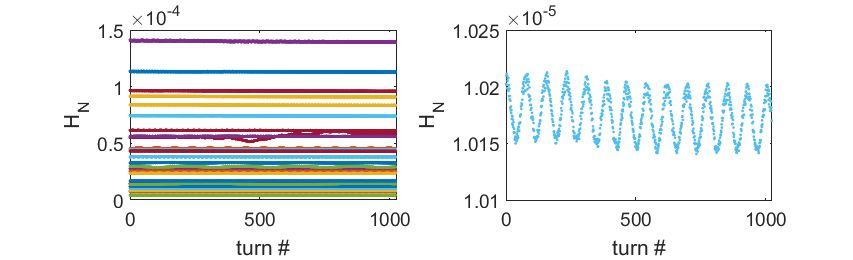}
 	\caption{Conserved invariant $H_N$ for simple quasi-integrable octupole lattice, calculated in WARP simulation with no space charge. }
   \label{fig:warp-invar}
\end{figure}

I use the simple octupole model in WARP to track the invariant quantity $H_N$ (Eq. \ref{eq:H-norm-2}) for lattice parameters $L = 64$ cm and $k=3.3264 m^{-1}$ ($\beta_* = 0.3$ m and $\nu_{drift} = 0.2603$).
For 1024 passes through this octupole channel, a particle of $\langle H_N\rangle =1.02 \times 10^{-5}$ m experiences RMS variation of $2.8\times 10^{-10}$ m ($0.003\%$) without octupoles and variations of $1.9\times 10^{-8}$ ($0.19\%$) for maximum octupole current of 1 A ($G_{3,max} = 50 T/m^3$). 
Despite the small-amplitude oscillation, the particle energy appears to be well-bounded as expected. Invariant evolution for a selection of particles at various amplitudes is shown in Fig. \ref{fig:warp-invar}. Different orbits are distinguished by color. The right plot is a close-up of the left.

\begin{figure}[tb]
   \centering
    \includegraphics[width=\textwidth]{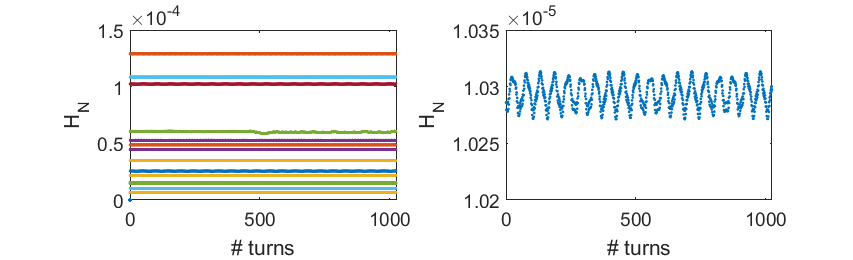}
 	\caption{Conserved invariant $H_N$ for simple quasi-integrable octupole lattice, calculated with Elegant. }
   \label{fig:ele-invar}
\end{figure}

An identical test was run with the Elegant model. A particle of amplitude $\langle H_N\rangle =1.03 \cdot 10^{-5}$ has an RMS variation of $1.3 \cdot 10^{-8}$ ($0.13\%$) for maximum octupole current 1 A ($G_{3,max} = 50 T/m^3$). Accompanying plots are in Fig. \ref{fig:ele-invar}. 

Discrepancy between true $H_N$ conservation and the observed low-level variation is likely due to numerical noise and/or higher-order terms in the Hamiltonian that are truncated by small angle approximations. As Elegant uses matrix-based tracking, the variation is likely due to truncation errors. Additionally, the octupole profile $G_3(s)$ is approximated by flat-topped octupole elements, which may contribute to the variation. The smoothly-varying octupole element in WARP is closer to the target function. However, the percent variation is comparable to the Elegant results, suggesting that interpolation across finite step size may contribute as well.

\begin{figure}[]
   \centering
    \includegraphics[width=0.8\textwidth]{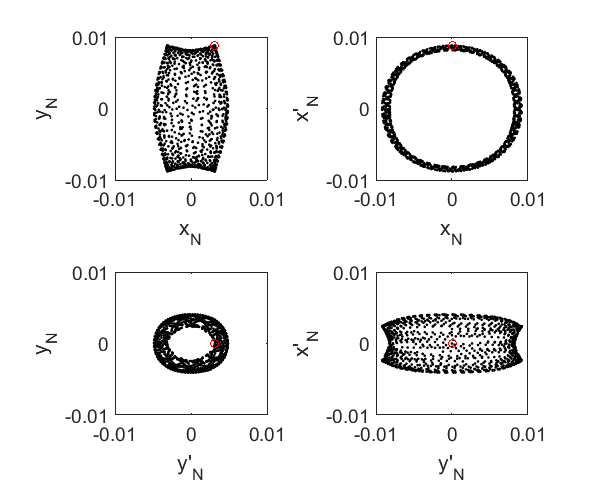}
 	\caption{Particle motion in normalized coordinates in a WARP simulation with $\kappa = 3940$ ($\beta_*=0.3$ m and $G_{3,max} = 50\ T/m^3$). Launch position is indicated with red circle. }
   \label{fig:toy-poincare}
\end{figure}

Poincar\'{e} plots typical of a single stable particle from the WARP simple model simulation are shown in Fig. \ref{fig:toy-poincare}. In a perfectly linear system, we expect motion in the normalized frame to be simple harmonic motion, tracing out a spherical surface in $[x_N,y_N,x'_N,y'_N]$ space with amplitudes $J_x = \frac{1}{2}\left(x_N^2 + p_{x,N}^2\right)$, $J_y = \frac{1}{2}\left(y_N^2 + p_{y,N}^2\right)$. Instead, the particle lives on a surface defined by Eq. \ref{eq:H-norm-2}, which is highly x-y coupled and therefore has no simple projection in any plane.

\subsection{Frequency Map Analysis of simple model} \label{sec:qio:fma}

\begin{figure}[]
\centering
\subfigure[Configuration space.]{
\includegraphics[width=0.8\textwidth]{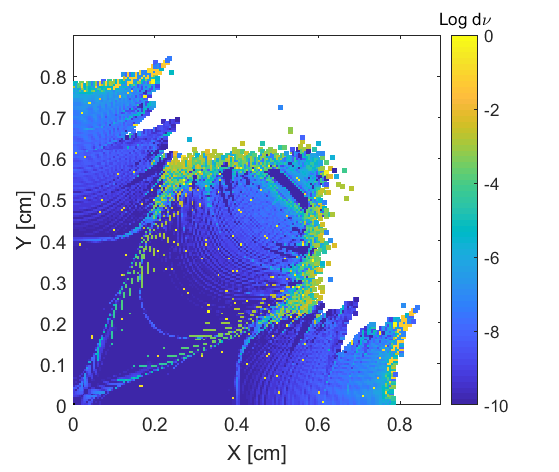}
\label{fig:warp-fma-xy}
}
\hspace{.5in}
\subfigure[Tune space with up to 3rd order resonance lines.]{
\includegraphics[width=0.8\textwidth]{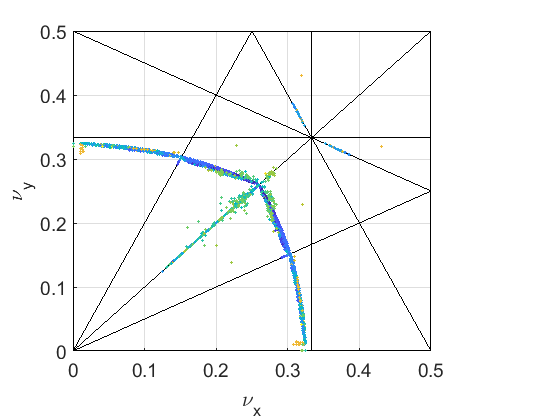}
\label{fig:warp-fma-tune}
}
\caption{Frequency map analysis of simple octupole lattice from WARP model, for operating point $\beta_* = 0.3$ m, $\nu_{dr} = 0.2603$. }
\label{fig:warp-fma}
\end{figure}

\begin{figure}[]
\centering
\subfigure[Configuration space.]{
\includegraphics[width=0.8\textwidth]{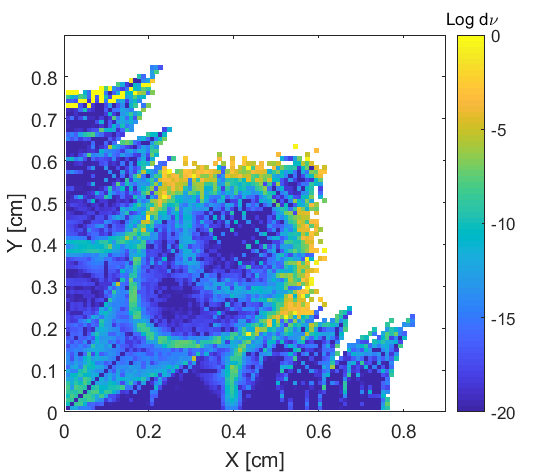}
\label{fig:ele-fma-xy}
}
\hspace{.5in}
\subfigure[Tune space with up to 3rd order resonance lines.]{
\includegraphics[width=0.8\textwidth]{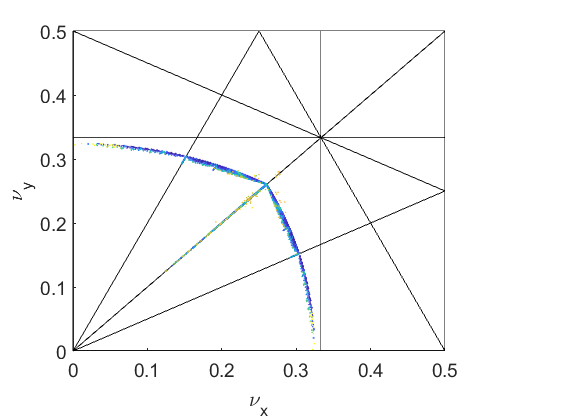}
\label{fig:ele-fma-tune}
}
\caption{Frequency map analysis  of simple octupole lattice from Elegant model for operating point $\beta_* = 0.3$ m, $\nu_{dr} = 0.2603$. }
\label{fig:ele-fma}
\end{figure}

Frequency map analysis (FMA) \cite{Laskar2003} is applied to the simple model of the nonlinear channel, as described in Section \ref{sec:numeric:fma}. This a method to simultaneously sample dynamic aperture and quantify spatially-varying nonlinearities in the lattice. Here it is applied to the simple model to estimate the ``best-case" aperture and tune spread for the quasi-integrable octupole lattice. 

FMA results from both WARP and Elegant models are shown in Fig. \ref{fig:warp-fma} and Fig. \ref{fig:ele-fma} respectively. For both these simulations, a zero-current particle distribution is initiated over a square grid between $x,y=\left[0,2\right]$ cm and tracked for 1024 passes. Particles are launched at the center of the octupole channel, at the matched solution waist $\beta_x = \beta_y = \beta_*$. Therefore, aperture size should be interpreted relative to the expected beam size at this location in the lattice, $x_{RMS} = \sqrt{\epsilon_{RMS} \beta_*}$. The simulation parameters are identical to the previous section. Here the ``thin-lens" external focusing has a strength of $k=3.3264\ m^{-1}$, for $\beta_* = 0.3$ m and $\nu_{dr} = 0.2603$,  and peak octupole gradient is $G_{3,max} = 50\ T/m^3$.

\begin{figure}[]
\centering
\includegraphics[width=0.8\textwidth]{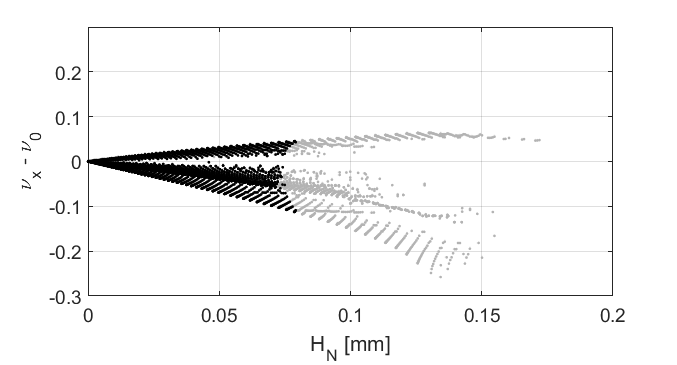}
\caption{Particle tune distribution versus amplitude from Elegant calculation. Particles outside the round dynamic aperture limit are shown in light gray. }
\label{fig:ele-tune-vs-amp}
\end{figure}

Figures of merit are dynamic aperture and tune spread, quantified as both $\max{|\delta \nu|}$ and RMS tune spread RMS$(\delta \nu)$ for $\delta \nu = \nu - \nu_{dr}$. 
For the given octupole channel parameters, Elegant predicts $ DA = 0.62 \pm 0.01 $ cm. The error-bar is taken to the the resolution of the simulated gridded distribution. In Elegant, the maximum tune spread imprinted on all stable particles within the 1024 pass aperture is $\max \delta \nu = 0.256$, close to the predicted maximum $\max{\delta \nu} = \nu_{dr}$.
However, $\max \delta \nu = 0.113$ and RMS$\delta \nu = 0.034$ for particles within the largest  circular dynamic aperture.  The distribution of tune versus initial radial position is shown in Fig. \ref{fig:ele-tune-vs-amp} for operating point $\beta_* = 0.3$ m, $\nu_{dr} = 0.2603$.

In WARP, the maximum radial aperture is $ DA = 0.62 \pm 0.005 $ cm. The maximum tune spread imprinted on all stable particles within the 1024 pass aperture is $\max \delta \nu = 0.250$, while $\max \delta \nu = 0.108$ and RMS$\delta \nu = 0.032$ for particles within a circular dynamic aperture.  These numbers are very similar to the Elegant prediction and demonstrate good agreement for the particle dynamics between the two codes.

Dynamic aperture is related to a maximum accepted emittance as:
\begin{equation}
\epsilon = \frac{DA^2}{\beta_*}.
\end{equation}
In this case, maximum emittance for the edge of the beam to be within the stable aperture is $\epsilon_{edge} \approx 130\ \mu$ m.  
The proposed low-current beam for UMER experiments has measured edge emittance $\epsilon_{edge} \sim 100 \mu m$.\footnote{Discussed in Chapter \ref{ch:design}.} 
At the waist the matched beam edge is $r(s_*) = \sqrt{\epsilon_{edge} \beta_*} \approx 0.55$ cm, comparable to the aperture limit $DA = 0.62$ cm. Another proposed low-current beam has calculated emittance $\epsilon=0.13\ \mu$m.\cite{BernalAAC2017}
The octupole strength required for $DA=r_{edge}$ would be $G_{3,max}\sim 47000\ T/m^3$, outside the capabilities of the UMER octupole magnets.

\section{Choosing an operating point for the octupole lattice  }
\label{sec:qio:op}

When choosing an octupole strength for the quasi-integrable lattice, we want to maximize amplitude-dependent tune spread for the strongest possible damping of resonances. At the same time, as a general rule of lattice design we should maximize dynamic aperture to avoid particle losses. In the quasi-integrable lattice, the tune shift is largest for particles near the aperture limit. The aperture limit is inversely proportional to octupole strength due to unstable fixed points in the quasi-integrable potential.  

\begin{figure}[tb]
\centering
\subfigure[Dynamic aperture]{
\includegraphics[width=0.45\textwidth]{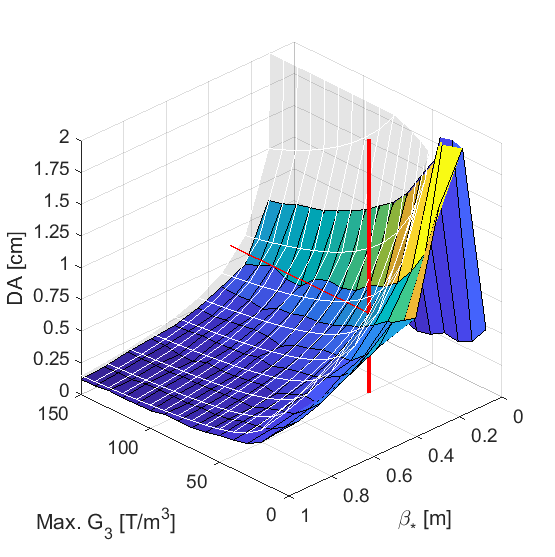}
\label{fig:fma-op-aperture}
}\hspace{.25in}
\subfigure[maximum tune spread]{
\includegraphics[width=0.45\textwidth]{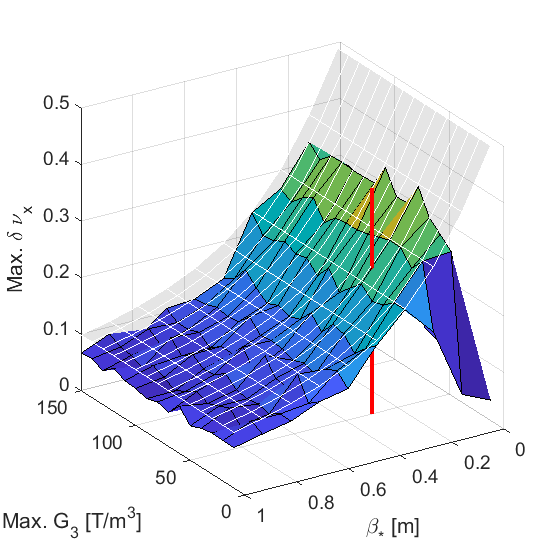}
\label{fig:fma-op-tune}
}
\caption{Parameter space landscape of simple octupole lattice, generated with Elegant FMA for 1024 turns. Figures of merit are plotted vs. peak octupole strength and waist size. 
}
\label{fig:fma-op}
\end{figure}

Examination of the Hamiltonian in \ref{eq:H-norm-2} reveals one stable fixed point at $\left( x_N, y_N \right) = \left(0,0\right)$ and four unstable fixed points at $\left( x_N, y_N \right) = \left(\pm \sqrt{\frac{1}{2\kappa}},\pm \sqrt{\frac{1}{2\kappa}}\right)$.\cite{Webb2012} 
This should result in unstable orbits beyond a radius $r_{max} = \sqrt{\frac{\beta_*}{2\kappa}}$.
Meanwhile, the tune spread is bounded by $\delta \nu \le \nu_{dr}$, as the integer resonance for fractional tunes $\nu = 0$ is very strong. 
This relationship between aperture and $\kappa$, tune spread and $\nu_{drift}$ is shown in Fig. \ref{fig:fma-op}. 
In Fig. \ref{fig:fma-op-aperture} the shaded surface is the estimated aperture limit based on fixed point location, $r_{max} = \sqrt{\frac{\beta_*}{\kappa}}$. 
In Fig. \ref{fig:fma-op-tune} the shaded surface is the bare lattice tune. Red bars indicate nominal operating point chosen for the channel octupole lattice at UMER.
Elegant predictions agree well with the analytic results, with tune spreads and radial aperture only slightly less than the predicted values.

\begin{figure}[tb]
\centering
\subfigure[RMS beam radius, $\epsilon = 100 \mu m$.]{
\includegraphics[width=0.45\textwidth]{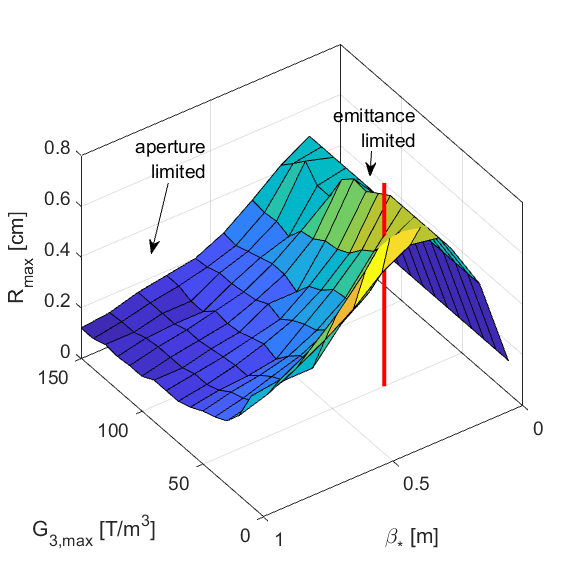}
\label{fig:fma-aperture-100micron}
}\hspace{.25in}
\subfigure[Tune spread, $\epsilon = 100 \mu m$.]{
\includegraphics[width=0.45\textwidth]{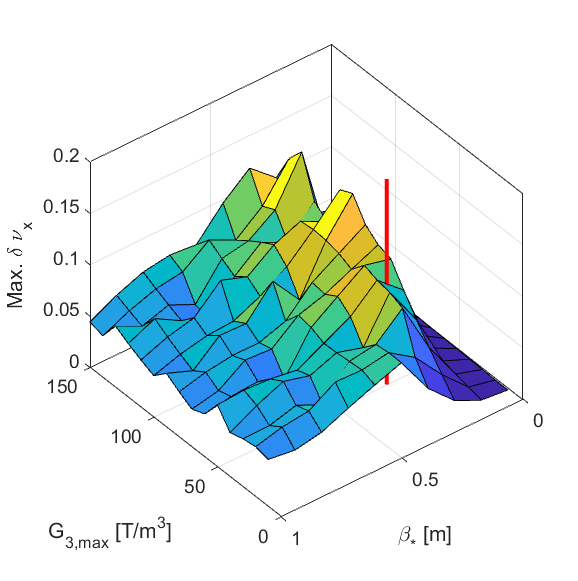}
\label{fig:fma-tune-100micron}
}
\subfigure[RMS beam radius, $\epsilon =7.6 \mu m$.]{
\includegraphics[width=0.45\textwidth]{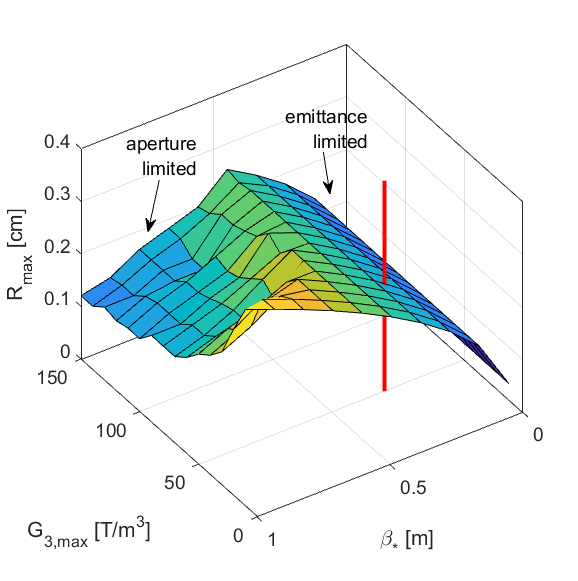}
\label{fig:fma-aperture-7microns}
}\hspace{.25in}
\subfigure[Tune spread, $\epsilon = 7.6 \mu m$.]{
\includegraphics[width=0.45\textwidth]{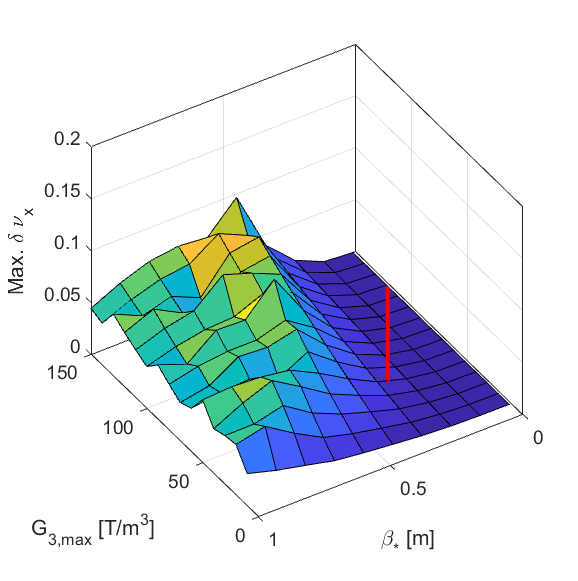}
\label{fig:fma-tune-7microns}
}
\caption{Parameter space landscape of simple octupole lattice, generated with Elegant FMA for 1024 turns. 
} 
\label{fig:fma-eps}
\end{figure}

Choosing a design point for nonlinear experiments requires balancing octupole strength with external focusing to maximize observable tune spread without exceeding the aperture and driving transverse beam loss. 
Fig. \ref{fig:fma-eps} shows maximum beam radii and tune spreads for two different emittance values: $\epsilon = 100 \mu$m and $\epsilon = 7.6 \mu$m, corresponding with values for two UMER beams (low-current DC beam described in in Chapter \ref{ch:design} and the 0.6 mA pencil beam in Chapter \ref{ch:apparatus}). Here maximum beam radius is calculated to be $r_{max}=2\sqrt{\epsilon_{RMS} \beta_*}$ when the beam is within the stable aperture and $r_{max} =$DA when it is not. As expected, tune spread is maximized for operating points where the beam size is limited by the dynamic aperture. The red line in Fig. \ref{fig:fma-eps} indicates nominal operating point used in most calculations.
The goal of initial experiments is to operate near the aperture limit for maximum observable spread.

The parameters $G_{3,max} = 50\ T/m^3$ and $\beta_* = 0.3$ m as identified as a promising operating point for planned experiments with low-current, high-emittance beams. This point is marked in the plots as a red line. Less octupole current is required for lattices with smaller $\beta_*$ and reliance on existing UMER optics makes it challenging to achieve a very small $\beta_*$. However, predicted tune spreads increase with smaller $\beta_*$. $\beta_* = 0.3$ m is chosen as a reasonable compromise, and most calculations in this chapter are computed near this operating point.  
Experiments with smaller emittance beams (with small $r_edge$) require either larger $\beta_*$ or stronger octupole fields. Fields with $G_{3,max} > 150\ T/m^3$ may require more than passive cooling to avoid magnet damage, as $I_{oct}$ exceeds the usual threshold of $3$ A for UMER PCBs.


\clearpage
\section{Damping of mismatch oscillations and halo suppression} \label{sec:qio:halo-studies}

An important limitation of high-intensity beam transport is the phenomenon of beam halo. 
A demonstrated mechanism for halo generation is resonant interaction between individual particles and the beam core. 
A beam that is “mismatched” oscillates about the optimum transverse size.
This coherent oscillation can drive particles to the halo region through parametric particle-core resonance.\cite{Gluckstern1994},\cite{Wangler1998} 
Although halo is low-density, generally containing $<0.01\%$ of the total current, interception on internal boundaries is sufficient for radio-activation of the surrounding environment in a high intensity machine. 

NLIO is predicted to suppress halo formation by quickly damping coherent core oscillations. This effect was thoroughly investigated in simulation studies supporting the IOTA ring.\cite{Webb2012},\cite{WebbIPAC2013}. For a strong nonlinear lattice in both the integrable and quasi-integrable (octupole lattice) cases with weak space charge, the previous study showed no visible halo after 500 passes through a similar reduced model. This work was repeated for parameters appropriate to UMER in order to understand the interplay of space charge tune shift with external octupole-induced tune spread. This is important for UMER experiments, as even the lowest charge (0.6 mA) beam used in normal operation has a significant space charge tune shift of $\nu - \nu_0 = 0.94$.

Simulations were done in the WARP model, with parameters $L=0.774$ m and external focusing strength $k =4.200\ m^{-1}$, which gives $\beta_* = 0.185$ m and $\nu_{drift} = 0.358$. Octupole strengths of $G_{3,max} = 53.8\ T/m^3$ ($\kappa = 1020$) were compared to the linear ($G_{3,max} = 0$) case. 
The initial distribution is KV-like: particles are seeded on a surface in $\left[x_N,y_N,x'_N,y'_N\right]$ space with value $H_N = H_0$. In this case, the effective emittance value was set to $H_{edge} = 5 \times 10^{-6} m$, which gives a beam size in the FOFO lattice equivalent to edge emittance $\epsilon = 10 \times 10^{-6} m$. This is similar to the UMER 0.6 mA (pencil) beam, which has initial RMS emittance $\epsilon \sim 7.6\times 10^{-6}$ m. This type of distribution is used to stay consistent with the work in \cite{Webb2012}. 

\begin{figure}[!tb]
\centering
\subfigure[Initial particle distribution]{
\includegraphics[width=0.45\textwidth]{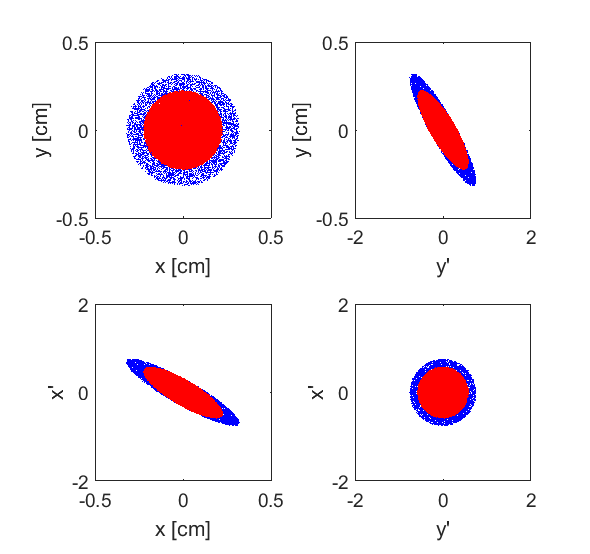}
}\label{fig:initial-distr}
\subfigure[Final distribution, $G_{3,max} = 0$]{
\includegraphics[width=0.45\textwidth]{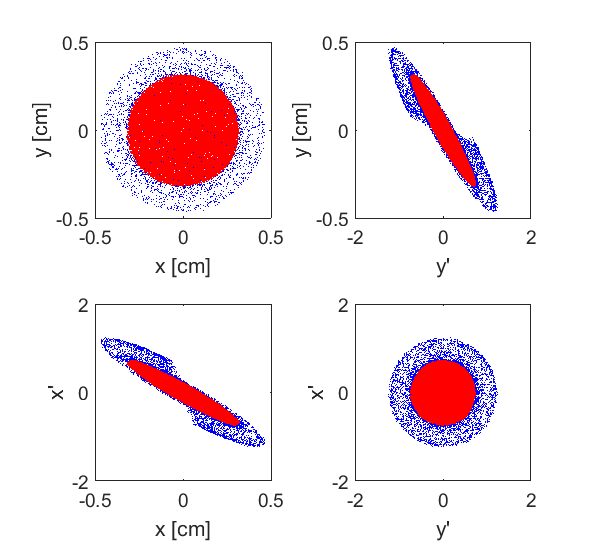}
}\label{fig:final-distr-T0}
\subfigure[Final distribution, $G_{3,max} = 53.8 T/m^3$]{
\includegraphics[width=0.45\textwidth]{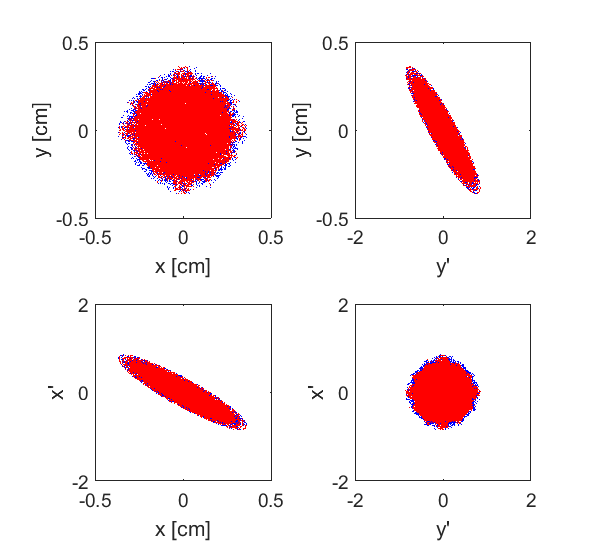}
}\label{fig:final-distr-T10}
\subfigure{
\includegraphics[width=0.45\textwidth]{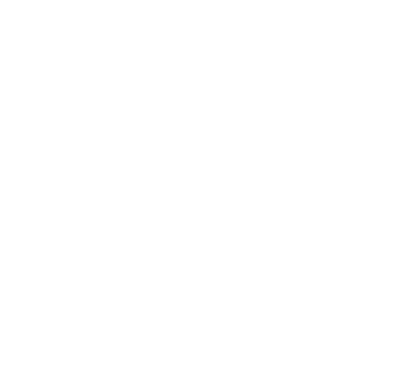}
}
\caption{(a) Initial and (b,c) final distributions after 500 ``turns" for core/halo particles with and without octupole fields. }
\label{fig:halo-distr}
\end{figure}

\begin{figure}[]
\centering
\subfigure[Linear lattice]{
\includegraphics[width=0.75\textwidth]{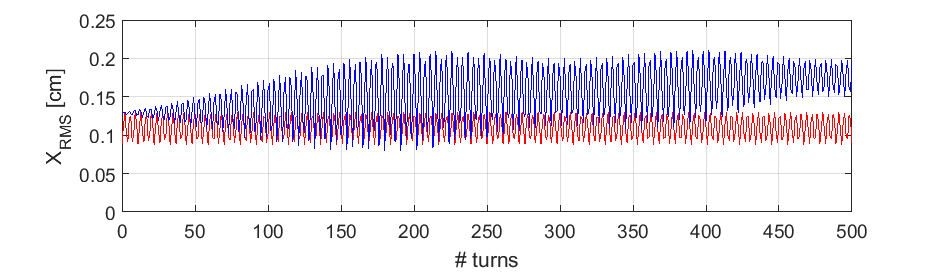}
}
\subfigure[Octupole lattice, $G_{3,max}=53.8 T/m^3$]{
\includegraphics[width=0.75\textwidth]{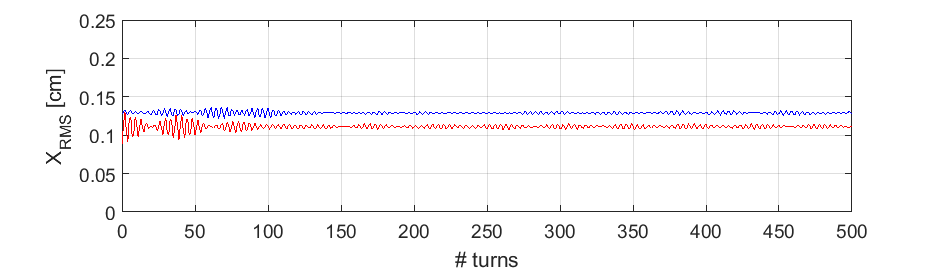}
}
\caption{Evolution of halo (blue, upper) and core (red, lower) populations for $60 \mu$A beam current without (a) and with (b) octupole fields. }
\label{fig:halo-rms-60muA}
\end{figure}

\begin{figure}[]
\centering
\subfigure[Linear lattice]{
\includegraphics[width=0.75\textwidth]{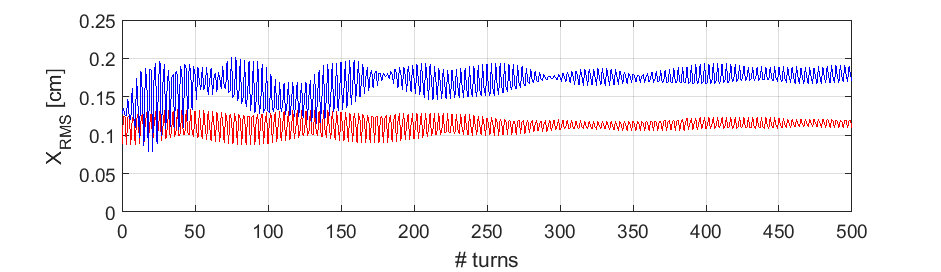}
}
\subfigure[Octupole lattice, $G_{3,max}=53.8 T/m^3$]{
\includegraphics[width=0.75\textwidth]{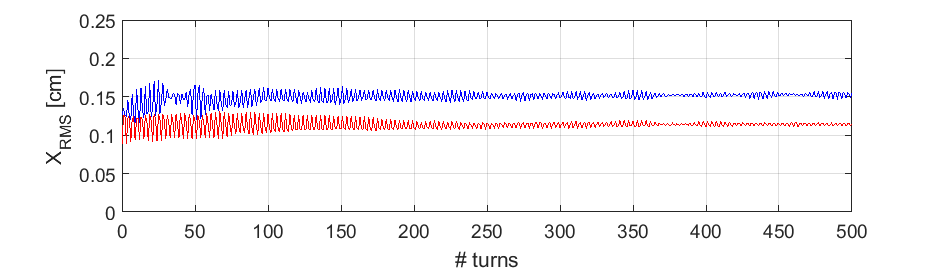}
}
\caption{Evolution of halo (blue, upper) and core (red, lower) populations for 0.6 mA beam current without (a) and with (b) octupole fields. }
\label{fig:halo-rms-06mA}
\end{figure}

The particle distribution is seeded at the start of the channel (downstream of the thin lens kick) with an initial mismatch (the beam core is $30\%$ smaller than the matched solution in both planes). 
In addition to the mismatched distribution, which I refer to as the ``core", I seeded a witness distribution of 
particles with zero current weight to map out the halo dynamics. This ``halo" distribution is initiated with the same ``emittance" $H_{edge}$ as the core but without mismatch. This is similar to the core/pre-halo approach used in \cite{Webb2012}. The initial distributions are plotted in Fig. \ref{fig:halo-distr}(a). For this case, ``edge emittance" $H_{edge} = 5\times 10^{-6}$. Core particles are plotted in red, overlaid on the halo distribution in blue.

The other two plots in Fig. \ref{fig:halo-distr} show the final distribution of a low-charge ($0.03$ mA) beam core and halo after propagating 500 turns through the octupole channel. 
In a linear FOFO lattice with $G_{3,max}=0$, halo particles are driven to higher amplitudes and the beam core oscillates about the matched beam size. Note that the core appears to have relaxed to the matched condition but this is an artifact of examining a single snapshot. In fact, the core oscillations continue out to 500 turns in this case. In comparison, for the case $G_{3,max}=53.8\ T/m^3$, the core and halo populations quickly filament and equilibrate to a steady-state. There is small emittance growth of the core when compared to the linear case, but there are no high-amplitude halo particles.

Simulations were run for a variety of current densities for fixed effective emittance. Beam currents from 0.6 mA to 0.01 mA were tested. Fig. \ref{fig:halo-rms-60muA} compares the time-evolution of the core and halo populations for the 0.06 mA beam with and without an octupole insertion. In the linear FOFO lattice the maximum halo extent grows linearly until saturating at about twice the RMS core size, at which the halo particle frequency is no longer resonance with the core mismatch oscillation frequency. The amplitude of the core mismatch oscillations remains constant out to 500 turns. When octupole fields are added, the halo population does not grow at all and the core oscillations are quickly damped within the first 100 turns. 

Fig. \ref{fig:halo-rms-06mA} shows the same plot in the 0.6 mA case. At higher current density, the efficacy of the octupole insert is reduced. In the linear lattice, the halo still saturates at approximately twice RMS beam radius, but the growth is more rapid (within 50 turns). Additionally, the space charge force also acts to decohere the mismatch oscillation, and by turn 500 the oscillation has been mostly damped out (although the damage has already been done in terms of halo growth). With the octupole insert turned on, halo is partially damped, although the population still grows to $\sim 1.5$x the core radius. Core mismatch oscillations are also completely damped, but this takes longer ($\sim 200$ turns) than in the low-charge case.

\begin{figure}[]
\centering
\includegraphics[width=0.7\textwidth]{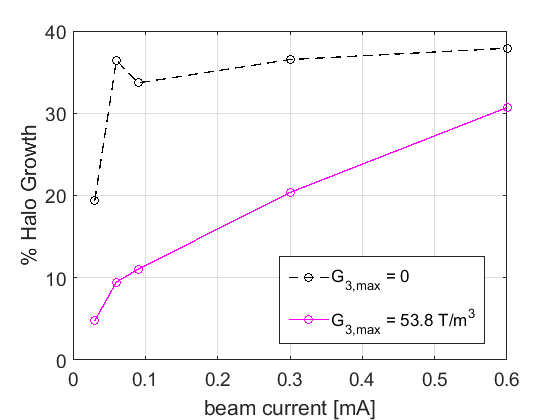}
\caption{Percent halo growth between initial and final halo distribution after 500 turns in simple octupole lattice, with and without octupole fields. }
\label{fig:halo-growth}
\end{figure}

Fig. \ref{fig:halo-growth} plots the percent growth of the RMS halo size with and without the octupole insert after 500 turns. 
In all cases the octupole insert reduced halo growth. However, the relative effect when compared to the linear cases was diminished in the presence of space charge. Even the 0.6 mA beam, which has the lowest current of all ``standard" UMER beams, suffers from significant space charge tune spread (0.85 depression in the standard FODO lattice). 

Extrapolating the curve in Fig. \ref{fig:halo-growth}, at sufficiently high space charge density we expect the insertion will have no observable effect on halo growth.  
Naively, one might assume that dialing up the octupole strength will improve damping for higher charge beams. However, as discussed in Section \ref{sec:qio:op}, the maximum available octupole tune spread is frozen-in. If the beam size is much smaller than the available stable aperture, increasing octupole strength will increase the damping rate \textit{up to a certain point}. Once beam size starts to exceed the area of stable aperture, dialing up the octupole strength will not increase the imprinted tune spread but, instead, high amplitude particles will become unstable and be lost.

As we intend to study mismatch-driven halo evolution and suppression in the octupole lattice, there is compelling reason to extend UMER operation to lower space charge density than the standard low-charge beam at 0.6 mA. For nonlinear optics experiments, this will be accomplished through the double-apertured and DC-mode beams, which both have beam currents in the $\mu$A range.

\section{Chapter summary}

In this chapter I used a reduced model of a quasi-integrable octupole lattice to investigate particle dynamics and set tolerances  for planned single-channel nonlinear experiments. In the zero-charge, single particle limit, the octupole-induced tune shift has a characteristic ``arrow" shape in tune space. The dynamic aperture is sharply defined by fixed points in the transverse potential. I choose a nominal operating point at $\beta_*=0.3$ m and $G_{3,max}=50\ T/m^3$ that should allow for operation near maximum tune spread with a low-current, high-emittance beam.

The effect of octupole potentials on halo formation due to collective mismatch oscillation is studied. For low beam currents, octupole potentials quickly damp the mismatch oscillation and no halo is observed.  Two effects are observed when space charge is included. The space charge force acts to decohere the core oscillations but also increases the rate of halo growth. The effect of the octupole-induced damping is reduced in the presence of space charge. For initial testing we will operate at low space charge density, where the octupole nonlinearity will have a stronger effect. 

Space charge tune shift complicates the dynamics. Due to the space charge tune spread, only an infinitely small sample of the particle distribution will meet the integer phase advance condition exactly. It is assumed that tune-depressed particles will need to be sufficiently close to the quasi-integrable condition to have stable motion. 
This further motivates the use of the low-current, high-emittance beam, which is predicted to have very small space-charge tune shift $\nu-\nu_0=0.005$ (to be discussed in Chapter \ref{ch:design}).

%% file: 6.Chapter.tex

\renewcommand{\thechapter}{6}

\chapter{UMER Experimental Apparatus and Diagnostics}
\label{ch:apparatus}

The University of Maryland Electron Ring (UMER) is a 10 keV electron storage ring used to study the dynamics of intense electron beams over long time scales. This section describes the layout of UMER, as well as primary diagnostics and general measurement techniques. UMER is a flexible machine supporting a range of optics and beam intensity. The design supports FODO transport at different space charge densities for the purpose of studying intensity-dependent beam physics.

Section \ref{sec:apparatus:umer} describes the overall layout of UMER. Section \ref{sec:apparatus:beams} describes the electron gun and properties of the apertured beams. Section \ref{sec:apparatus:diag} covers available diagnostics and Section \ref{sec:apparatus:meas} covers two methods for measuring orbit distortion and resonant structure that are used in later chapters.\footnote{Chapter \ref{ch:steering} and Chapter \ref{ch:res}, respectively}

\section{UMER hardware} \label{sec:apparatus:umer}

\begin{table}[htb]
\centering
\caption{Nominal parameters for the UMER lattice. }
\label{tab:umer-params}
\vspace{10pt}
\footnotesize
\begin{tabular}{l l}
\hline
Beam energy & 10 keV \\
Beam current & 0.6 - 100 mA \\
Circumference & 11.52 m \\
Pipe diameter & 5.08 cm \\
Bunch length &  100 ns  \\
Revolution period & 197.2 ns \\
Repetition rate & 60 Hz  \\
FODO cell length & 32 cm \\
Bare tune & 6.7		\\
\hline
\end{tabular}
\end{table}

\begin{figure}[!tb]
   \centering
   \includegraphics[width=\textwidth]{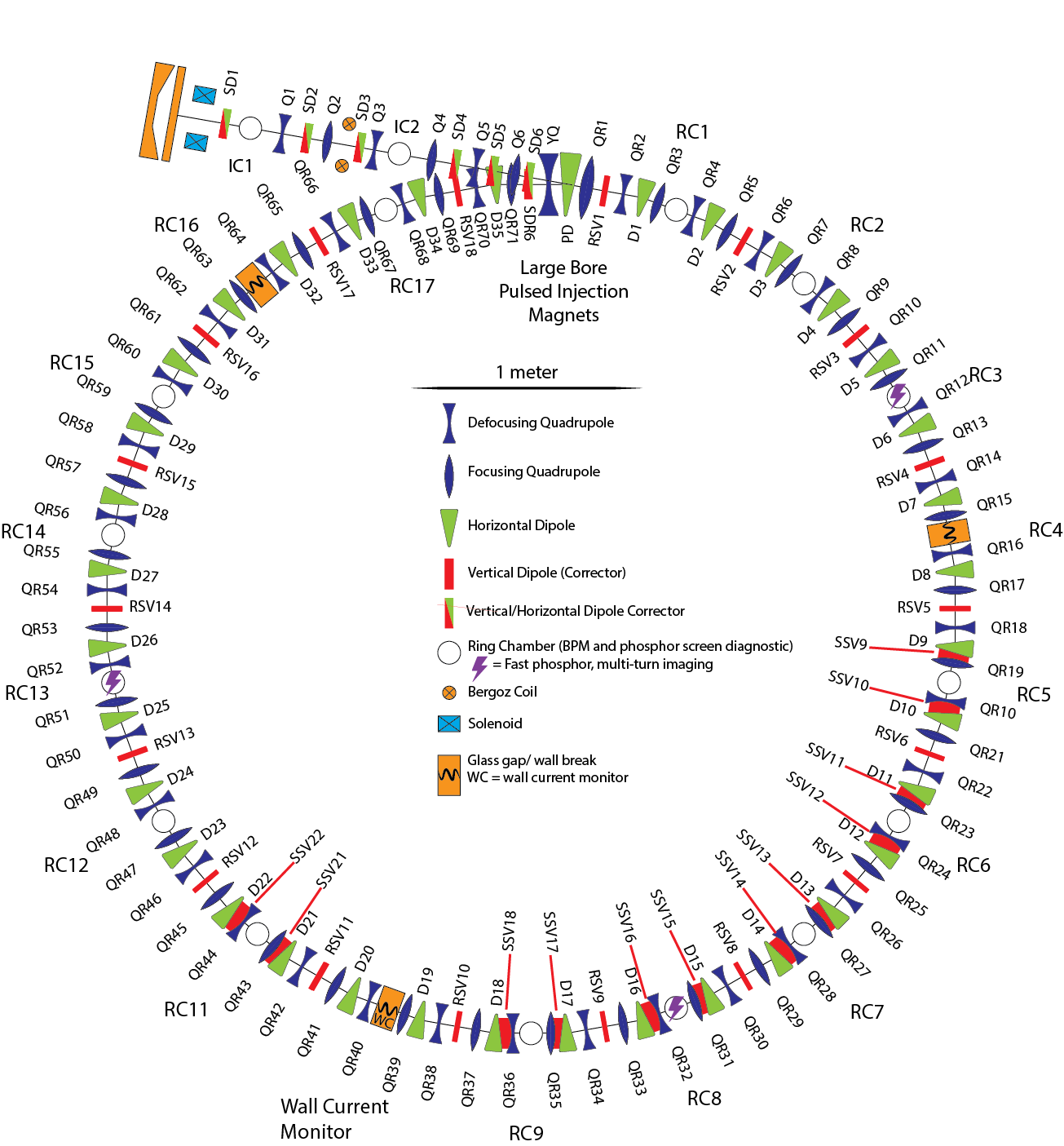}
   \caption{UMER ring, with all magnets labeled. }
   \label{fig:umerring}
\end{figure}

\begin{figure}[]
   \centering
   \includegraphics[width=0.7\textwidth]{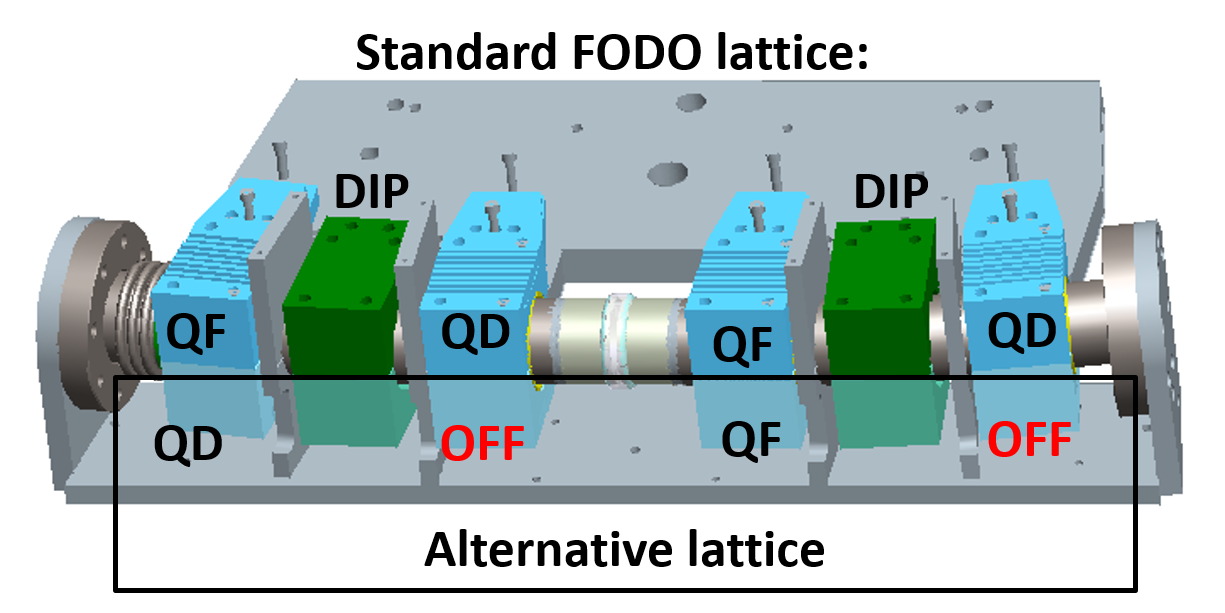}
   \caption{Two standard UMER FODO cells (blue quadrupoles and green dipoles). }
   \label{fig:FODOcell}
\end{figure}

A diagram of the ring is shown in Figure \ref{fig:umerring}. 
The nominal UMER operating parameters are listed in Table \ref{tab:umer-params}. Most parameters can easily be varied, including bunch length, energy, and lattice tune.
UMER is laid out as a 36-sided polygon, comprised of 18 modular $20^o$ sections. A single $20^{\circ}$ section is shown in Fig. \ref{fig:FODOcell}. Each ring section contains two dipole magnets placed over $10^{\circ}$ pipe bends and four quadrupole magnets spaced at 16 cm intervals. Additionally, each section contains room for diagnostics directly between dipoles. Fourteen of the eighteen ring sections contain ring chambers, which house beam position monitors and transverse-imaging phosphor screens. Three have glass breaks in the pipe, to allow for coupling of electromagnetic fields to/from the beam. The last ring section is the ``Y section," which connects the source/injection line to the ring.

\subsection{Injection line and Y section}

The injection line transports the 10 keV beam from the source to the injection at the ``Y section." The injection line contains a solenoid and six quadrupoles for control of injection match as well as six horizontal and vertical steering dipoles (SDs).
Injection is achieved through a pulsed magnetic dipole (PD in Fig. \ref{fig:umerring}). PD is switched between the injection polarity needed to deflect beam into the ring and the recirculating polarity which acts to keep beam in the ring. Prior to bunch arrival from the source, PD is switched to the injection polarity then switched back before the head of the injected beam has completed one turn (approximately 197.2 ns). 

\subsection{Focusing optics}

The UMER ring quadrupoles are arranged the primarily be used in a FODO configuration, with each of the 72 quadrupoles alternating in polarity for a total of 36 FODO cells.
The default configuration is the most efficient orientation for quadrupole focusing, with lowest average beam size for a given emittance, and is ubiquitous in accelerator experiment and theory.

Although nominally a FODO lattice, UMER optics have significant built-in flexibility, as each quadrupole magnet is powered by an independent power supply. 
Previous work has described alternative arrangements of the UMER quadrupoles.\cite{Bernal2012} An alternative arrangement with a longer FODO cell (64 cm instead of 32 cm) is considered as the basis for a strongly nonlinear lattice as described in Chapter \ref{ch:distr}.

\subsection{Printed circuit magnets}

\begin{figure}[tb]
\centering
\includegraphics[width=\textwidth]{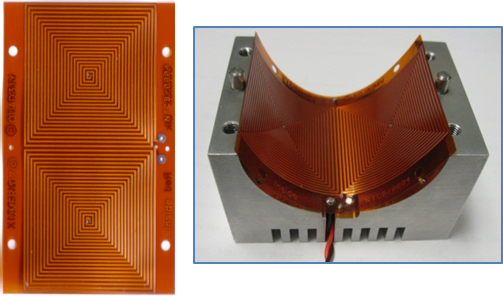}
\caption{Quadrupole PCB and bottom half of an assembled quadrupole element. }
\label{fig:quad-pic}
\end{figure}

All UMER dipole and quadrupole magnets are flexible printed circuit boards (PCBs). 
A PCB quadrupole circuit and mount are pictured in Fig. \ref{fig:quad-pic}. Due to low beam energy, pole field strengths on the order of 10-20 G are sufficient to confine the beam. Required magnet currents are typically 0.5 - 3 A, with higher currents possible if the heat load is compensated by active cooling or pulsed operation. 
The magnets are designed to maximize the purity of the axially-integrated transverse field, with undesired harmonic content $\le 1\%$.\cite{Zhang2000}

\section{Beam generation} \label{sec:apparatus:beams}

\subsection{Electron source}

\begin{figure}[tb]
\centering
\includegraphics[width=0.7\textwidth]{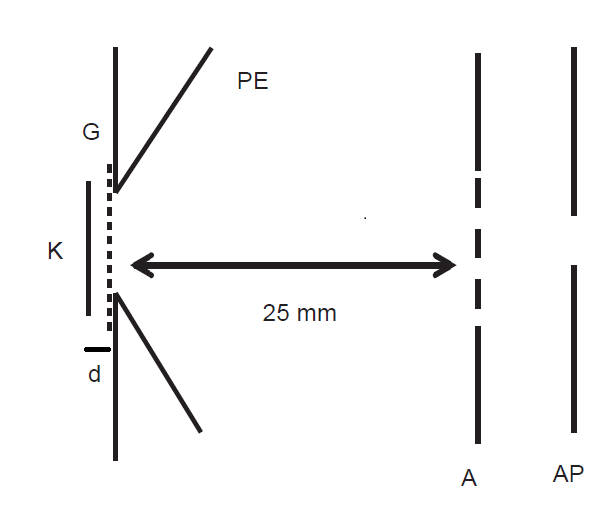}
\caption{Schematic of the UMER gridded gun showing cathode K, cathode grid G, cathode-grid gap d, focusing Pierce electrode PE, gridded anode A and rotating aperture plate AP. }
\label{fig:egun}
\end{figure}

The UMER electron source is a 10 keV gridded triode with a hot dispenser cathode. A schematic is shown in Fig. \ref{fig:egun}. The anode (A) is held at ground and the cathode assembly, including cathode (K), cathode grid (G) and focusing (Pierce) electrodes, is floated at -10 keV. 

A bias voltage is applied across K-G gap d to suppress/draw current.
Typically UMER operates with a $V_b = -30 \to -20$ V bias applied to the cathode when the beam is on. $V_b = +30 \to +40$ V suppresses dark current emission when the cathode is off. 
The longitudinal bunch shape is formed by a square wave pulse on the bias voltage.
The nominal pulse length is 100 ns, for an initial bunch length equal to half the ring circumference. A single bunch is produced at a repetition rate of $60$ Hz. UMER is a storage ring, meaning coasting beam is stored at a fixed energy in the ring.

The gun is operated in saturation mode, meaning the current drawn between the cathode and the grid is the maximum transportable current across the A-K gap. This limit is defined by the Child-Langmuir relation for A-K separation $g$ and potential different $V \sim 10$ keV,

\begin{equation}
I_{beam} = F \epsilon_0 \frac{2q}{m}^{1/2} \frac{V^{3/2}}{g^2}
\end{equation}

where F is a geometry factor.
In saturation, current ``piles up" in the A-K gap. At equilibrium, a \textit{virtual cathode} emitting surface is formed downstream of the grid. 
Because of the virtual cathode, running in saturation mode reduces beam distribution dependence on spatial variation in the cathode surface work function.

\subsection{Selection of space charge density}

The full current produced by the gun is nominally 100 mA. 
Different currents (and therefore different space charge regimes) are selected by collimating the beam with a circular aperture. A rotatable aperture plate, located directly downstream from the anode (see Fig. \ref{fig:egun}), allows selection from a variety of aperture radii. The beam emittance has a linear relationship with the transverse size (defined by the aperture radius $r_0$), while the apertured beam current $I_b$ goes as $r_0^2$. As described by Eq. \ref{eq:tune-spread}, the space charge tune shift $\nu - \nu_0$ has dependence $I/\epsilon$. This gives the space charge tune shift a linear dependence on aperture radius. Table \ref{tab:umer-beams} gives parameters for the five available beam apertures.  

\begin{table}
\centering
\caption{Parameters for all apertured UMER beams. Reproduced from \cite{BeaudoinHB2010} and \cite{HaoThesis}. Emittance $\epsilon$ is unnormalized RMS emittance. Values in three right columns are estimated for FODO lattice with operating point $\nu \sim 6.6$. }
\label{tab:umer-beams}
\vspace{10pt}
\resizebox{\textwidth}{!}{%
\begin{tabular}{l c c c c c c}
\hline
Current [mA] &$r_0$ [mm] & $\epsilon$ [mm-mrad] & Avg. radius [mm] & $\nu / \nu_0$ & Incoh. shift \\ 
\hline 
 0.6 & 0.25&  7.6 & 1.6 & 0.85 & 0.94 \\
 6.0 &0.875&  25.5& 3.4 & 0.62 & 2.4  \\
 21  &1.5  &  30.0& 5.2 & 0.31 & 4.5  \\
 80  &2.85 &  86.6& 9.6 & 0.17 & 5.5  \\
 100 & 3.2 &  97.3& 11.1& 0.14  & 5.7  \\
\hline
\end{tabular}}
\end{table}
\normalsize

\section{Diagnostics} \label{sec:apparatus:diag}

\subsection{Beam Position Monitors}

Multi-turn beam position is measured on capacitive beam position monitors (BPMs), located at the 14 ring chambers in Fig. \ref{fig:umerring}. 
The BPM is a non-interceptive diagnostic consisting of four curved-plate pick-up electrodes, one for each of the transverse directions: top, bottom, left and right. The BPM plates subtend a circular aperture with the same radius as the nominal pipe radius (2.54 cm). As the beam passes, image current runs along each BPM plate and inducing a voltage drop across a resistor proportional to the relative distance of the beam. Comparing differences (top $-$ bottom and left $-$ right) gives a measurement of the beam centroid position with respect to the BPM center.  
The BPMs have a spatial resolution of 0.1 mm with 100 mA beam (0.4 mm for 20 mA).\cite{Quinn2003}
The repeatability of BPM measurements is measured to be 1 micron for the 6 mA beam with 16 averages on the scope.\cite{LevonRCDS}

\subsection{Wall Current Monitor}

A resistive wall current monitor (WCM) is used to measure the temporal beam current profile non-interceptively. It is located at RC10 (labeled in Fig. \ref{fig:umerring}). The WCM consists of a glass gap in the metal pipe boundary and a resistor that electrically connects the two sides of the gap. Voltage drop across the resistor is proportional to the beam's image current, with conversion $I_{beam} = V_{scope} / 4.545\ \Omega$. A ferrite core around the beam pipe adjacent to the glass gap forces the image current to run through the resistor rather than the surrounding environment. The equivalent circuit for the WCM has parallel inductive originating primarily from the ferrite core.
The inductive term causes an apparent drifting baseline in the WCM signal and effectively makes the WCM ``DC blind," as it cannot be relied upon to measure absolute current.
If this inductance is known, it can be integrated out of the measured voltage, 

\begin{equation}
I_{beam} = \frac{V_{scope}}{4.545\ \Omega} + \frac{1}{L} \int_0^t V_{scope}dt
\end{equation}

\noindent as described in \cite{irvwcmnote}.

\subsection{Transverse imaging}

The transverse beam profile is imaged with beam-intercepting phosphor screens, which produce photon intensity proportional to the number of incident electrons. These screens measure the XY projection of the transverse beam distribution. They can also be used measure beam emittance using the quad scan technique. In this method, the Courant-Snyder parameters are reconstructed by measuring the dependence of beam size on the strength of an upstream quadrupole.\footnote{this method is only valid for low-current beams with weak space charge}

The phosphor screens are housed in each of the fourteen ring chambers. 
A screen consists of 31.75 mm diameter glass coated with P43 phosphor (Gd$_2$O$_2$S:Tb), which has a 1 ms decay time. At some ring locations, fast-phosphor ZnO:Ga is used with decay times $< 3$ ns. The fast-phosphor screen locations are indicated in Fig. \ref{fig:umerring}.
Both types of phosphor screens are patterned with fiducial marks for the calibration of beam size.

Image capture is done with GigE Vision Flea3 cameras. These cameras have 12-bit ADC, with max. rate 120 FPS and resolution 648 $\times$ 488. While shutter speed is fast enough to capture a single bunch at $60$ Hz repetition rate, there is not enough time resolution to examine slices within the bunch. Each image represents an integration along the bunch length.

\section{Measurement techniques} \label{sec:apparatus:meas}

This section describes specific measurement techniques utilizing the above diagnostics. These methods are applied for work described in Chapters \ref{ch:res} and \ref{ch:steering}. Section \ref{sec:apparatus:quad-as-bpm} describes the quad-as-BPM method. This is a tool for measuring transverse orbit distortion at a higher longitudinal resolution than the BPM spacing. The method has been refined and incorporated in the UMER control system as a result of this work. Section \ref{sec:apparatus:knock-out} described a beam knock-out method for multi-turn imaging and reconstruction of DC current components. Section \ref{sec:apparatus:tune-scan} discusses tune scans, which measure of beam transmission as a function of transverse focusing strength. This method is applied to identify good operating points and observe transverse frequency-dependent resonant phenomena.

\subsection{Quadrupole as BPM technique} \label{sec:apparatus:quad-as-bpm}

Measuring transverse beam offset at the BPM locations is not sufficient to fully resolve the beam trajectory. At the nominal operating point the wavelength of transverse beam oscillations is $\lambda \sim 170$ cm. With 14 BPMs spaced at 64 cm intervals, the beam can only be sampled every third of a wavelength. Higher resolution for trajectory data is achieved by measuring beam centroid offset in the quadrupoles (which have a spacing of 16 cm), using a technique called ``quad as BPM" or ``virtual BPM." On the first turn only, beam centroid position can be reconstructed from quadrupole response data. This technique is applied in an orbit correction algorithm described in Chapter \ref{ch:steering}.

\begin{figure}[!tb]
\begin{center}
\includegraphics[width=\textwidth]{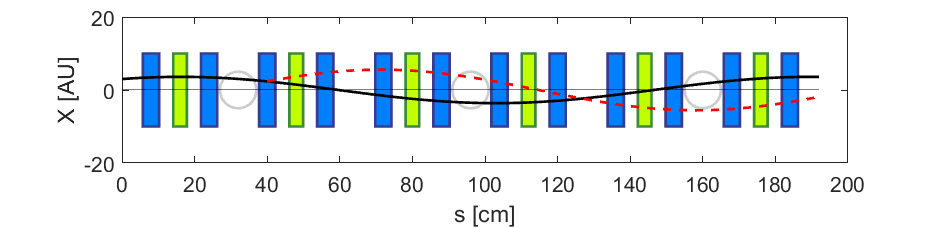}
\end{center}
\renewcommand{\baselinestretch}{1}
\small\normalsize
\begin{quote}
\caption{Diagram of UMER beam line, with quadrupoles (blue), dipoles (green) and BPMs (circles). Black curve is possible centroid trajectory, dashed-red curve is perturbed centroid due to change in quadrupole strength at 40 cm. Phase advance per cell (32 cm) is $66.4^o$. Beam trajectory is shown in a smooth focusing limit. }
\label{fig:simple-beamline}
\end{quote}
\end{figure}

A beam passing through a quadrupole off-axis experiences a dipole force acting on the centroid (in addition to the quadrupole force on the beam envelope). The strength of the dipole kick will depend on the position of the centroid in the quadrupole as well as strength of the quadrupole. Variation of the quadrupole strength will cause variation of the centroid proportional to the position in the quadrupole as detected on a downstream BPM. This concept is illustrated in Fig. \ref{fig:simple-beamline}. As the beam centroid evolves according to single particle equations of motion, knowledge of the single particle transformation between the quadrupole and BPM allows reconstruction of the beam position.

\renewcommand{\baselinestretch}{2}
\small\normalsize

The centroid position in a downstream BPM has a linear dependence on the position in the quadrupole, $x_Q$: 

\begin{equation} 
\frac{d \tilde{x}(s_{BPM})}{d I_Q} = x_Q \frac{L}{\sigma} \frac{G}{B\rho} \sin{\frac{\sigma}{L}s_{BPM}}. 
\label{eq:quad-response} 
\end{equation}

Here, $I_Q$ is the quad current, $L$ is the cell length, $\sigma$ is the phase advance per cell and $G$ is the linear quad gradient.
The full derivation for Equation \ref{eq:quad-response} is covered in Appendix \ref{ap:quadasbpm}.

In the approach implemented in this thesis, we first measure the quad response slope $\frac{d \tilde{x}(s_{BPM})}{d I_Q}$. The Matlab-based in-house tracking code VRUMER\cite{vrumer} is used to calculate the constant $\frac{L}{\sigma} \frac{G}{B\rho} \sin{\frac{\sigma}{L}s_{BPM}}$. The result is a measurement of $x_Q$.
This approach is similar to a previous implementation in UMER \cite{KPRnote:2012}. The main difference is that the approach described here uses a tracking code to calibrate position, rather than a transfer matrix calculation.

As a result of this work, the quad-as-BPM method has been integrated into the UMER controls interface. VRUMER orbit tracking is calculated responsively during the data collection process. This not only streamlines the quad-as-BPM method, but also dramatically increases the flexibility of this method to accommodate different ring optics without increasing the complexity for the user. 
Details of the implementation, including error analysis, are discussed in Appendix \ref{ap:quadasbpm}.

\subsection{Knock-out method} \label{sec:apparatus:knock-out}

\begin{figure}[!tb]
\centering
\includegraphics[width=\textwidth]{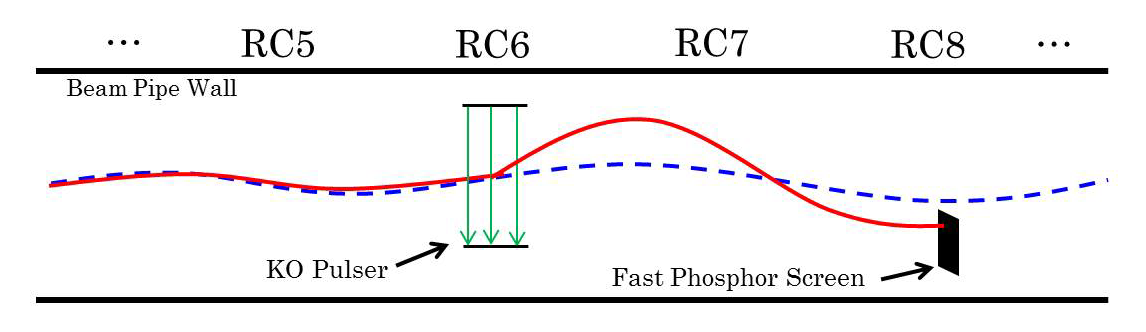}
\caption{Schematic of knock-out method. An electrostatic kick is applied on BPM electrodes at RC6 location. The blue dashed line indicates the previous turn and the red solid is the kicked turn. This drawing assumes beam motion in a smooth focusing limit, and magnets are not shown. }
\label{fig:knockout}
\end{figure}

Beam knock-out is a method devised to allow multi-turn imaging of the transverse beam profile. A BPM is converted from a set of four independent pickup electrodes into a pulsed electrostatic kicker with a vertically positive kick. 
A fast pulse is synchronized to perturb the desired turn. A pulse voltage $V = 1 \to 3$ kV is sufficient to deflect the beam such that it is intercepted on a  downstream, off-axis phosphor screen. The downstream screen is positioned below the closed orbit such that it only intercepts the perturbed turn. The screen is longitudinally separated from the kick location by approximately $2/3$ of a betatron oscillation, where the perturbed centroid oscillation is at a maximum. This is illustrated in Fig. \ref{fig:knockout}\footnote{Figure is taken from \cite{Stem2015}.}.

The knock-out technique is also used to restore AC signal when space charge forces act to elongate the bunch, which fills the ring and gains a DC current component. The full bunch is knocked out and the resulting AC structure measured with the WCM. This is used for current loss rate measurements, which are shown in Chapter \ref{ch:res}.

\subsection{Tune scans}	
\label{sec:apparatus:tune-scan}

\begin{figure}[htb]
\centering
\includegraphics[width=0.7\textwidth]{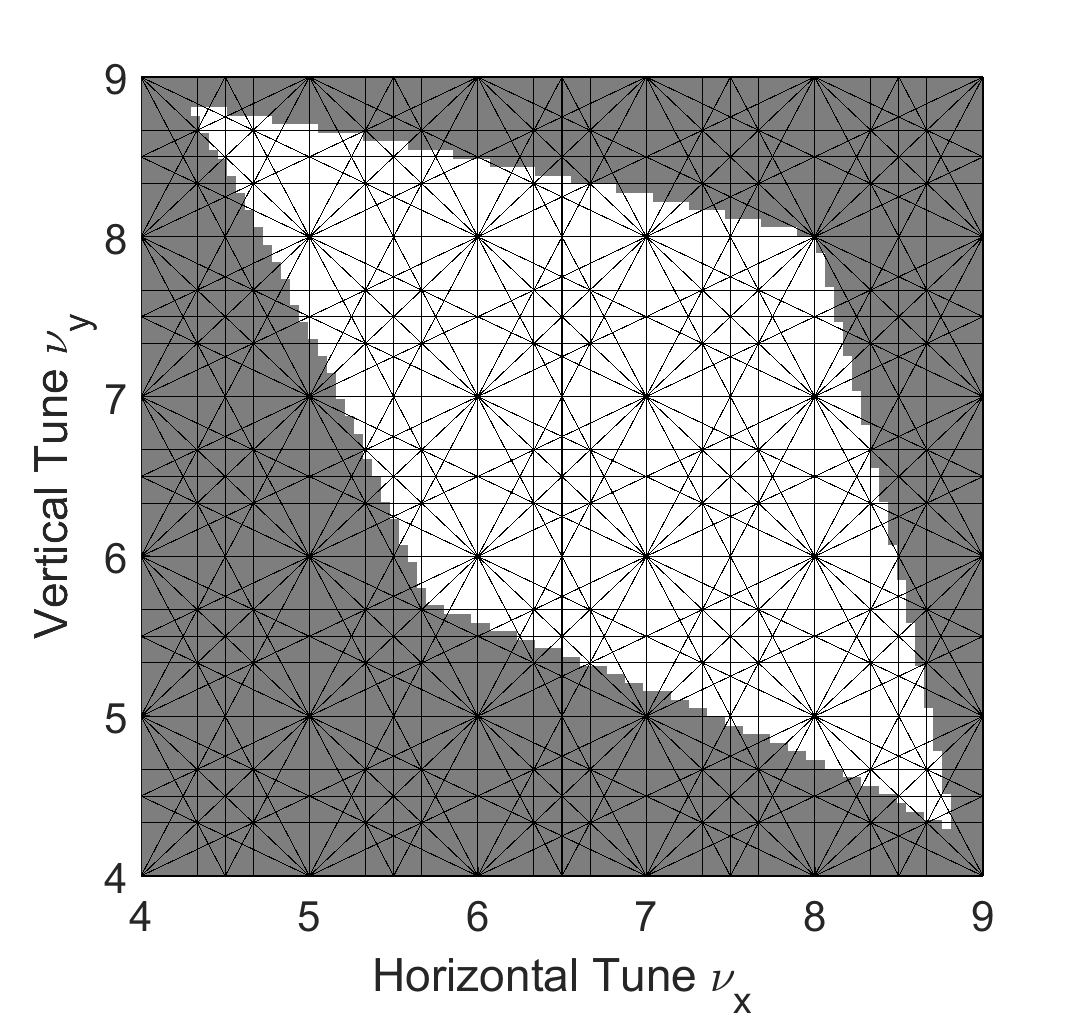}
\caption{Schematic of area in tune space covered in typical UMER tune scan, varying $I_F,I_D = 1.6 \to 2.1$ A. Resonance lines up to third order are included. }
\label{fig:tune-scan-schematic}
\end{figure}	

Tune scans (also called quad scans) are used to measure beam transmission as a function of lattice tune in a FODO lattice. In this method, ring quadrupoles are divided into two families, identified by their horizontal polarity as focusing (F) or defocusing (D). The currents in the two families, $I_F$ and $I_D$, are varied in a 2D raster scan. The WCM is used to measure beam transmission at each operating point. From this a survival plot can be generated, which shows the surviving beam current after a given number of turns as a function of operating point. This can be used to characterize the resonant response of the beam over a wide range of frequencies.

All survival plots are shown in tune space. The transformation from quad currents $\left(I_F,I_D\right)$ to tune $\left(\nu_x,\nu_y\right)$ is analytically calculated based on a smooth-focusing approximation of the UMER FODO lattice.\cite{Santiago} 
The typical range covered in a tune scan is $I_F,I_D = 1.6 \to 2.1$ A with step size 0.01 A. In tune space this resolution is roughly 0.07. The maximum possible resolution (based on power supply resolution) is 0.001 A, increasing tune space resolution to $\sim0.007$.
A diagram showing the tune space region covered in a typical scan is shown in Fig. \ref{fig:tune-scan-schematic}. 

\subsection{Measuring tune in UMER} \label{sec:apparatus:meas-tune}

The smooth focusing model used to convert ring operating point from quad currents to bare tune is known to be an over-simplification. For example, it does not include geometric and edge-focusing in the dipoles, which is known to split the horizontal and vertical tunes in a FODO lattice. By the smooth model prediction, $\nu_x = \nu_y$ when $I_F=I_D$. 

\begin{figure}[!tb]
\centering
\subfigure[Typical BPM signal used for tune calculation with 6 mA beam.]{
\includegraphics[width=\textwidth]{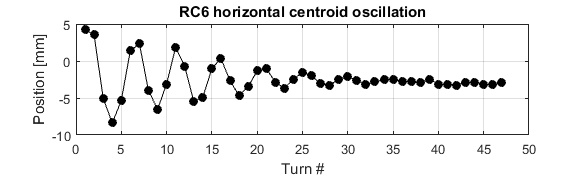}
\label{fig:tune-meas-a}}
\hspace{.5in}
\subfigure[Smooth model prediction compared to measured tunes.]{
\includegraphics[width=\textwidth]{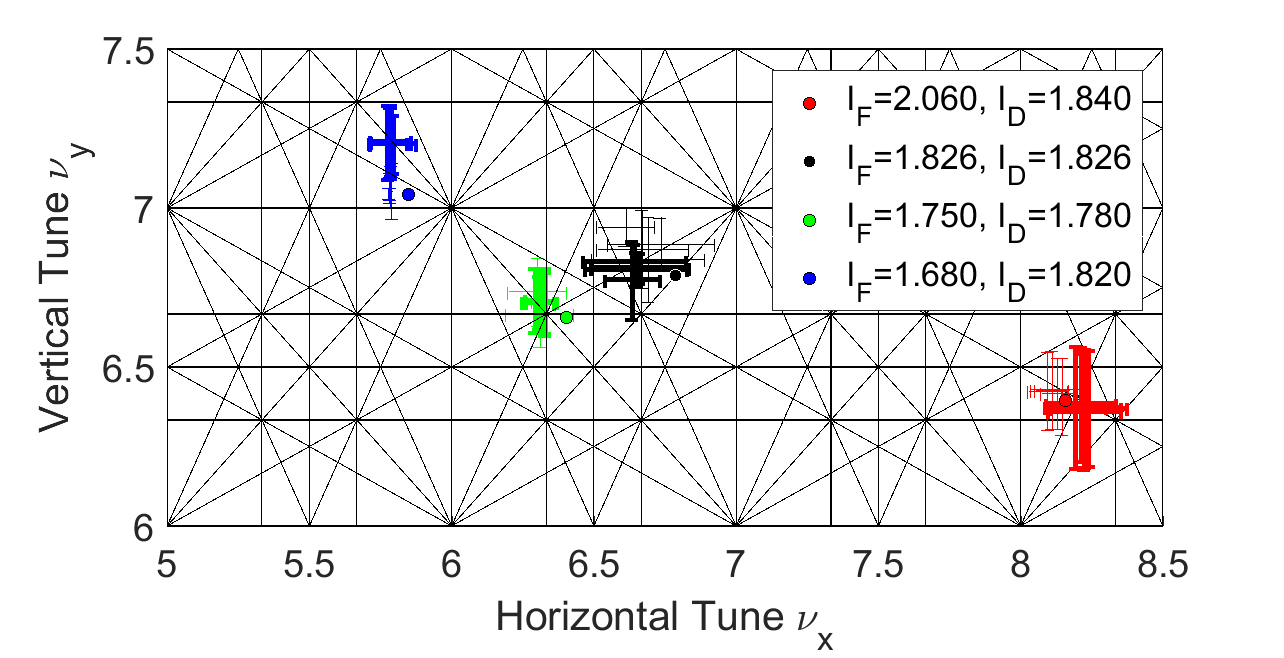}
\label{fig:tune-meas-b}}
\caption{Comparison of tune measurement with smooth model analytic predictions. Color indicates operating point. Dots mark the smooth model prediction, heavy hashes indicate predictions from the four-turn formula \cite{Koutchouk} and light hashes are NAFF calculations \cite{Laskar1992}. Multiple measurements were taken for different centroid injection errors. }
\label{fig:tune-meas}
\vspace{.5in}
\end{figure}

A set of ring tune measurements using the 6 mA beam is plotted in Fig. \ref{fig:tune-meas}. Fig. \ref{fig:tune-meas-a} shows a typical beam trajectory as sampled at a single BPM location. The decay of the oscillation is due to peak signal loss as the beam debunches under space charge driven expansion.
Fig. \ref{fig:tune-meas-b} compares tune measurements at four different operating points with the smooth model prediction. 
Two methods for calculating tune from position data are shown: Numerical analysis of fundamental frequency (NAFF) \cite{Laskar1992} and a four-turn formula.

The four-turn formula uses a geometric argument to return fractional tune from four subsequent turns of BPM data \cite{Koutchouk}: 

\begin{equation}
\label{}
\cos \nu = \frac{x_n - x_{n+1} + x_{n+2} - x_{n+3}}{2(x_{n+1}-x_{n+2})}
\end{equation}

At the $I_F = I_D = 1.826$ A operating point, measurements indicate a tune splitting of $\nu_y - \nu_x \approx 0.2$, with predicted tunes (averaged over measurements from all 14 BPMs) $\nu_x = 6.647$ and $\nu_y = 6.768$. 
The smooth focusing model predicts $\nu_x = \nu_y = 6.786$.
In general, the measured fractional tune agrees with the smooth model tune to within 0.2 (with four-turn) or 0.1 (with NAFF) for the four operating points considered.

A caveat is that any method which reconstructs tune based on a single BPM can only return the fractional tune. The measured fractional tune is related to the full tune as $\nu = n \pm \nu_f$ for some unknown integer n. There is a degeneracy in the ``direction" of the fractional tune, whether it is a fraction above or below the integer part. A large model-measurement discrepancy might appear smaller when comparing only fractional tunes. For the predicted tunes, the integer part was assumed based on the model prediction.

Understanding the source of discrepancy between measured and predicted tunes will allow more accurate modeling of the  quasi-integrable experiment, which requires accurate control of the lattice tune. 
Moving to tracking codes rather than analytic predictions will make it easier to include edge effects. 
Additionally, a well-benchmarked numerical method will allow flexibility for different UMER configurations, including the 64 cm FODO ``alternative lattice" and non-FODO optics for the octupole lattice.

\section{Chapter summary}

This chapter reviewed the operational parameters and principle diagnostics available at UMER. 
In additional to the primary diagnostics of beam position monitors (BPMs) and wall current monitor (WCM), the quad-as-BPM, knock-out and tune scan methods are described. Although the methods were refined in the context of nonlinear optics experiments, they are generally useful for day-to-day operations in UMER.

%% file: 7.Chapter.tex
 \renewcommand{\thechapter}{7}

\chapter{Design and Apparatus for Nonlinear Optics Experiments at UMER}
\label{ch:design}

This chapter extends the previous description of the UMER apparatus to include the planned nonlinear optics experiments. This includes a re-design of the UMER ring to incorporate octupole fields, as well as design of octupole magnets and production of a low-current beam for nonlinear optics experiments.

Section \ref{sec:design:framework} reviews two different options for experiment design: many short octupole inserts or a single long octupole channel. Section \ref{sec:design:octu} covers design of the octupole printed circuit board (PCB) windings and a multi-PCB octupole channel. Finally, Section \ref{sec:design:lowcurrent} details a voltage-amplification mode for electron source operation that produces a low-current, high-emittance beam with very low space charge.
In Section \ref{sec:design:error}, the simple model of Chapter \ref{ch:qio} is applied to study octupole lattice performance as a function of various errors. Errors in the form of orbit distortion, linear lattice tune advance and realistic octupole fields (based on Biot-Savart solution of the PCB windings) are considered.

\section{Layout for nonlinear optics experiments}
\label{sec:design:framework}

\begin{figure}[htb]
\centering
\includegraphics[width=0.7\textwidth]{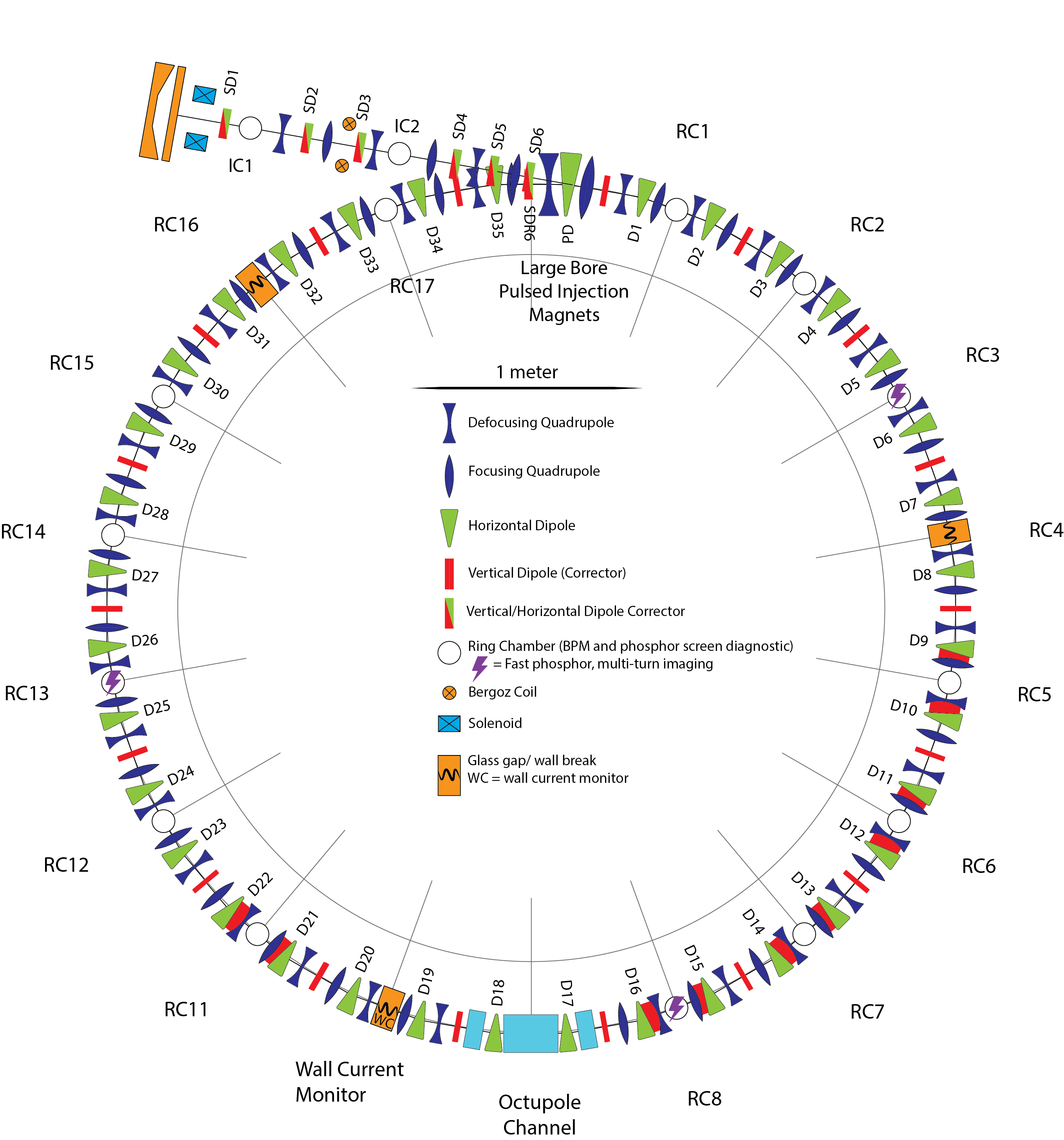}
\caption{Proposed layout for single-channel octupole lattice. }
\label{fig:exp-layout-a}
\end{figure}

\begin{figure}[htb]
\centering
\includegraphics[width=0.7\textwidth]{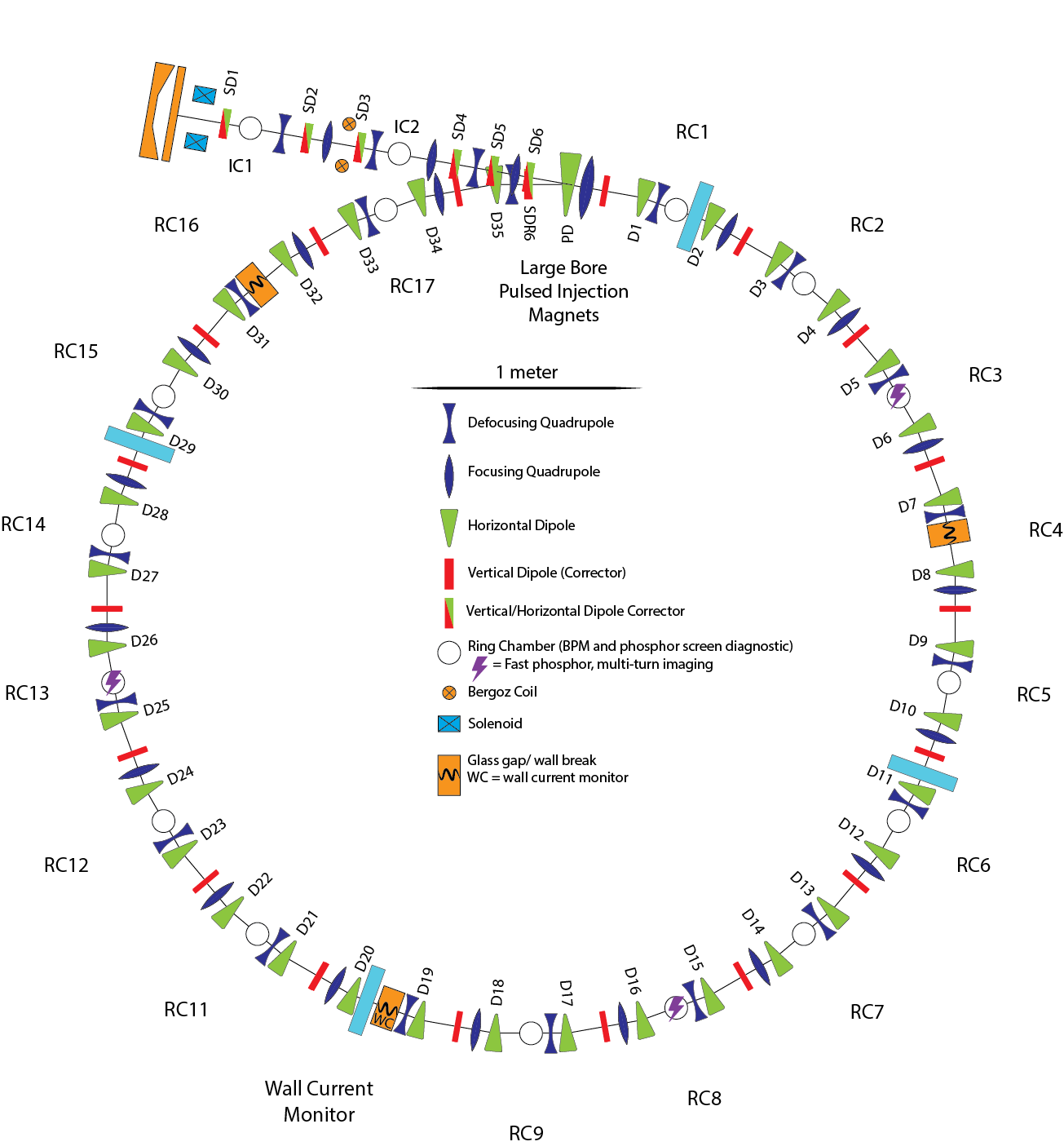}
\caption{Proposed layout for distributed octupole lattice with 4 octupole inserts. }
\label{fig:exp-layout-b}
\end{figure}

For the proposed experiments in nonlinear optics, the existing UMER configuration (described in Chapter \ref{ch:apparatus}) will be modified to include octupole elements. This requires modifying the linear focusing optics to accommodate the octupole inserts and meet the conditions for quasi-integrability. As presented in Chapter \ref{ch:theory}, for particle motion with one conserved invariant the conditions on the linear focusing are: (1) the beam envelope solution must be $XY$ symmetric inside octupole inserts, $\beta_x(s) = \beta_y(s)$, and (2) the phase advance between inserts must be equal to $n \pi$ for integer $n$. Two different philosophies for lattice design are identified. 

The first design, referred to as the single channel octupole lattice, is a non-FODO lattice solution that accommodates a single long octupole insertion (relative to other beam elements). This approach is similar to the IOTA ring design \cite{Antipov2017}, which includes a long nonlinear insert element. In UMER, the long channel will occupy a custom-designed mount in a single $20^{\circ}$ ring section, while the rest of the mechanical ring structure will be undisturbed. A diagram of the mechanical configuration is shown in Fig. \ref{fig:exp-layout-a}. To maintain the beam profile $\beta_x = \beta_y$ through the octupole channel, the linear ring optics will be modified to create a symmetric beam waist at the center of the channel. The basic behavior of this long octupole channel were explored in a reduced model in Chapter \ref{ch:qio}, and tune spreads up to $~\sim 0.25$ are expected. The required linear focusing optics to support quasi-integrable motion is much different from the standard UMER FODO lattice and designing the optics is a complex task. Linear lattice solutions are discussed in Chapter \ref{ch:lattice}. 

The second approach, the distributed octupole lattice, takes advantage of the UMER ``alternative lattice," which utilizes half of the available quadrupoles in a FODO arrangement with periodicity 18 (FODO cell length 64 cm).\cite{Bernal2012} In this arrangement, short octupole insertions are placed at the mid-point of the FODO cell, over symmetry points in the periodic envelope solution where $\beta_x \approx \beta_y$. The hardware layout is shown in Fig. \ref{fig:exp-layout-b}. As the tune of the alternative lattice is adjustable in the approximate range 2-4, a design with four short octupole elements (the ``N4 lattice") was identified as a candidate for testing the quasi-integrable theory. For this lattice, focusing and defocusing quad strengths can be adjusted so that the phase advance between elements is $2\pi$. Simulations and experimental measurements on the distributed octupole lattice are presented in Chapter \ref{ch:distr}. The predicted tune spreads are $\sim0.07$. Although easier to implement in terms of lattice design, the outlook for observable effects is not as promising when compared to the single-channel design.

\section{Printed circuit board octupoles} \label{sec:design:octu}

\begin{figure}[!tb]
\centering
\subfigure[Half of a UMER printed circuit octupole magnet]{
   \centering
   \includegraphics[width=0.8\textwidth]{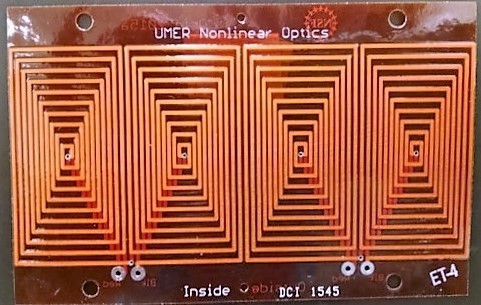}
   \label{fig:octupcb}
   } 
\subfigure[FFT of octupole fields from rotating coil measurement.]{
   \centering
   \includegraphics[width=\textwidth]{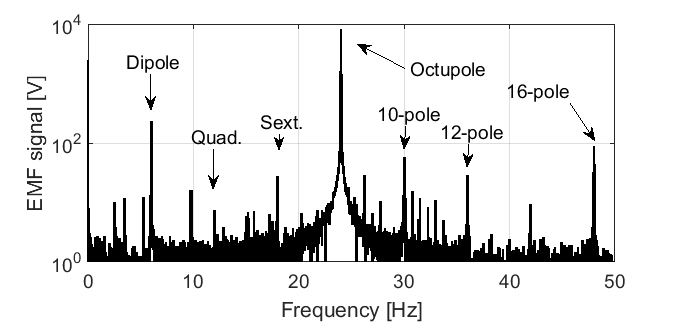}
   \label{fig:octucoil}
   }
\caption{1st generation octupole PCB designed for nonlinear optics experiments and measured field harmonics. }
\end{figure}

Printed circuit  board (PCB) octupole magnets were designed for use in the UMER nonlinear optics experiments.
The approach to octupole PCB design follows the same philosophy used in the design of the UMER quads and dipoles. \cite{Godlove2004}, \cite{Zhang2000} 
The guiding design principle was that the axially integrated longitudinal component of the current density $\vec{K}$ on the cylindrical surface should have an azimuthal dependence $\int K_z dz = \cos{n\theta}$ for desired multipole order n.
For a discrete number of conductors with longitudinal length $z_i$, the conductor length must be related to azimuthal position according to the relationship
\begin{equation}
| \sin{n\theta}| = 1 -\left( \frac{2z_i}{al}\right)^2
\end{equation}
where $l$ is the longitudinal magnet length and $a$ is an adjustable parameter near one. 
 
A set of printed circuit octupoles has been manufactured and characterized.\cite{Matthew2015} The PC circuits, pictured in Fig \ref{fig:octupcb}, are made in two double-layered halves, which fit inside the standard UMER quadrupole mount. Based on the similarity to existing UMER PC quadrupoles and dipoles, each magnet should easily be able to sustain 3 A DC with the existing mounts and up to 10 A with the addition of water cooling. Solutions generated with commercial code Maxwell 3D predict peak field gradient per amp as $G_3/I = 66.5\ T/m^3/$A with the 16-pole as the next significant, unwanted multipole.\cite{Matthew2015} 
Biot-Savart integration done with the in-house code MAGLI \cite{magli} predicts $49\ T/m^3/$A.

Gauss probe measurements of the assembled octupole magnet confirm $G_3/I = 51.6 \pm 1.5\ T/m^3/A$.\cite{BaumgartnerNAPAC2016}
The magnet was also characterized using an integrated-field rotating coil measurement. 
A radial coil, with longitudinal length exceeding the magnet length, rotates at 6 Hz, generating an induced current in the coil. The method is described in \cite{Zhang2000}. 
The FFT of the rotation coil measurement is shown in Fig. \ref{fig:octucoil}. The primary peak is the octupole component. The next largest contribution is the dipole. Despite $\mu$-metal shielding of the measurement, some component of the background Earth field is included in the measurement and contributes to the dipole component.
The next largest harmonic impurity is the 16-pole, as predicted in simulation.
The harmonic purity of the octupole element is within the $<1\%$ tolerance met by the quadrupole and dipole elements.

\section{Design of octupole channel}
\label{sec:design:channel}

\begin{figure}[tb]
\centering
\includegraphics[width=\textwidth]{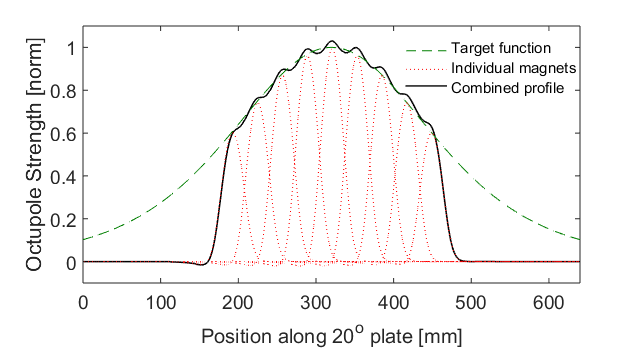}   
\caption{Composite octupole channel made of over-lapping, evenly-spaced discrete short PCB octupoles. }
\label{fig:octchannel}
\end{figure}

This section describes the design of a composite element for the single-channel experiments. A requirement for quasi-integrability is that the octupole potential is constant in the normalized frame.\footnote{See Section \ref{sec:theory:norm-coord}} In the lab frame, the strength of the octupole fields must scale as $G_3(s) \propto \beta(s)^{-3}$. 
For all UMER PCB magnets, which have a short aspect ratio and are fringe-field dominated, the effective magnet length is shorter than the physical length. To obtain a smoothly-varying octupole profile $G_3(s)$, short octupole circuits are placed to overlap significantly, as illustrated in Fig. \ref{fig:octchannel}. This is mechanically feasible up to two layers, as the printed circuits are very thin. 
For two layers of PCBs of length $4.65$ cm, the minimum spacing between magnet centers is $2.33$ cm.

Using field solutions from MAGLI, the profile $G_3(s)$ was examined for a 32 cm channel with evenly spaced octupole elements. Residuals were calculated with respect to the desired profile, as plotted in Fig. \ref{fig:octchannel}. 
Only odd numbers were considered to maximize smoothness at the channel center. While seven elements (center-to-center spacing 4.56 cm) was an improvement over five, nine elements (spacing 3.42) did not improve the amplitude of residuals. Seven elements in a 32 cm channel is determined to be the optimal spacing for a composite channel insert.


\section{Generation and detection of a low-current beam} \label{sec:design:lowcurrent}

\begin{figure}[bt]
\begin{center}
\includegraphics[width=\textwidth]{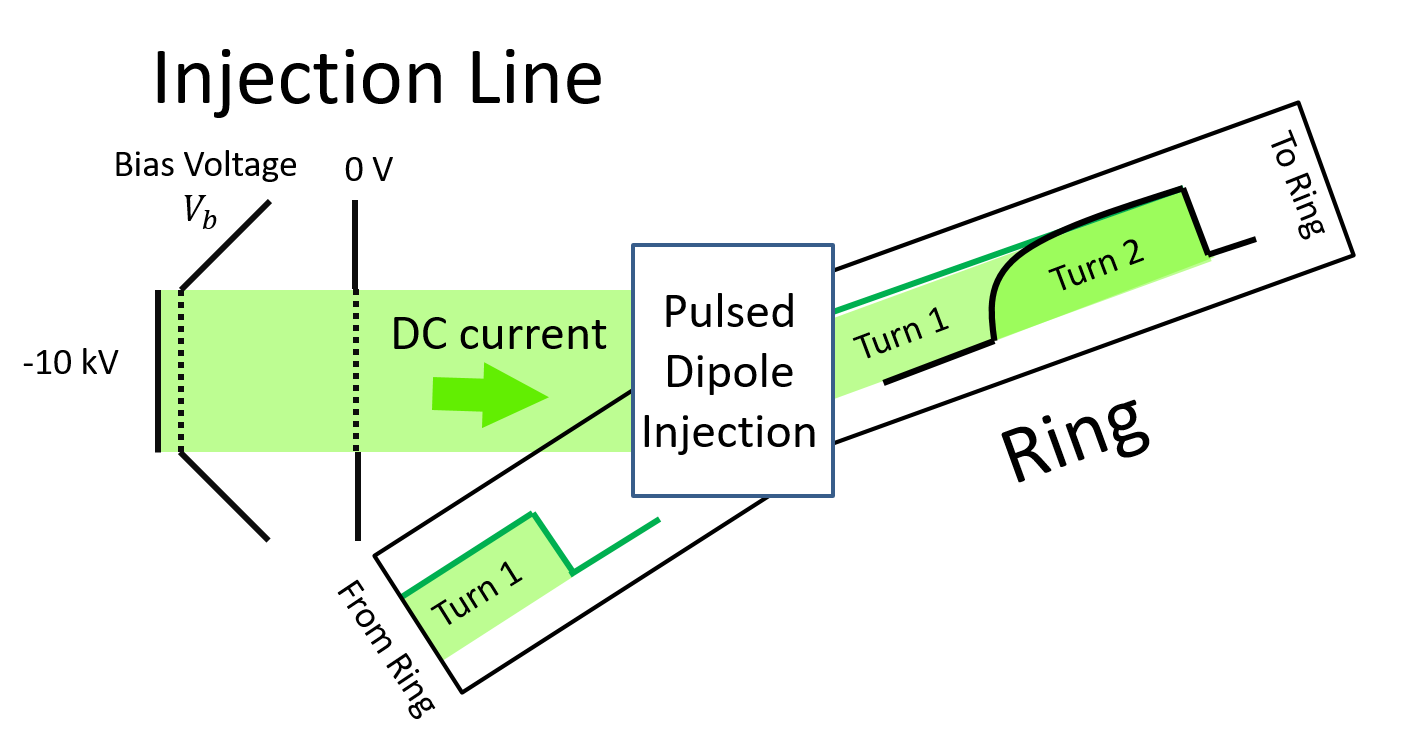}
\end{center}
\renewcommand{\baselinestretch}{1}
\small\normalsize
\begin{quote}
\caption{Generation of variable current DC beam in UMER gun. Cathode-grid bias is lowered until DC current leaks through and pulse formation is done with injection dipole. }
\label{fig:DCbeamcartoon}
\end{quote}
\end{figure} 
\renewcommand{\baselinestretch}{2}
\small\normalsize

For experiments with quasi-integrable optics, it is desirable to start near the zero-current limit where space charge effects are minimal. This will allow observation of the effects of octupole-induced tune spreads without complication by space charge spreads. The least intense apertured UMER beam, 0.6 mA, still has a very large tune shift, $\delta \nu = 0.94$. Predicted octupole-induced spreads will be, at most, 0.25 (this will be discussed in Chapter \ref{ch:qio}).
Generating and detecting a ``low space charge" beam is not straightforward, as UMER is designed as a high-intensity machine. Operating with minimal space charge requires low beam current, but not so low that beam current is undetectable. Space charge tune shift is also inversely proportional to beam emittance, so increasing emittance further reduces the space charge effect.

A low current, high emittance beam can be produced by running the triode electron gun in voltage amplification mode (rather than power amplification). 
This has been demonstrated by turning off the cathode grid pulse and reducing the positive bias to $\sim4$ V, which allows $\mu A$-level leakage current. This mode of operation is referred to as the ``DC beam" due to the method of operation. The positively-biased gun could be operated in pulsed mode, but any ripple on the cathode pulse will be amplified in the longitudinal bunch structure. For now, longitudinal structure is created through the pulsed magnetic injection. The time dependence of the injector dipole PD only allows a $\sim150$ ns slice of the DC current into the ring at a 60 Hz rate. The injection schematic is illustrated in Fig. \ref{fig:DCbeamcartoon}.

\begin{figure}[bt]
\centering
\includegraphics[width=\textwidth]{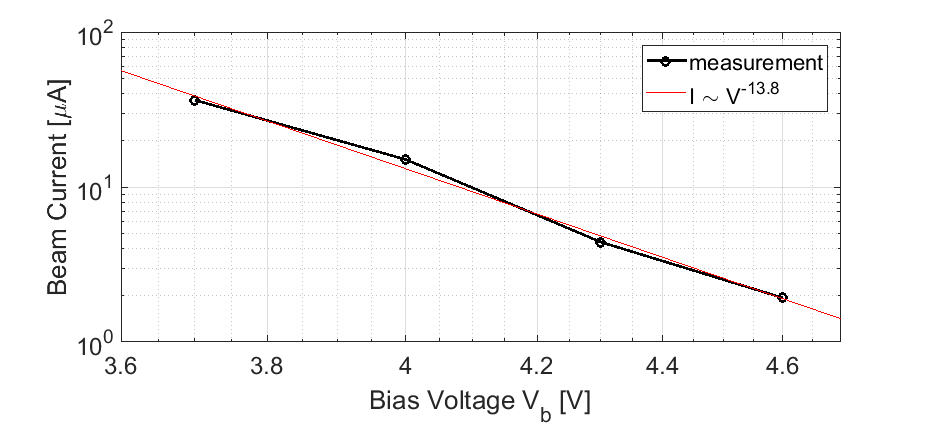}
\caption{Measured output of electron gun in DC mode as a function of cathode-grid bias voltage. Current is measured at the WCM. }
\label{fig:DC-IV-curve}
\end{figure}

\begin{figure}[!tb]
\centering
\includegraphics[width=\textwidth]{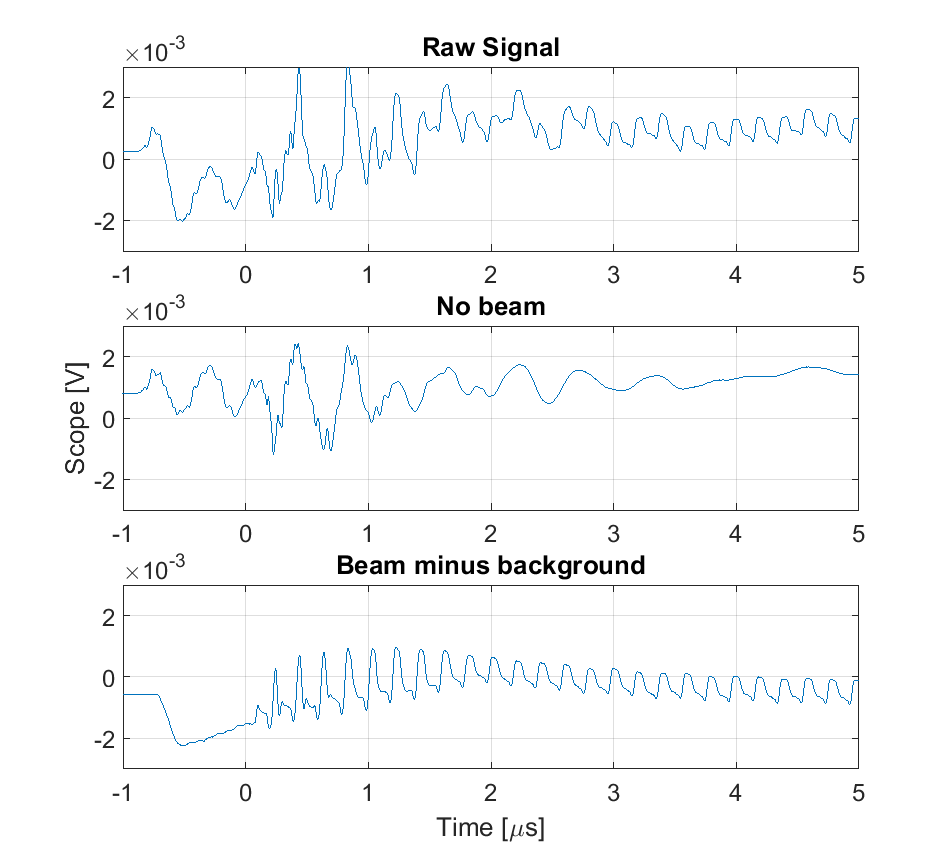}
\caption{First 10 turns of DC beam signal on WCM. (a) Raw WCM signal, (b) background signal and (c) background-subtracted signal for DC beam. }
\label{fig:microA-WCM-trace}
\end{figure}

\begin{figure}[!tb]
\centering
\includegraphics[width=0.8\textwidth]{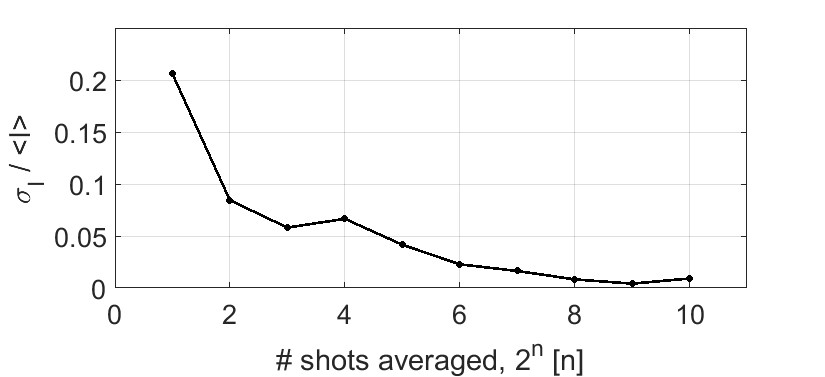}
\caption{Error in beam current measurement as a function of number of shots averaged. Standard deviation is calculated for 20 measurements and averaged over 500 turns. }
\label{fig:microA-stderror}
\end{figure}

The current density depends very sensitively on the bias voltage. A 3.70 V bias produces a $40\ \mu$A beam. The measured current-voltage relationship is plotted in Fig. \ref{fig:DC-IV-curve}.
The transverse edge emittance (4 RMS, unnormalized) was measured with the quadrupole-scan technique for an output current of $40\ \mu$A. 
Measured values are $\epsilon_x = 100\pm 20\ \mu$m, $\epsilon_y = 300\pm 20\ \mu$m.\cite{Bernal2017} 
The reason for the large asymmetry is unknown, although the large transverse beam size complicated the measurement.
From Equation \ref{eq:tune-spread}, for I = $60\ \mu$A and $\epsilon=100\ \mu$m, the predicted space charge tune shift (for the equivalent KV distributions) is $\delta \nu = 0.005$.

The time-resolved beam current signal, picked up by the wall current monitor (described below in Section \ref{sec:apparatus:diag}), is plotted in Fig. \ref{fig:microA-WCM-trace}. This signal is shown for an initially $\sim 40\ \mu$A beam. As the DC beam current is an order of magnitude below the lowest current apertured beam (which already has poor signal to noise ratio), a +20 dB amplifier was used to boost the low-level signal. 

The background subtracted trace (bottom plot) clearly shows the effect of the injection-gated pulse formation. Approximately $1 \mu s$ after the injection dipole is switched to allow beam into the ring, the pulsed dipole switches polarity and only a single $\sim 150$ ns temporal slice of the DC current is recirculated. An artifact of this switching is an increase in current from bunch head to bunch tail that reflects the PD pulse shape.

The UMER environment is noisy, particularly near injection. The magnitude of the pulsed circuit noise picked up on the wall current monitor is comparable to the beam signal. In addition to this ``frozen-in" noise at 60 Hz repetition, which is background subtracted, there are also statistical fluctuations due to noise at other frequencies. The signal to noise ratio can be improved by taking longer averages on the scope. Fig. \ref{fig:microA-stderror} shows the uncertainty in beam current measurement due to statistical fluctuations as a function of the number of averaged waveforms. While low errors ($<1\%$) can be achieved by averaging over at least 256 bunches, this becomes time consuming when collecting many data points. Characterizing measurements discussed in Chapter \ref{ch:res} were averaged over 16 waveforms, which results in a statistical error of $\sigma_I / I \sim 7\%$. 

One concern while operating with the DC beam is increased average beam current drawn from the cathode and additional heat/current load due to current intercepted by the cathode grid.
In typical UMER operation, 100 mA is drawn from the cathode when the bias voltage is negative. For a 100 ns pulse length that fires at 60 Hz, the duty factor is $6 \times 10^{-6}$. Therefore the average power drawn from the gun is 6 mW. However, in DC mode, for a $40\ \mu$A beam, the average power is 400 mW, nearly $70\times$ above the design value. 

The UMER gun has been operated very sparingly in the DC mode except for two lengthy data collection cycles ($\geq 12$ hours) run one month apart. The data was collected to characterize DC beam transmission and resonant behavior over a range of tunes, shown in Chapter \ref{ch:res}. Shortly afterwards, the cathode was observed to fail and was replaced. The cathode was near the end of its ten year life span and it is not clear to what extent the extended DC operation of the gun may have hastened its demise. During the experimental run, a slow decrease in gun output current was observed (also described in Chapter \ref{ch:res}). This may be related to increased power load across the 10 kV stand-off or warping of the cathode geometry under heat. Pulsing of the bias voltage is recommended to reduce average power for cathode safety, possible using a long pulse ($>150$ ns) to reduce ripple being amplified on the bunch structure.

\section{Error analysis using simple model of octupole lattice} \label{sec:design:error}

In this section the reduced model of the quasi-integrable octupole lattice, introduced in Chapter \ref{ch:qio}, is used to study the effect of lattice errors. Performance is quantified in terms of dynamic aperture and tune spread. The simple model is comprised of two elements: a thin lens FOFO lattice of cell length $L=64$ cm and a long octupole insertion with longitudinal strength profile $G_3(s) \propto \beta(s)^{-3}$. More description of this model is given in Chapter \ref{ch:qio} and Appendix \ref{ap:simple-model}. Properties of the FOFO lattice, including  matched solution $\beta(s)$, are described in Appendix \ref{ap:fofo}. 

\subsection{Sensitivity to closed orbit distortion} \label{sec:design:steering}

The effect of background fields on the low-rigidity UMER beam is significant.  The vertical field is roughly constant along the ring with an average strength of $400$ mG. This gives approximately $2.2^o$ of bend per horizontal dipole. The radial field has a sinusoidal dependence on $s$ with amplitude $\sim 200$ mG, for maximum bend angle of $2.2^o$ per vertical corrector. Due to these background fields, the beam orbit trajectory traces arcs between corrector magnets. 
For a long element like the octupole insert, the orbit will not be well-centered. Correction is possible through shielding or field-canceling Helmholtz coils.  
It is therefore necessary to define alignment tolerances for acceptable closed orbit distortions from magnetic centers. This effect is discussed in more detail in Chapter \ref{ch:steering}.

To quantify lattice performance, frequency map analysis (FMA) is applied to measure dynamic aperture and octupole-induced tune spread. All simulations are performed in a zero-current limit. This study considered distortions of the closed orbit but neglects oscillations about the closed orbit. This discussion also does not extend to  consider misalignment of octupole elements in the channel. A gross mis-alignment of the octupole element can be considered equivalent to an orbit distortion, while misalignments between individual circuits in the long octupole channel requires a separate treatment.

\begin{figure}[tb]
\centering
\subfigure[Case 1]{
\includegraphics[width=\textwidth]{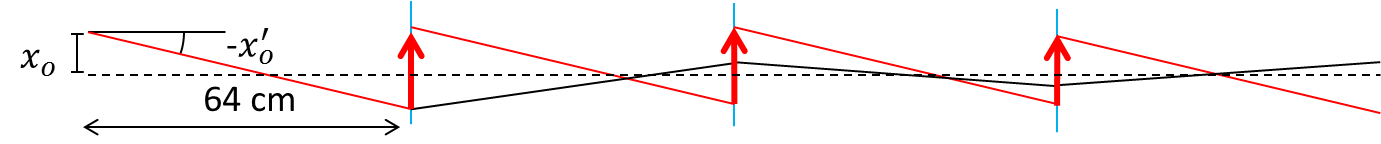}
\label{fig:straightorbitdistortion}
}
\subfigure[Case 2]{
\includegraphics[width=\textwidth]{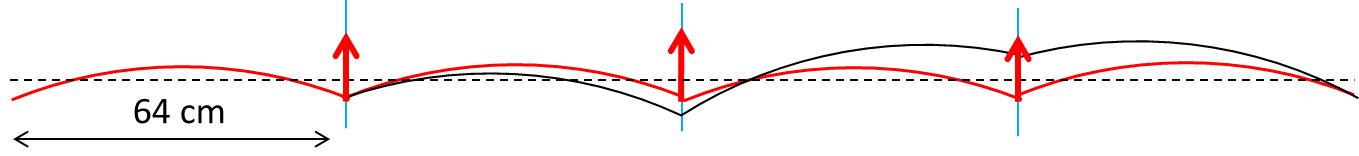}
\label{fig:vertcurvedorbitdistortion}
}
\caption{
Single particle trajectories in the two orbit distortion cases considered. 
Red line is centroid motion of a particle with applied thin lens centroid transformation (red arrows). Black line is centroid motion without appropriate transformation. }
\end{figure}

Tolerance simulations use the WARP PIC model of an $L= 64$ cm octupole channel in an ideal linear FOFO thin-lens transformation of $k=2.92\ m^{-1}$ with $\beta_*=0.3417$ m. Octupole fields are set at $G_{3,max}=50\ T/m^3$. Dynamic aperture is calculated over 1024 turns. In these simulation, a constant closed orbit distortion term is included in the thin lens transformation for each particle. Two cases are examined:

\begin{enumerate}
\item Orbit distortion in otherwise shielded 64 cm section (centroid has straight trajectory between steering elements), depicted in \ref{fig:straightorbitdistortion}.
\item Curved orbit distortion due to constant background field, depicted in Fig. \ref{fig:vertcurvedorbitdistortion}.
\end{enumerate}

\subsubsection{Case 1: Straight/shielded closed orbit errors}

\begin{figure}[tb]
\centering
\includegraphics[width=0.8 \textwidth]{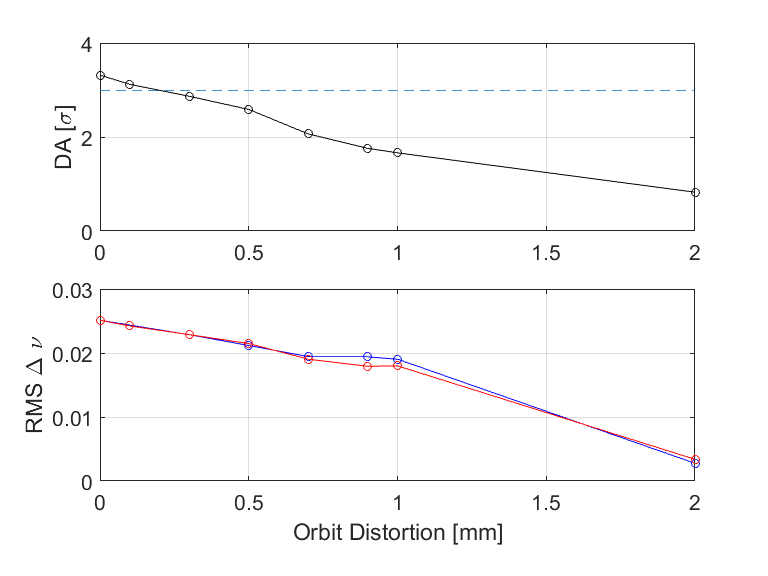}
\caption{Dependence of dynamic aperture and tune spread on orbit distortion. On dynamic aperture (DA) plot, horizontal dashed line indicates $90\%$ of ideal aperture. }
\label{fig:DAvsorbitdistort}
\end{figure}

Fig. \ref{fig:straightorbitdistortion} shows the distorted closed orbit in the case where particle trajectories are straight between magnetic elements. The distortion is defined by initial conditions $x_0$ and $x'_0$, which represent the initial offset in beam centroid. At the same location as the FOFO thin lens transformation (that represents focusing in the linear part of the ring), a centroid transformation is also made. Details on centroid transformation are covered in Appendix \ref{ap:simple-steering}.

Fig. \ref{fig:DAvsorbitdistort} shows the dynamic aperture and tune spread dependence on orbit distortion $x_0$. Here $x_0'=L/2x_0$. There is quite a stringent requirement on distortion, compared to usual UMER operations, with $x_0 < 0.2$ mm desired for less than 10\% loss of dynamic aperture. At $x_0 = 0.2$ mm, decrease of RMS tune spread from ideal case of 0.025 is $\sim 6\%$.

\subsubsection{Case 2: Orbit distortion due to background field}

To model the effect of orbit distortion caused by immersion in ambient background fields, a uniform background field is added to the model. Fig. \ref{fig:vertcurvedorbitdistortion} depicts the path of single particles in the immersed field, with appropriate thin-lens centroid transformation (see Appendix \ref{ap:simple-steering}).

\begin{figure}[tb]
\centering
\includegraphics[width=0.8\textwidth]{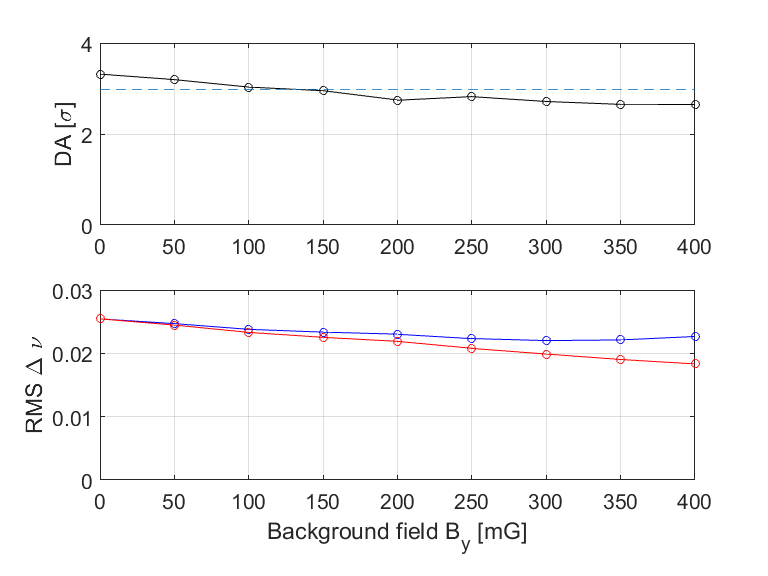}
\caption{Dependence of dynamic aperture and tune spread on vertical background field. On dynamic aperture (DA) plot, horizontal dashed line indicates $90\%$ of ideal aperture. }
\label{fig:DAvsBGfield}
\end{figure}

\begin{figure}[!tb]
\centering
\subfigure[Dynamic aperture contours for beam immersed in vertical field]{
	\centering
	\includegraphics[width=0.6\textwidth]{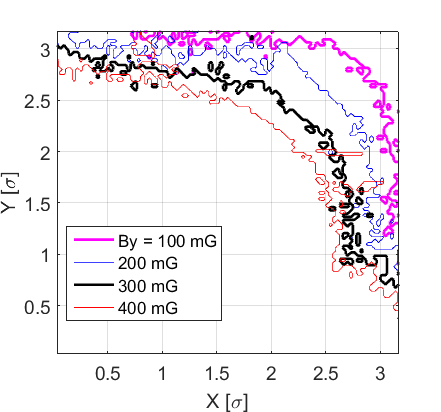}
	\label{fig:FMADA1}}
\hfill
\subfigure[Dynamic aperture contours for beam immersed in vertical and horizontal fields.]{
	\centering
	\includegraphics[width=0.6\textwidth]{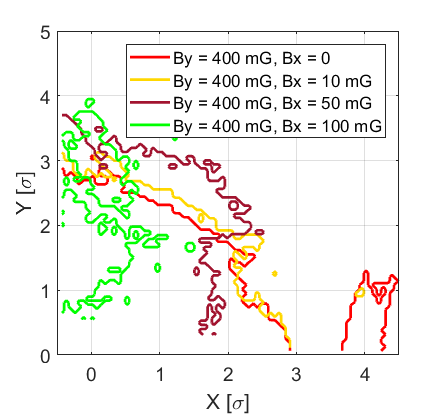}
	\label{fig:FMADA2}}
\caption{Shape of dynamic aperture for steering errors introduced by ambient fields. In all cases, $G_{3,max}=50\ T/m^3$. }
\label{fig:FMADA}
\end{figure}

As seen in Fig. \ref{fig:DAvsBGfield} and Fig. \ref{fig:FMADA1}, the presence of a vertical background field at 400 mG  incurs a $20\%$ loss of radial dynamic aperture when compared to the $0$ mG case. The addition of horizontal/radial field, shown in Fig. \ref{fig:FMADA2}, causes severe loss of stability. No particles are stable when $B_y=400$ mG, $B_x=200$ mG.

For reasonable dynamic aperture, the single-channel experiment requires control of orbit distortions inside the octupole insert to within $0.2$ mm. The $\sim 400$ mG vertical background field should be shielded or compensated to $< 100$  mG. Compensation for the horizontal background field, with peak $\sim 200$ mG should also be made.
Measurement of the orbit control in the UMER lattice is discussed in Chapter \ref{ch:steering}. The octupole insert in the single-channel experiment can be placed in a region of low measured distortion, such as RC9.

\clearpage
\subsection{Sensitivity to phase errors in external focusing} \label{sec:design:phase-error}

A quasi-integrable solution for the single-channel octupole lattice has been shown to exist when the phase advance in the linear portion of the lattice (between octupole inserts) is an integer multiple of $\pi$. In reality, the phase advance of any given particle will be $n\pi + \delta$, with $\delta$ representing errors in external focusing. $\delta$ could also be the result of tune depression (due to space charge) or chromatic tune shift (due to energy spread). The simple model is used to examine dynamic aperture and tune spread dependence on phase error. The phase error is introduced in the thin lens focusing transformation through use of a more general expression for the ``T-insert" matrix. For tune error $\Delta \nu_z$,

\begin{equation}
T_z = 
\begin{bmatrix} 
\cos{2\pi\Delta\nu_z} - \alpha_z \sin{2\pi\Delta\nu_z} & \beta_z \sin{2\pi\Delta\nu_z} \\
-\frac{1-\alpha_z^2}{\beta_z}\sin{2\pi\Delta\nu_z}-\frac{2\alpha_z}{\beta_z}\cos{2\pi\Delta\nu_z} & \cos{2\pi\Delta\nu_z}-\alpha_z\sin{2\pi\Delta\nu_z}\\
\end{bmatrix}
\label{eq:tune-error-matrix}
\end{equation}

Origin of Eq. \ref{eq:tune-error-matrix} is shown in Appendix \ref{ap:matrix}.
Simulations with phase error were carried out in Elegant, for an octupole lattice of length $L=64$ cm. Tracking was done for 1024 passes, set up as described above with peak octupole gradient $G_{3,\max} = 50\ T/m^3$. For these simulations, external focusing strength is $k=3.3264\ m^{-1}$ for lattice tune advance $0.260$ and waist size $\beta_* = 0.3$ m.

\begin{figure}[tb]
\centering
\includegraphics[width=0.7\textwidth]{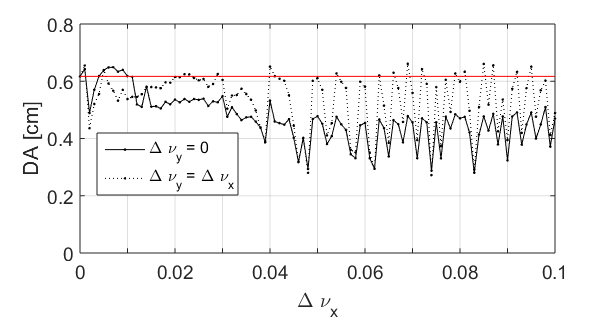}
\caption{Dependence of dynamic aperture on tune error, for equal (dashed line) and unequal (solid line) errors. Value for ideal case ($\Delta \nu_x = \Delta \nu_y$ = 0) is indicated by red line. }
\label{fig:DA-vs-phase-error}
\end{figure}

\begin{figure}[!htb] \centering
\subfigure[Max. tune spread for unequal phase error.]{\includegraphics[width=0.45\textwidth]{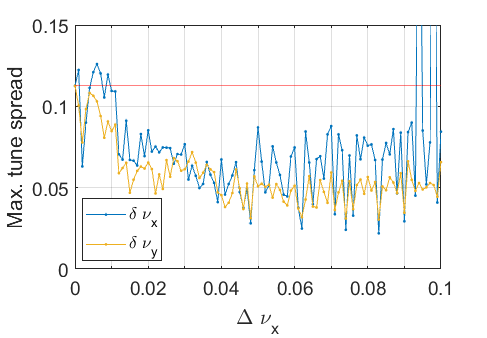}}
\subfigure[Max. tune spread for equal phase error.]{
	\includegraphics[width=0.45\textwidth]{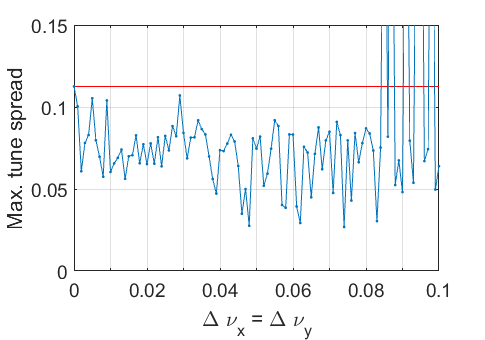}}
\subfigure[RMS tune spread for unequal phase error.]{
	\includegraphics[width=0.45\textwidth]{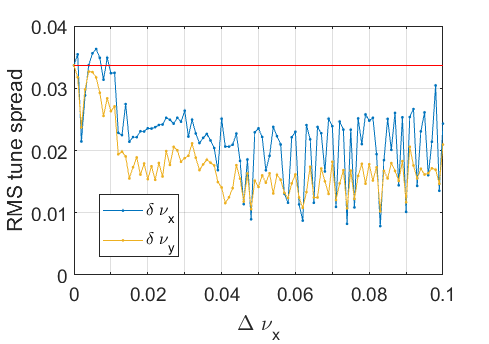}}
\subfigure[RMS tune spread for equal phase error.]{
	\includegraphics[width=0.45\textwidth]{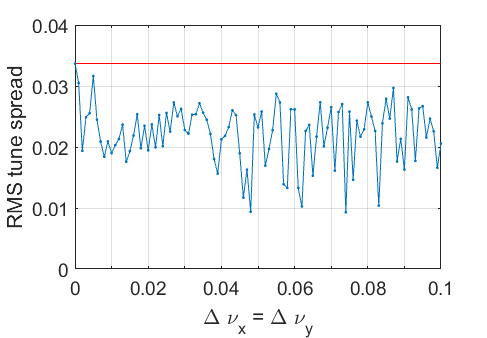}}
\caption{Dependence of tune spread on phase error, for both equal (top row) and unequal (bottom row) tune errors. Value for ideal case ($\Delta \nu_x = \Delta \nu_y$ = 0) is indicated by red line. } 
\label{fig:tune-vs-phase-error}
\end{figure}

\begin{figure}[tb] \centering
\subfigure[Subset of particles at r=0.18 cm]{
	\includegraphics[width=0.35\textwidth]{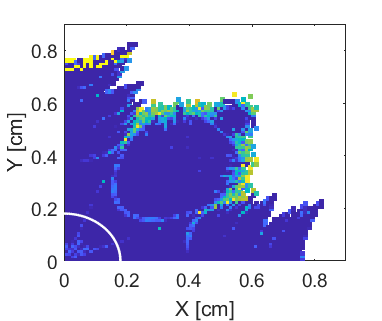}\label{fig:H-pts}}
\subfigure[$\sigma H / <H>$ versus tune error]{
	\includegraphics[width=0.55\textwidth]{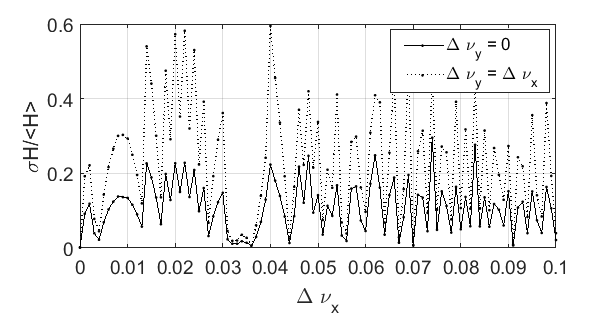}\label{fig:H-vs-tune}}
\caption{Standard deviation of invariant H over 1024 passes in simple octupole lattice. $\sigma H / <H>$ in \ref{fig:H-vs-tune} is averaged over subset of particles in \ref{fig:H-pts}. } 
\label{fig:H-vs-phase-error}
\end{figure}

\begin{figure}[tb] \centering
	\includegraphics[width=0.7\textwidth]{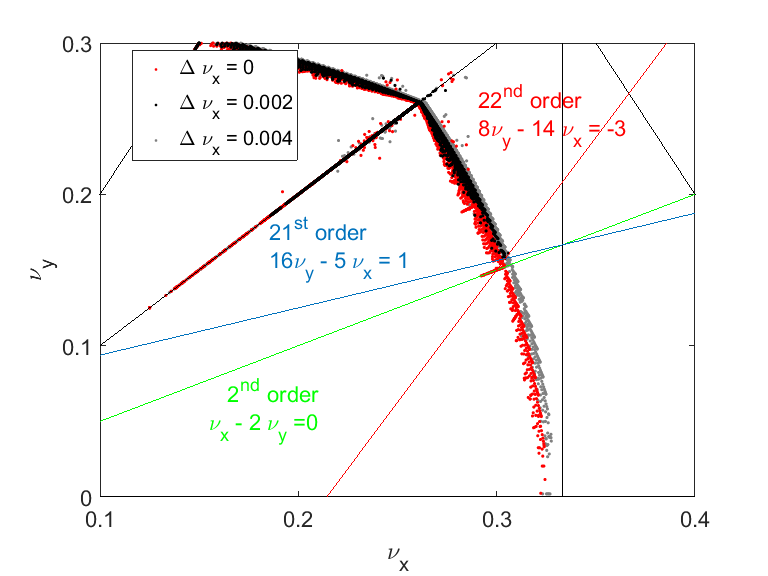}
\caption{Tune footprint from Elegant calculation with tune errors $\Delta \nu_x = 0, 0.002, 0.004$. } 
\label{fig:bump-at-002}
\end{figure}

Two cases are examined: $\Delta \nu_x = \Delta \nu_y$ (equal tune errors) and $\Delta \nu_x \neq 0, \Delta \nu_y = 0$ (unequal tune errors). 
From Fig. \ref{fig:DA-vs-phase-error}, it is apparent that the dynamic aperture is reduced in the presence of phase errors. In the case of equal tune errors, however, the lattice was relatively forgiving for tune errors $\Delta \nu < 0.03$. The dependence for $\Delta \nu > 0.03$ is very irregular, with alternating bands of near-ideal and reduced aperture. The outlook is slightly worse in the case of unequal errors. A large loss of aperture is seen around $\Delta \nu \sim 0.01$, and likewise beyond $\Delta \nu = 0.03$ the dependence is irregular, but in general much lower than the ``best-case" aperture size. 

The dependence of RMS and maximum tune spread on errors in the bare lattice tune are shown in Fig. \ref{fig:tune-vs-phase-error}. Similar behavior is seen when compared to the dynamic aperture measurement. For unequal tune errors there is a sharp loss in tune spread around $\Delta \nu \sim 0.01$. The dependence on equal tune errors is comparable, with less of a margin for acceptable small tune errors. 

Finally, the invariant $H_N$ (Eq. \ref{eq:H-norm-2}) is shown as a function of tune error in Fig. \ref{fig:H-vs-phase-error}. As $H_N$ is a single particle property (rather than aperture and tune spread, which are sampled in the whole transverse space), $H_N$ conservation is calculated by averaging over a subset of particles indicated in Fig. \ref{fig:H-pts}. Averaging is meant to reduce the chance of interpreting a single resonantly excited particle to mean lack of $H_N$ conservation over the entire space. In general, $H_N$ conservation is very sensitive, with fractional variation up to $0.6$ in the case of equal tune errors.

A common feature in all metrics is the sharp dip at small tune error $\Delta \nu = 0.002$. This appears to be related to resonant structure; Fig. \ref{fig:bump-at-002} shows that the tune footprint for the $\Delta \nu = 0.002$ case is cut-off near the 3rd order resonance line $\nu_x-2\nu_y = 0$. However, the neighboring cases $\Delta \nu = 0$ and $\Delta \nu = 0.004$ are not similarly affected. On closer inspection, there is a crossing of higher order resonance lines $8 \nu_y - 14 \nu_x = -3$ and $16 \nu_y - 5 \nu_x = 1$ that may be allowing particles to escape the stable region. These resonances are very high order, and in general tune spreads in the distribution (due to space charge and chromaticity) tend to damp out the effects of very high order resonances. However, if the cause is more closely related to the third-order resonance, we would expect to observe this effect in experiment.

\begin{figure}[tb] \centering
\subfigure[$\Delta\nu_x = 0.001, \Delta\nu_y=0$]{
	\includegraphics[width=0.45\textwidth]{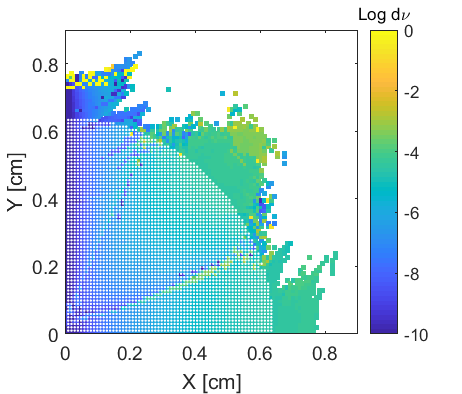}}
\subfigure[$\Delta\nu_x = \Delta\nu_y = 0.001$]{
	\includegraphics[width=0.45\textwidth]{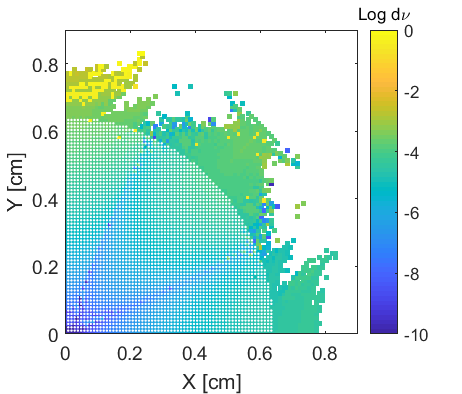}}

\subfigure[$\Delta\nu_x = 0.01, \Delta\nu_y=0$]{
	\includegraphics[width=0.45\textwidth]{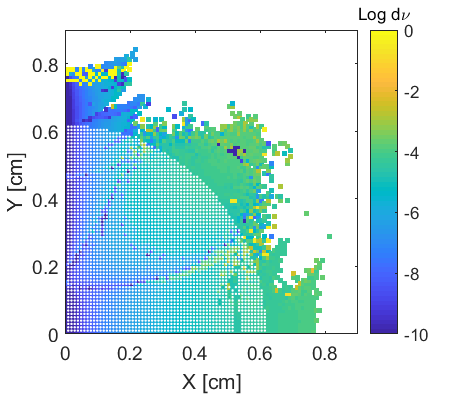}}	
\subfigure[$\Delta\nu_x = \Delta\nu_y = 0.01$]{
	\includegraphics[width=0.45\textwidth]{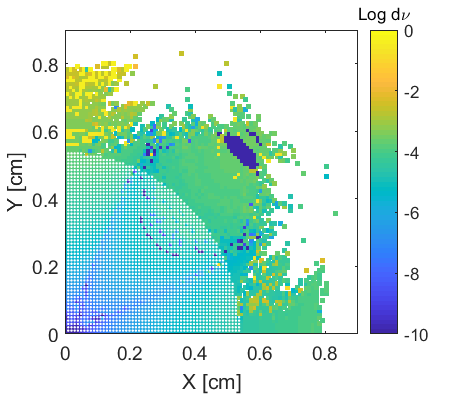}}	

\subfigure[$\Delta\nu_x = 0.1, \Delta\nu_y=0$]{
	\includegraphics[width=0.45\textwidth]{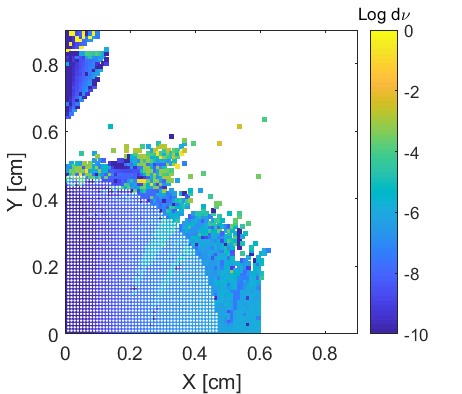}}
\subfigure[$\Delta\nu_x = \Delta\nu_y = 0.1$]{
	\includegraphics[width=0.45\textwidth]{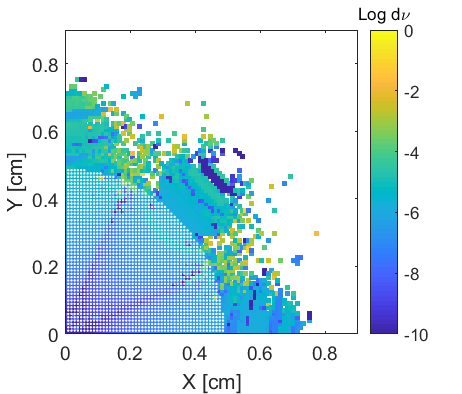}}
\caption{Elegant frequency maps (in configuration space) for 1024 turns for different $\Delta \nu_x$ and $\Delta \nu_y$. } 
\label{fig:phase-error-xy}
\end{figure}

\begin{figure}[tb] \centering
\subfigure[$\Delta\nu_x = 0.001, \Delta\nu_y=0$]{
	\includegraphics[width=0.4\textwidth]{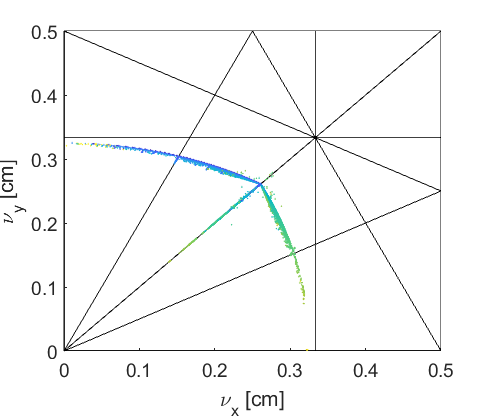}}
\subfigure[$\Delta\nu_x = \Delta\nu_y = 0.001$]{
	\includegraphics[width=0.4\textwidth]{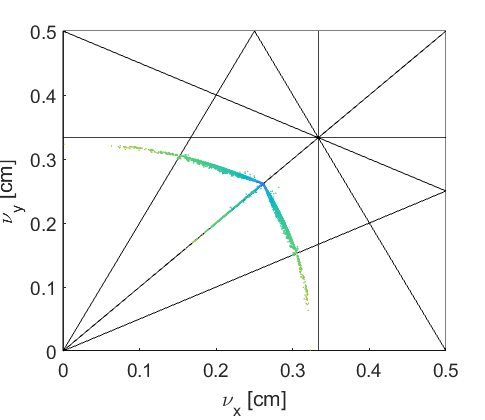}}

\subfigure[$\Delta\nu_x = 0.01, \Delta\nu_y=0$]{
	\includegraphics[width=0.4\textwidth]{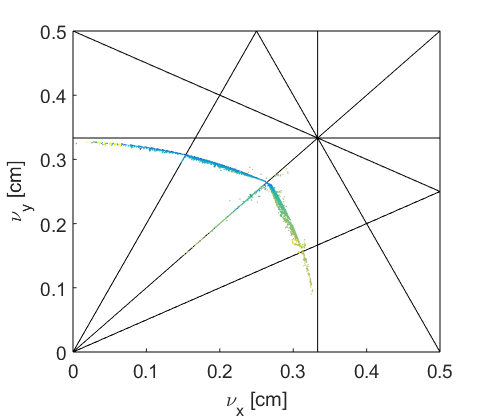}}	
\subfigure[$\Delta\nu_x = \Delta\nu_y = 0.01$]{
	\includegraphics[width=0.4\textwidth]{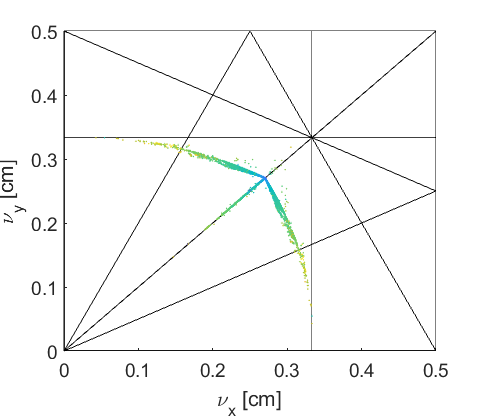}}	

\subfigure[$\Delta\nu_x = 0.1, \Delta\nu_y=0$]{
	\includegraphics[width=0.4\textwidth]{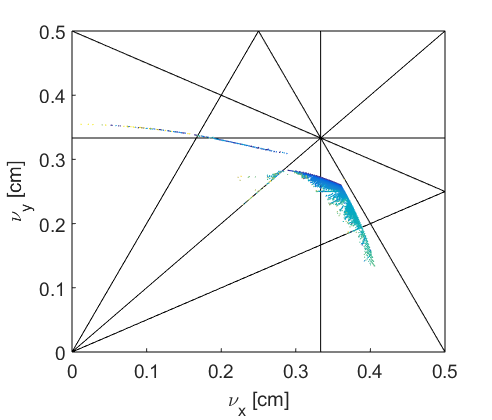}}
\subfigure[$\Delta\nu_x = \Delta\nu_y = 0.1$]{
	\includegraphics[width=0.4\textwidth]{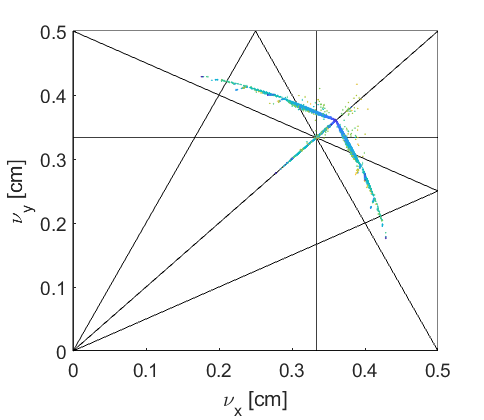}}
\caption{Elegant frequency maps (in tune space) for 1024 turns for different $\Delta \nu_x$ and $\Delta \nu_y$. Color axis is identical to Fig. \ref{fig:phase-error-xy}. } 
\label{fig:phase-error-tune}
\end{figure}

The shape of the dynamic aperture and tune footprints for a range of tune errors can be seen in Fig. \ref{fig:phase-error-xy} and \ref{fig:phase-error-tune} respectively. These cane be compared to the case of no errors ($\Delta \nu_x = \Delta \nu_y$ = 0) in Fig. \ref{fig:ele-fma}.
For the dynamic aperture plots, the largest round stable region is shown in the lightly shaded region. For tune footprints, the entire tune shift imprinted for all configuration space is shown (no cut is made based on the assumption of a round beam). In the case of equal tune errors, the characteristic octupolar shape of the dynamic aperture is preserved although the size of the stable region shrinks. With it, regions of high  nonlinear tune spread are also lost. In the case of unequal tune errors, the shape of the aperture skews to one side or the other.
It appears that the $\nu_x = \nu_y$ coupling resonance is destructive in this case, allowing particles to wander from the stable region and be lost. 

In general the octupole lattice is fairly robust to small ($<0.1$) tune errors. Errors above this value were not considered, as matched lattice solutions have been found in this range. However, in the experiment, the beam optics may be sufficiently different from the model that larger tune errors are possible. This analysis could be extended to see if any structure exists beyond $\Delta \nu = 0.1$. For now, while tune errors should be as low as possible, $\Delta \nu < 0.1$ should result in reasonable aperture and tune spread for experimental purposes. Avoiding differences in tune error $\Delta \nu_x - \Delta \nu_y$, appears to be more important.

\clearpage
\subsection{Effect of octupole profile and fringe fields} \label{sec:design:real-octus}

Up to this point a simplified model of the longitudinally varying octupole potential has been assumed. Details of the WARP and Elegant implementation can be found in Appendix \ref{ap:simple-model}, but both cases assume near-ideal profile $G_3(s)$ and pure octupole fields.
In reality, for the UMER experiments the octupole channel will have to work around the restrictions of space and flexibility of existing optics. A 64 cm, $20^o$ ring section will be replaced, but this ring section must preserve two $10^o$ bending dipoles (BD's) at locations $s={16,48}$ cm. This means there must be breaks in the octupole channel to accommodate the BD mounts. 
As discussed above in Section \ref{sec:design:channel}, the smoothness of the $G_3(s)$ profile from overlapping short PCBs is limited. Additionally, the UMER octupoles contain additional, unwanted multipole terms, due to fringe fields as well as additional harmonics in the PCB surface current distribution. 

\begin{figure}[htb]
\centering
\subfigure[19 octupoles spaced at 3.2 cm center-to-center.]{
\includegraphics[width=0.7\textwidth]{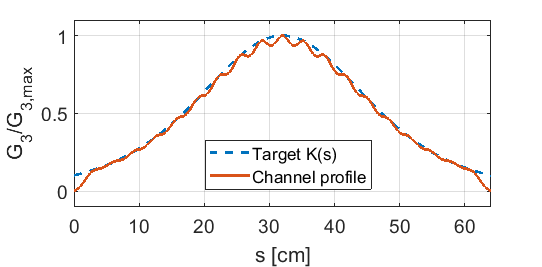}
\label{fig:octu-64cm}}
\subfigure[9 octupoles spaced at 3.2 cm center-to-center, occupying only 32 cm between dipole locations.]{
\includegraphics[width=0.7\textwidth]{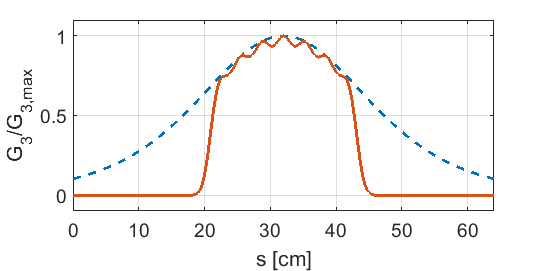}
\label{fig:octu-32cm}}
\subfigure[Same as (b), with additional (single) octupoles in existing dipole and quad mounts.]{
\includegraphics[width=0.7\textwidth]{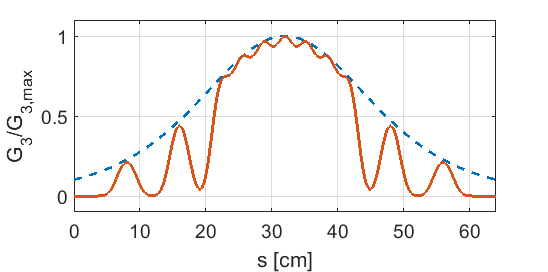}
\label{fig:octu-hybrid}}
\caption{Various arrangements of over-lapping short octupole circuits in 64 cm channel. Only cases (b) and (c) are compatible with UMER's configuration. }
\label{fig:octu-bgrd-channel}
\end{figure}

This section examines the effect of realistic magnetic fields for three three configurations shown in Fig. \ref{fig:octu-bgrd-channel}. The WARP simple model is used with gridded field elements (BGRD) generated from Biot-Savart solutions of the PCB windings in MAGLI \cite{magli}. Simulation parameters were $k=3.3264\ m^{-1}$ and $\beta_* = 0.3$ m. Peak octupole strength is fixed at $G_{3,max}=50\ T/m^3$.

\clearpage

\begin{figure}[!htb]
\centering
\subfigure[BGRD octupoles, case(a)]{
\includegraphics[width=0.4\textwidth]{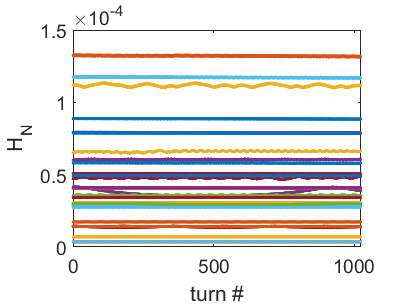}
}
\subfigure[BGRD octupoles, case(b)]{
\includegraphics[width=0.4\textwidth]{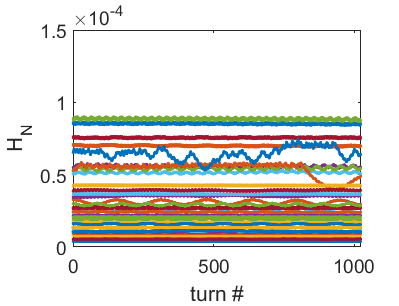}
}
\subfigure[BGRD octupoles, case(c)]{
\includegraphics[width=0.4\textwidth]{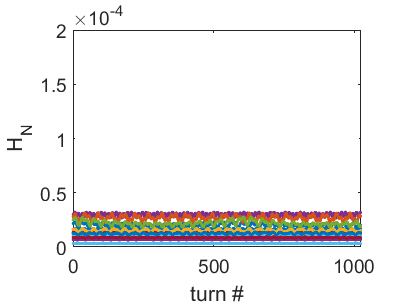}
}
\subfigure[BGRD octupoles, case(b)$^\dagger$]{
\includegraphics[width=0.4\textwidth]{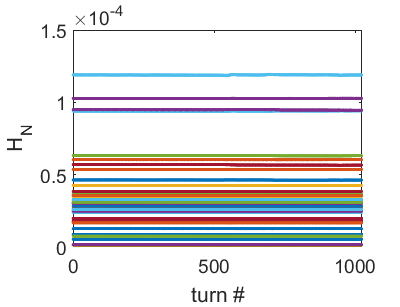}
}
\caption{Evolution of $H_N$ for a subset of particles in a WARP simulation of the octupole channel. Colors indicate unique orbits. }
\label{fig:Hcons-warp-bgrd}
\end{figure}

Case (a) has 19 octupoles evenly spaced at intervals of 3.2 cm center-to-center. This is unrealistic for the experiment, as there is no accommodation for the dipoles, but serves as an intermediate step between ideal ``flat-top" elements and a more realistic $G_3(s)$ with unwanted multipole components. Case (b) has 9 octupoles evenly spaced at 3.2 cm center-to-center, occupying only the center of the channel. Case (c) is identical to case (b), with octupole circuits added in existing mounts at the dipole locations $s={16,48}$ cm and vacated quad locations $s={8,56}$ cm. As this analysis is done with the simple model, dipole fields are not included and therefore neither is the fact that the beam takes a curved path through that elements at $s={16,48}$ cm even in the case of perfect steering without background fields.

\begin{figure}[!tb]
\centering
\subfigure[Dynamic aperture contours]{
\includegraphics[width=0.6\textwidth]{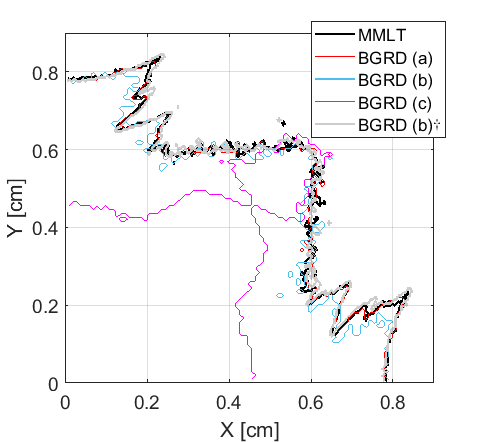}
}
\subfigure[Tune footprints]{
\includegraphics[width=0.6\textwidth]{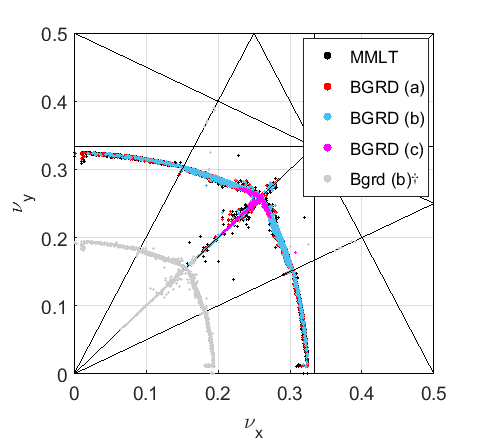}
}
\caption{Dynamic aperture and tune footprints for different octupole field models and configurations.$\dagger$ Results for short 32 cm channel. }
\label{fig:FMA-warp-bgrd}
\end{figure}

\begin{table}
\centering
\begin{threeparttable}
\caption{Figures of merit for different octupole configuration in simple model of quasi-integrable lattice. Metrics are taken for particle distribution inside the largest circular dynamic aperture. }
\label{tab:warp-bgrd}
\vspace{10pt}
\begin{tabularx}{\textwidth}{X L{2cm} L{2cm} L{2cm} L{2cm}}
\hline
Case & DA [cm] & RMS $\Delta \nu$ & $\max \Delta \nu$ & $\overline{\frac{\sigma_H}{\langle H_N \rangle}}$ \\
\hline
MMLT     & 0.62 & 0.033 & 0.108 & 0.005 \\
BGRD (a) & 0.62 & 0.033 & 0.108 & 0.005 \\
BGRD (b) & 0.55 & 0.024 & 0.077 & 0.015 \\
BGRD (b) $^\dagger$& 0.62 & 0.020 & 0.067 & 0.001 \\ 
BGRD (c) & 0.44 & 0.008 & 0.025 & 0.045 \\
\hline
\end{tabularx}
\begin{tablenotes}
\footnotesize
\item $\dagger$ Results for short 32 cm channel.
\end{tablenotes}
\end{threeparttable}
\end{table}

Results for $H_N$ invariance are plotted in Fig. \ref{fig:Hcons-warp-bgrd}. Results from FMA are compared in Fig. \ref{fig:FMA-warp-bgrd}. Values for $H_N$ variation, dynamic aperture limit and tune spreads are given in Table \ref{tab:warp-bgrd}. 
In general, there is little difference by any metric when comparing the idealized 64 cm channel with MMLT elements (interpolated between 100 discrete points) and 19 overlapping BGRD octupoles. 
In the more realistic case of 9 overlapping BGRD octupoles between dipole locations, there is a significant reduction of aperture ($11\%$) and tune spread ($28\%$), although the level of $H_N$ variation only increases by a small fraction. 
Finally, the third case, which is maximally-populated given space constraints for the magnet mounts, is clearly unsuitable for experiment, due to dramatic loss of aperture and the associated tune spread. 

The most promising configuration for experiment is therefore case (b). We can possibly improve on the aperture and tune spread limitations by considering a mechanically-identical setup where the linear lattice is tailored to provide an integer phase advance transformation across a 32-cm channel (rather than a 64-cm channel). Holding the beam size and octupole strength constant ($\beta_* = 0.3$ cm, $G_{3,max} = 50\ T/m^3$), the simulation parameters become $L = 0.32$ cm and $k = 2.7682\ m^{-2}$, for a channel tune of $\nu_{drift} = 0.1560$. While dynamic aperture is slightly improved in this configuration, tune spread is slightly lower, since shortening the channel length reduces the maximum possible tune spread. I expect that the two configurations may yield indistinguishable experimental results. If it is feasible to test both, this should be done. 

The physical length of the ``32-cm" multi-PCB element in Figs. \ref{fig:octu-32cm} and \ref{fig:octu-hybrid} is actually 30.25 cm, considering spacing and physical circuit length $4.65$ cm. As the BD mount is centered over each $10^o$ pipe bend (meaning the longitudinal axis of the magnet is tilted $\pm5^o$ in horizontal plane with respect to the upstream and downstream beam pipe) and the physical BD length is 4.44 cm, the actual clearance for the channel mount is at most 26.6 cm. Therefore, the multi-PCB element will be a bit shorter than the case studied here. For seven individual PCBs, the center-to-center spacing would be shortened from 3.20 to 2.74 cm. This is not expected to change the results significantly.

\clearpage
\section{Chapter summary}

This chapter reviewed the basic design of a quasi-integrable octupole lattice utilizing the UMER framework. Octupole magnets are designed using the same approach as the UMER quad/dipole magnets, and are measured to be within the same tolerances. Arrangement of several octupole PCBs in a long composite element is near optimal for a center-to-center spacing of 4.56 cm. 

Generation and detection of a high-emittance, low-current ``DC beam" is also discussed. This is a large deviation away from the nominal UMER design, in which the gun is run in saturation and the beam apertured downstream to access different intensity regimes. The estimated incoherent space-charge tune spread for the DC beam, $\delta \nu \sim 0.005$, is much lower than the least space-charge-dominated apertured beam (0.6 mA, $\delta \nu \sim 0.94$). This will be essential for testing operation of the octupole lattice, which can provide a maximum octupole-induced tune spread of $\delta \nu \sim 0.25$.

Precise orbit control and mechanical alignment in the octupole insert is critical, as the dynamic aperture is very sensitive to orbit distortions from magnetic center. Shielding or compensating the ambient fields in the ring is necessary to run the proposed experiments. Requirements for external focusing are less stringent, but tune spread and dynamic aperture suffer for unequal phase advance errors in the linear lattice section. These observables appear fairly robust to equal tune errors, but in this case $H_N$ is ``less-conserved." For experimental design $\Delta \nu < 0.1$ and $|\Delta \nu_x - \Delta \nu_y| < 0.01$ should be sufficient for stable beam transport with optimal tune spreads. 

The use of realistic, fringe-dominated octupole fields over ideal, hard-edged elements does not significantly affect dynamics, but mechanical restrictions on octupole placement have a large effect. Shortening the octupole channel to 32 cm seems more promising than trying to maximize octupole density within available space. This may be due to cancellation of fringe fields in overlapping elements, while large spaces between elements results in some resonances being more strongly driven.

%% file: 8.Chapter.tex

\makeatletter
\def\env@matrix{\hskip -\arraycolsep
  \let\@ifnextchar\new@ifnextchar
  \linespread{1}\selectfont
  \renewcommand{\arraystretch}{0.5}%
  \array{*\c@MaxMatrixCols c}}
\makeatother

\renewcommand{\thechapter}{8}

\chapter{Design of the Single-Channel Octupole Experiment}
\label{ch:lattice}

UMER has been proposed as a test-bed for demonstrating the concept of nonlinear integrable optics (NLIO), particularly in the regime where space charge is significant. 
With minimal modification to the ring hardware, UMER will be modified to operate as quasi-integrable octupole (QIO) lattice with a single long octupole insert. 
This scheme is similar to the design of the IOTA ring, which will test the fully integrable solution.\cite{Antipov2017} 
Modifying UMER for QIO involves both creating the octupole insert element as discussed in Section \ref{sec:design:channel} and designing a linear focusing profile utilizing existing ring quadrupoles that meets the conditions required for quasi-integrability.
While Chapter \ref{ch:qio} explored particle dynamics in a reduced model of the system consisting only of the octupole element, this chapter focuses on descriptions of the full ring, including optimization of quadrupole strengths in the linear focusing section (the ``T-insert" of Chapter \ref{ch:qio}) and simulations of particle transport in the full ring.

Section \ref{sec:lattice:setup} describes the approach to modifying the UMER linear optics to meet the quasi-integrable condition. The linear lattice section should have a transfer function with phase advance equal to $n\pi$ for integer $n$ with equal focusing strength in both planes. As UMER optics are very flexible (all 72 quadrupoles are powered independently), it was assumed that identifying a solution would be straightforward. However, there is enough complexity that finding a workable approach to the optimization problem was not trivial. This section describes both an approach to optimization, including reduction of the problem dimension, as well as identifying a promising lattice solution.

Section \ref{sec:lattice:solenoids} describes how solenoid focusing elements can be incorporated into the single-channel lattice. The use of solenoid lenses provides straightforward control of the lattice tune without disruption to the optimized quadrupole focusing profile. In Section \ref{sec:lattice:warp}, particle-in-cell (PIC) WARP simulations are used with a model of UMER in the QIO configuration with linear focusing as proposed in Section \ref{sec:lattice:setup}. Results both with and without the octupole insert and space charge forces are discussed. 
Finally, Section \ref{sec:lattice:another} describes another approach for linear focusing design and shows some results from beam transport in the modified lattice.

\section{Modification of existing UMER optics for single-channel experiment}\label{sec:lattice:setup}

Initial experiments include plans for a single nonlinear insertion element.
A single $20^o$ ring section will be modified to house a long octupole element composed of many short octupole PCB magnets. 
The bends will be preserved in this section, but there will be no quadrupole fields. As UMER is equipped with 72 quadrupoles at $5^{\circ}$ azimuthal spacing, the remaining 68 can be adjusted to provide the desired transverse focusing. 

Recall that the transfer function of the linear portion of the quasi-integrable lattice (referred to as the ``T-insert") should be equivalent to a symmetric, thin focusing lens: 

\begin{equation}
T = \begin{bmatrix} 1 & 0 & 0 & 0 \\ -k & 1 & 0 & 0  \\ 
0 & 0 & 1 & 0  \\ 0 & 0 & -k & 1   \\ 
\end{bmatrix}
\label{eq:t-insert}
\end{equation}

As described in Eq. \ref{eq:k}, $k$ is constrained by the choice of octupole channel length $L$ and beam waist size $\beta_*$. In the previous chapter, $\beta_* = 0.3$ m was identified as a suitable operating point for a channel of length $L=64$ cm. 
This requires the linear lattice to provide the transport function $T$ with focusing strength $k=3.3264\ m^{-1}$. Another constraint placed on the ``T-insert" is that the phase advance be $n\pi$ for any integer $n$ (this is equivalent to an integer or half-integer tune advance). This requirement ensures that particles experience quasi-continuous motion through the octupole potential, allowing the invariant $H_N$ to be conserved over the particle lifetime. 

The 68 ``free" quadrupoles are independently powered, with focusing strengths that may be varied between $\pm K_{1,max} = 10.8$ G/cm (determined by the maximum safe excitation of 3 A). The simplest posing of the problem has 6 target quantities: $\left[ X', X, Y, Y' \right]$ to ensure the beam is matched through the octupole insert and $\left[ \nu_x, \nu_y \right]$ to meet the phase advance requirement. The equations to be solved (in each plane) are the RMS envelope equation (Eq. \ref{eq:env-eq}) with the boundary condition $X(s+C) = X(s)$ and $X'(s+C) = X'(s)$ for ring circumference $C$. Given there are 68 free parameters, it should be trivial to meet the targets exactly, but there are additional constraints on the beam that limit the possible solution space.

A note on notation: 
Typically $\nu_x$ and $\nu_y$ denote full ring tune. 
Here, $\nu_{x,T}$ and $\nu_{y,T}$ are defined as tune advances in the linear, ``T-insert" portion of the ring. 
The linear lattice tune advance depends on ring tune as $\nu_{x,T} = \nu_{x}- \nu_{dr}$ (and similar for $y$), where $\nu_{dr}$ is the tune advance through the insertion region when the octupole fields are off.

\subsection{Additional constraints for lattice solution}

To create a transfer function equivalent to Eq. \ref{eq:t-insert}, we only need to consider the six-parameter target $\left[ X', X, Y, Y', \nu_x, \nu_y \right]$ where $X$ and $Y$ are the RMS beam edge.
However, meeting just these constraints does not result in a good solution. 
We require further constraints to find a well-behaved lattice function.

First, the beam edge must not exceed the pipe radius at $2.5$ cm.
Second, large values of $X'$ should be avoided. Steep gradients in the envelope function cause sensitivity to quad errors. In the extreme case, a quadrupole error in a sensitive, high-gradient location could cause the envelope to become unstable and particle motion unbounded. Third, large asymmetry in the beam $XY$ aspect ratio can disrupt the particle distribution through asymmetric space charge forces and can lead to unpredictable, nonlinear behaviors. Therefore we desire to maximize beam roundness (while maintaining sufficiently strong alternating-gradient focusing).

Energy spread in the particle distribution is another consideration. As UMER is a coasting beam (without synchrotron focusing), at the center of the bunch the energy spread can be assumed to be low. However, the lattice design should still consider energy-dependent effects to ensure off-momentum orbits are still near the quasi-integrable condition.\footnote{Off-momentum effects are discussed in Appendix \ref{ap:Hamiltonian}.}  
The octupole insert should occupy a zero-dispersion region, as dispersion distorts particle orbits. Ideally the single-channel lattice is achromatic, with chromaticity $C_{x} = 0$\footnote{Equation \ref{eq:chromaticity}}, so that off-momentum particles do not experience a tune shift from the design value. As the UMER dipole locations and strengths are fixed, it is impossible to exactly match this criterion. At present UMER is not equipped with chromaticity-correcting sextupoles, therefore we aim for a lattice design with low natural chromaticity. It is possible to implement this correction in the future with PCB sextupoles.

\subsection{Reducing dimensionality of the optimization problem}

Feeding all 68 free parameters into an optimization routine with $5+$ targets is inefficient and generally ineffective at finding a well-behaved solution. The dimensionality of the problem is reduced by assuming periodicity $N$ of the lattice solution and optimizing for one period. In the typical FODO configuration, the lattice solution therefore has a periodicity of 36.\footnote{In reality, the optics in the injection section are distinct from the other 35 cells, and the ring has a super-periodicity of one.}

Choice of $N$ reflects a balance between lattice complexity (high $N$ is more repetitive and therefore easier to tune) and flexibility in focusing optics (high $N$ has fewer free quads). The periodicity must be evenly divisible into $72$. In the single-channel lattice, the length of a cell will be longer than 32 cm. Considering there will be a drift space of length 64 cm (requiring the omission of 4 quadrupoles), the number of free quads per cell is $72/N-4$. At $N=18$, there are no quadrupoles between drift/insertion regions and therefore no external focusing. This limits the possible periodicities to the set $N \in \left[12,9,6,4,3,2,1\right]$. 

Existing UMER routines for matching injected beam shape into the lattice assumes symmetry at the diagnostic locations (for a well-tuned match the beam spot has an identical aspect ratio at all imaging screens).\cite{HaoThesis} The periodicity of RCs in UMER is 18. Therefore, for the most straightforward diagnostics of the beam match, $N$ should also be divisible into 18. This further limits $N$ to the set $\left[9,6,3,2,1\right]$. 

$N=9$ contains only 4 quads per cell. This is less than the minimum number of target parameters. Cases $N=2$ and $N=1$ do not significantly simplify the problem. Therefore the two most promising symmetries are $N=6$ and $N=3$, with eight and twenty free quads per cell, respectively. 

The following discussion assumes a lattice with periodicity $N=3$. A single cell with available quadrupoles is shown in Fig. \ref{fig:p3-cell}. In the full ring solution the matched beam comes to a waist at three azimuthal locations. For empirical tuning of beam match, it will be possible to image the beam spot at three symmetric locations per turn. The octupole insert will be placed over one of the quad-free waist regions and the bare tune advance between exit and entrance of the octupole element will be an integer or half-integer. In the $N=3$ lattice, dimensionality is further reduced by assuming forward-backward symmetry in each cell, as drawn in Fig. \ref{fig:p3-halfcell}. In other words, only lattice solutions for which the periodic solution is mirror-image symmetric about the middle of the cell are considered. This reduces the free parameters to ten. 

The optimization procedure described below was applied to the $N=6$ case for a beam waist $\beta_*=0.3$ m. No good solutions were found at the desired tune value, suggesting that there was not enough flexibility with only four free parameters. The $N=6$ lattice may be more tractable for a larger waist size $beta_*$ and may be revisited in this case.

\subsection{Approach for finding QIO lattice solutions} \label{sec:lattice:approach}

\begin{figure}[]
\centering
\subfigure[Half a lattice cell, with quads numbered according to location in cell.]{
\includegraphics[width=\textwidth]{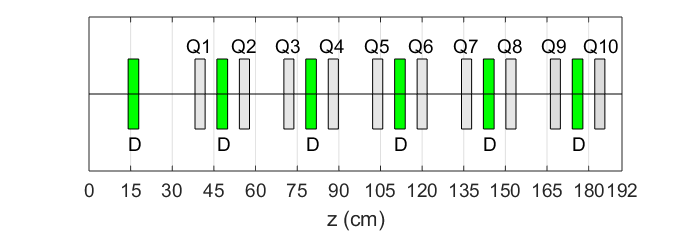}
\label{fig:p3-cell}}
\hspace{.5in}
\subfigure[One lattice cell, with quads numbered according to location within symmetric cell.]{
\includegraphics[width=\textwidth]{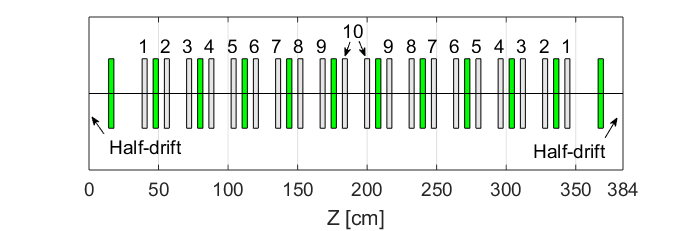}
\label{fig:p3-halfcell}}
\caption{Arrangement of magnetic elements in $N=3$ lattice. Quads are gray, dipoles are green. } 
\label{fig:p3-cells}
\end{figure}

In order to reduced the computational time when examining many solutions, the envelope equations are only solved over a half-cell, as drawn in Fig. \ref{fig:p3-halfcell}. $s=0$ is aligned with the center of the insertion/drift region.
I set the initial conditions based on the desired beam size at the waist, $X(0)=Y(0)=\sqrt{\epsilon \beta_*}$ and $X'(0) = Y'(0) = 0$. 
The free parameters are the currents in the ten free quads, $\vec{I} = \left[I_1,I_2,...I_9,I_{10}\right]$. 
Cell length is $L=3.84$ meters.

To find the periodic matched solution for one cell, over the half-cell calculation I minimize the target terms $\vec{T} = \left[ X'(L/2),Y'(L/2), \Delta \nu_x, \Delta \nu_y, \Delta XY \right]$. For single-objective optimizations, I define a minimization function $f = \sqrt{<T\cdot w >}$ for weight vector $w$. 
$X'(L/2)=Y'(L/2)=0$ for the matched, periodic solution. 
Meanwhile, $\Delta \nu_x = \nu_{x} - \nu_{dr} - \frac{n}{2}$ for any integer $n$, where $\Delta \nu_x = 0$ is the quasi-integrable condition (and similar in y). The term $\Delta XY$, defined $\Delta XY\equiv \max{|X - Y|}$ is an approximate measure of roundness. 

Another method for limiting asymmetry is the use of a ``reference trajectory," first described in the envelope code \verb|SPOT| \cite{Allen1998}. Prior to lattice function optimization, the user defines a reference trajectory $R(s)$, which is the desired average beam size as a function of $s$. Two additional terms are added to the minimization function: $<|X(s)-R(s)|>$ and $<|Y(s)-R(s)|>$. This should guide the optimized solution towards the desired average behavior with minimum asymmetry. Used in this context, the reference trajectory approach was not more effective than simply adding the roundness term $\Delta XY$ described above. 

Finding the appropriate weight factor to balance tune advance and envelope terms is difficult. Tune is generally more slowly-varying than $\alpha$ when adjusting quadrupole strength, as it is a global rather than local property. Additionally, the tune term contains the arbitrary integer $n$. My approach was to calculate the tune for the non-optimized initial condition (a non-periodic solution) and choose a tune target near the initial condition. 
In general the tune condition may not be met exactly. 
Or, in switching between simulation models with different magnet models, the tune may drift from the desired value. In these cases, I explore nearby solutions and parametrize tune in terms of quadrupole strength in order to move towards $\Delta \nu = 0$.

\clearpage
\subsection{Lattice solution in MENV} \label{sec:lattice:solution}

\begin{figure}[tb]
\centering
\subfigure[Lattice solution for $1/6$th of ring (half-cell in $N=3$ lattice). Quads are numbered.]{
\includegraphics[width=\textwidth]{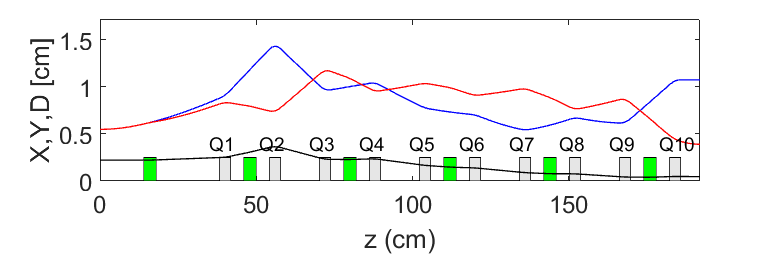}
\label{fig:p3-at7f-full:a}}
\hspace{.5in}
\subfigure[Lattice solution over whole ring. Numbers indicate RC position.]{
\includegraphics[width=\textwidth]{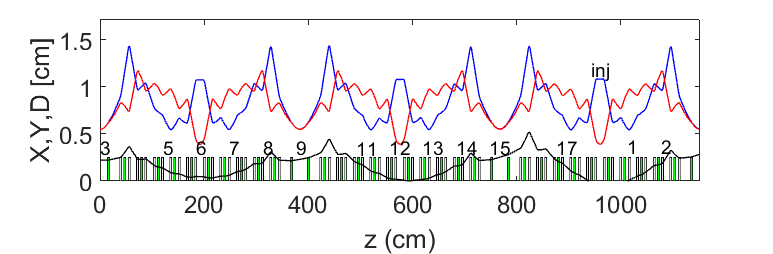}
\label{fig:p3-at7f-full:b}}
\caption{MENV solution for $100$ micron, $60 \mu$A beam. Dipoles are green, quads are gray. Y-axis shows RMS beam size in x (blue) and y (red) and dispersion (black). Numbers indicate RC position. }
\label{fig:p3-at7f-full}
\end{figure}

\begin{table}
\centering
\caption{Quadrupole currents (in Amps) for MENV $N=3$ lattice solution shown in Figure \ref{fig:p3-at7f-full}. } 
\label{tab:p3-at7f-quad-vals}
\vspace{10pt}
\begin{tabular}{r r r r r r r r r r}
\hline
Q1&Q2&Q3&Q4&Q5&Q6&Q7&Q8&Q9&Q10\\
\hline
-0.590& 1.247& -0.972& 0.616& -0.332& 0.414& -0.510& 0.738& -1.166& 0.762 \\
\hline
\end{tabular}
\end{table}

Initial solutions are found by integrating the RMS envelope equation (Eq. \ref{eq:env-eq}) using the in-house, Matlab-based code MENV \cite{menv}.\footnote{Described in Chapter \ref{ch:numeric} with model parameters in Table \ref{tab:menv-sim}.} 
Lattice optimization was performed using the built-in Matlab method GlobalSearch, part of the Global Optimization toolbox, which tests convergence over a range of initial conditions in order to locate a global minimum.\cite{Ugray2007} For this problem I applied the \verb|fmincon| minimization tool. I use the target function $\vec{T} = \left[ \alpha_{x,f}, \alpha_{y,f}, \Delta \nu_x, \Delta \nu_y, \Delta XY \right]$. All terms are defined in the previous section, with equal weights except the $\Delta XY$ term weighted at $1\%$ of the other contributions.

Figure \ref{fig:p3-at7f-full} shows a solution for a $60\ \mu$A, $\epsilon_x = \epsilon_y = 100\ \mu$m beam computed in MENV.\footnote{Comparable to the proposed ``DC beam" of Section \ref{sec:design:lowcurrent}.} Beam currents for this solution are listed in Table \ref{tab:p3-at7f-quad-vals}. Lattice tune and chromaticity are given in Table \ref{tab:p3-at7f-params}. Note that dispersion is not optimized or matched in Fig. \ref{fig:p3-at7f-full}. An alternate solution with non-dispersive drift sections is shown in Appendix \ref{ap:param}, but the tune values are further from optimal.

\begin{table}
\centering
\caption{Parameters of matched lattice solution shown in Figure \ref{fig:p3-at7f-full}. }
\label{tab:p3-at7f-params}
\vspace{10pt}
\begin{tabular}{l r r r}
\hline
Parameter &  \\
\hline
Full ring tune 		 & $\nu_x = 3.270 $ \\
					 & $\nu_y = 3.267 $\\
Drift/insertion tune & $\nu_x = 0.265 $\\
					 & $\nu_y = 0.272 $\\
Linear lattice tune  & $\nu_x = 3.004 $\\
					 & $\nu_y = 2.995 $\\
Half-cell tune		 & $\nu_x = 0.545$\\
					 & $\nu_y = 0.544$\\
Chromaticity         & $C_x = -4.270$ \\
					 & $C_y = -3.277$ \\
\hline
\end{tabular}
\end{table}

\clearpage
\subsection{Lattice sensitivity} \label{sec:lattice:sensitivity}
 
\begin{figure}[tb]
\centering
\subfigure[\% error in beam waist $\beta_*$ in insertion region as a function of injected mismatch.]{
\includegraphics[width=0.7\textwidth]{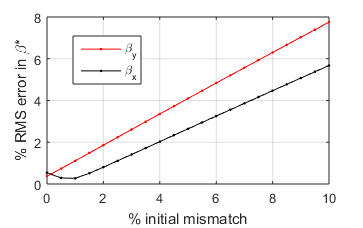}
\label{fig:p3-at5a-mismatch-a}}
\hspace{.5in}
\subfigure[RMS envelope evolution for $60 \mu$A, 100 $\mu$m beam over 10 turns with initial 10\% mismatch.]{
\includegraphics[width=\textwidth]{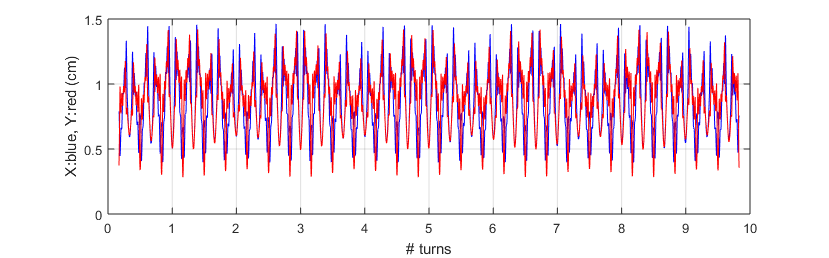}
\label{fig:p3-at5a-mismatch-b}}
\caption{Beam envelope evolution with injection mismatch error. }
\label{fig:p3-at5a-mismatch-env}
\end{figure}

\begin{figure}[tb]
\centering
\includegraphics[width=0.8\textwidth]{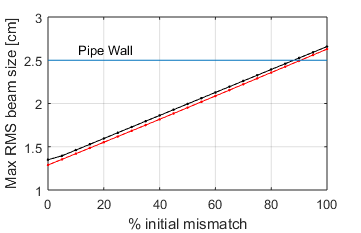}
\caption{Maximum RMS size versus initial mismatch over 10 turns. }
\label{fig:p3-at5a-mismatch-max}
\end{figure}

One of the concerns with designing a non-FODO lattice is finding a stable solution. Small errors in injected beam or quadrupole strength should not result in secular growth envelope. Instead the RMS beam extent should oscillate about the matched solution, with a large tolerance for injection error. To test lattice stability, MENV is used to study RMS envelope propagation with an ``injected beam" mismatch. Initial conditions are defined as $X(0) = (1-f)X_M$ and $Y(0) = (1-f)Y_M$ where $f$ is the fractional mismatch. 

Fig. \ref{fig:p3-at5a-mismatch-env} shows results up to $f=0.10$ for the $60 \mu$A lattice solution.
In general the solution is well-bounded for even large ($>10\%$) injected mismatch. Fig. \ref{fig:p3-at5a-mismatch-max} shows the maximum RMS beam size for a $60 \mu$A,  100 $\mu$m beam as a function of initial mismatch. The behavior is very predictable and bounded even up to very large injection errors. The results predict good stability for the proposed experiments, as the acceptance for injection errors will be limited by scraping on the pipe wall and not lattice sensitivity.


\clearpage
\subsection{Tuning and tune scans in the single-channel lattice} \label{sec:lattice:parameterization}

\begin{figure}[tb]
\centering
\includegraphics[width=0.5\textwidth]{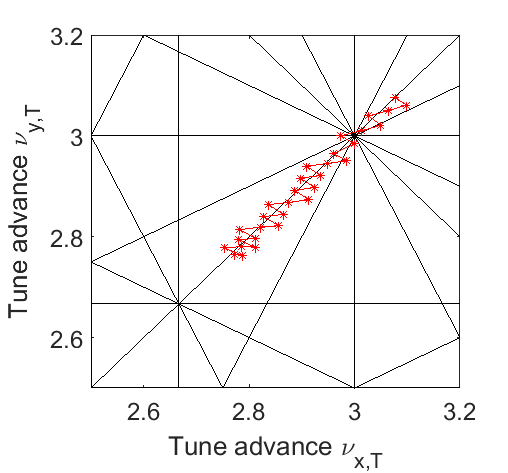}
\caption{Operating points $\nu_x\approx\nu_y$ for nearby matched solutions. }
\label{fig:tunespace-param}
\end{figure}

\begin{figure}[tb]
\centering
\subfigure[Linear fit to tune as a function of $i_{Q1}$.]{
\includegraphics[width=0.45\textwidth]{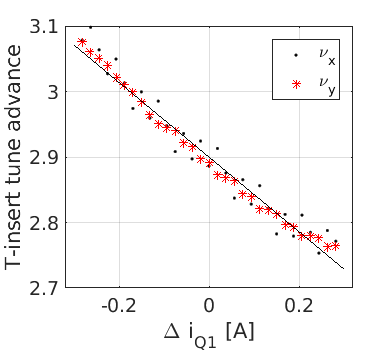}
}\label{fig:parameterization-a}
\subfigure[Linear fit to quad currents as a function of $i_{Q1}$.]{
\includegraphics[width=0.45\textwidth]{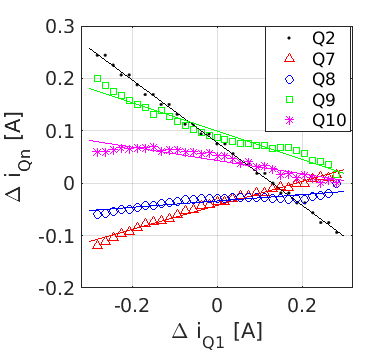}
}\label{fig:parameterization-b}
\caption{Linear fits that parameterize quad current and tune with respect to $i_{Q1}$ while maintaining a matched beam in the drift/insertion region. }
\label{fig:parameterization}
\end{figure}

In the above lattice solution, we desire independent control of the lattice tune without disrupting the beam match through the octupole insert. The tune dependence on quadrupole strength is slowly varying compared to the dependence of the envelope solutions $X(s)$, $Y(s)$. Therefore, it is possible to linearize the tune dependence of nearby matched lattice solutions in terms of quadrupole strengths. This technique is described in more detail in Appendix \ref{ap:param}. 

Linearizing the local solutions allows tuning of a lattice towards the optimal tunes. As long as the tune is known, a path can be drawn towards the desired tunes and the required changes in quadrupole strengths predicted. In Appendix \ref{ap:param}, this approach is applied both to correcting the MENV solution and tuning quadrupole currents for implementation in a WARP model.

Another application is to perform tune scans in the single-channel experiment. In Chapter \ref{ch:apparatus} the tune scan method is described and in Chapter \ref{ch:res} it is applied to resonance studies in the FODO lattice. Tune scans can be used in the single-channel experiment to examine performance as a function of distance from the quasi-integrable condition as well as a way of inducing resonant losses to demonstrate octupole-driven damping. Tune scans should be performed in such a way that the tune is shifted without mis-matching the beam through the octupole insert. Fig. \ref{fig:tunespace-param} shows the operating points of a handful of matched solutions near the line $\nu_x = \nu_y$. Fig. \ref{fig:parameterization-a} shows the dependence of lattice tune on the current in Q1, while Fig. \ref{fig:parameterization-b} shows how Q2, Q7-10 must be varied to maintain a matched solution. Further discussion of this technique, including linear fits for behavior along the $\nu_x = \nu_y$ line, is in Appendix \ref{ap:param}.

\section{Solenoids for flexibility of optics}
\label{sec:lattice:solenoids}

\begin{figure}[!tb]
\centering
\includegraphics[width=\textwidth]{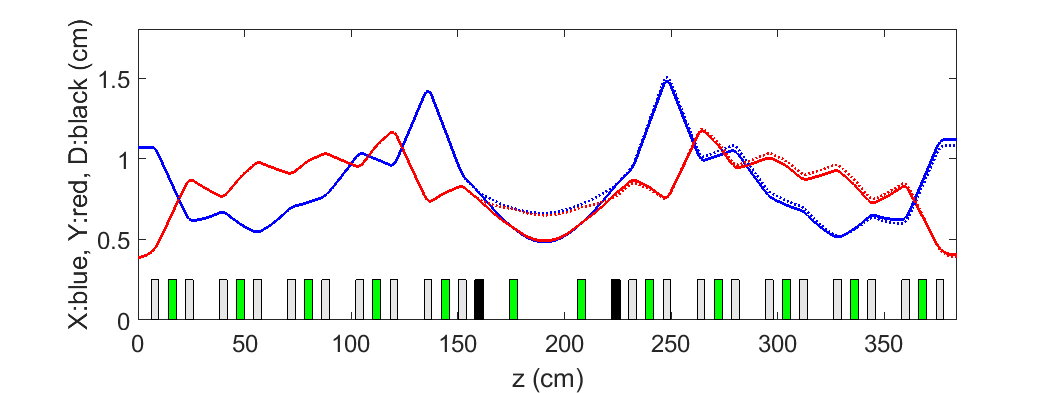}
\caption{Ring lattice solution with the addition of solenoids (in black). Dashed line indicates solenoids with $B_{||}=100$ G, solid line has $B_{||}=70$ G in opposite polarity. }
\label{fig:at7f-solenoids}
\end{figure}

A lattice solution of the type suggested in Chapter \ref{ch:lattice} is based on the assumption of fixed beam profile in the octupole insertion. The design values assumed for the above solutions are $\beta_0 = 0.3$ m, $\nu_{dr} = 0.27$. Flexibility in the size of the beam waist would be useful for experiments, as this has direct correlation with dynamic aperture. This can be achieved using solenoid focusing elements on either side of the channel. With the same focusing strength in each solenoid, the size of the beam waist can be adjusted without creating mismatch, as illustrated in Fig. \ref{fig:at7f-solenoids}. 

MENV calculations show beam waist can be adjusted in the range $\beta_0 = 0.236 \to 0.442$ m, which corresponds to fractional lattice tunes $\nu = 0.40 \to 0.10$. This assumes a four cm-long element with on-axis field $B_z=26$ Gauss. This requires approximately $84$ Amp-turns, which is feasible for a UMER magnet.
Independent control of the beam waist with solenoids as described allows a ``single knob" to adjust lattice tune, without relying on an optics model of the ring as described in Chapter \ref{ch:lattice}. This capability would be very useful for examining beam transport near different resonant conditions, as is done for the FODO lattice in Chapter \ref{ch:res}.

\section{WARP simulations of single-channel experiment} \label{sec:lattice:warp}

This section uses a WARP model of the lattice solution found above to predict single-channel experiment performance. 
Initial lattice solutions were found in both Elegant and MENV, using envelope integrators in a linear lattice. However, the dynamics that can be probed in these codes are limited. MENV is strictly an envelope integrator, allowing only linear fields (quadrupole magnets and linear (KV) space charge). Elegant is more powerful, with a variety of magnet models including symplectic models of nonlinear fields. However, Elegant is most trustworthy in the low-charge regime as it does not include a fully self-consistent space charge model. As the goal is to perform low-charge experiments before moving into a high-charge regime, building a lattice model in WARP will allow for self-consistent simulations. Additionally, the WARP deck for UMER includes gridded field elements based on MAGLI calculations of the PCB circuits, while all other models use hard-edged approximations.\footnote{See Appendix \ref{ap:model} for hard-edged model parameters.}

\subsection{Implementing MENV lattice solution in WARP}

\begin{table}
\centering
\caption{Quadrupole currents for WARP implementation of $N=3$ lattice. }
\label{tab:warp-quad-vals}
\vspace{10pt}
\begin{tabular}{r r r r r r r r r r}
\hline
Q1&Q2&Q3&Q4&Q5&Q6&Q7&Q8&Q9&Q10\\
\hline
-0.538&1.161&-0.973&0.616&-0.434&0.410&-0.458&0.925&-1.109&0.650\\
\hline
\end{tabular}
\end{table}

The MENV-optimized solution is implemented in the WARP model using gridded field elements. As the transfer function of the hard-edged model is not in perfect agreement with the gridded elements, additional optimization and tuning of the lattice in WARP is required. This includes applying the tune parameterization method described above to shift the WARP lattice towards the desired operating point. A more thorough discussion is given in Appendix Section \ref{sec:param:warp}.

\begin{figure}[tb]
\centering
\includegraphics[width=\textwidth]{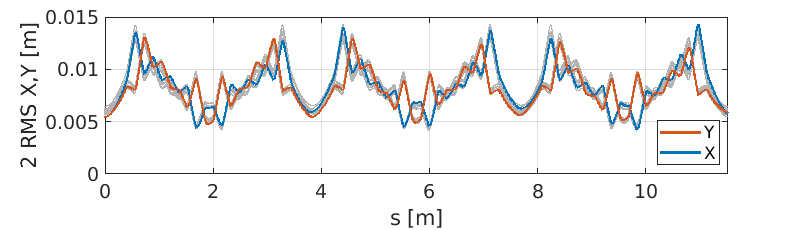}
\caption{Transverse beam evolution in WARP model. First turn is shown with bold lines; subsequent turns 2-10 are shown in gray to show magnitude of mismatch/envelope oscillation. }
\label{fig:warp-linear}
\end{figure}

Lattice tune is measured using frequency analysis of low-amplitude probe particle orbits. Full ring tunes are measured to be $\nu_{x} = 0.240$ and $\nu_{y} = 0.259$. From this measurement, tune errors are $\Delta \nu_x = -0.023$, $\Delta \nu_y = -0.013$ and $\Delta \nu_x-\nu_y = -0.019$. 
Figure \ref{fig:warp-linear} shows RMS beam size evolution over 10 turns in the optimized WARP linear lattice. The average twice-RMS beam size in the drift/insertion regions is $\beta_x = 0.36 \pm 0.03$ m, $\beta_y = 0.34 \pm 0.03$ m measured over 10 turns from an initial condition of $\beta_x = \beta_y = 0.3$. Equivalently, $2x_{RMS} = 6 \pm 0.3$ mm and $2y_{RMS} = 5.8 \pm 0.2$ mm compared to the desired matched condition of $2x_{RMS} = 2y_{RMS} = 5.5$ mm.

\subsection{Frequency map analysis of octupole lattice in WARP} \label{sec:lattice:fma}

In this section frequency map analysis\footnote{Described in Chapter \ref{ch:numeric}.} 
is applied to the full ring WARP model. 
To map the space, probe particles (with zero-current) are launched on an $N \times N$ grid in $x \in [0,0.8]$ cm and $y \in [0,0.8]$ cm. 
To stay consistent with reduced model simulations, the distribution is launched at the $s_*$ waist location, so the configuration space plot Fig. \ref{fig:warp-lin-fma-xy} corresponds with the drift/insertion center. At this location, for a $60 \mu$A, $100\mu$m beam, the beam edge is expected to be $\sim 0.55$ cm, so the gridded distribution over-fills the space. The pure-multipole \verb|MMLT| element is used to model an octupole insert of length 32 cm (dipole-center to dipole-center) and gridded field \verb|BGRD| elements as described above. Unless otherwise stated, frequency maps are calculated for 512 turns.

Simulations are run both with and without space charge. In the latter, it is only necessary to initialize probe particles. For the former, I also initiate a KV distributed beam with $60 \mu$A current and $\epsilon = 100 \mu$m emittance that is matched to the linear lattice. While the KV distribution is high-correlated and thermally unfavorable state, it is a commonly used test case for lattice dynamics with space charge. Additionally, in a self-consistent PIC model the particle distribution decorrelates very quickly. 

Simulations are presented for the lattice solution in Table \ref{tab:warp-quad-vals}, which has fractional tunes $\nu_{x} = 0.240$,$\nu_{y} = 0.259$ near the desired operating point $\nu_{x} = 0.265$, $\nu_{y} = 0.272$. An analysis of dynamics in a less-optimal operating point $\nu_{x} = 0.426$, $\nu_{y} = 0.185$ is presented in Appendix \ref{ap:warp-nonopt}. Dynamics are similar in both cases.

\subsubsection{Linear WARP lattice}

\begin{figure}[tb]
\centering
\subfigure[Dynamic aperture over 512 turns.]{
\includegraphics[width=0.6\textwidth]{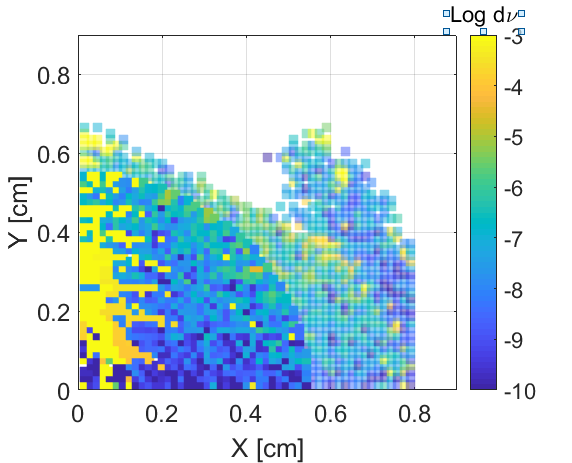}
\label{fig:warp-lin-fma-xy}
}
\hspace{.5in}
\subfigure[Tune footprint, with up to 3rd order resonance lines.]{
\includegraphics[width=0.6\textwidth]{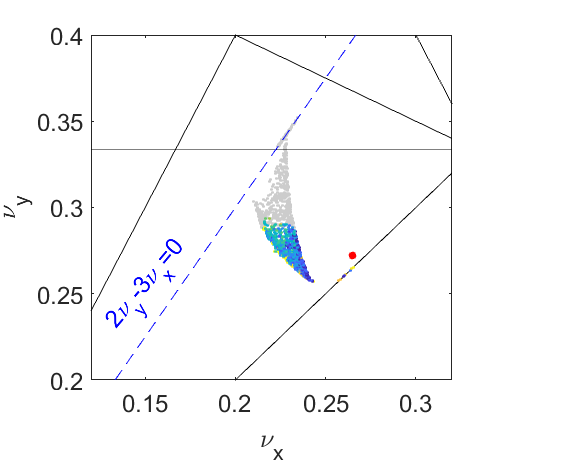}
\label{fig:warp-lin-fma-tune}
}
\caption{Frequency map analysis of full ring linear lattice in WARP for lattice solution in Table \ref{tab:warp-quad-vals}. }
\label{fig:warp-lin-fma}
\end{figure}

In the linear case, the octupole insert is not powered.
As the maximum transverse beam edge is quite large ($\sim 1.5$ cm), the beam samples non-linear regions of the UMER PCB magnets. This nonlinearity causes tune spread even in the ``linear" lattice. A zero-current distribution of probe particles was used to sample the proposed lattice in the WARP model, the resulting frequency map analysis is shown in Fig. \ref{fig:warp-lin-fma}. The ideal quasi-integrable operating point is indicated by a red dot. The full interrogated space is shown, but a cut at $r=0.55$ cm is indicated by pixel saturation (this corresponds with the expected beam edge). The dynamic aperture for a round beam is 0.56 cm, and the aperture appears to be limited by the $2\nu_y-3\nu_x$ resonance. 

As seen, there is significant tune spread even in the absence of octupole fields and space charge. 
As the the nonlinearity in the quadrupoles is supralinear, we expect high-amplitude particles to occupy a higher tune. This is true for the vertical spread, but not the horizontal.
For particles within $r=0.55$ cm, $\max \delta \nu_x = 0.044$ and $\max \delta \nu_y=0.039$ with RMS $\delta \nu_x=0.029$ and RMS $\delta \nu_y=0.019$. These values are comparable to the ``best-case" values in the reduced model ($\max \delta \nu = 0.113$ and RMS $\delta \nu = 0.034$), while the amplitude-dependence of the tune shift resembles that of the octupole lattice (high amplitude particles at larger tune shifts). This may act to obscure the effect of the octupoles in the proposed experiments.

\begin{figure}[tb]
\centering
\subfigure[Dynamic aperture for 128 turns.]{
\includegraphics[width=0.6\textwidth]{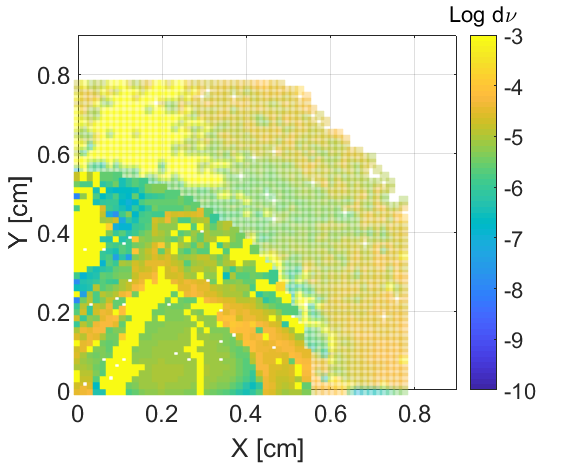}
\label{fig:warp-lin-sc-fma-xy}
}
\hspace{.5in}
\subfigure[Tune footprint, with up to 3rd order resonance lines.]{
\includegraphics[width=0.6\textwidth]{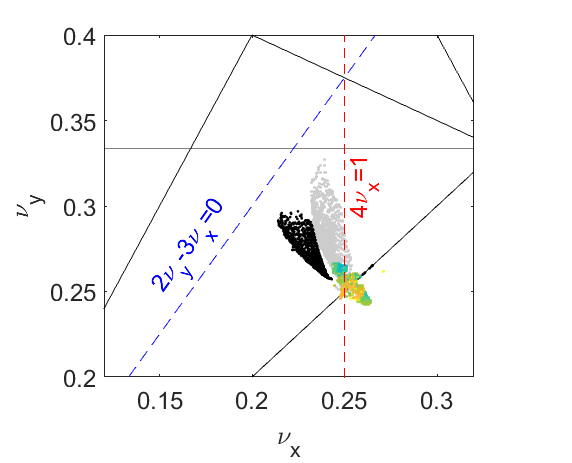}
\label{fig:warp-lin-sc-fma-tune}
}
\caption{Frequency map analysis of full ring linear lattice in WARP with $60\mu$A beam. }
\label{fig:warp-lin-sc-fma}
\end{figure}

\begin{figure}[tb]
\centering
\includegraphics[width=0.7\textwidth]{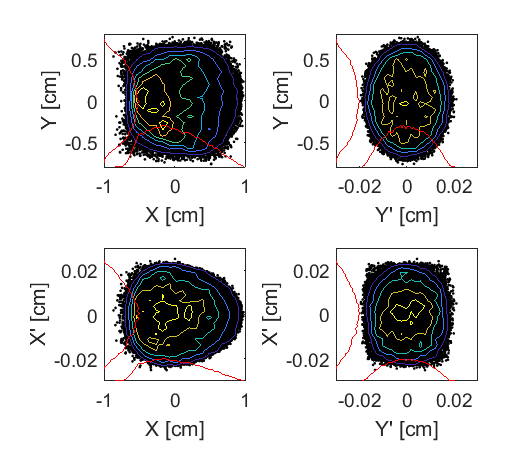}
\caption{Projections of particle distribution after 128 turns in the linear WARP lattice ($G_{3,max}=0$) for $60\mu$A beam. }
\label{fig:warp-sc-0-pdist}
\end{figure}

Fig. \ref{fig:warp-lin-sc-fma} shows the same case with $60\ \mu$A current. Again, a radial cut is made at $r=0.55$ cm. The footprint for the zero-charge case (Fig. \ref{fig:warp-lin-fma-tune}) is shown in black.
The tune footprint of the particle distribution is shifted from the zero-current bare tune $\nu_0$. We expect that the tune is depressed with space charge. As seen in Fig. \ref{fig:warp-lin-sc-fma}, the vertical tune is depressed but the horizontal tune experiences a small positive shift. The partial tunes for a low-amplitude particle are $ \nu_{x,ring} = 0.263$ and $\nu_{y,ring}=0.242$ (compare to $\nu_{x,ring} = 0.243$ and $\nu_{y,ring} = 0.257$ for the zero-charge case). A negative tune depression is unexpected, so it is possible the method of tune reconstruction is not correct. 

Comparing the configuration space to the no-charge case (Fig. \ref{fig:warp-lin-sc-fma} with Fig. \ref{fig:warp-lin-fma}), we see stronger nonlinear behavior with the inclusion of space charge, with higher $d\nu$ on average. Interestingly, the dynamic aperture is increased in the presence of space charge, to $r=0.79$ cm. This seems to be because the space charge tune shift moves the distribution away from the $2\nu_y-3\nu_x$ resonance. For particles within $r=0.55$ cm, $\max \delta \nu_x = 0.017$ and $\max \delta \nu_y=0.020$ with RMS $\delta \nu_x=0.008$ and RMS $\delta \nu_y=0.010$. This is a smaller spread than in the zero-charge limit, suggesting that even small amounts of space charge act to shield particles from external nonlinearity.

Projections of the beam distribution in planes $[X,Y,X',Y']$ are shown in Fig. \ref{fig:warp-sc-0-pdist}. There is some distortion in the $X-Y$ beam projection (which should be round, as this image is taken at the waist). 
The initial, seeded distribution is uniform in all projections. After 128 turns there is some charge redistribution, with tails developing and the profile tending towards a more Gaussian-type distribution.

\clearpage
\subsubsection{Nonlinear WARP lattice}

\begin{figure}[]
\centering
\subfigure[Dynamic aperture for 512 turns.]{
\includegraphics[width=0.6\textwidth]{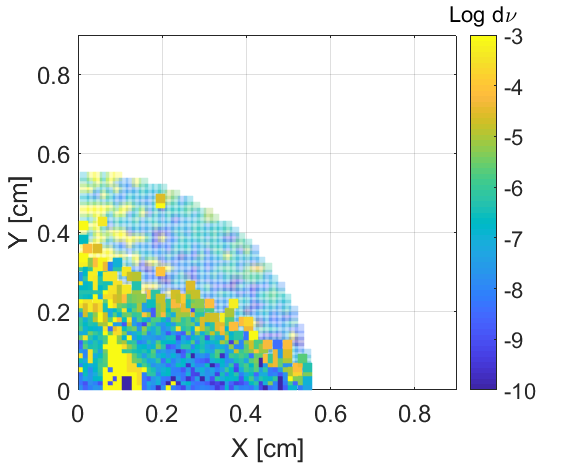}
\label{fig:warp-50-fma-xy}
}
\hspace{.5in}
\subfigure[Tune footprint with up to 3rd order resonance lines shown. $G_{3,max}=0$ case is shown in gray.]{
\includegraphics[width=0.6\textwidth]{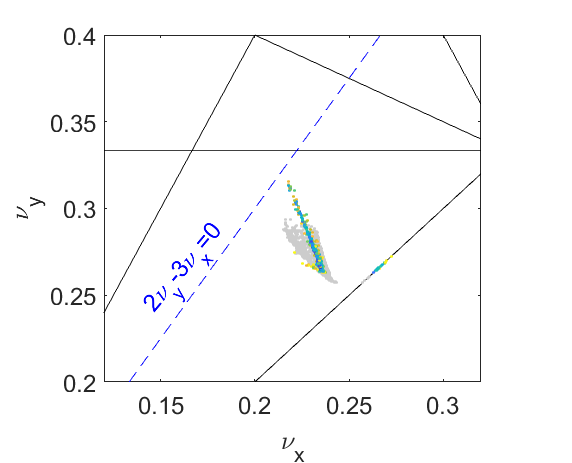}
\label{fig:warp-50-fma-tune}
}
\caption{Frequency map analysis of full ring octupole lattice at $G_{3,max}=50\ T/m^3$ in WARP with zero current. }
\label{fig:warp-50-fma}
\end{figure}

Simulations were also run including the octupole element.
Figure \ref{fig:warp-50-fma} shows the frequency map when the octupole insertion is powered at $G_{3,max}=50\ T/m^3$ ($\sim 1$ A peak). The unsaturated pixels are the configuration space and tune footprint from the linear case, Fig. \ref{fig:warp-lin-fma}. Compared to the linear case, the dynamic aperture is decreased to $r=0.22$ cm. As the aperture is much smaller, it is hard to directly compare tune spreads, since both the octupole-induced spreads and the dipole/quadrupole nonlinearities are greater at high amplitudes. 
However, for the distribution of stable particles in Fig. \ref{fig:warp-50-fma}, $\max \delta \nu_x = 0.038$ and $\max \delta \nu_y=0.050$ while RMS $\delta \nu_x=0.008$ and RMS $\delta \nu_y=0.014$. 
The tune spreads are almost identical to the linear case, although the total area of the tune footprint is smaller.

\begin{figure}[]
\centering
\subfigure[Dynamic aperture for 128 turns. Note the color axis is shifted when compared to most other plots.]{
\includegraphics[width=0.6\textwidth]{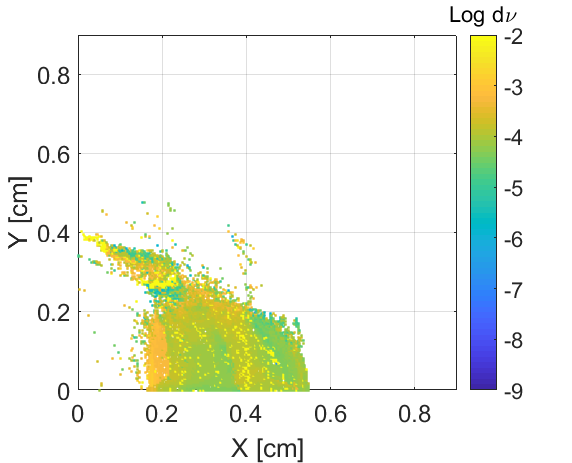}
\label{fig:warp-sc-50-fma-xy}
}
\hspace{.5in}
\subfigure[Tune footprint, with up to 3rd order resonance lines. Black points represent the same case without space charge.]{
\includegraphics[width=0.6\textwidth]{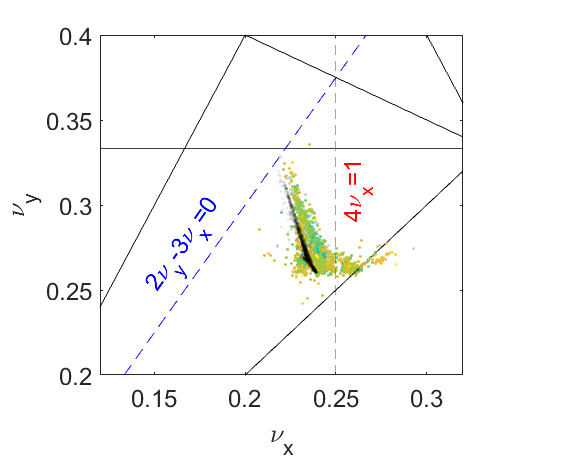}
\label{fig:warp-sc-50-fma-tune}
}
\caption{Frequency map analysis of full ring octupole lattice at $G_{3,max}=50\ T/m^3$ in WARP with $60 \mu$A, $100 \mu$m beam. }
\label{fig:warp-sc-50-fma}
\end{figure}

\begin{figure}[tb]
\centering
\includegraphics[width=0.7\textwidth]{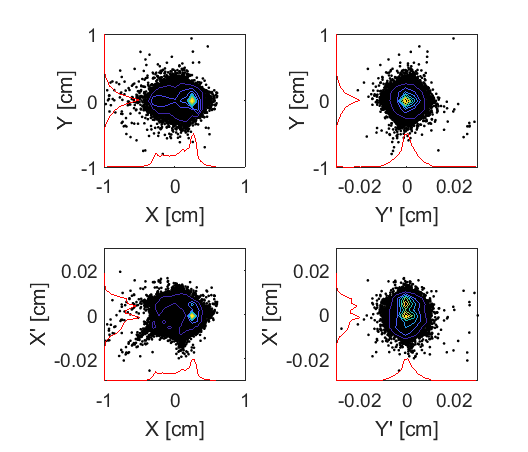}
\caption{Projections of particle distribution after 128 turns in the WARP octupole lattice at $G_{3,max}=50\ T/m^3$ for $60\mu$A beam. }
\label{fig:warp-sc-50-pdist}
\end{figure}

With space charge introduced as described above, the tune spreads increase. The frequency map is shown in Fig. \ref{fig:warp-sc-50-fma} for $G_{3,max}=50\ T/m^3$. The tune footprint at zero charge is also plotted (in black) for comparison.
For particles stable particles, $\max \delta \nu_x = 0.051$ and $\max \delta \nu_y=0.077$ while RMS $\delta \nu_x=0.022$ and RMS $\delta \nu_y=0.016$. Both the tune spread increases and the central tune shifts, as seen in Fig. \ref{fig:warp-sc-50-fma-tune}. However, there is essentially no dynamic aperture, due to particle losses that appear to primarily be along the fourth order resonance $4\nu_x = 1$. The beam distribution after 128 turns is shown in Fig. \ref{fig:warp-sc-50-pdist}. Compared to the linear case, the loss of dynamic aperture is apparent. The transverse beam shape starts to reflect the shape of the octupole fields: the $XY$ projection gains ``wings." Horizontally, the bunch appears to hollow due to the loss of unstable particles near the core.

The WARP results predict poor performance when implementing the proposed lattice experimentally. The maximum octupole-induced tune spread in a $60\ \mu$A beam is only approximately twice that of the linear focusing lattice, which is still a relatively weak effect
For the chosen operating point $\nu_x=0.240$ and $\nu_y = 0.259$, space charge appears to strongly drive particle losses along a fourth order resonance when octupole fields are included. 
Tuning to an operating point above the $\nu_x=0.25$ resonance might yield better results.

\clearpage
\section{Another strategy for finding lattice solutions} \label{sec:lattice:another}

\begin{figure}[tb]
\centering
\includegraphics[width=\textwidth]{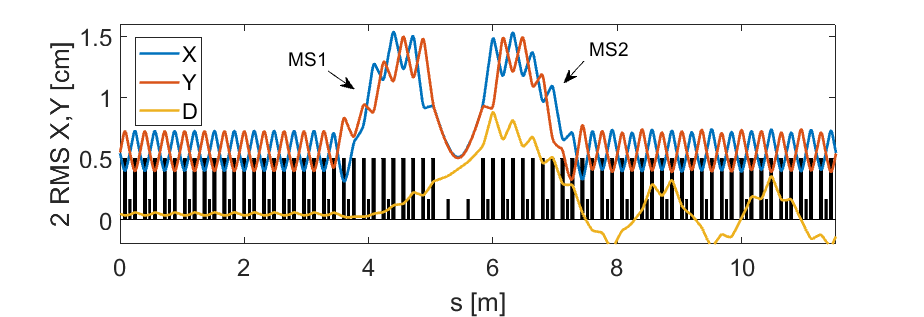}
\caption{Another possible ring solution, with custom solution over 1/3 of ring and standard FODO lattice for remaining 2/3 of lattice. }
\label{fig:ele-other-solution}
\end{figure}

\begin{table}[tb]
\centering
\caption{Quadrupole currents for Elegant solution shown in Figure \ref{fig:ele-other-solution}. }
\label{tab:ele-quad-vals}
\vspace{10pt}
\begin{tabular}{c r r r r r r r r r }
\hline
&Q1&Q2&Q3&Q4&Q5&Q6&Q7&Q8&Q9\\
&Q10&Q11&Q12&Q13 &&&&&\\
\hline
FODO &-1.418  & 1.418 &&&&&& \\
MS1 & -1.370 &1.581&-0.997& 0.701&-0.504&0.648&-0.587&0.606&-0.623\\
&0.741& -0.751&-0.751&0.376 &&&&& \\
MS2 & -0.376& 0.751&0.751&-0.741 &0.623 &-0.606&0.587&-0.688& 0.701\\
& -0.821& 1.036& -1.558& 1.556 &&&&& \\
\hline
\end{tabular}
\end{table}

Another approach to matching, which also yielded promising results, follows a completely different philosophy of ring structure. In this approach, the super-periodicity of the ring is assumed to be one. The ring is divided into 4 regions. the quad-free drift region is one, while the ``T-insert" is composed of three distinct regions: two ``matching sections" on either side of the drift region, and a region with a FODO solution.

Fig. \ref{fig:ele-other-solution} shows an example lattice solution (with current values in Table \ref{tab:ele-quad-vals}), in the zero-current limit. 
In this case, the drift length is 64 cm with tune advance $\nu_{dr} \sim 0.28$. The matching sections each contain 13 free quadrupoles over 2.5 $20^{\circ}$ ring sections. The quads in the matching section are optimized in order to match the initial condition at the beam waist to the matched condition for a FODO cell. After an initial matched solution is found, the FODO cell is adjusted to move the tune advance closer to the desired value, and the matching section quads re-tuned to maintain a match between the drift region and the FODO region. This solution is matched for beam envelope but no effort is made to match or minimize dispersion.

For the solution in Fig. \ref{fig:ele-other-solution}, the tune in the FODO region is $\nu_x = 3.890$, $\nu_y = 3.926$. In the matching sections, $\nu_{x,1}=0.640$, $\nu_{y,1} = 0.441$ and $\nu_{x,2}=0.501$, $\nu_{y,2} = 0.642$. The tune advance through the drift/insertion regions is $\nu_{x,drift}=0.283$, $\nu_{y,drift} = 0.288$. This gives a linear lattice tune advance of $\nu_{x,T}=5.031$ and $\nu_{y,T} = 5.000$. Tune errors in this lattice are $\Delta \nu_x = 0.031$ and $\Delta \nu_y = 0.000$.

The appeal of this approach is that the beam is well-contained during transport in the FODO section, with stronger focusing reflected in the higher tune, and the average beam size is well within the physical pipe aperture and linear quadrupole region. The beam is only large near the insertion region. This relaxes steering requirements for $2/3$ of the ring, as compared to the previously discussed solution where the beam is large in many sections. The downside is the super-period of one, which may make it difficult to characterize the beam match. This also complicates optimization of the lattice solution, as there are three distinct regions that must be optimized instead of one. 
The exact solution shown in Fig. \ref{fig:ele-other-solution} will not translate well to WARP or lab implementation, as the built-in Elegant edge-focusing model, using sector-dipoles with $\zeta = 5^{\circ}$ edge angles, is known to over-estimate vertical edge focusing when comparing FODO lattice tunes to experimental measurements.

\begin{figure}[]
\centering
\subfigure[]{
\includegraphics[width=0.46\textwidth]{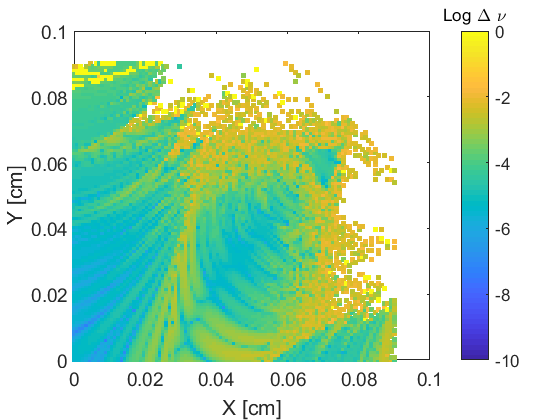}
\label{fig:ele-old-matching-xy}
}
\subfigure[]{
\includegraphics[width=0.46\textwidth]{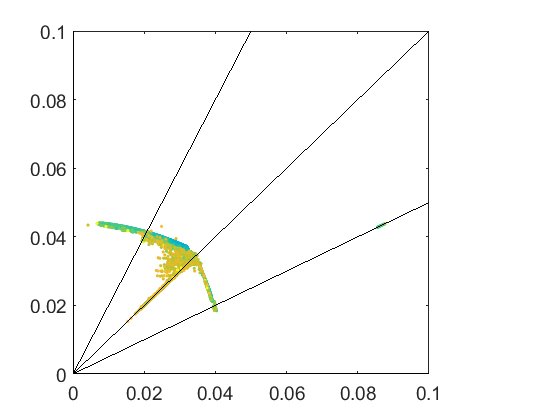}
\label{fig:ele-old-matching-tune}
}
\caption{Frequency map analysis of full ring octupole lattice from Elegant model for lattice solution similar to Fig. \ref{fig:ele-other-solution}. Octupole lattice is at operating point $\beta_* = 0.45$ m, $\nu_{dr} = 0.03$. (a) Dynamic aperture (survival vs. particle initial position). (b) Tune footprint, with up to 3rd order resonance lines. }
\label{fig:ele-old-FMA}
\end{figure}

Results from frequency map analysis on a similar solution is shown in Fig. \ref{fig:ele-old-FMA}. This simulation uses 8 discrete hard-edged octupole \verb|MULT| elements as the insertion. The octupole channel is 28.24 cm long, situated in the straight section between dipoles. $G_{3,max} = 84.7\ T/m^3$ for normalized strength $\kappa = 4394\ m^{-1}$. This lattice has phase errors of $\Delta \nu_x = 0.002$ and $\Delta \nu_y = 0.007$.
The shape of the dynamic aperture reflects the distortion we expect in the case of unequal phase advances in the linear lattice, as described in Section \ref{sec:design:phase-error}.

\section{Chapter summary}

This chapter discussed development of a workable approach for finding a lattice solution to meet the requirements of a single-channel quasi-integrable lattice using the existing UMER framework. I justified the choice of a lattice with $N=3$ symmetry, which has ten free quad strengths for lattice optimization. Using the envelope integrator MENV, I demonstrate an effective approach for optimization and show three possible lattice solutions. 
Nearby matched solutions are explored to find a working point closer to the quasi-integrable condition. With knowledge of the surrounding tune landscape, I show a parameterization of lattice solutions around the $\nu_x = \nu_y$ line. This will be useful for tuning of the simulated and experimental lattices to operate at the desired tune, as well as provide a framework for performing tune scans in the octupole lattice. 
The use of solenoid elements is considered for additional control of the second order moments near the insertion region. Required solenoid strength is relatively weak ($<30$ Gauss on-axis field) and the implementation would be a powerful tool for examining resonant losses in the quasi-integrable lattice.
Finally, in the same vein of lattice optimization, Section \ref{sec:lattice:another} described a completely different approach for matching using a ``half-FODO" lattice. Solutions of this have a smaller average beam size, but many more free quadrupole strengths and in general are less approachable than the $N=3$ type. 

This Chapter also reviewed efforts to implement the proposed lattice in the WARP PIC code. This will be a powerful tool for both numerical experiments and comparison with measurements, as WARP allows for self-consistent treatment of space charge effects. 
A WARP model has been assembled for the single-channel octupole experiment using realistic gridded field elements. This model uses the lattice solution found in MENV for a $N=3$ periodic lattice, with small adjustment to maintain a matched beam profile in the drift/insertion regions. The tune of this lattice is outside the bounds of the desired tolerance $\Delta \nu_x, \Delta \nu_y < 0.1$ and $\Delta \nu_x - \Delta \nu_y< 0.01$ found with the reduced model. While this may be comparable to what we expect in a first experimental pass, further optimization of the WARP lattice function should be done for a more robust model nearer the integrable condition. 

Moving beyond the reduced model into full ring simulation of the quasi-integrable octupole lattice with the proposed lattice function, poor behavior is observed when all nonlinearities (quad/dipole fields, octupole insert and space charge) are included. However, the tested working point is far from the quasi-integrable condition, with tune errors $\Delta \nu_x = 0.163 $, $\Delta \nu_y = -0.085$ and $\Delta \nu_x-\nu_y = 0.240$ beyond the range that was tested in the reduced model. The tune footprint of this working point spans a third order resonance, which appears to be driven by space charge and not strongly mitigated by the octupole-induced tune spread. The WARP PIC calculations should be repeated at a more favorable working point.

%% file: 9.Chapter.tex

\renewcommand{\thechapter}{9}

\chapter{Distributed Octupole Lattice}
\label{ch:distr}

This chapter discusses design of and preliminary measurements on a distributed octupole lattice. As discussed in Chapter \ref{ch:design}, the distributed octupole lattice is an alternative design for the QIO experiment which is simpler to implement in UMER. This design requires less custom hardware and uses a FODO arrangement for linear focusing, in comparison to the single-channel experiment which requires a custom $20^{\circ}$ ring section and a non-FODO lattice solution. However, the predicted nonlinear detuning effect is predicted to be much weaker and the quasi-integrable condition identified by Danilov and Nagaitsev \cite{Danilov2010} cannot be met exactly. This work was originally presented as a conference proceedings.\cite{RuisardHB2016}

\section{Motivation}

\begin{figure}[htb]
\centering
\includegraphics[width=\textwidth]{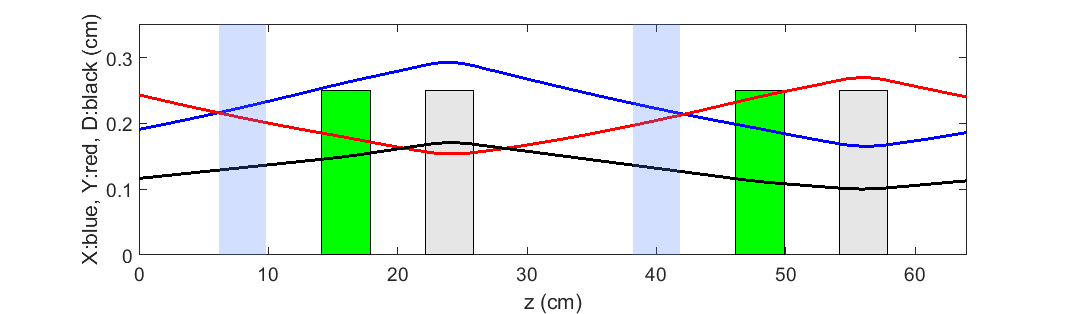}
\caption{Diagram of alternative lattice FODO cell, including possible locations for distributed octupole elements (blue). }
\label{fig:altfodo-cell}
\end{figure}

The Danilov-Nagaitsev condition for quasi-integrability is that nonlinear insert is placed where the beam is round ($\beta_x = \beta_y$) and the nonlinear potential scales longitudinally as $V(s) \propto \beta(s)^{-3}$. Additionally, the phase advance between nonlinear inserts (determined by the linear focusing function) should be $\phi_T = n\pi$ for integer $n$. The distributed octupole lattice proposes using FODO-like linear focusing and placing short octupole inserts at the locations where $\beta_x \sim \beta_y$.

The linear focusing function in the distributed octupole lattice takes the form of the alternative configuration of the UMER FODO lattice. Here the other length of the FODO cell is extended from 32 to 64 cm and 36 of the 72 available ring quadrupoles are left un-powered. A diagram of a single FODO cell is shown in Fig. \ref{fig:altfodo-cell} with lattice elements indicated in green (dipoles) and gray (quads). The envelope solution assumes the $I=0.6$ mA pencil beam with emittance $\epsilon=7.6\ \mu$m.  

In the alternative lattice, the mid-cell location where $\beta_x = \beta_y$ corresponds roughly with the location of un-used quadrupole elements. These locations are indicated in Fig. \ref{fig:altfodo-cell} in lightly shaded blue. As the octupole PCBs were designed with the same aspect ratio as the UMER ring quadrupoles (Section \ref{sec:design:octu}), they can be installed at this location using existing magnet mounts. 

A key liberty taken with the quasi-integrable theory is the requirement that $\beta_x=\beta_y$ throughout the nonlinear element. In this case, $\beta_x \approx \beta_y$. The error $\beta{x,y} - \left<\beta_x,\beta_y\right>$ is approximately 15\%. Even assuming $\beta_x =\beta_y=$ constant, the PCB octupole field is fringe-dominated. The longitudinal profile is not flat-top and therefore the magnet does not meet the requirement that $V_{oct} = 1/\beta^3 =$ constant. Theoretical calculations of the UMER magnets predict that fringe fields cancel due to the relatively short magnet length.\cite{VenturiniThesis} It is unclear if this cancellation will help preserve the nonlinear invariant. 
Simulations presented here utilize a hard-edged approximation. 

The alternative FODO lattice is designed for a bare tune of $\nu_0 = 3.36$.\cite{Bernal2012} The tune operating point can be adjusted in a wide range by varying currents $I_F$ and $I_D$ in the focusing and defocusing quad families. 
To meet the $n-\pi$ phase advance requirement, the linear focusing ring must have a tune of $\frac{n}{2}+ \delta$ where $\delta$ indicates the tune advance through the octupoles. $\delta$ is approximated as

\begin{equation}
\delta = 2\pi \psi_{oct} = \int_{oct}\frac{ds}{\beta(s)} \approx \frac{N L_eff}{\left<\beta(s)\right>}
\end{equation}

\noindent for effective octupole length $L_{eff}$ and number of octupole inserts $N$. With four octupole inserts length $L_{eff}=5$ cm and average beta function $\left<\beta(s)\right> \sim 0.5$ m, the tune advance through the octupole is $\delta \approx 0.06$. 

As discussed in Chapter \ref{ch:qio}, the maximum octupole induced tune spread $\delta \nu$ is roughly equal to the tune advance through the octupoles. Therefore, the distributed lattice tune spread scales with the number of octupole inserts $N$. 
$N=18$ and $N=36$ lattices have tune spreads comparable to the single-channel design, with maximum $\delta \nu = 0.25$. However, as the linear lattice is only tunable in the range $\nu \approx 3.36 \pm 1$, the number of inserts should be restricted to $N\leq9$. Therefore, the resonance suppression effect in the distributed lattice is expected to be weaker than in the single-channel experiment.

\section{Simulations of distributed octupole lattice}

\subsection{Frequency map analysis} 

\begin{figure}[tb]
\subfigure[N36 octupole lattice]{ \centering
	\includegraphics[width=.3\textwidth]{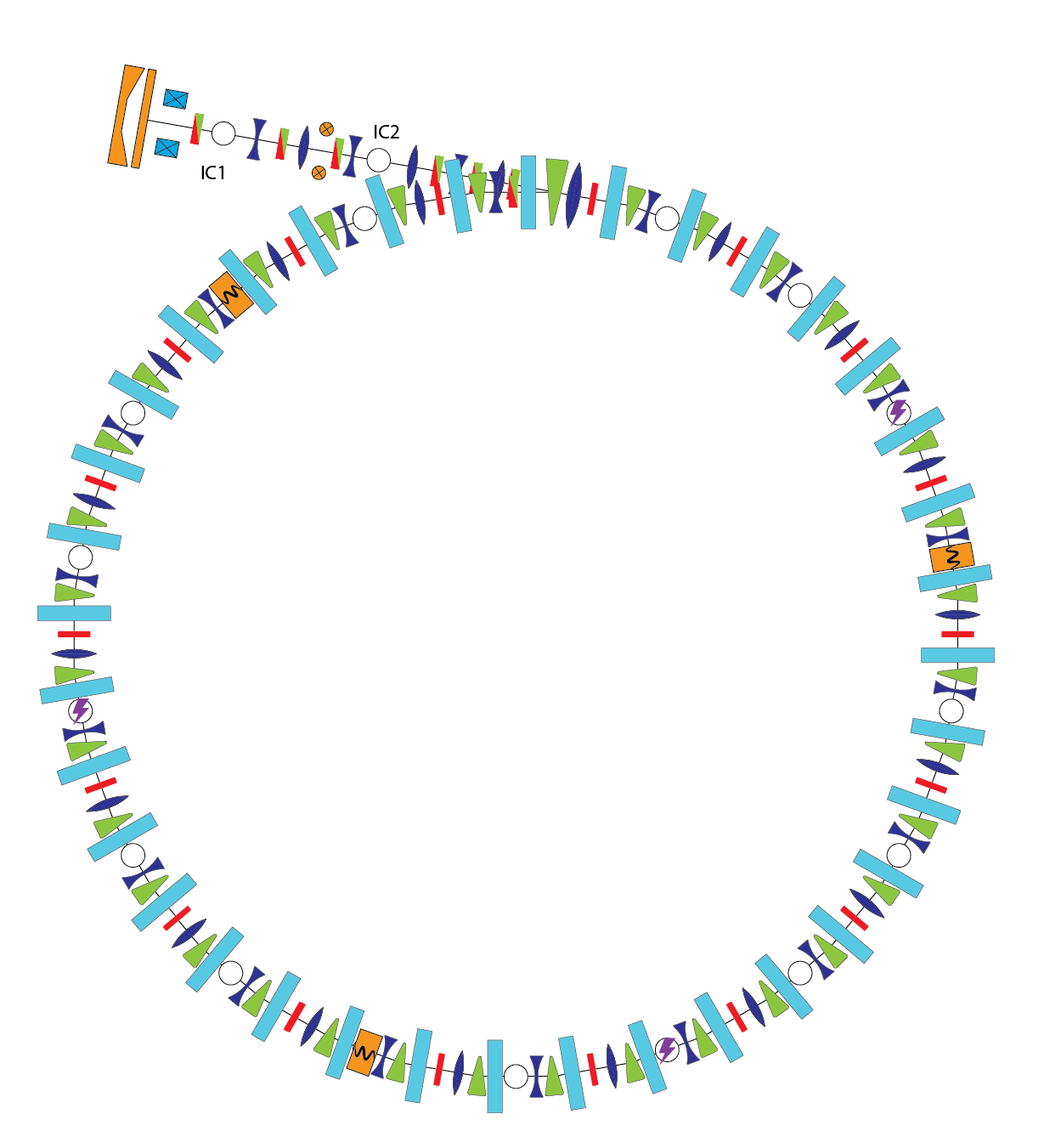}
	}
\subfigure[N9 octupole lattice]{ \centering
	\includegraphics[width=.3\textwidth]{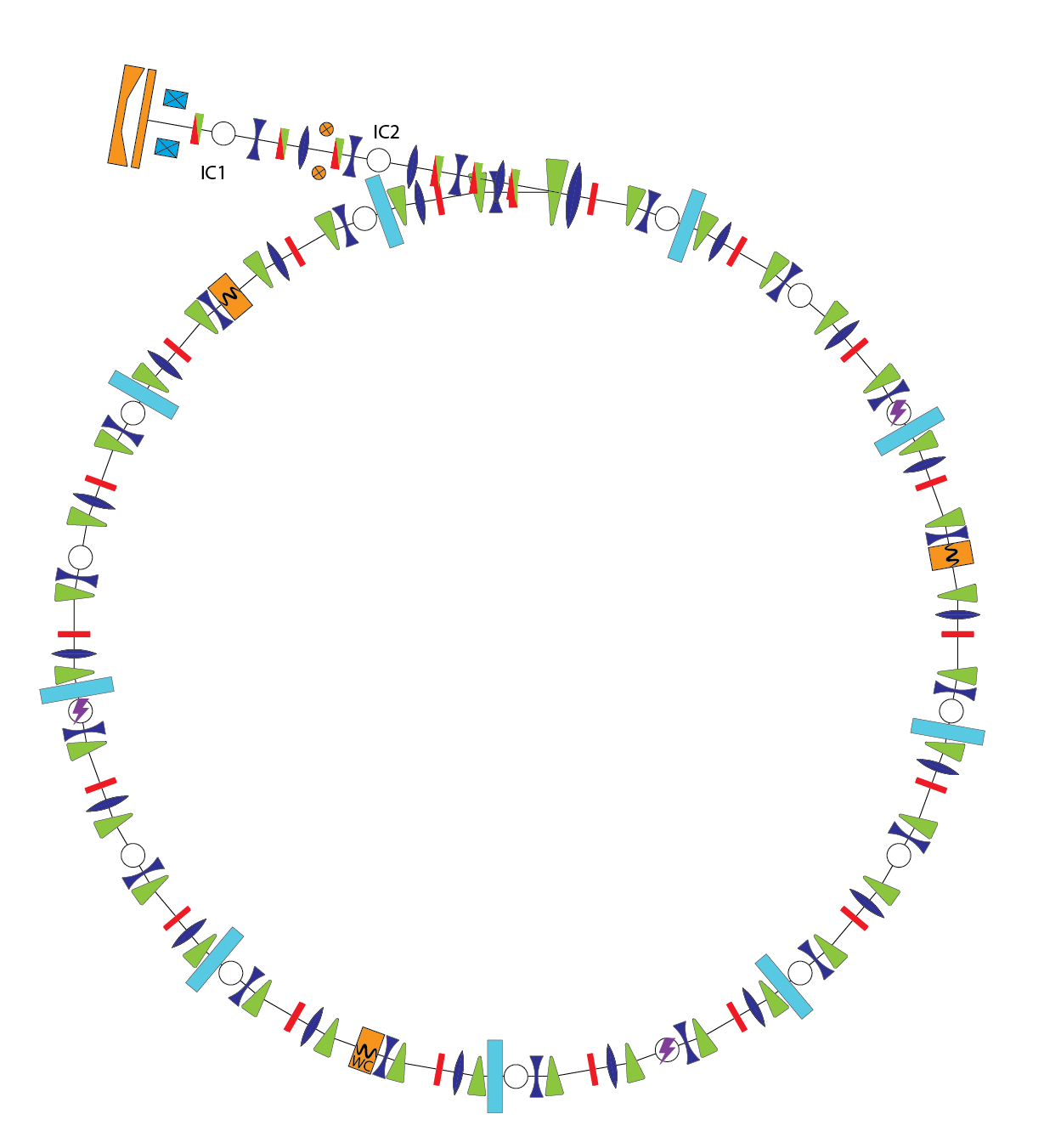}
	}
\subfigure[N4 octupole lattice]{ \centering
	\includegraphics[width=.3\textwidth]{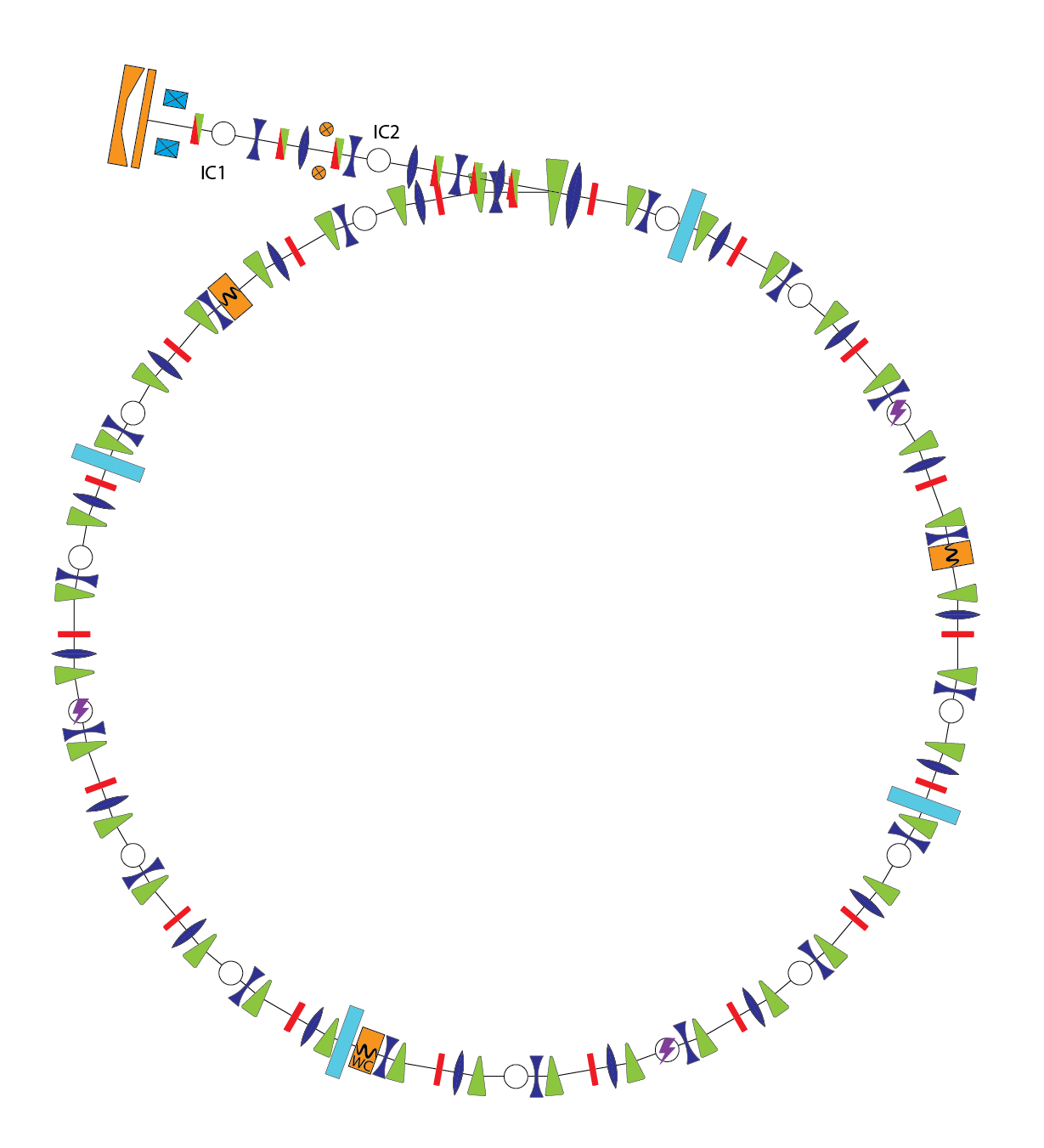}
	}	
\caption{A selection of distributed octupole lattices tested for dynamic aperture and tune spread. }
\label{fig:Noctulattices}
\end{figure}

\begin{table}[!tb]
\centering
\caption{Parameters and results for three configurations of distributed octupole lattice. $\nu_T$ is tune advance between octupole inserts. }
	\vspace{10pt}
	\begin{tabular}{ccccc c}
	\hline
	\# octupoles & Separation [m] & $\nu_{x,T}$ & $\nu_{y,T}$ & DA [mm] & tune spread $\delta \nu$ \\
	\hline
	36 & 0.32 & 0.11 & 0.11 & 7.0 & 0.1 \\
	9 & 1.28 & 0.48 & 0.49 & 1.5 & 0.05 \\
	4 & 2.88 & 1.10 & 1.12 & 2.5 & 0.1\\
	\hline
	\end{tabular}
\label{tab:Noctlattices}
\end{table}

There are several configurations of distributed octupoles that result in a periodic lattice. Three cases are shown in Fig. \ref{fig:Noctulattices}. Here $N\#$ is used to identify lattices by the number of octupole inserts. For the three cases shown, frequency map analysis was used to predict maximum dynamic aperture and tune spread for each case. Integrated octupole gradient and lattice tune was held fixed between cases. These simulations uses the Elegant \cite{elegant} model of the ring (see Section \ref{sec:numeric:elegant} and Appendix \ref{ap:model} for more details).

\begin{figure}
	\subfigure[Configuration space, in units of RMS beam size $\sigma$]{\centering
    \includegraphics[width=0.5\textwidth]{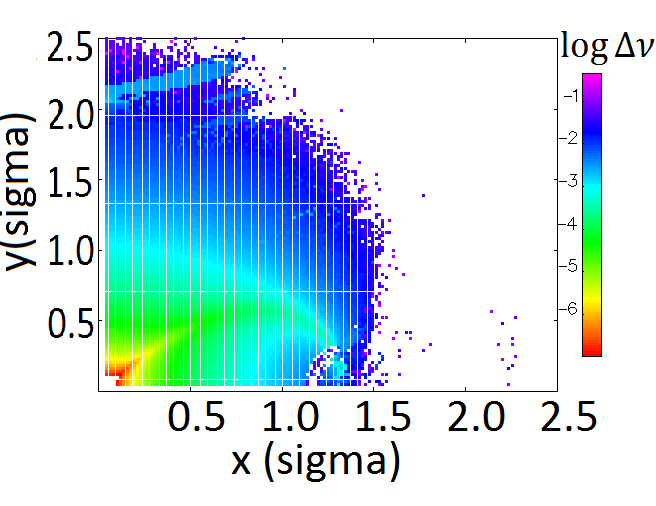}
	}
	\subfigure[Tune footprint in fractional tune space]{\centering
	\includegraphics[width=0.5\textwidth]{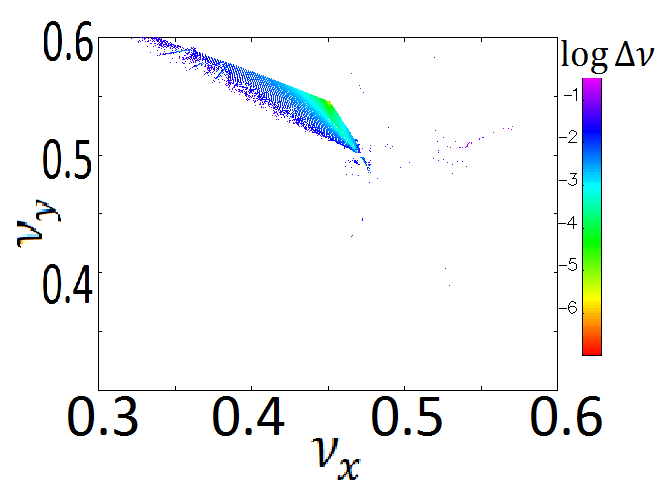}
	}
 	\caption{Frequency map analysis of N4 lattice in configuration and tune space. }
   \label{fig:N4fma}
\end{figure}

The operating point for Elegant calculations was at ring tunes $\nu_x=4.45$, $\nu_y=4.54$. The resulting tune advance between centers is given in Table \ref{tab:Noctlattices}. The $N4$ lattice possessed a larger dynamic aperture than the $N9$ case. The $N36$ case gave large tune spread and large dynamic aperture, but was not investigated further for the reason stated above. In all cases, tune spreads were comparable to the expected $\sim 0.06$.

The results for the $N4$ are plotted in Fig. \ref{fig:N4fma}. 
Here, results are plotted in terms of RMS beam size $\sigma$ for the 0.6 mA pencil beam. 
The region of stability is limited by the half-integer band $\nu_y = 4.5$ (seen in Fig. \ref{fig:N4fma} in lower left corner of stable region). The ideal operating point at $\nu_x = \nu_y = 4.06$ is farther away from this resonance.

\subsection{Tracking $H_N$ invariant}

\begin{figure}[tb]
   \centering
    \includegraphics[width=0.7\textwidth]{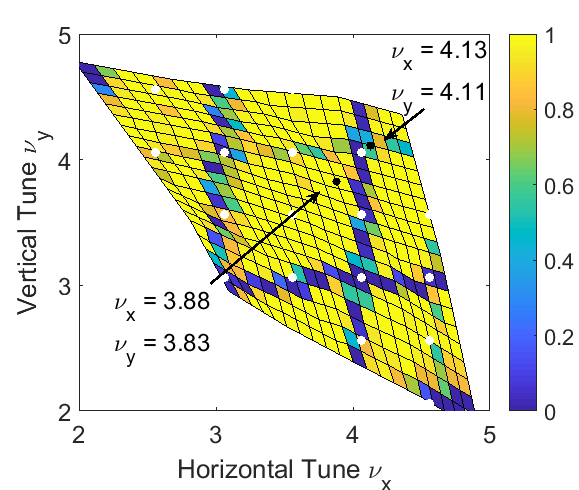}
 	\caption{Beam survival plot simulated in WARP, using hard-edged elements and thin multipole kicks as driving terms. }
   \label{fig:warpscan}
\end{figure}

Studies of the $N4$ lattice were followed up with quantification of the invariant $H_N$ conservation (Eq. \ref{eq:H-norm}) using a WARP model of the ring. Two operating points both nearby and far from quasi-integrable conditions were compared.
The WARP model uses a hard-edged approximation for ring elements, described in Appendix \ref{ap:model}. Hard-edged octupoles of length 5.2 cm and peak strength $75 T/m^3/A$ are placed at 2.88 m intervals.
Fig. \ref{fig:warpscan} shows a survival plot for a simulated tune scan in the WARP model. As the smooth-focusing approximation (used to transform quadrupole current to tune for the 32 cm FODO lattice, Section \ref{sec:apparatus:tune-scan}) is less valid for a less-dense FODO cell, the tune map is generated using probe particles in the WARP model. 

Two cases are considered:  The historically utilized alternative lattice operating point $I_F=I_D=0.87$ A, which has a tune (as calculated in WARP) of $\nu_x=3.88$, $\nu_y=3.83$ and $I_F=0.938 A$, $I_D=0.944 A$, with tunes $\nu_x=4.13$, $\nu_y=4.11$. The two operating points are marked in Fig. \ref{fig:warpscan}. $H_N$ is tracked over 50 turns.

\begin{figure}[]
   \centering
	\subfigure[$I_{oct}=0$ A]{ \centering
    \includegraphics[width=\textwidth]{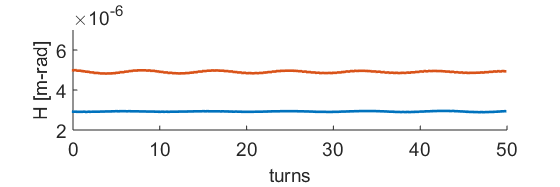}
	}
	\subfigure[$I_{oct}=2$ A]{\centering
    \includegraphics[width=\textwidth]{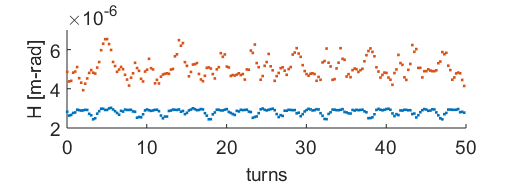}
	}
	\subfigure[$I_{oct}=4$ A]{ \centering
    \includegraphics[width=\textwidth]{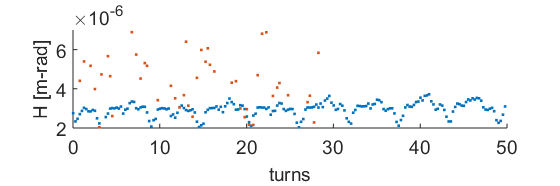}
	}
 	\caption{Invariant $H_N$ for N4 distributed octupole lattice at $\nu_x=3.88$, $\nu_y=3.83$. }
   \label{fig:N4invar1}
\end{figure}

\begin{figure}[]
	\subfigure[$I_{oct}=0$ A]{ \centering 
    \includegraphics[width=\textwidth]{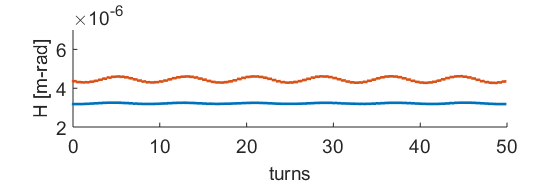}
	}
	\subfigure[$I_{oct}=2$ A]{ \centering
    \includegraphics[width=\textwidth]{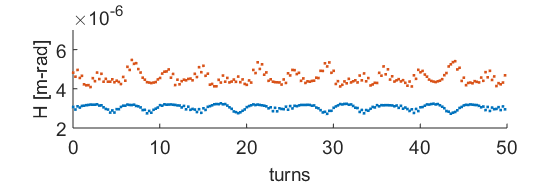}
	}
	\subfigure[$I_{oct}=4$ A]{ \centering
    \includegraphics[width=\textwidth]{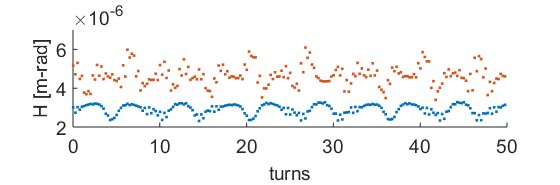}
	}
 	\caption{Invariant $H_N$ for N4 distributed octupole lattice at $\nu_x=4.13$, $\nu_y=4.11$. }
    \label{fig:N4invar2}
\end{figure}

The variation in $H_N$ is reported in Table \ref{tab:N4invar} for both operating points. Invariant tracking for two sample particles, launched at fixed initial phase space coordinates, are plotted in Fig. \ref{fig:N4invar1}  and Fig. \ref{fig:N4invar2}.
One expects the invariant to be perfectly conserved in the linear case ($I_{oct}=0$). Low-level variation on the order of $2\%$ is seen at both operating points.
As the octupole strengths are increased, $H_N$ becomes ``less-conserved" and larger oscillations are observed.
However, $H_N$ variation is lower for the case nearer the $\nu_x=\nu_y=4.06$ quasi-integrable condition. This suggests that improved stability is gained from operating near integrability in the distributed lattice. 
Simulation studies at the $\nu_x=\nu_y=4.06$ were not fruitful, as the proximity to integer resonances led to instability.

A natural extension of this work is to extend consideration to a wider range of operating points. Fig. \ref{fig:warpscan} indicates all the quasi-integrable operating points with white dots. Unfortunately, due to the small $\delta$, all ideal operating points are very near integer and half-integer resonance bands.

\begin{table}[tb]
   \centering
   \caption{Invariant $H_N$ conservation in N4 distributed octupole lattice. }
   \begin{tabular}{lccc} \hline
	   $\nu_x=4.13$ & $\nu_y=4.11$\\ \hline
       \textbf{$I_{oct}$ [A]} & \textbf{$\langle H_N \rangle$}        &RMS variation             & \% peak-to-peak variation   \\
           0        & 3.22E-6    &2.3E-8      & 2.4       \\ 
          0.5       & 3.17E-6     &4.2E-8     & 6.2       \\ 
           2.0        & 3.05E-6    &1.1E-7     &17.6       \\ 
           4.0      & 2.91E-6      &2.1E-7   & 33.5        \\ \hline
	   $\nu_x=3.88$ & $\nu_y=3.83$\\ \hline
       \textbf{$I_{oct}$ [A]} & \textbf{$\langle H_N \rangle$}          &RMS variation          & \% peak-to-peak variation \\
           0        & 2.92E-6      &1.5E-8    & 2.3       \\ 
          0.5       & 2.90E-6      &3.5E-8    & 6.8       \\ 
           2.0        & 2.82E-6     & 1.0E-7    & 20.8       \\ 
           4.0      & 2.93E-6       & 1.4E-7   & 59.9        \\

   \end{tabular}
   \label{tab:N4invar}
\end{table}

\clearpage
\section{Preliminary measurements}

To create the $N4$ lattice, octupole PCBs were installed at the locations of QR4, QR22, QR40 and QR58. Only odd-numbered quads were used in the alternative FODO lattice configuration.
The injection quadrupole currents were set according to a matched solution found via envelope integration in \cite{Bernal2012}. 
Prior to data collection, orbit corrections were made using the algorithm described in Chapter \ref{ch:steering}. 
The 0.6 mA pencil beam was used for the initial measurements. This beam is predicted to have tune spread $\delta \nu = 0.87$ in the alternative FODO lattice.\cite{Bernal2012} 

Transmission in an $N4$ distributed octupole lattice was characterized. As there are many quasi-integrable conditions in the available tune range, a tune scan was performed using the method described in Section \ref{sec:apparatus:tune-scan}. 
A tune scan measures beam loss/survival as a function of ring tune. Tune is varied by adjusting currents in two families of quadrupoles (horizontally focusing and defocusing, notated as $I_F$ and $I_D$). 
For this experiment, ring quadrupoles were varied in the range $I=0.7\to1.0$ A, which covers a tune range $\nu = 2.5 \to 4.5$. Measurements were repeated for a variety of octupole strengths $I_{oct}=0\to1.5$ A.

\subsection{Tune scan results}

\begin{figure}[]
	\centering
	\subfigure[Alternative lattice tune scan ]{
	\centering
    \includegraphics[width=0.7\textwidth]{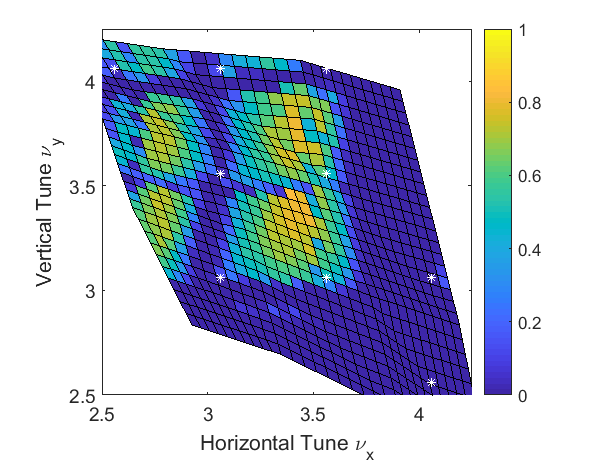}
	\label{fig:oct-tune-scan-pencil-a}}
	\subfigure[N4 lattice tune scan with octupoles powered at 0.5 A]{
	\centering
    \includegraphics[width=0.7\textwidth]{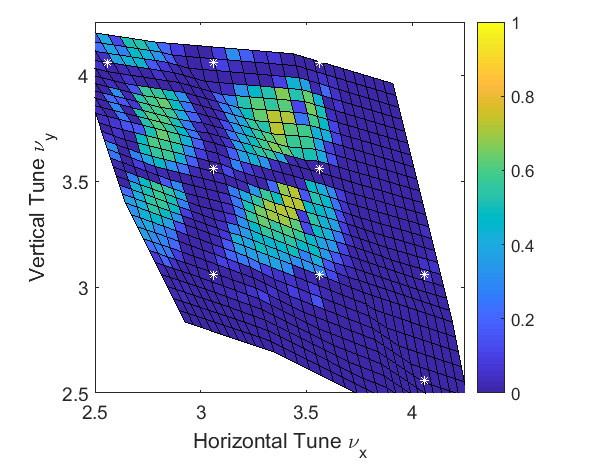}
	\label{fig:oct-tune-scan-pencil-b}}
 	\caption{Beam survival plot for 0.6 mA ``pencil" beam  at turn 25. Color axis is current normalized to 10th turn. }
   \label{fig:oct-tune-scan-pencil}
\end{figure}

Results from the linear lattice ($I_{oct} =0$ A) are compared with an octupole lattice ($I_{oct} =0.5$ A) in Fig. \ref{fig:oct-tune-scan-pencil}.
The lattice tune is different from the model prediction, so the integer resonance bands are used to orient the measurement in tune-space. A correction of $\nu_x \to \nu_x -0.45$ and $\nu_y \to \nu_y -0.35$ from the WARP prediction is applied. 

The tune scan of the linear lattice (Fig. \ref{fig:oct-tune-scan-pencil-a}) shows broad stop-bands around the horizontal integer and vertical integer and half-integer resonances. As the $N4$ lattice is intended to be run at $\nu_x=\nu_y=n/2 + \delta$, most ideal operating points are blocked by stop-band losses. With octupoles on (Figs. \ref{fig:oct-tune-scan-pencil-b}), no increase in beam survival is seen at any operating point. Quasi-integrable operating points are indicated with white asterisks.

\subsection{Errors in beam matching and steering}

The beam matching quadrupoles and steering correctors were optimized to a single operating point, at $I_F=I_D = 0.87$ A. It is expected that the accrued errors in the match and the steering grow with greater distance from the orbit-corrected operating point. 
First-turn horizontal orbit distortions at the ``tuned-up" operating point were measure to be RMS$x=0.5$ mm and $\max |x|= 1.3$ mm. In the vertical plane, RMS$x= 3.2$ mm and $\max |y|=8.5$ mm. The contribution of these steering errors can be seen in the width of the integer resonance bands. More accurate steering corrections will likely reduce the size of the integer bands. Section \ref{sec:steering:altlat} discusses orbit correction in the alternative lattice.

\begin{figure}[htb]
\centering
\subfigure[Transverse beam distribution from phosphor screens]{\centering
\includegraphics[width=\textwidth]{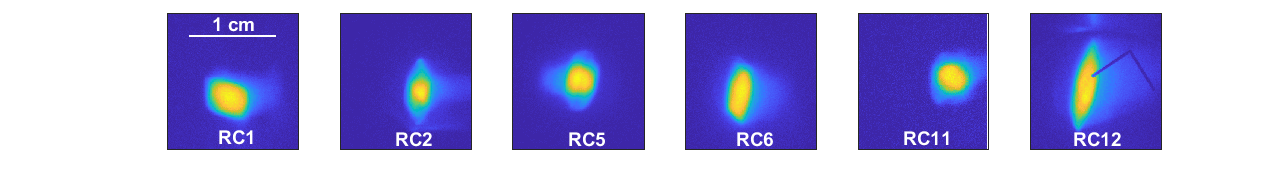}
\label{}}
\hspace{.5in}
\subfigure[Measured RMS beam sizes]{\centering
\includegraphics[width=0.8\textwidth]{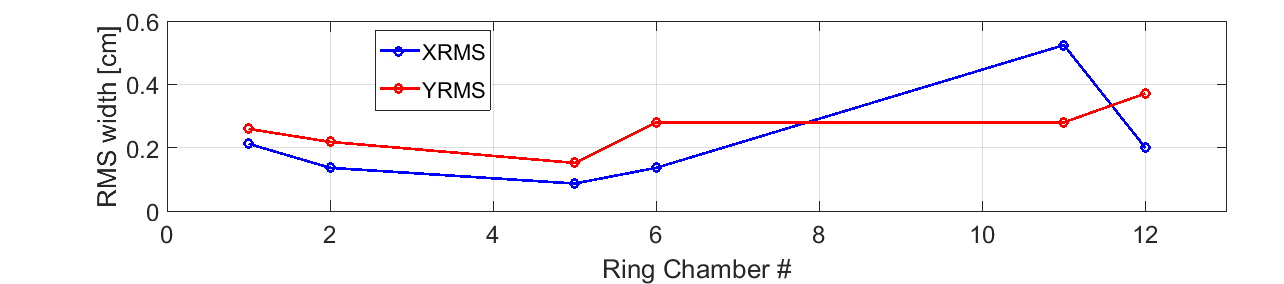}
\label{}}
\caption{Mismatch oscillations observed for measured beam profile on first turn in alternative FODO lattice. }
\label{fig:alt-match}
\end{figure}

The beam match was also not very accurate. The first-turn beam profile measurements are shown in Fig. \ref{fig:alt-match}. The RMS beam size varies 33\% in the horizontal and 28\% in the vertical plane. More accurate matching solutions have been demonstrated in UMER, up to standard deviations of 0.17 mm horizontally and 014 mm vertically for the 6 mA beam. \cite{HaoThesis} Large mismatch oscillations increase beam scraping and generally increase beam loss. 

\section{Chapter summary}

The distributed octupole lattice is natively suited to the UMER structure, allowing the installation of octupoles with minimal disruptions to the ring (utilizing existing mounts and power supplies). However, it only approximately satisfies the quasi-integrable condition. It is expected that these approximations will limit the extent to which the Hamiltonian $H_N$ is conserved and the lattice is stable even for strong octupole potentials. Tracking of $H_N$ in an $N4$ lattice shows that the invariant is less conserved (experiences large and irregular oscillations) when octupole fields are included. However, this diffusion of the invariant is less strong when the operating point is near the quasi-integrable condition. 

Characterization of the 0.6 mA pencil beam in the $N4$ distributed octupole lattice via tune scan shows that, for all operating points, current loss is increased when octupole fields are included. However, initial measurements were limited in scope. Better results may be obtained through more careful tuning of the beam match and orbit correction to reduce scraping losses and shrink the width of integer resonance bands. 
 
The 0.6 mA beam tune spread ($\delta \nu = 0.87$) is large compared to the predicted octupole-induced spread in the N4 lattice ($\delta \nu = 0.06$). The $40\ \mu$A DC beam is a better candidate for experiments, although orbit tolerances will be tighter due to the large average beam size.
Additionally, the experiment was limited to 25 turns due to space-charge driven bunch erosion and inter-penetration. Losses over short path lengths are dominated by scraping and low-order centroid resonances. Increasing path length by operating with longitudinal confinement or at lower current will allow for observation of slower-acting resonances.
Finally, lattices with a higher density of octupoles should be considered, as they will have a stronger and more measurable effect.

%% file: 10.Chapter.tex
\renewcommand{\thechapter}{10}

\chapter{Steering and Orbit Correction}
\label{ch:steering}


Precise control of the beam centroid is necessary for good recirculation with low scraping losses and small integer stop-bands. 
In general we desire the beam to be centered with respect to the quadrupole magnets.
This results in the least amount of coupling between focusing strength and orbit distortion. This is essential for measurements like tune scans, in which beam transmission is measured over a wide range of tune operating points. 

In the context of nonlinear lattice experiments, the beam must be well-centered through the octupole insert.
As discussed in Section \ref{sec:design:steering} the dynamic aperture of the quasi-integrable octupole lattice is greatly reduced when the orbit distortion from magnetic center exceeds $0.1$ mm. 
Additionally, the QIO experiments will use a low-current, high-emittance beam with relatively weak focusing (tune $\nu \sim 3.3$ compared to the nominal UMER operation at $\nu \sim 6.7$). 
A weakly-focused, large emittance beam corresponds to a large beam cross-section in configuration space and therefore tighter orbit tolerances are required to minimize scraping.

Conventionally in accelerator design, a \textit{reference trajectory} is defined as the ideal path of an on-energy design particle. This trajectory is typically centered within the focusing (quadrupole) elements. 
The \textit{closed} or \textit{equilibrium orbit} is a continuous, closed path around the ring that is defined by the strength of the steering dipoles. Ideally, the closed orbit is close to the reference trajectory and centered in the quadrupoles. 
The closed orbit only exists for ``good" choice of steerer strengths and in general may have large excursions from the reference trajectory.
A beam that is not injected on the closed orbit will oscillate about it with an amplitude that depends on the injection error. \textit{First-turn orbit} is defined as the path of the centroid on the first turn only, which is only identical to the closed orbit in the case of perfect injection.
In general we desire lowest-possible deviation of the first-turn orbit from the quadrupole centers and low oscillation amplitude in subsequent turns.

This chapter describes an approach to steering that minimizes centroid position in the quads (referred to as \textit{quad-centering}) and presents results using the 6 mA beam as the test case. Section \ref{sec:steering:rigidity} discusses the effect of ambient fields on the UMER beam. 
Section \ref{sec:steering:mags} describes the magnets used for steering and orbit correction. 
Section \ref{sec:steering:steeringalgorithm} describes the general approach to quad-centering in the first turn and uses particle-tracking code VRUMER\footnote{See Section \ref{sec:numeric:vrumer}} to test the algorithm and predict ``best case" results in the limit of no mechanical mis-alignments. 
The approach to minimizing closed orbit oscillations is relegated to Appendix \ref{ap:steering}.
Section \ref{sec:steering:results} shows measured results from lab implementation and Section \ref{sec:steering:SSV} shows improvement in the vertical plane after additional correctors are installed. 
Section \ref{sec:steering:nlumer} discusses orbit control in the context of QIO experiments, including both the single-channel and distributed octupole designs. 
Finally, Section \ref{sec:steering:annealing} presents an application of a global stochastic optimization method to orbit correction.

All centroid data is shown in the co-moving beam frame coordinates $(x,y,s)$.\footnote{Frenet-Serret coordinate system, as discussed in Section \ref{sec:theory:linear}.} In all plots, $+x$ is radially outwards and $+y$ is (naturally) in the upwards direction. 

\section{Considerations for low-rigidity electron beam} \label{sec:steering:rigidity}
\begin{figure}
\centering
\includegraphics[width=\textwidth]{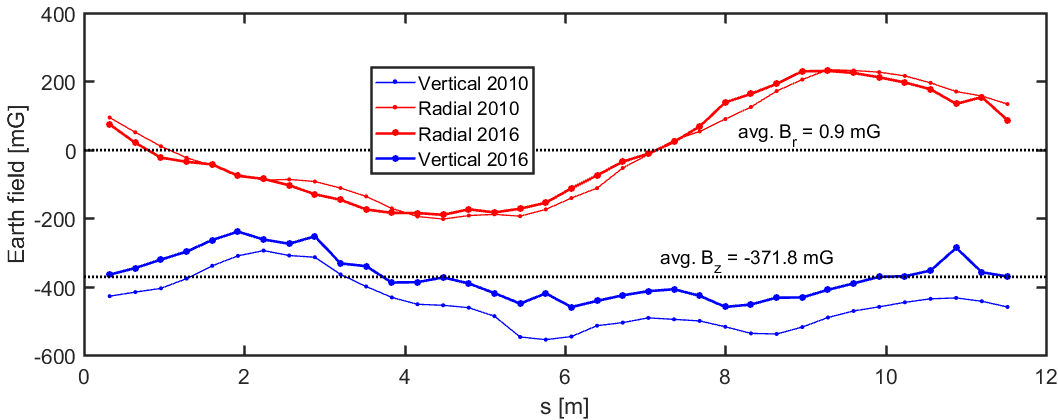}
\renewcommand{\baselinestretch}{1}
\small\normalsize
\caption{Ambient fields measured at UMER dipoles, from Dave Sutter measurements 6/1/2010 and 7/22/2016. }
\label{fig:earthfield}
\end{figure}

\renewcommand{\baselinestretch}{2}
\small\normalsize
\begin{figure}[th]
\centering
\includegraphics[width=0.8\textwidth]{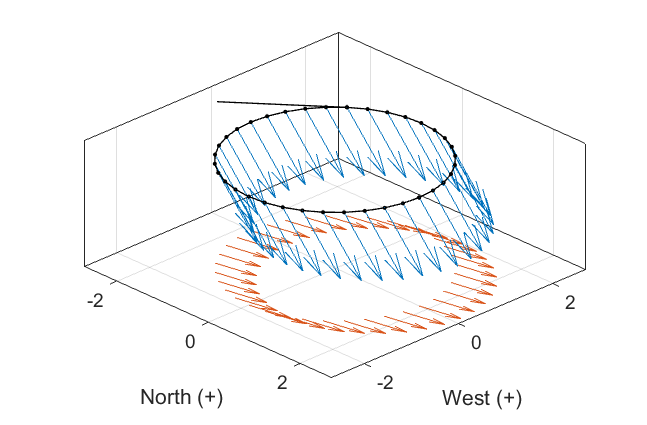}
\renewcommand{\baselinestretch}{1}
\small\normalsize
\caption{Earth field vectors, including $xy$ projection. $x$,$y$ units are meters. }
\label{fig:earthfield3D}
\end{figure}
\renewcommand{\baselinestretch}{2}
\small\normalsize

In typical high-energy rings the beam travels in straight lines between steering elements, tracing an N-sided polygon. At low energies, the beam is significantly affected by ambient background fields and there are no straight lines. At UMER, the background field that complicates steering solutions mainly originates from the Earth's magnetic field.  Measurements of the ambient field at the location of ring dipoles is plotted in Fig. \ref{fig:earthfield}.\footnote{Note that the convention  for radial fields is opposite that of centroid position. Here negative is radially-outwards, positive is radially-inwards.} 
A simplified diagram of the Earth field orientation with respect to the UMER lattice is shown in Fig. \ref{fig:earthfield3D}.
The strongest component is the vertical field. The beam is immersed in a near-constant vertical field of average strength $-372$ mG. The radial component of the field is low, with an average  strength $<1$ mG, but has a sine-like dependence in $s$ with a peak amplitude of $\approx  210$ mG. 

A 10 keV electron beam has magnetic rigidity $B\rho = 338.859$ G-cm. The required integrated field to bend the beam $10^o$ is $59.154$ G-cm. Given an average ambient vertical field of $\approx 372$ mG per cell, the integrated field is $ 32 * 0.400 = 11.9$ G-cm. Therefore, $\sim 20\%$ of the total horizontal bending in the ring is provided by the ambient vertical fields. The ambient horizontal field is weaker and gives only a small average orbit distortion over each turn that can be compensated for with a weak corrector. However, local closed orbit distortions due to the horizontal fields can be large and, as a result, relatively strong vertical corrections are necessary to maintain a vertically ``flat" orbit. 

\begin{figure}
\centering
\includegraphics[width=0.5\textwidth]{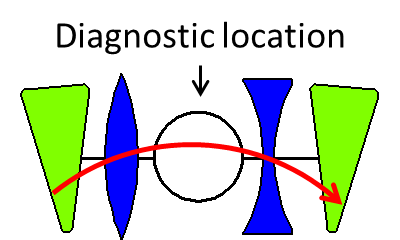}
\renewcommand{\baselinestretch}{1}
\small\normalsize
\caption{Diagram of beam trajectory in BPM diagnostic (circle) for an orbit that is horizontally centered in the quadrupoles. Dipole elements are green, quads are blue. }
\label{fig:BPMcartoon}
\end{figure} 
\renewcommand{\baselinestretch}{2}
\small\normalsize

Because the beam is immersed in a bending field, in the perfectly aligned case a beam orbit that is centered in the quadrupoles is required to be displaced in the BPMs, demonstrated for the horizontal trajectory in Fig. \ref{fig:BPMcartoon}. Simple calculations with a constant background field show that we expect the ideal orbit in the BPMs to be radially displaced by $+0.93$ mm.

\section{UMER steering magnets} \label{sec:steering:mags}

\begin{figure}
\begin{center}
\includegraphics[width=\textwidth]{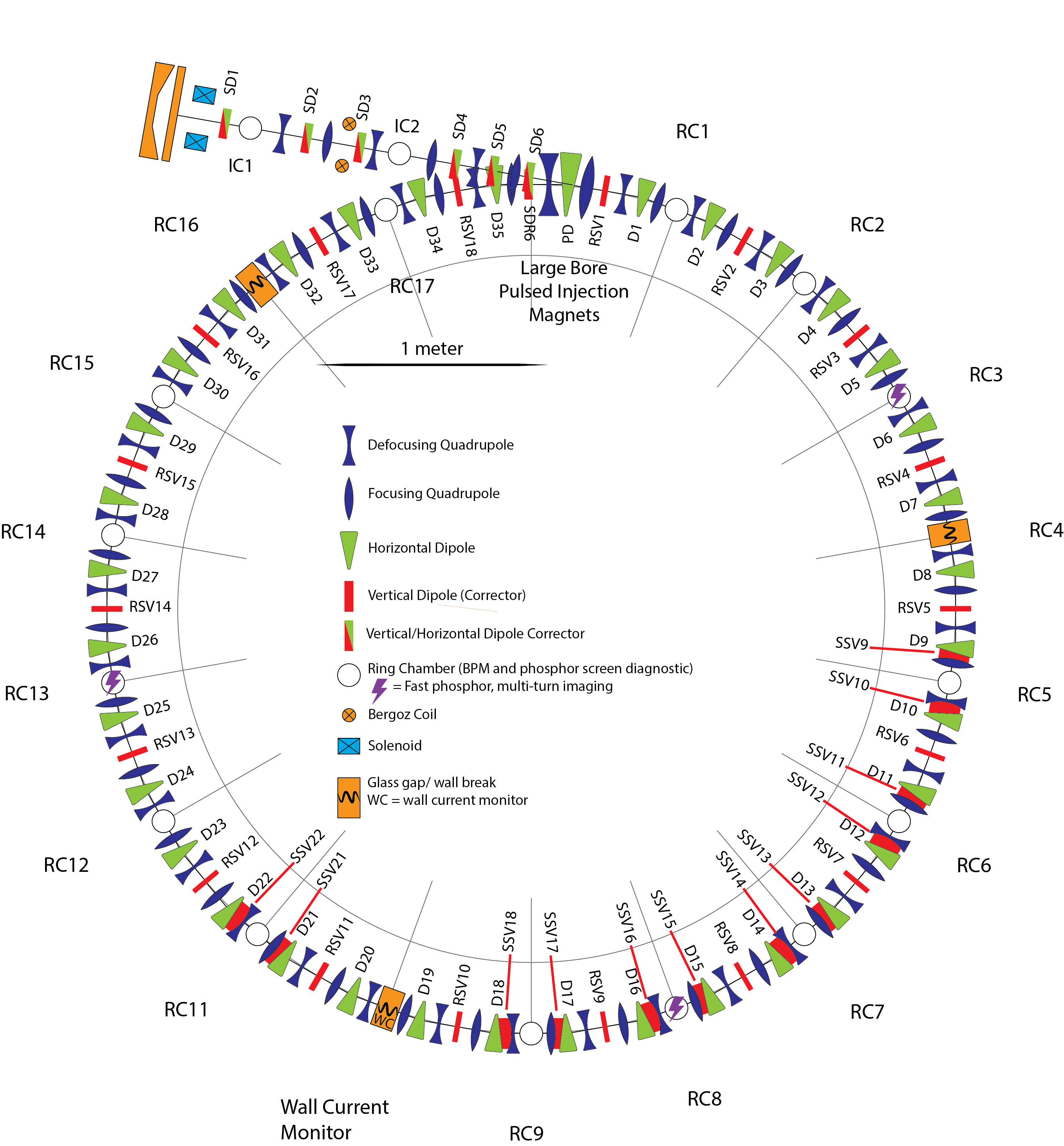}
\end{center}
\renewcommand{\baselinestretch}{1}
\small\normalsize
\begin{quote}
\caption{UMER Diagram with all steerers labeled. Quads are indicated in dark blue. SSV family is discussed in Section \ref{sec:steering:SSV}. }
\label{fig:steerers-labelled}
\end{quote}
\end{figure}

Horizontal steering in UMER is controlled by 36 bending dipoles (BD) which can be independently adjusted for optimization. 
There is 1 independent horizontal dipole for every 2 quads.  
Vertical correction is made with 18 vertical ring steerers (RSV) located at the flanges between $20^o$ sections. There is 1 RSV corrector for every 4 quads. Additional short vertical correctors (SSV) have been installed, as described in Section \ref{sec:steering:SSV}. 
Location of all steerer magnets in the ring (not including injection line steerers) is shown in Fig. \ref{fig:steerers-labelled}. To avoid heat-damage, steerer set-points are limited to $<2$ A. For bending dipoles, in which the aluminum mount acts as a heat-sink, this limit is extended to $<3$ A.

\renewcommand{\baselinestretch}{2}
\small\normalsize

\section{Orbit correction algorithm} \label{sec:steering:steeringalgorithm}

Historically, on-line orbit optimization in UMER has been made on the basis of minimizing centroid offset from the centers of the beam position monitors (BPMs). A global correction can be made using a response matrix technique, in which the effect of each corrector is measured at each BPM and the resulting matrix inverted to find an optimized steering solution.\cite{KPRnote:2011} However, orbit corrections based exclusively on BPM data utilize 14 data points per turn. With the quad-as-BPM method, \footnote{Described in Section \ref{sec:apparatus:quad-as-bpm}.} much more information is available. In addition to the 14 BPM locations, position can be measured at 71 quadrupole locations. I have developed an approach to steering that uses this additional information to ensure that a closed orbit is found with minimal excursion in the quadrupoles.

The general approach for setting all steerers is to start with the first steerer after injection and minimize a target function that depends only on the local orbit distortion. After the first is set, the algorithm proceeds one-by-one until all ring steerers have been optimized. Here the definition of ``local" is limited to quad locations between the current steerer and the next downstream steerer. As an example, QR3 and QR4 are between D1 and D2, so the target function for D1 has the form

\begin{equation}
f(I_{D1}) = F\left(z_{QR3}(I_{D1}),z_{QR4}(I_{D1})\right)
\end{equation}

\noindent where $z \in (x,y)$. The set-point for D1 is decided according to

\begin{equation}
I_{D1} = \min{f(I_{D1})}.
\end{equation}

A simple model of the UMER ring in VRUMER was used to evaluate suitability of different target functions $F$ for both the horizontal and vertical planes. The VRUMER model includes background field measurements from 2016 (Fig. \ref{fig:earthfield}) and hard-edged models of the steering magnets and ring quadrupoles. Parameters are given in Section \ref{sec:model:vrumer}. Simulations and measurements were compared for quadrupole set-point $I_Q=1.826$ A.

\subsection{Horizontal steering target}
\begin{table}[tb]
\centering
\caption{Different horizontal steering targets and their simulated performance for initial condition $x=1$ mm, $x'=0$. All position measurements are in millimeters. }
\label{tab:Hsteeringalgorithm}
\begin{tabular}{lllll}
target & min. function & RMS($x_Q$) & Max($x_Q$) & RMS($x_Q$)  \\
& & $\sigma=0$ & $\sigma=0$& $\sigma=5$mm \\
\hline\hline
$x_F$         & $\| x_F \|$                 & 5.27 & 15.36&$>$25 \\
$x_D$         & $\| x_D \|$                 & 0.26 & 0.52 & 7.60 \\
$x_F$, $x_D$  & $\sqrt{x_F^2+x_D^2}$        & 2.01 & 6.55 &10.71 \\
$x_F$, $x'_F$ & $\sqrt{x_F^2+(x_D-x_F)^2}$  & 0.26 & 0.52 & 7.76 \\
$x_D$, $x'_F$ & $\sqrt{x_D^2+(x_D-x_F)^2}$  & 0.28 & 0.64 & 7.38 \\
\hline
\end{tabular}
\end{table}

\begin{figure}[!htb]
\centering
\subfigure[No misalignments]{
	\includegraphics[width=\textwidth]{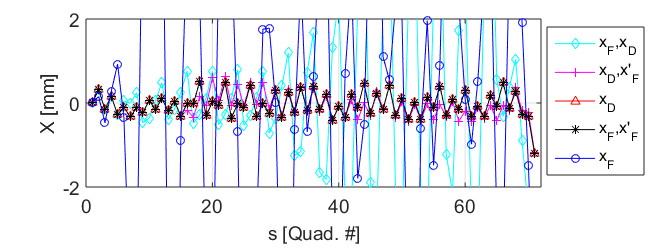}}
\hspace{.05in}
\subfigure[Random quad misalignments from Gaussian distribution, $\sigma=5$mm]{
	\includegraphics[width=\textwidth]{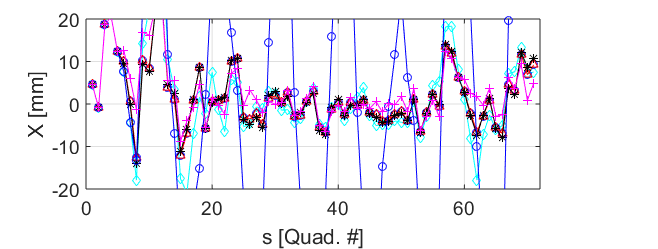}}
\renewcommand{\baselinestretch}{2}
\small\normalsize
\caption{First-turn VRUMER orbits with applied quad-centering correction for initial condition $x=1$mm, $x'=0$. }
\label{fig:steeringalgorithm}
\end{figure}
\renewcommand{\baselinestretch}{2}
\small\normalsize

The bending dipoles provide sufficient correction for local variation of the Earth field. At $3$ A the dipoles can provide $10.1^o$ of bend, but typical operating points are closer to $I=2.4$ A for $8^o$ of bend. The ambient fields bend the beam $1.3^o \to 2.5^o$ per $20^o$ section. The needed corrections are well within the safe range of dipole current.

Simulations were run for the test case of the standard UMER FODO lattice at operating point $I_Q=1.826$ A. The target function $F$ only depends on the two quads immediately downstream of a given dipole and upstream of the next dipole. The 2 downstream quadrupoles are indicated as focusing ``F" or defocusing ``D" based on polarity in the horizontal plane (QR1 is a focusing quad). In the standard UMER FODO configuration the nearest downstream quadrupole is focusing, as sketched in Fig. \ref{fig:BPMcartoon}.
 
The target function can depend on position $x$ and angle $x'$. The set of targets considered are listed in Table \ref{tab:Hsteeringalgorithm}. In the thin lens approximation for two quads separated by a drift of distance $L$, $x'_F \equiv \frac{x_D-x_F}{L}$. The term $x'_F \propto x_D-x_F$ is to include $x'_F$ in the RMS minimization term. 

To evaluate target performance, two cases are tested: a lattice with perfect horizontal alignment of the quads, and a lattice with random quad misalignments sampled from a Gaussian distribution with standard deviation $\sigma=5$ mm (well above the estimated misalignments).
The results of multiple steering targets are shown in Table \ref{tab:Hsteeringalgorithm}. Simulated orbits with and without misalignment errors are shown in Fig. \ref{fig:steeringalgorithm}. 

Qualitatively, the best performers were $\left(x_D\right)$, $\left(x_F,x'_F\right)$, $\left(x_D,x'_F\right)$, resulting in almost identical orbits (sub-millimeter differences) that converge very quickly towards the center of the quads given an injection error or quad misalignment. In general, algorithms that give higher weight to $x_D$ perform better. There appears to be a ``lever arm" effect, where choosing a farther-away target like $x_D$ converges toward a flatter trajectory, while choosing a too-near target like $x_F$ leads to over-correction and large orbit offsets. Before performing these tests it was assumed that equal weighting of all the available information (i.e., taking the RMS of $\left(x_D,x_F\right)$) would lead to a good orbit, but better results are found by simply aiming for the center of the downstream defocusing quad. 

A single variable target function like $(x_D)$ is attractive as the data collection time is much faster. Fitting is also simpler as the measured response is linear.
However, early measurements indicate the $\left(x_F,x'_F\right)$ target may yield better results. 
This may be because of its ability to handle relative misalignments between the focusing and defocusing quadrupoles, as well as reduced sensitivity to nonlinearities in the quad fields and BPM response for large centroid positions. This effect was observed prior to ring re-alignment, which reduced the magnitude of alignment errors. A follow-up using the $\left(x_F,x'_F\right)$ target was not done. 
For implementation in the lab (described in Appendix \ref{ap:steering}) and all results shown here, I chose to use $(x_D)$ as the target function due to the shorter time needed per iteration.

Based on VRUMER predictions, the ``best-case" orbit tolerances for a perfectly aligned ring are $\max{(x)} \sim 0.5$ mm, rms$(x) \sim 0.3$ mm. Reducing these tolerances requires shielding the ambient fields or trying a different target function other than the options listed here.

\subsection{Vertical steering target}

\begin{table}[!tb]
\centering
\caption{Performance of vertical steering targets. }
\label{tab:vert_algorithm}
\vspace{.5in}
\begin{tabular}{llcccccc}
target& min. function & RMS($y_Q$)  & $\max{y_Q}$ & RMS($y_Q$) & $\max{y_Q}$\\
      &               & $I\le2.5$ A &             & $I\le2$ A  &             \\
\hline \hline
none  & RSV current=0 & 6.54       & 19.95       & --         & --    \\ 
\hline
$y_1$ & $\| y_1 \|$   & 9.47        & $>25$       & 7.74       & 24.00 \\
$y_2$ & $\| y_2 \|$   & 6.03        & 18.00       & 5.06       & 15.14 \\
$y_3$ & $\| y_3 \|$   & 2.51        & 8.44        & 2.46       & 8.23  \\
$y_4$ & $\| y_4 \|$   & 1.82        & 4.81        & 2.02       & 6.44 \\
\hline
$y_1$,$y_2$  & $\sqrt{\frac{1}{2}\left(y_1^2 +y_2^2\right)}$ & 8.93        & $>25$       & 7.24       & 23.88 \\
$y_2$,$y_3$  & $\sqrt{\frac{1}{2}\left(y_1^2 +y_3^2\right)}$ & 2.84        & 10.16       & 2.68       & 9.55  \\
$y_3$,$y_4$  & $\sqrt{\frac{1}{2}\left(y_3^2 +y_4^2\right)}$ & 2.07        & 6.22        & 2.16       & 6.30 \\
$y_1$,$y_3$  & $\sqrt{\frac{1}{2}\left(y_1^2 +y_3^2\right)}$ & 3.05        & 11.28       & 2.81       & 10.28 \\
$y_2$,$y_4$  & $\sqrt{\frac{1}{2}\left(y_2^2 +y_4^2\right)}$ & 1.88        & 5.02        & 2.07       & 6.20 \\
\hline 
$y_1$,$y'_1$ & $\sqrt{\frac{1}{2}\left(y_1^2 +(y_2-y_1)^2\right)}$ & 6.10        & 18.63        & 5.09       & 14.71 \\
$y_2$,$y'_2$ & $\sqrt{\frac{1}{2}\left(y_2^2 +(y_3-y_2)^2\right)}$ & 2.25      & 6.87       & 2.28      & 6.99 \\
$y_3$,$y'_3$ & $\sqrt{\frac{1}{2}\left(y_2^2 +(y_4-y_3)^2\right)}$ & 3.28        & 11.78        & 3.03       & 11.08 \\
$y_1$,$y_2$,$y_3$,$y_4$ &  $\sqrt{\frac{1}{4}\left(y_1^2 +y_2^2+ y_3^2 +y_4^2\right)}$ & 2.50        & 8.83        & 2.44       & 8.22 \\
\hline
\end{tabular}
\vspace{.3in}
\end{table}
\renewcommand{\baselinestretch}{2}
\small\normalsize

Vertical steering in UMER is accomplished by 18 vertical correctors (RSVs) located at the pipe flanges every $20^o$. All of these magnets are fairly weak (see Table \ref{tab:UMERsteererstrength}) and there are half the number as horizontal dipoles (for a density of 1 corrector for every 4 quadrupoles).  The largest source of vertical alignment errors is the radial component of the Earth's field, plotted in Fig. \ref{fig:earthfield}. At most, the radial field bends $\sim 2.5^o$ over $20^o$ of the ring, while the corrector at 2 A excitation supplies $1.2^o$ of correction. It is already apparent that it is not possible with the existing RSV correctors to fully compensate for the ambient radial field at certain ring locations.

In an approach similar to the horizontal plane, VRUMER is used to test various vertical steering algorithms and place a lower bound on the best-possible first-turn orbit using existing vertical steerers. Two cases with perfect alignment are considered for steerer currents limited to $< 2.0$ A and $<2.5$ A.
The results are summarized in Table \ref{tab:vert_algorithm}.\footnote{Note that these results are valid for the stated test case only. Changes in the focusing lattice and background field can significantly change the statistics, although the better-performing targets tend to do well in all cases.}
All units are in millimeters and subscript indicates quad \# counting downstream from vertical steerer ($y_1$ is position in first quad downstream from each RSV).

\begin{figure}[!tb]
\centering
\includegraphics[width=\textwidth]{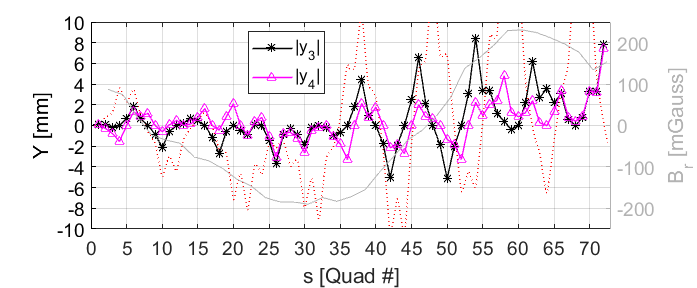}
\renewcommand{\baselinestretch}{2}
\small\normalsize
\caption{Simulated first-turn vertical orbits, using $\| y_3 \|$ and $\| y_4 \|$ as the targets for setting RSV magnets. Uncorrected orbit is shown with red-dashed line and Earth field with gray. }
\label{fig:vert_sim}
\end{figure}
\renewcommand{\baselinestretch}{2}
\small\normalsize

The most successful targets for steering when the RSVs are current-limited to $\le2.5$ A are $\left(y_4\right)$, and $\left(y_2,y_4\right)$. 
One observes the same ``lever arm" effect as in the horizontal plane - a greater distance between the steerer and the target results in a flatter orbit. Choosing a target too close to the steerer leads to over-correction, and the steered solution might be worse than the uncorrected solution!

Simulated results, applying steering with the targets $\left(y_3\right)$ and $\left(y_4\right)$, are plotted in Fig. \ref{fig:vert_sim}. With perfect alignment and corrector currents up to 2 A, we can obtain vertical steering with $\max{(y)} \sim 6.4$ mm, rms$(y) \sim 2.0$ mm. Increasing the current limit slightly to 2.5 A reduces these values to $\max{(y)} \sim 4.8$ mm, rms$(y) \sim 1.8$ mm. Interestingly, for all algorithms tested, allowing the current limit to increase to 10 A did not reduce the orbit statistics. For the well-performing targets, currents up to 4 A were found to be optimal, but the RMS and maximum orbit excursions also increased (slightly). 

Steering was also tested in the presence of vertical misalignments, in the cases $\sigma = 0.1$ mm and $\sigma = 1$ mm. The performance of the quad-centering algorithm was almost identical to the aligned case for $\sigma = 0.1$ mm (which is close to the expected mechanical tolerance). Large excursions were seen in the $\sigma = 1$ mm case (RMS $\sim10$ mm, $\max{\sim20}$ mm).

As predicted, the ``best-case" for vertical orbit control has larger offsets, with $\max{(y)} \sim 4.8$ mm, rms$(y) \sim 1.8$ mm, than in the horizontal plane, with $\max{(x)} \sim 0.5$ mm, rms$(x) \sim 0.3$ mm. Existing vertical correcters are too weak and spread out to provide orbit control equivalent to the horizontal plane, despite the average radial field being much smaller than the vertical field. VRUMER results suggest that increasing the strength of the RSV's (up to 4 A) will not result in better control. Instead, more correctors are required. 
An upgrade in the form of additional weak vertical correctors is described in Section \ref{sec:steering:SSV}.


\section{Corrected beam orbit} \label{sec:steering:results}

\begin{figure}[t]
\centering
\subfigure[Horizontal position in quads for 1st turn. Data is from 10/23/17 steering solution.]{\label{fig:horz_quads}
	\includegraphics[width=\textwidth]{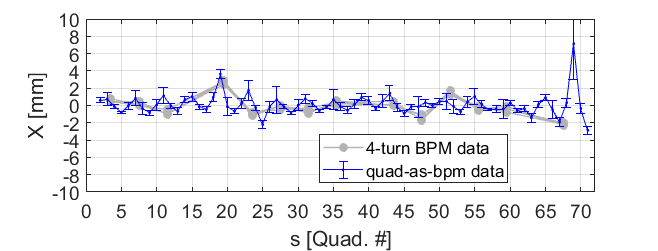}}
\hspace{.05in}
\subfigure[Horizontal position in BPMs for first 4 turns. Data is from 10/23/17 steering solution.]{\label{fig:horz_BPMs}
	\includegraphics[width=\textwidth]{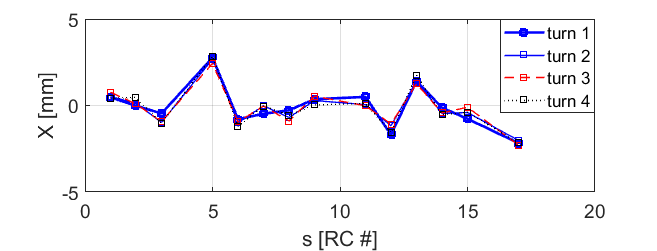}}
\renewcommand{\baselinestretch}{2}
\small\normalsize
\caption{Example of good horizontal orbit obtained for 6 mA beam with quad-centering method. }
\label{fig:horz-result}
\end{figure}

\begin{figure}[tb]
\centering
\subfigure[Vertical position in quads for 1st turn. Data is from 9/1/17 steering solution.]{\label{fig:vert_quads}
	\includegraphics[width=\textwidth]{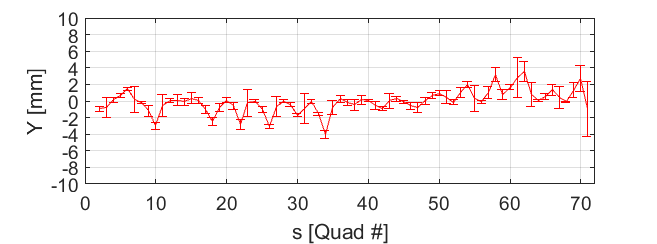}}
\hspace{.05in}
\subfigure[Vertical position in BPMs for first 4 turns. Turn-to-turn amplitude minimized with a raster scan approach. Data is from 12/16/16 steering solution.]
	{\label{fig:vert_BPMs}
	\includegraphics[width=\textwidth]{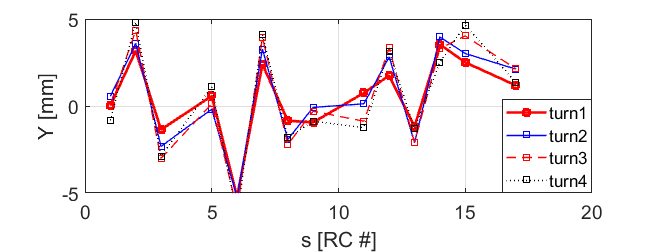}}
	\hspace{.05in}
\renewcommand{\baselinestretch}{2}
\small\normalsize
\caption{Example of good vertical orbit for 6 mA beam obtained with quad-centering method. }
\label{fig:vert_result}
\end{figure}

Results of applying the horizontal and vertical quad-centering algorithm are shown in Figures \ref{fig:horz-result} and \ref{fig:vert_result}\footnote{Minimization of turn-to-turn oscillation amplitude is achieved with 2D raster scan.}. 
The target used for the horizontal plane was $\|x_D\|$. Both $\|y_3\|$ and $\|y_4\|$ were tested for vertical steering. Both performed well, but $\|y_3\|$ seemed to yield more dependable results. 
Multi-turn orbit solutions are obtained by applying the quad-centering algorithm for the first turn, then the last two steerer magnets before the Y-section (D34/D35 for horizontal, RSV17/RSV18 for vertical) are used to minimize turn-to-turn deviation from the first-turn injected orbit.
Statistics for the plotted data sets are given in Table \ref{tab:steeringresults}. 

In general, control over the amplitude of oscillations about the closed orbit to $\sim0.5$ mm is possible. Closed orbit oscillations are quantified by calculating the turn-to-turn deviation in the BPM data. This is defined here as

\begin{equation}
\Delta_n = \|x_n-x_1\|
\label{eq:delta}
\end{equation}

\noindent for $n>1$. In the horizontal solution in Fig. \ref{fig:horz-result} the maximum turn-to-turn deviation in the first four turns is $\max \Delta = 0.67$. The RMS value for turns $2 \to 4$ is RMS$\Delta = 0.36$ mm.  
The turn-to-turn control shown in Fig. \ref{fig:vert_BPMs} is less impressive ($\max{\Delta} = 2.00$ mm, RMS $\Delta = 0.93$ mm), but improved control is possible.
The technique applied in Fig. \ref{fig:vert_BPMs} uses a brute-force raster-scan search for minimum RMS$\Delta$ as a function of the two corrector strength. The resolution is limited by the step size for the raster scan, and the data collection time is lengthy. In comparison, the Robust Conjugate Direction Search (RCDS) algorithm, used for minimization of RMS$\Delta$ in Figs. \ref{fig:horz_BPMs}, has proven to be both more effective and time-efficient.\cite{Huang2013, LevonRCDS} Results with RCDS were $\max{\Delta} = 0.53$ mm, RMS $\Delta = 0.22$ mm.

\begin{table}[!tb]
\begin{threeparttable}
\caption{Orbit statistics for quad-centering steering method, using 6 mA beam as test case. }
\label{tab:steeringresults}
\centering
\begin{tabularx}{\textwidth}{X R{4cm} R{4cm}}
Statistics & Horz. [mm] & Vert. [mm] \\
\hline \hline
First turn RMS in quads& 1.24$\ast$& 1.32$\dagger$ \\
First turn Max in quads& 3.64      & 4.05          \\
Four turn RMS in BPMs  &1.17       & 2.81$\ddagger$\\
Four turn Max in BPMs  &2.85       & 6.32          \\
RMS $\Delta$ (Eq. \ref{eq:delta})& 0.36      & 0.93\\
Max. $\Delta$          & 0.67      & 2.00          \\
\hline
\end{tabularx}
\renewcommand{\baselinestretch}{2}
\small\normalsize
\begin{tablenotes}
\footnotesize
\item $\ast$ Horizontal data is from 10/23/17 steering solution, with recirculation tuned by RCDS method.
\item $\dagger$ Vertical quad-as-BPM data is from 9/1/17 steering solution for first turn orbit.
\item $\ddagger$ Vertical BPM data is from 12/16/16 steering solution, with recirculation tuned by 2D raster scan method.
\end{tablenotes}
\end{threeparttable}
\end{table}
\renewcommand{\baselinestretch}{2}
\small\normalsize

\begin{figure}[!tb]
\subfigure[]{
\includegraphics[width=\textwidth]{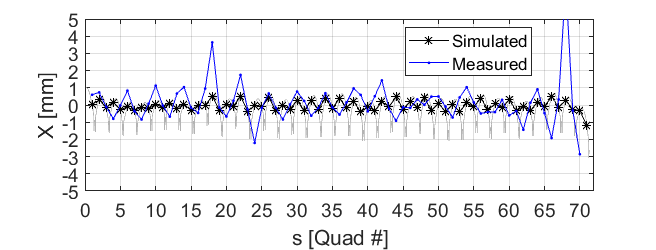}
\label{fig:horz_quads_vs_sim}}
\subfigure[]{
\includegraphics[width=\textwidth]{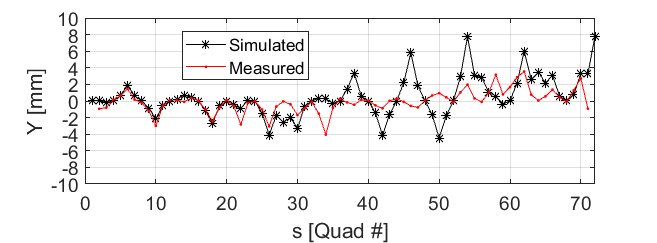}
\label{fig:vert_quads_vs_sim}}
\renewcommand{\baselinestretch}{2}
\small\normalsize
\caption{Simulation/measurement comparison for best-case orbit obtained via quad-centering. Simulated orbit is shown in quads (black) and (in \ref{fig:horz_quads_vs_sim}) at higher resolution along s (light gray). Spikes indicate dipole locations, where a coordinate transformation is applied to unwind ring. }
\label{fig:quads_vs_sim}
\end{figure}
\renewcommand{\baselinestretch}{2}
\small\normalsize

\clearpage
While we can control oscillations about the closed orbit relatively well, the present limiting factor for orbit tolerances is the closed orbit offset caused by background fields and mechanical misalignments. The data in Fig. \ref{fig:horz_quads} shows a near-best-case solution for the first turn horizontal closed orbit. Fig. \ref{fig:horz_quads_vs_sim} compared the simulated ``best-case" results against this solution. In the lab, we measured RMS$x_Q\sim1.2$ mm and $\max{x_Q}\sim 3.6$ mm, while the simulation predicts RMS$x_Q\sim0.3$ mm and $\max{x_Q}\sim 0.5$ mm.

However, the large excursions in Fig. \ref{fig:horz_quads_vs_sim} are localized. Between these locations, the horizontal distortions are close to smallest-possible. The maximum horizontal mis-alignment of the quadrupoles, based on survey data from 2016, is $<0.1 mm$. As this is much smaller than the observed orbit distortion, the dominant ``misaligning" must be due to the background fields. This is apparent by comparing the orbit statistics for the simulation results when considering orbit as measured in the quads (RMS$x_Q\sim0.3$ mm and $\max{x_Q}\sim 0.5$ mm) and as measured at all points in s (RMS$x(s)\sim0.6$ mm and $\max{x(s)}\sim 2.1$ mm). The quad-centered orbit has minimal position at the quad locations and largest offset right between quads. 

Vertical orbit statistics are close to VRUMER predictions. Fig. \ref{fig:vert_quads_vs_sim} shows excellent agreement for the first half of the ring (up to QR28) before the trajectories start to diverge. In fact, the lab implementation appears to perform slightly better than predicted in simulation.
In the lab, we measured RMS$y_Q\sim2.8$ mm and $\max{y_Q}\sim 6.3$ mm while the simulation predicts RMS$y_Q\sim2.5$ mm and $\max{y_Q}\sim 8.4$ mm. 
The vertical orbit is also limited by the effect of background fields. In Fig. \ref{fig:vert_quads}, characteristic arcs are visible with a periodicity of four quads (for example, between QR18 and QR22). This is due to the constant beam bending in the background field. Additionally, one observes that the regions with largest deviation from quad centers corresponds roughly with the peak radial field.

\section{Decreasing vertical orbit distortion}\label{sec:steering:SSV}

\begin{figure}[t]
\centering
\includegraphics[width=0.7\textwidth]{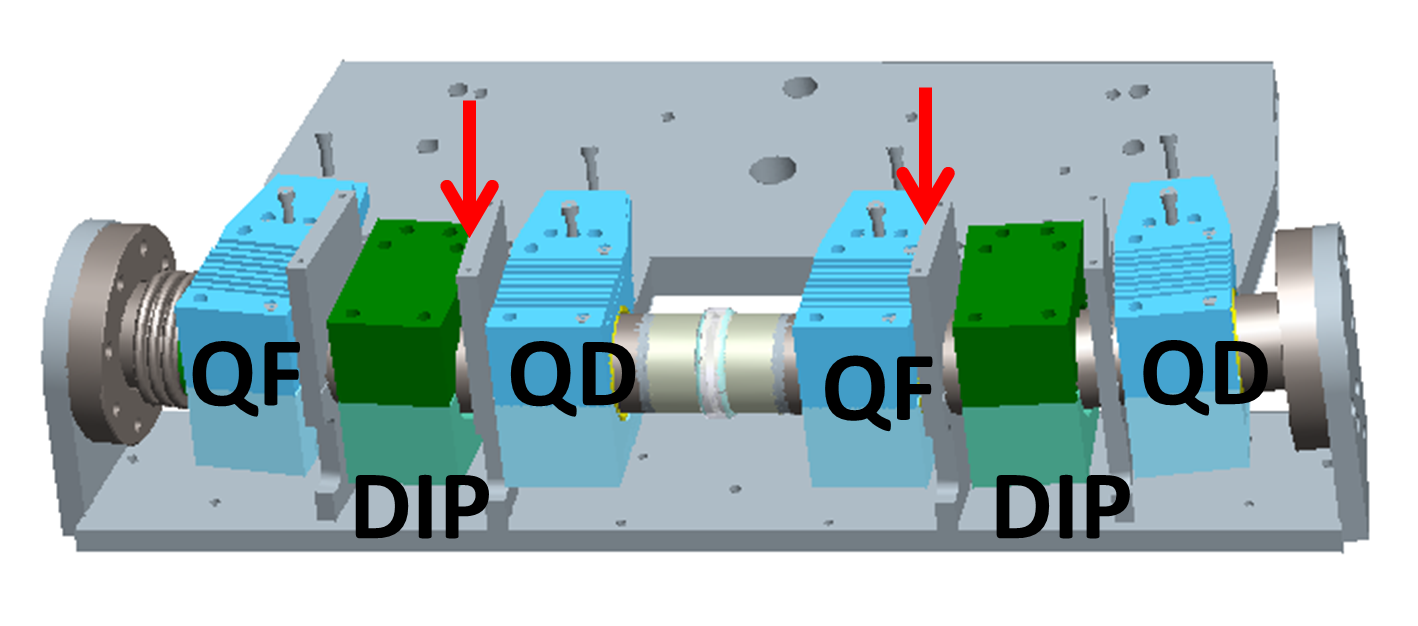}
\renewcommand{\baselinestretch}{2}
\small\normalsize
\caption{Locations of SSV's on UMER $20^o$ plate indicated with arrows. RSV's are located at vacuum flanges at ends of $20^o$ section. }
\label{fig:SSVlocation}
\end{figure}
\renewcommand{\baselinestretch}{2}
\small\normalsize

As discussed in the previous section, vertical orbit correction is inherently limited by the strength and number of the RSV correctors.
Previous assumption was that sparsely populated, low-field vertical correctors were sufficient to correct for the low average radial field, but the above results show that large local offsets will be present when the correctors are too weak to fully compensate for the effects of local fields. 

I tested the effectiveness of increasing steerer density by adding short vertical steerers (SSVs) to the ring. These are short PCBs of physical length $1.54$ cm identical to magnets already in use in the injection line.
There is space in the dense UMER lattice for two additional thin SSVs per $20^o$ plate, as shown in Fig. \ref{fig:SSVlocation}. From Table \ref{tab:UMERsteererstrength}, the available correction of each SSV is $\sim1.2^o$ per amp, comparable to RSV strength. With twice the density of the RSV and a similar limitation of $\le2$ A, there should be enough strength to compensate for the maximum ambient field bend of $\sim 2.4^o$ over $20^o$.

\begin{figure}[]
\centering
\includegraphics[width=\textwidth]{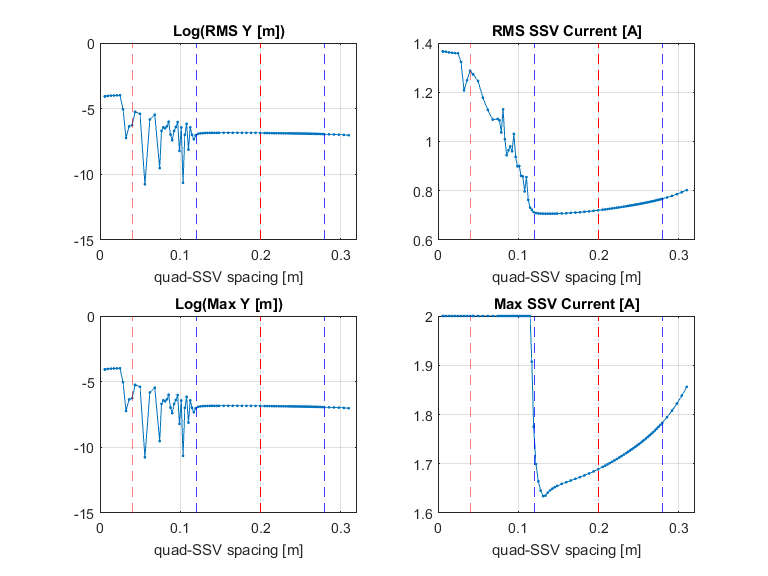}
\renewcommand{\baselinestretch}{2}
\small\normalsize
\caption{Steering statistics for varying steerer-target distance. Quad center locations are indicated by dashed lines. Red dashes indicate distance from 1st SSV on $20^o$ plate, blue from 2nd SSV on $20^o$ plate. }
\label{fig:lever_arm}
\end{figure}
\renewcommand{\baselinestretch}{2}
\small\normalsize

There are two quads between each SSV, similar to the BDs in the horizontal plane. Applying the same approach as in the horizontal plan, the steering algorithm aims for the center of the second downstream quad, $\|y_2\|$. Because of the uneven spacing of the SSV magnets in the $20^o$ plate, the SSV-target spacing varies from cell to cell. Fig. \ref{fig:lever_arm} shows the dependence of SSV strength and correction on SSV-target separation when applying the quad-centering algorithm. A choice of target that is too close (such as the first downstream quad) leads to over-correction and large SSV currents. Longer distances between steerer and target are necessary for corrections within the available strength, and there is a sharp transition from ``good steering" to over-correction for separations $< 12$ cm. The target quadrupoles for the two SSV locations are at 20 cm and 28 cm respectively, within the range of good correction.

\begin{figure}[tb]
\centering
\includegraphics[width=\textwidth]{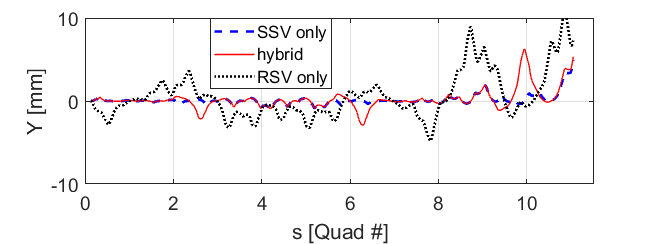}
\renewcommand{\baselinestretch}{2}
\small\normalsize
\caption{VRUMER simulated orbits for quad-centered steering with 18 RSVs (black), hybrid (red, 28 SSVs and 4 RSVs), and 36 SSVs (blue). }
\label{fig:SSVsimorbit}
\end{figure}
\renewcommand{\baselinestretch}{2}
\small\normalsize

\begin{figure}[bt]
\centering
\includegraphics[width=\textwidth]{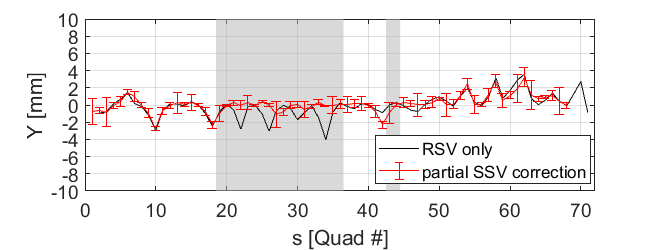}
\renewcommand{\baselinestretch}{2}
\small\normalsize
\caption{Measured first turn orbit for vertical steering with SSV's on 30\% of the ring (locations indicated by gray shading). Data taken on 8/31/17. }
\label{fig:SSVexporbit}
\end{figure}
\renewcommand{\baselinestretch}{2}
\small\normalsize

The resulting first turn orbit with SSV correction from VRUMER calculations is plotted in Fig. \ref{fig:SSVsimorbit}. There are three results shown: orbit correction using only 18 RSVs (black dot), orbit correction using only 36 SSVs (blue dash), and orbit correction using 8 RSVs and 28 SSVs. In the ring, there are four $20^o$ sections with welded glass gap breaks in the pipe (including the injection Y-section). Extra supports needed to protect the glass occupy the space needed for SSV placement. In these sections, the two RSV's bookending the $20^o$ plate are also utilized. As seen in \ref{fig:SSVsimorbit}, there are relatively large local deviations at the four glass gap sections.

Simulation results were very promising when SSVs were added to the UMER model. To test their effectiveness in the lab, 11 SSVs were installed on ring sections 5 - 11 (skipping section 10 due to the wall current monitor diagnostic).\footnote{The SSV numbering system corresponds with the nearest horizontal dipole (SSV9 is immediately downstream of dipole D9, SSV10 upstream of D10, etc).} The resulting first-turn orbit 
is plotted in Fig. \ref{fig:SSVexporbit}. Orbit statistics are in Table \ref{tab:SSVexporbit}. These can be directly compared to the orbit results without SSVs in Table \ref{tab:steeringresults}. The addition of SSVs reduces the vertical orbit deviation by a factor of $\sim 2$, almost to a tolerance of $\pm 1$ mm.   

\begin{table}[t]
\centering
\caption{First-turn measured orbit statistics with SSV correction. }
\label{tab:SSVexporbit}
\vspace{2mm}
\begin{tabular}{lc}
\hline
First turn RMS & 0.98 mm\\
First turn Max. & 3.25 mm\\
Shaded RMS & 0.45 mm\\
Shaded Max. & 1.11 mm \\
\hline
\end{tabular}
\end{table}
\renewcommand{\baselinestretch}{2}
\small\normalsize

\section{Steering for QIO UMER experiments} \label{sec:steering:nlumer}

Improved orbit control is essential for nonlinear experiments. Nonlinear inserts will imprint an amplitude-dependent tune spread on the beam, where amplitude is measured from the magnetic center of the octupole element. Ideally, the magnet center is identical to the beam center. For offset orbits, the nonlinear kick will not be symmetric about the beam center and the tune footprint will likewise be asymmetric. 
Generally, we expect smaller induced tune spreads and smaller acceptance in the case of an un-centered centroid orbit, resulting in a smaller beneficial effect and larger beam loss. This section describes the measured orbit distortion during distributed octupole tests and possible means for improvement. In light of the best-possible results shown here, the single-channel experiment can be oriented to coincide with a low-distortion region. 

\subsection{Steering for alternative FODO lattice} \label{sec:steering:altlat}

\begin{table}[htb]
\centering
\begin{threeparttable}
\caption{Orbit statistics for alternative lattice steering with 6 mA beam. }
\label{tab:steering-altlat}
\begin{tabularx}{\textwidth}{X R{4cm} R{4cm}}
Statistic & X [mm] & Y [mm] \\
\hline \hline
First turn RMS in quads $\ast$ & 2.04 & 6.31 \\
First turn Max in quads & 5.02 & 18.13 \\
Four turn RMS in BPMs $\dagger$ & 2.55 &4.40\\
Four turn Max in BPMs   & 7.52 & 8.09\\
RMS $\Delta$ (Eq. \ref{eq:delta})& 3.29 & 1.31  \\
Max. $\Delta$            & 6.50 & 2.89 \\
\hline
\end{tabularx}
\renewcommand{\baselinestretch}{2}
\small\normalsize
\begin{tablenotes}[flushleft]
\footnotesize
\item $\ast$ Quad-as-BPM data taken 1/21/17 after orbit tuning.
\item $\dagger$ BPM data taken 2/9/16 prior to distributed octupole lattice tune scans.
\end{tablenotes}
\end{threeparttable}
\end{table}

\begin{figure}[tb]
\centering
\includegraphics[width=\textwidth]{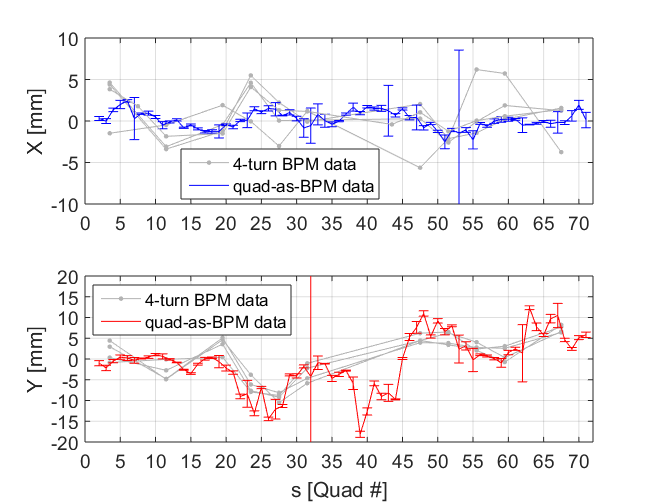}
\renewcommand{\baselinestretch}{2}
\small\normalsize
\caption{First-turn orbit for alternative lattice measured on 1/21/16. }
\label{fig:quadasbpm-altlat}
\end{figure}
\renewcommand{\baselinestretch}{2}
\small\normalsize

\begin{figure}[htb]
\centering
\includegraphics[width=\textwidth]{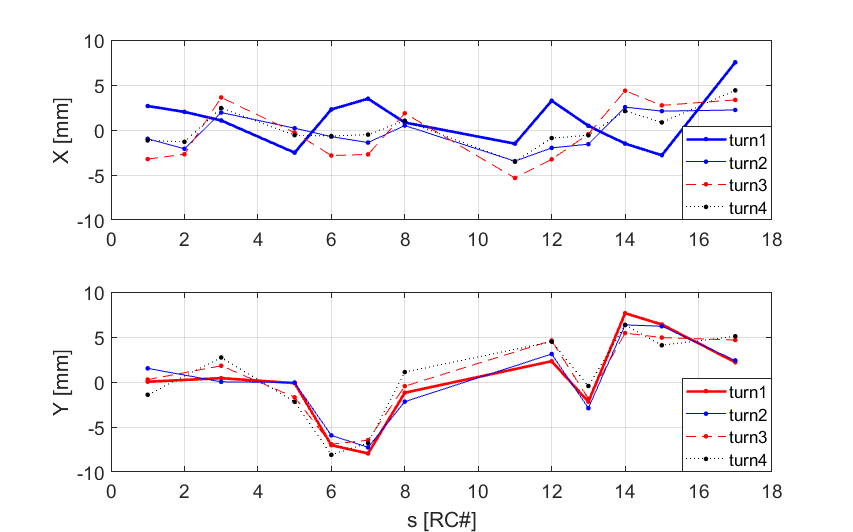}
\renewcommand{\baselinestretch}{2}
\small\normalsize
\caption{Four-turn BPM data for the alternative lattice measured on 2/9/16. 
}
\label{fig:bpmdata-altlat}
\end{figure}
\renewcommand{\baselinestretch}{2}
\small\normalsize

The distributed octupole lattice uses the alternative lattice configuration of UMER, in which half the quadrupoles are removed and the length of the FODO cell is doubled.\footnote{Described in detail in Chapter \ref{ch:distr}.} This lattice has a tune of roughly half the nominal UMER operating point, $\nu \sim 3.3$. The quad-centering technique was applied to find a steering solution for the alternative lattice prior to characterizing beam transmission in the $N4$ distributed lattice. The orbit statistics are given in Table \ref{tab:steering-altlat}.

Quad-as-BPM data for the alternative lattice is plotted in Fig. \ref{fig:quadasbpm-altlat}. The quad-centering technique was less successful in this case, most noticeably in the vertical plane. There are several reasons for this. The large vertical excursions can be attributed to the fact that RSV currents were held to $< 2.0$, rather than $<2.5$ A allowed in later tests. Also, RC9 did not house a BPM at this time and RC11 had a short on the lower BPM plate. Quad-centering between RC8 and RC12 (QR32 to QR47) was not very effective in both planes due to the high likelihood of scraping at or before RC11.

\begin{figure}[!htb]
\centering
\subfigure[PD waveform before, during and after sinple pulse lifetime.]{
	\includegraphics[width=\textwidth]{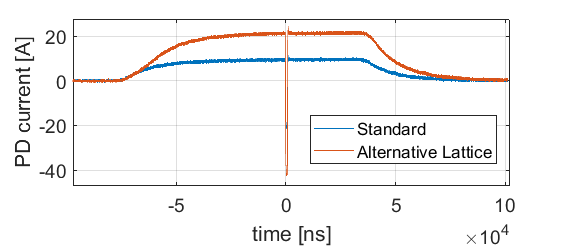}
	\label{fig:PD-far}}
\hspace{.05in}
\subfigure[Close-up of PD waveform at injection. Shaded regions indicate when beam is passing through PD.]{
	\includegraphics[width=\textwidth]{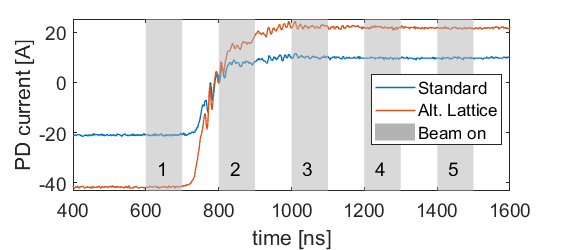}
	\label{fig:PD-closeup}}
\renewcommand{\baselinestretch}{2}
\small\normalsize
\caption{Pulsed dipole (PD) waveform is shown for both standard and alternative lattice operation.  }
\label{fig:PDtrace}
\end{figure}
\renewcommand{\baselinestretch}{2}
\small\normalsize

The horizontal orbit control for the alternative lattice is also worse than in the standard lattice. While the first turn orbit statistics are only about twice as large, the control over turn-to-turn orbit oscillations, as seen in Fig. \ref{fig:bpmdata-altlat}, is much worse. The extent to which the oscillation amplitude measured by RMS$\Delta$ (Eq. \ref{eq:delta}) can be minimized is limited by the pulsed dipole (PD) waveform. 

For every beam injection, the injection dipole PD is powered on with enough time to settle before the beam arrival. Prior to injection, PD is fast pulsed to the opposite polarity in order to kick the beam into the ring. After injection it is switched to the original polarity in order to keep the recirculating beam in the ring. This is shown in Fig. \ref{fig:PD-far}. In standard UMER operation, the pulsed injection quad YQ is set to the appropriate polarity (horizontally defocusing) so that the dipole kick imparted by YQ is the proper polarity for injected and recirculating beams. This reduces the required strength in both PD injection and recirculating kicks. However, in the alternative lattice YQ is turned off and the strength of PD must be increased. The fast PD polarity switch is not quite settled within the 100 ns ``no beam" window between injection and recirculation. This results in a weaker kick on the beam between turns 1 and 2 when compared to steady state.  

In standard operation, the injection pulse height, PD-Inj, is set to 31 A and the steady-state recirculating current, PD-rec, is set to 9.8 A. Using the parameters in Table \ref{tab:UMERsteererstrength}, this imparts a kick of $\theta = -6.8^o$ on injection and $\theta = 3.2^o$ on recirculation. However, as illustrated in Fig. \ref{fig:PD-closeup}, the difference in kick strength between turn $1 \to 2$ and the steady state is 13\% (considering the average applied kick about the center of the bunch, $\pm 20$ ns). The kick from PD between turn $1 \to 2$ is $\Delta \theta = -0.4^o$ weaker than the steady state.
In the alternative lattice, due to the absence of YQ, PD-Inj is set to 64 A and PD-rec, is set to 22 A. The strength of the PD kicks are $\theta = -13.6^o$ on injection and $\theta = 7.1^o$ on recirculation. The difference between turn $1 \to 2$ and the steady state is 16\%, which translates to $\Delta \theta = -1.1^o$, almost three times the discrepancy seen in the standard lattice. This effect is clearly seen in both the large $\Delta$ values in Table \ref{tab:steering-altlat} and the four-turn data in Fig. \ref{fig:bpmdata-altlat}. While the second through fourth turns appear to oscillate about a shared closed orbit, the first turn clearly follows a different trajectory. 

Control over horizontal orbit distortions is limited to the values given here unless there are upgrades to either hardware or tuning algorithms. As there is not much room near PD for additional correctors, a faster-switching pulser is needed. Adjusting the multi-turn tuning algorithm to optimize over turns $2 \to 5$ instead of $1 \to 4$ may be successful, as long as large distortions in the first turn orbit are acceptable. This approach may work well for smaller-emittance beams, but lead to scraping at high emittance. 

\subsection{Steering for single-channel octupole lattice} \label{sec:steering:octu}

Section \ref{sec:design:steering} sets the threshold for the orbit deviation within the long octupole element should be $<0.2$ mm. At first glance, the horizontal and vertical orbit statistics are well outside this range. However, as this tolerance is for the insertion region only and looser tolerances may be tolerated for the rest of the ring, it will be sufficient to demonstrate precision control over a single $20^o$ section.

Orbit correction results suggest RC9 (location of quadrupole magnets QR34-37) as a likely candidate for the octupole channel. The radial field is locally low in this section. The vertical orbit control is already demonstrated to be within $\pm 0.1$ mm measured in the quadrupoles on the first turn with addition of SSV corrector magnets (Table \ref{tab:RC9orbit}). However, leaving room for SSV correctors in the octupole section limits the length of the octupole channel, which should be as long as possible to maximize tune spread. Without the SSV correctors, local orbit deviations will likely be larger. Field-canceling radial Helmholtz coils at this ring section could provide this additional correction. 

The horizontal orbit deviation in RC9 is measured to be larger than $0.2$ mm. However, $\mu$-metal shielding or field-canceling vertical Helmholtz coils will be necessary for the beam to remain straight through the octupole insert. This should reduce distortion of the local horizontal orbit significantly, although the horizontal dipoles in this section will have to be run at a higher current ($\sim 3$ A) to account for the lower vertical background field.

\begin{table}[h]
\caption{Measured centroid position (in millimeters) in RC9 quads on first turn. Vertical data is from 8/31/17, horizontal data is from 10/23/17. }
\label{tab:RC9orbit}
\centering
\begin{tabular}{lcccc}
Axis & QR34 & QR35 & QR36 & QR37 \\
\hline \hline
Horizontal & $-0.18\pm0.16$ & $0.69\pm1.11$ & $-0.14\pm0.49$ & $-0.56\pm0.41$ \\
Vertical   & $ 0.10\pm0.03$ & $0.05\pm0.99$ & $-0.08\pm0.54$ & $-0.06\pm0.88$ \\
\hline
\end{tabular}
\end{table}
\renewcommand{\baselinestretch}{2}
\small\normalsize

\section{Global optimization of closed orbit} \label{sec:steering:annealing}

While the algorithm described above relies on local orbit corrections, this section describes a global approach based on BPM data. A global approach in general is faster and can be more robust to nonlinearities, although, as discussed below, we find that corrections based on BPM data only allow for large local distortions between BPMs.

While quad-centering method outlined above is successful and reproducible, there are some drawbacks. It is time consuming, taking more than one day of operator time to improve both vertical and horizontal orbits. Depending on the distance of the initial configuration to the corrected state, multiple iterations between horizontal and vertical passes may be necessary. The method is also very sensitive to and behaves poorly in the presence of nonlinearities. While the equations of motion for centroid trajectory in a lattice of quadrupoles and dipoles are linear, nonlinearities arise due to scraping of the beam, higher order magnet terms and geometric effects in the BPM. A highly nonlinear response curve results in poor choice of steerer strength and a local orbit bump. 
This issue also effects the response matrix method of tuning, which assumes linearity. An optimization algorithm that is relatively fast and requires little operator intervention to converge, while being robust to nonlinearities, would be a very powerful tool for regular use in ring tuning. 

Search for a global orbit solution is equivalent to solving an unknown nonlinear function with a large number ($N = 18-36$) of independent variables. There may be a unique ``best" solution, but within the limits of measurement noise and machine imperfections there are many ``good-enough" solutions existing in the N-dimensional parameter space. This problem appears suited for global stochastic optimization methods. Stochastic methods seek to balance sampling of a wide parameter space with efficient convergence on an optimized solution. The stochastic search reduces the probability of mistakenly converging on a local, rather than global minimum, due to the non-zero probability at each iteration that a solution will randomly jump between valleys in parameter space.

Global optimization using stochastic search methods is already applied to a variety of accelerator tuning problems. Genetic algorithms have been used to optimize parameters for individual beam-line elements, as well as tuning accelerator working points, such as skew corrections and maximum dynamic aperture at CEBAF \cite{Hofler2013} using simulation models. The application to chromaticity correction for light sources has been demonstrated experimentally at SPEAR3.\cite{Tian2013} Multi-objective genetic algorithm and particle swarm optimization were also tested for setting matching quadrupoles and RF bunching cavity operating points at the LANSCE linear accelerator.\cite{pang2014}

This section applying the principle of simulated annealing to tune the UMER orbit. Simulated annealing is a stochastic optimization technique, based on the principles of statistical mechanics.\cite{Kirkpatrick2015} It is an analog to the physical process of annealing, in which the degrees of freedom of a system are ``heated" and slowly cooled, allowing for self-organization to the lowest energy state. At high temperatures, the system has a high probability of jumping to a very different state (that is, one that is far away in parameter space). As the temperature cools, the distance between one state and a possible new state decreases, as does the probability that the system will move to that new state. 

\subsection{Procedure for simulated annealing of closed orbit}

\begin{figure}[!tb]
\centering
\subfigure[Temperature schedule.]{
\includegraphics[width=0.47\textwidth]{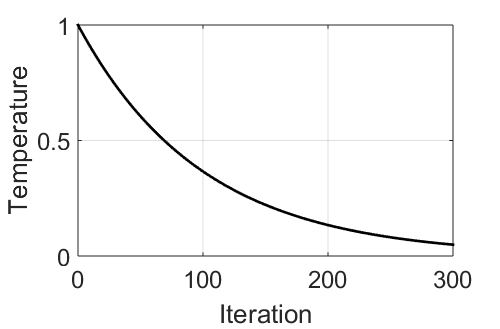}
\label{fig:temperature}}
\subfigure[Probability of jumping to new state.]{
\includegraphics[width=0.47\textwidth]{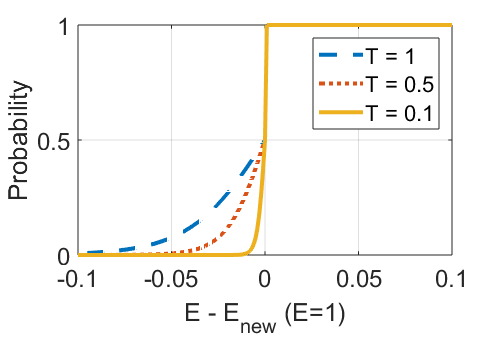}
\label{fig:probability}}
\renewcommand{\baselinestretch}{2}
\small\normalsize
\caption{Temperature and probability functions used in simulated annealing of UMER orbit. }
\label{fig:annealingtemp}
\end{figure}
\renewcommand{\baselinestretch}{2}
\small\normalsize

In this application, there are N degrees of freedom for each of the N steerer magnets in use. The ``state" of the system is simply the N-length vector of steerer currents. A time-dependent temperature schedule is defined. In this case, time is the number of iterations and temperature determines the maximum step size that the system can take when moving to another state. Energy is equivalent to a fitness (or minimization) function used to define the ``goodness" of a state. 

The state is defined as the current values for a subset of the available steerers. An initial state was chosen with good transmission in the first four turns and a centered orbit on the first turn, but otherwise not well-tuned for minimum turn-to-turn amplitude. Horizontal and vertical corrections were optimized separately.  Multiple passes of the simulated annealing algorithm were run, each with 300 iterations. Before the first step, the steerer currents of the initial state were each randomly perturbed by $\delta I$ chosen from a uniform distribution $\delta I \in [-0.1,0.1]$ A. 

The temperature schedule was set to
\begin{equation}
T = 0.990^i
\end{equation} 
\noindent where $i$ is the iteration number. The temperature versus iteration number is plotted in Fig. \ref{fig:temperature}. There is no established method for choosing an appropriate temperature schedule. Rather, this must be considered a tunable variable that affects algorithm performance. In this case, the temperature was tuned for convergence within 300 iterations. A more gradual schedule could be used, which would explore a larger range of possible states. 

For each iteration a neighboring state is randomly generated, with perturbations $\delta I_N$ selected from a Gaussian distribution with $\sigma = 0.1 \cdot T$. The ``energy" of the new neighbor state is evaluated. The fitness function used here is the RMS position of the beam in the BPMs over the first four turns:

\begin{equation}
E = RMS\left(x_{BPM,n}\right)
\label{eq:fitness}
\end{equation}

This is a very simple fitness function that overlooks some of the more subtle points of steering. More sophisticated functions are possible. Trivially, as described above in Section \ref{sec:steering:rigidity}, the center of the BPMs is not an accurate target if one is aiming for the center of the quads. 
An additional term could quantify scraping as measured on BPM sum signals or at the wall current monitor. A multi-objective optimization with loss as well as beam position may be more appropriate, as the lowest loss orbit may have local orbit bumps (particularly near injection). It would be ideal to include first-turn quad-as-BPM data in the fitness, as the BPM locations are sparse compared to quadrupole density and betatron frequency. However, this would make each iteration impossibly lengthy. 

After the energy of both states is known, the algorithm chooses whether the system stays in the current state or jumps to the neighboring state. The probability that the system jumps to the new state is
\begin{equation}
P(T,E,E_{new}) = 
\left\{
	\begin{array}{ll}
		1  & \mbox{if } E_{new} < E \\
		\left(1+e^{\frac{50(E_new - E)}{T}} \right)^{-1} & \mbox{if } E_{new} \geq E
	\end{array}
\right\}
\end{equation} 
\noindent If the neighboring state has a lower energy, the system always transitions to this state. If the neighboring state has a higher energy, there is still a probability that the system transitions. This is more likely at high temperatures or for small $\Delta E$. 

With the BPM scope averaging over eight waveforms, the evaluation of the fitness function (reading all 14 scope channels) takes $\sim 16$ seconds. 300 iterations takes 75 minutes to complete. The current bottleneck is switching multiplexer channels to read each BPM signal. 

\subsection{Orbit correction results for simulated annealing}

Simulated annealing was applied to find a steering solution in the ring. The data shown here was collected on 8/15/17 and 8/17/17. Vertical optimization was done prior to horizontal, as there is significant vertical-horizontal coupling due to skew rotation in the RSV steerers. 

\begin{figure}[!tb]
\centering
\subfigure[Fitness as a function of iteration.]{
\includegraphics[width=0.39\textwidth]{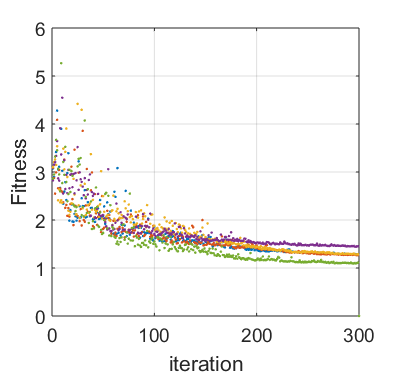}
\label{fig:vert-Ehist}}
\subfigure[Best states found on each pass.]{
\includegraphics[width=0.51\textwidth]{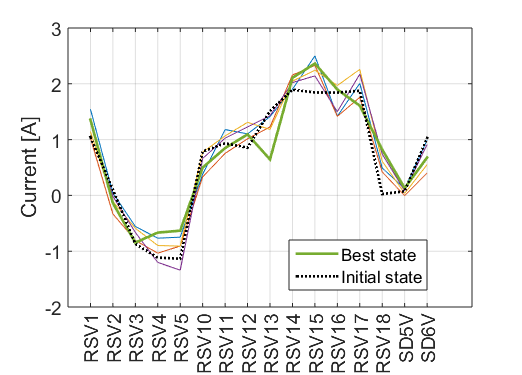}
\label{fig:vert-beststate}}
\renewcommand{\baselinestretch}{2}
\small\normalsize
\caption{Results of 5 unique passes of simulated annealing algorithm on vertical orbit in ring. }
\label{fig:vert-annealing}
\end{figure}
\renewcommand{\baselinestretch}{2}
\small\normalsize

\begin{figure}[!tb]
\centering
\includegraphics[width=\textwidth]{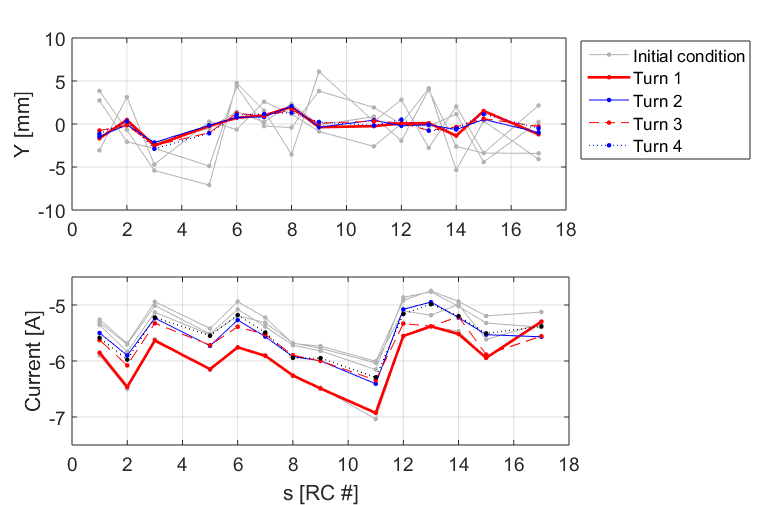}
\renewcommand{\baselinestretch}{2}
\small\normalsize
\caption{Four turn BPM data for vertical orbit after a simulated annealing pass, in comparison to orbit in initial state. Top plot is vertical position, bottom plot is signal sum (top+bottom+left+right plates). }
\label{fig:vert-pass-bpms}
\end{figure}
\renewcommand{\baselinestretch}{2}
\small\normalsize

For vertical correction, five independent trial runs were started from the same initial condition with a small random perturbation. Each pass ran for 300 iterations. The state contained $N=16$ steerers. 
Other vertical steerers were held at fixed values. The relaxation of the fitness function is shown in Fig. \ref{fig:vert-Ehist}, while the alignment of the degrees of freedom (RSV settings) is shown in Fig. \ref{fig:vert-beststate}. The four-turn orbit data for the best state found during the five passes (green curve in Fig. \ref{fig:vert-annealing}) is plotted in Fig. \ref{fig:vert-pass-bpms}. The fitness function (Eq. \ref{eq:fitness}) relaxed from a value of $E = 2.969$ to $E=1.096$ mm. This value is much lower than the RMS value found in the previous sections, RMS $y_{BPM}=2.81$ mm. For the initial state, the difference between the first and subsequent 3 turns has an RMS value of RMS$\Delta = 3.61$ mm, with a maximum of $\max \Delta_n=7.39$ mm. After annealing, these values are reduced to RMS$\Delta = 0.49$ mm and $\max \Delta_n=1.03$ mm. This is about comparable to $\Delta$ values achieved in the previous sections.

\begin{figure}[!tb]
\centering
\subfigure[Fitness as a function of iteration.]{
\includegraphics[width=0.39\textwidth]{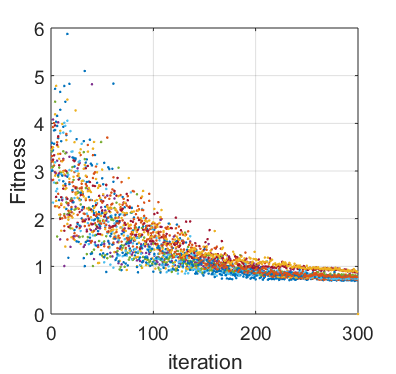}
\label{fig:horz-Ehist}}
\subfigure[Best states found on each pass.]{
\includegraphics[width=0.51\textwidth]{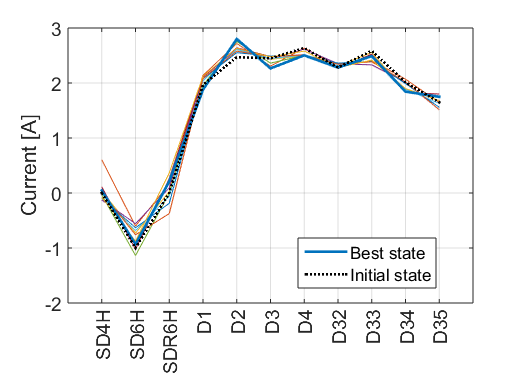}
\label{fig:horz-beststate}}
\renewcommand{\baselinestretch}{2}
\small\normalsize
\caption{Results of 10 unique passes of simulated annealing algorithm on horizontal orbit in ring. }
\label{fig:horz-annealing}
\end{figure}
\renewcommand{\baselinestretch}{2}
\small\normalsize

\begin{figure}[!tb]
\centering
\includegraphics[width=\textwidth]{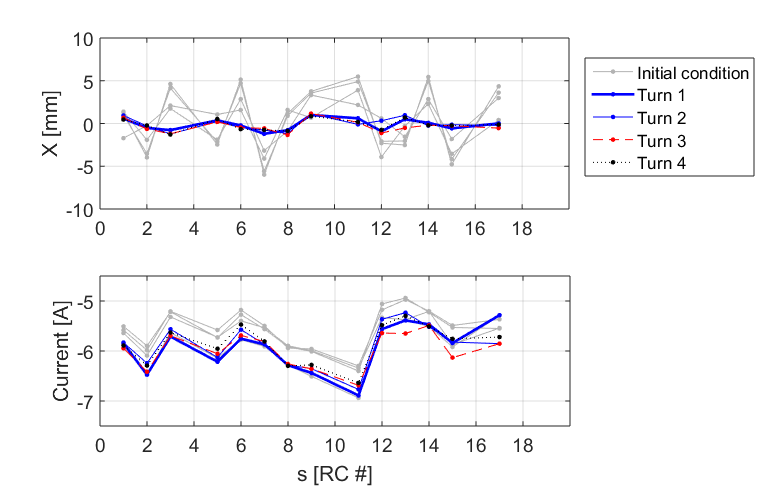}
\renewcommand{\baselinestretch}{2}
\small\normalsize
\caption{Four turn BPM data for horizontal orbit after a simulated annealing pass, in comparison to orbit in initial state. Top plot is horizontal position, bottom plot is signal sum (top+bottom+left+right plates). }
\label{fig:horz-pass-bpms}
\end{figure}
\renewcommand{\baselinestretch}{2}
\small\normalsize

Parameters for the horizontal run were the same as in the vertical, except the algorithm was allowed to run for ten independent passes. Only 11 horizontal steering magnets are used to define the state, concentrated around the Y-section.

The relaxation of the fitness function is shown in Fig. \ref{fig:horz-Ehist}, while the variation in final states (current settings) is shown in Fig. \ref{fig:horz-beststate}. The four-turn orbit data for the best state found in all ten passes is plotted in Fig. \ref{fig:horz-pass-bpms}. The fitness function relaxed from a value of $E = 3.203$ to $E=0.698$ mm in the best case. This is significantly smaller than the RMS value achieved through quad-centering, RMS$x_{BPM}=1.17$ mm. For the initial state, the difference between the first and subsequent 3 turns has an RMS value of RMS$\Delta = 1.77$ mm, with a maximum of $\max \Delta_n=3.94$ mm. After annealing, these values are reduced to RMS$ \Delta = 0.37$ mm and $\max \Delta_n=1.24$ mm. Again, this is comparable to values achieved with the quad-centering approach.

\begin{figure}[!tb]
\centering
\includegraphics[width=\textwidth]{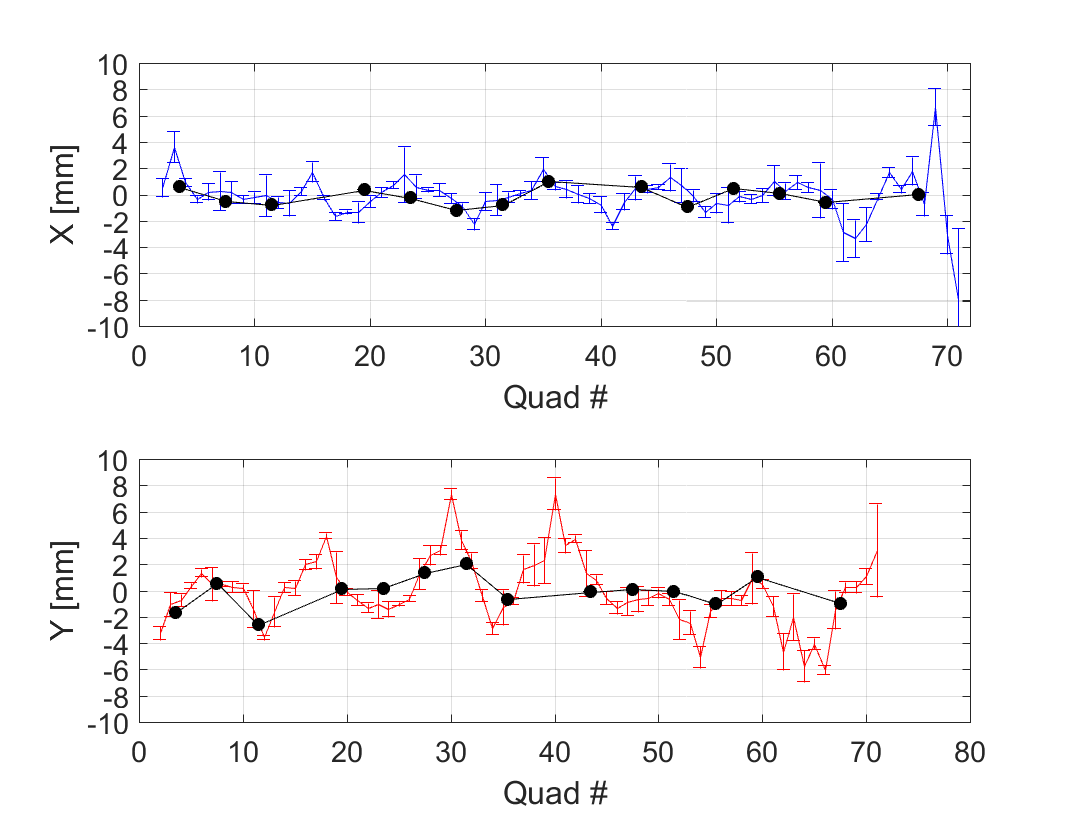}
\renewcommand{\baselinestretch}{2}
\small\normalsize
\caption{Measured first turn orbit for horizontal (upper) and vertical (lower) orbits after simulated annealing on both planes. Data taken on 8/17/17. }
\label{fig:quadasbpm-annealed}
\end{figure}
\renewcommand{\baselinestretch}{2}
\small\normalsize

\begin{table}[tb]
\caption{Orbit statistics for simulated annealing steering method, using 6 mA beam as test case. Can be directly compared to values in Table \ref{tab:steeringresults} for quad-centering method. }
\label{tab:annealing}
\centering
\begin{tabularx}{\textwidth}{X R{4cm} R{4cm}}
Statistic & Horz. [mm] & Vert. [mm] \\
\hline \hline
First turn RMS in quads & 1.70 & 2.49 \\
First turn Max in quads & 6.68 & 7.34 \\
Four turn RMS in BPMs   & 0.70 & 1.10 \\
Four turn Max in BPMs   & 1.16 & 2.07 \\
RMS $\Delta$ (Eq. \ref{eq:delta})& 0.37 & 0.49 \\
Max. $\Delta$           & 1.24 & 1.03 \\
\hline
\end{tabularx}
\end{table}
\renewcommand{\baselinestretch}{2}
\small\normalsize

The annealing algorithm initially appears to be very successful, converging to very low orbit distortions in the BPMs when compared to the quad-centering method of Section \ref{sec:steering:results}. 
This success, however, is only an apparent improvement. As the fitness function depends only on the position of centroid at the BPM locations, orbit excursions between BPMs is not constrained. Fig. \ref{fig:quadasbpm-annealed} shows the quad-as-BPM measurement for the best vertical and horizontal orbits found during annealing passes. Table \ref{tab:annealing} shows the orbit statistics, that can be compared directly to values in Table \ref{tab:steeringresults} for the quad-centering method.

The usefulness of this approach for ring tuning is questionable. For the temperature schedule and probability function used here, a single pass of the algorithm appears to converge on local minima and multiple passes are needed to effectively search the space of solutions. Lower values of the RMS four-turn position have been found using other methods such as RCDS.\cite{LevonRCDS} 

More fundamentally, any algorithm that aims to reduce orbit excursions \textit{only} at the BPMs will be under-constrained and allow large excursions between the BPMs (particular for large phase advance). The BPM spacing of 0.64 cm is close to half the betatron wavelength, $\sim 0.85$ m for the $I_Q=1.826$ A operating point, and steering can be found so that the zero-crossings of the transverse motion occur near the BPMs. This is clearly seen in Fig. \ref{fig:quadasbpm-annealed}, where largest orbit offsets appear in-between BPM locations. This effect is alleviated for lattice with much longer betatron wavelengths (weaker focusing) and therefore may still be useful for steering corrections in the single-channel octupole lattice. 

\section{Chapter summary}

Precise control over the horizontal and vertical orbits is necessary for the nonlinear optics experiments. A systematic approach for setting ring steerers to center the beam in the quadrupoles has been developed and demonstrated in the lab. Different targets for finding the centered orbit have been tested via simulation and the most effective were implemented in the lab. This quad-centering approach will be essential for tuning nonlinear UMER, as global optimization techniques that rely only on BPM data are not well-constrained. The quad-centering approach will also reduce beam loss due to scraping for all UMER experiments. Based on VRUMER predictions at the $I=1.826$ A operating point, the ``best-case" orbit distortion in a perfectly aligned ring are $\max{(x)} \sim 0.5$ mm, rms$(x) \sim 0.3$ mm and $\max{(y)} \sim 4.8$ mm, rms$(y) \sim 1.8$ mm. The best demonstrated orbit control is $\max{(x)} \sim 3.6$ mm, rms$(x) \sim 1.3$ mm and $\max{(y)} \sim 1.1$ mm, rms$(y) \sim 0.5$ mm.

Presently, orbit control limited by the effect of the ambient fields. For nonlinear experiments, additional correction of the vertical orbits is essential (with either additional SSV steerers or field-canceling Helmholtz coils). To achieve horizontal tolerances, shielding or cancellation of the vertical background field will be necessary. However, once these are in place, this approach should bring the orbit tolerances to an acceptable level both in and outside the octupole insert.

%% file: 11.Chapter.tex
\renewcommand{\thechapter}{11}

\chapter{Experimental Characterization of Beam Transmission and Transverse Resonances in the UMER FODO Lattice}
\label{ch:res}

This chapter discusses characterizing measurements of beam transmission in the UMER FODO lattice over a range of tune operating points. This is done to establish a baseline for resonance sensitivity and beam transmission prior to nonlinear optics experiments. 
Loss current loss rates are important for the proposed experiments which aim to test that the QIO lattice is stable over many turns.
Additionally, as the octupole fields are predicted to damp resonant particle growth and mitigate losses, establishing  sensitivity to resonant losses of different orders will guide experiment plans.

Single particles with tunes that meet the condition $m \nu_x + n \nu_y = p$ for integer $n,\ m,\ p$ will experience resonant excitation if the driving term $\exp{i \left( n + m \right) \Omega t}$ is present for revolution frequency $\Omega$. In practice, linear and nonlinear field errors can be minimized but not eliminated, and it can be assumed all order resonances are driven and can lead to losses. The observability of these losses depends on the strength of the driving term, the growth rate of the resonance (higher for lower orders) and whether any nonlinearities act to mitigate the resonance.\footnote{More careful discussion of resonance dynamics can be found in Section \ref{sec:theory:resonance} and many resources including \cite{Reiser,Ruth1986}.}

This chapter explores dynamics in a conventional FODO lattice, the UMER configuration that is simplest to operate, best-characterized and for which well-tuned solutions exist. The FODO lattice is not only a commonly used design, it is also a common test case for theory and numerical calculations. 
Measurements are done using the tune scan technique, where transmission is measured using the wall current monitor (WCM) over a range of tune operating points.\footnote{See Section \ref{sec:apparatus:tune-scan}.} Measurements are taken at the three lowest space charge density UMER beams: the 6 mA beam, the 0.6 mA ``pencil" beam and the $\sim 40\ \mu$A ``DC-beam." Specifics of beam generation were covered in Section \ref{sec:apparatus:beams} and Section \ref{sec:design:lowcurrent}. 

The 6 mA and 0.6 mA apertured beams have previously been well-characterized as a function of tune \cite{BeaudoinHB2010}, \cite{SutterPAC2009}. Although these are the lowest current beams designed for UMER, and the 0.6 mA beam is considered ``emittance dominated," space charge is still  a significant driver of dynamics. Section \ref{sec:res:linear} shows latest transmission measurements (with updated orbit correction, as discussed in Chapter \ref{ch:steering}) for comparison with the DC beam results. 

This chapter includes the first systematic study of transmission and resonant structure of the low-current DC beam. 
The DC beam, with $\mu$A-level currents and space charge tune spreads $\sim0.005$, is intended for use in initial tests of the quasi-integrable octupole lattice. 
Section \ref{sec:res:DCbeam} presents tune scan results for a nominally $40\ \mu$A beam in the FODO lattice. This work is the first attempt at examining the multi-turn behavior of the DC beam. 
Measurements of the transmission and resonant sensitivity of this beam will serve as the baseline for planning quasi-integrable experiments.

Finally, Section \ref{sec:res:octu} shows beam response to systematically increased nonlinearity. A single octupole element is added to the FODO lattice to act as a nonlinear driving term. With this method additional structure up to third order resonance is observed at low currents, although the higher space charge beams are less affected.

\section{Beam transmission measurements in the linear lattice} \label{sec:res:linear}

The section reviews beam transmission and loss rates over a range of tunes for the 6 mA and 0.6 mA beams in the linear FODO lattice.  
As the charge distribution is not KV (and the space charge force not linear), one might expect the space charge force to increase the width of observed resonance stop-bands as a result of the space-charge induced tune spread. However, the tune spread may also provide some amount of resonance detuning.
In this section comparison is made of the resonant structure between all three beams. 

\subsection{Experimental procedure}

\begin{figure}[tb]
\centering
\includegraphics[width=\textwidth]{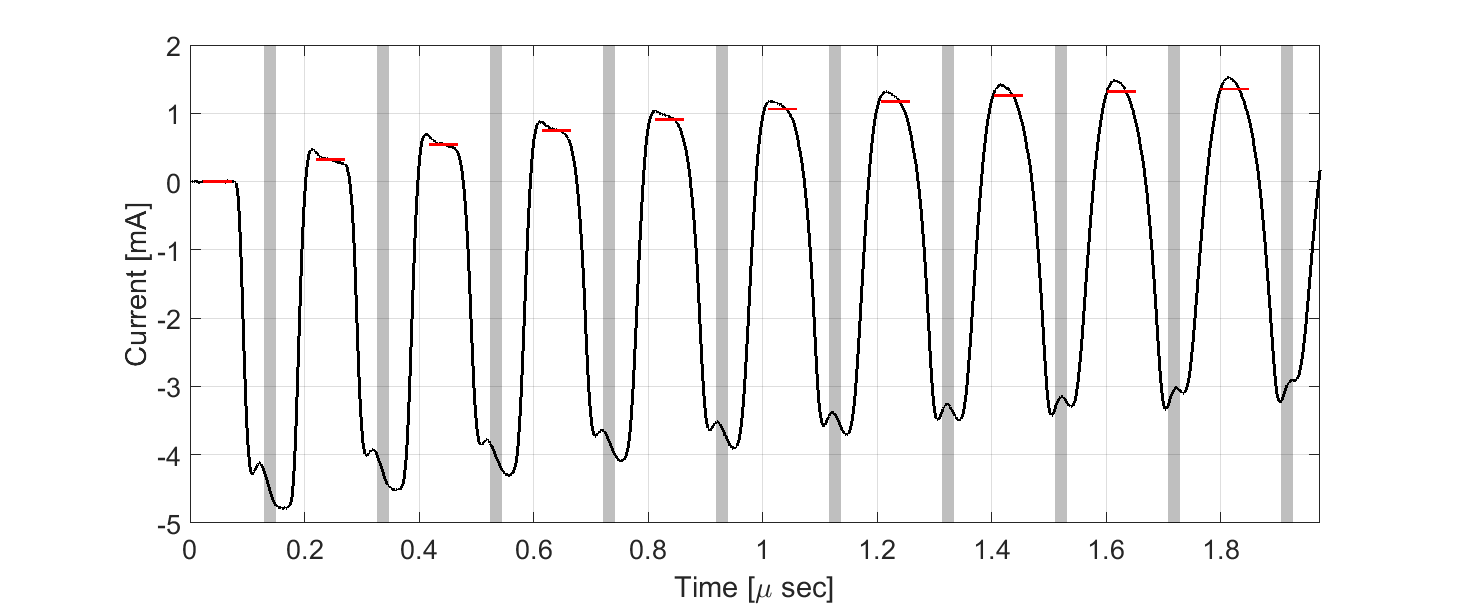}
\caption{Typical WCM signal, shown for first 10 turns of 6 mA beam. $\pm 10$ ns of beam are averaged around bunch center, indicated by gray boxes. The baseline level is averaged over $50$ ns between turns, the average for each turn is indicated by the red hash lines. }
\label{fig:6mA-beam-slices}
\end{figure}

The beam survival measurement uses the wall current monitor (WCM) to measure beam current per turn. This set-up is explained in more detail in Section \ref{sec:apparatus:tune-scan}. Current per turn is calculated by averaging over a 20 ns window about the bunch center. The beam center is identified in the first turn and subsequent turns are counted by assuming a fixed revolution frequency. For a beam at 9.97 keV (10 kV - 30 V bias voltage) on a closed path of $L=11.52$ m, one revolution period is $T=197.2$ ns. This value agrees well with observations. An example WCM signal is shown in Fig \ref{fig:6mA-beam-slices}.
As the baseline is not constant, the ``beam-off" level is calculated for each turn in a 50 ns window centered directly between the current turn and the preceding turn.\footnote{The drifting baseline is discussed in Section \ref{sec:apparatus:diag}. No inductive correction is made for the data in this chapter.}  
Measured current per turn is normalized to the peak current in the first turn at each operating point. This should account for any drift in output current over the multi-hour long data collection routines. 

For all the measurements below, injection quadrupole values are set near the matched condition for the 6 mA beam (identified in \cite{HaoThesis}). No additional tuning is made for the 0.6 mA beam, so the results include losses due to injection mismatch. Additionally, no match compensation is made for different lattice settings in the scan. The injection quadrupoles are held at the known matched condition for quadrupole currents $I_F = I_D = 1.826$ A. Therefore, there is an injection mismatch error that increases with departure  from the 1.826 A operating point.

All survival plots are shown in tune space. The conversion from quad excitations $\left(I_F,I_D\right)$ to tune $\left(\nu_x,\nu_y\right)$ is analytically calculated based on a smooth-focusing approximation of the UMER FODO lattice. This model does not include dipole edge-focusing terms, therefore $\nu_x = \nu_y$ when $I_F=I_D$.\footnote{See discussion of edge-focusing in Appendix \ref{ap:model}.} This is known to be an over-simplification, as measurements at the 1.826 A operating point show a tune splitting of $\nu_x = 6.636$, $\nu_y=6.752$, while the smooth-focusing model predicts $\nu_x = \nu_y=6.787$.
The typical range covered in a tune scan is $I_F,I_D = 1.65 \to 2.1$ A with stepsize 0.01 A. In tune space this resolution is roughly 0.07. The maximum possible resolution (based on power supply resolution) is 0.001 A, increasing tune space resolution to $\sim0.007$.

\subsection{6 mA beam}

\begin{figure}[!tb]
\centering
\includegraphics[width=\textwidth]{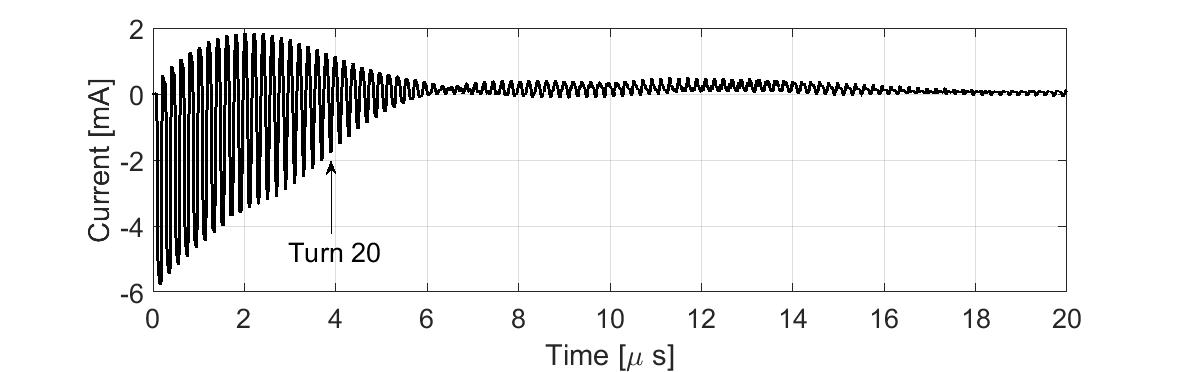}
\caption{Typical WCM signal for 6 mA beam, measured at operating point $I_F=1.830$ A, $I_D=1.930$ A (predicted bare tunes $\nu_x = 6.599$, $\nu_y = 7.413$). }
\label{fig:6mA-WCM-signal}
\end{figure}

\begin{figure}[tb]
\centering
\includegraphics[width=\textwidth]{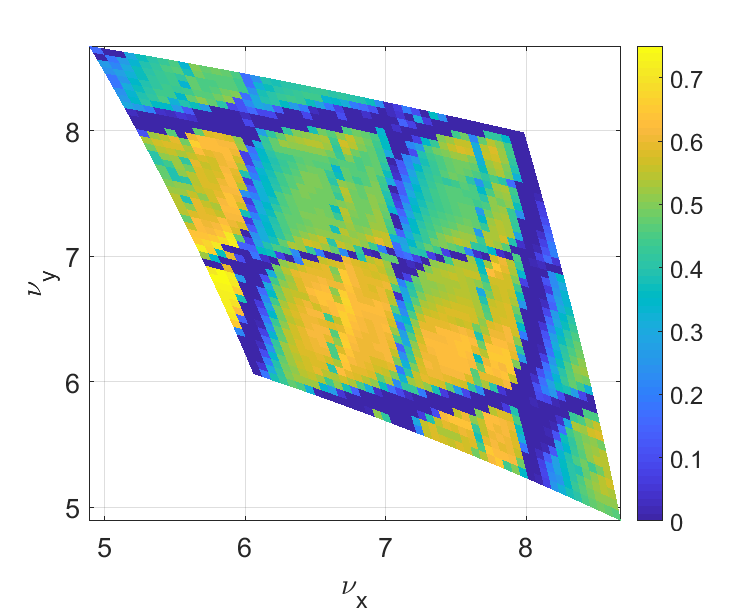}
\caption{Beam transmission for 6 mA beam on turn 20 ($\sim 4\ \mu$s), plotted as fractional survival (color axis) as a function of predicted tune. }
\label{fig:6mA-turn20}
\end{figure}

Results from the 6 mA beam are shown in Fig. \ref{fig:6mA-turn20}, and a sample WCM signal from this run is shown in Fig. \ref{fig:6mA-WCM-signal}. The distinct shape of the WCM signal is due to two phenomena. First, the WCM diagnostic has an inductive reactance that causes a drifting baseline, as discussed in Section \ref{sec:apparatus:diag}. Second, the initially 100 ns-long beam pulse elongates due to longitudinal space charge forces. Without longitudinal confinement, the bunch eventually interpenetrates and uniformly fills the ring. Peak current is reduced as charge redistributes and eventually the beam seems to disappear (at $\sim7\ \mu$s in Fig. \ref{fig:6mA-WCM-signal}). The end erosion effect is discussed in more detail in \cite{BeaudoinAAC2010}. For the purposes of this study, the bunch erosion limits the ``flat current" region of the beam (before erosion waves meet) to $\sim9$ turns, and the region of usable WCM signal to $\sim 25$ turns. 

Examining the tune scan results in Fig. \ref{fig:6mA-turn20}, integer resonance bands for $\nu_x = p$ and $\nu_y=p$ are clearly visible. Half-integer resonances $2\nu_x = p$ are present in the horizontal plane, but barely visible in the vertical. In some regions the second order sum resonance $\nu_x + \nu_y=p$ is visible. There is no visible resonances above order two. In regions of good transmission (say around $\nu_x = 6.7$, $\nu_y=6.5$), there is $\sim 30\%$ peak current loss from the first to $25^{th}$ turn.

\clearpage
\subsection{0.6 mA ``pencil" beam}

\begin{figure}[!tb]
\centering
\includegraphics[width=\textwidth]{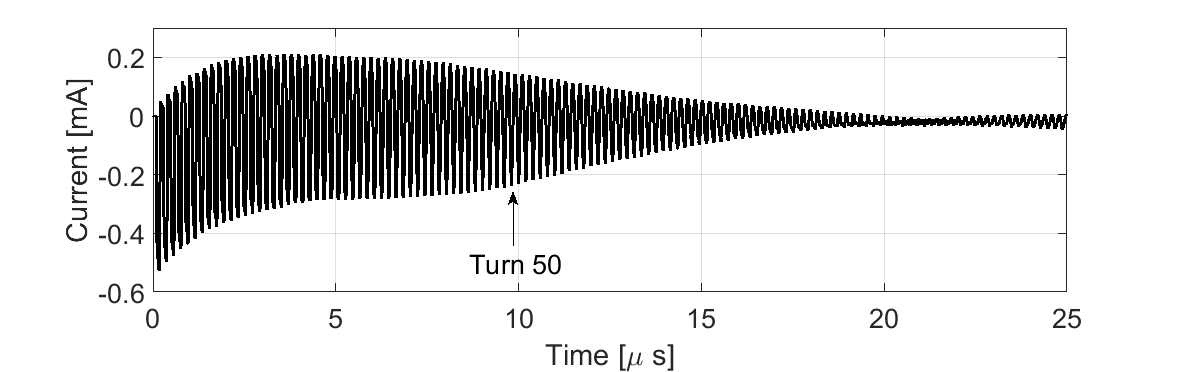}
\caption{Typical WCM signal for 0.6 mA pencil beam, measured at operating point $I_F=1.830$ A, $I_D=1.830$ A (predicted bare tune of $\nu_x=\nu_y=6.803$). }
\label{fig:pencil-WCM-signal}
\end{figure}

\begin{figure}[tb]
\centering
\includegraphics[width=\textwidth]{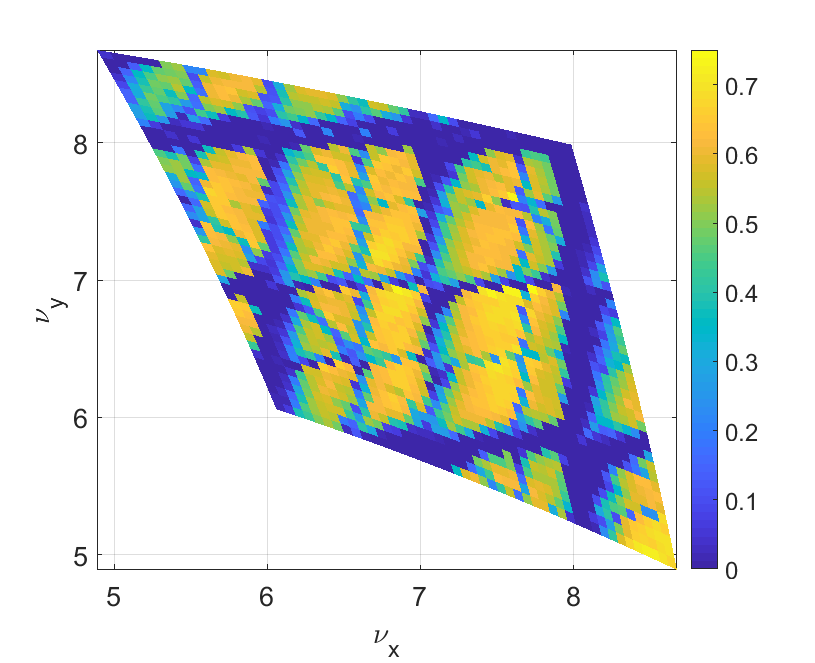}
\caption{Beam transmission for 0.6 mA pencil beam on turn 50 ($\sim 10\ \mu$s), plotted as fractional survival (color axis) as a function of predicted tune. }
\label{fig:pencil-turn50}
\end{figure}

The results from the 0.6 mA beam, which has a predicted space charge tune spread less than half that of the 6 mA beam, are shown in Fig. \ref{fig:pencil-turn50}. In the WCM signal (Fig. \ref{fig:pencil-WCM-signal}), the same characteristic shape is observed, although the DC point appears much later. The ``flat current" region extends to turn $\sim25$, with usable signal out to $\sim90$ turns.

\begin{figure}[tb]
\centering
\includegraphics[width=\textwidth]{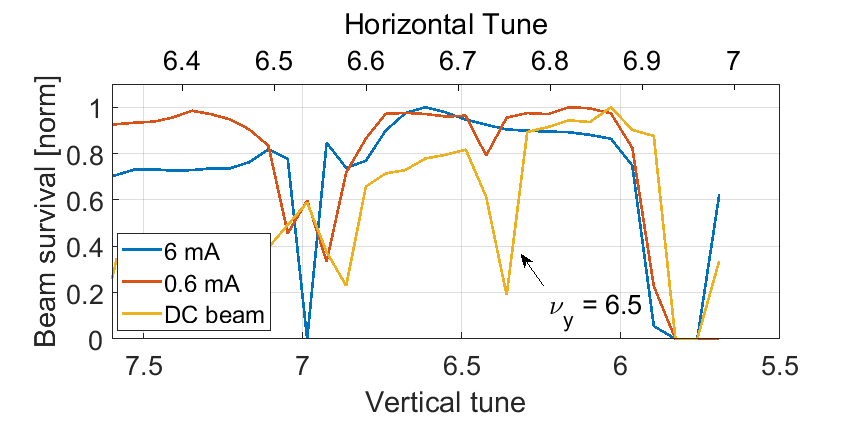}
\caption{Beam transmission along the line$I_F=1.8$ A ($\nu_y = -2.88\nu_x+25.9$) for different space charge densities. Curves are normalized to the maximum current in this turn along this line. The apparent $\nu_y=6.5$ stop-band is indicated. }
\label{fig:resonances-compare}
\end{figure}

Beam survival rates at turn 50 are plotted in \ref{fig:pencil-turn50}. Compared to the 6 mA results, second order resonant structure appears more clearly, with the half-integer bands more prominently visible. The second order sum resonance $\nu_x + \nu_y=p$ are also more apparent. Their increased prominence could either be due to the increased turn number (as the second order resonance is slower growing) or the decreased space charge concentration (as higher charge may act to detune resonances). Comparing transmission at turn 20 for all three beams, plotted in Fig. \ref{fig:resonances-compare}, reveals greater relative loss in the half-integer resonance band when there is less space charge. This suggests that the space charge detuning acts to mitigate the half-integer losses. 
Finally, in the regions of good transmission, the 0.6 mA beam  experiences $\sim 30\%$ loss when from the first to $50^{th}$ turn.

\section{Measurements of low-current DC beam} \label{sec:res:DCbeam}

While integrable optics have been proposed as a method for mitigating resonant instability, the theory is based on single particle dynamics and may not extend well to beams with significant collective effects. Initial exploration of nonlinear lattices of this type should be done as close to the zero-charge limit as possible. For this reason, the low current, high emittance ``DC beam" is proposed as a candidate for initial operation of the quasi-integrable lattice. 
This section explores beam transmission in this new regime in the linear UMER FODO lattice.

\subsection{Experimental procedure}

\begin{figure}[!tb]
\centering
\subfigure[125 turns at operating point $I_F = 1.800$ A, $I_D = 1.900$ A.]{
\includegraphics[width=\textwidth]{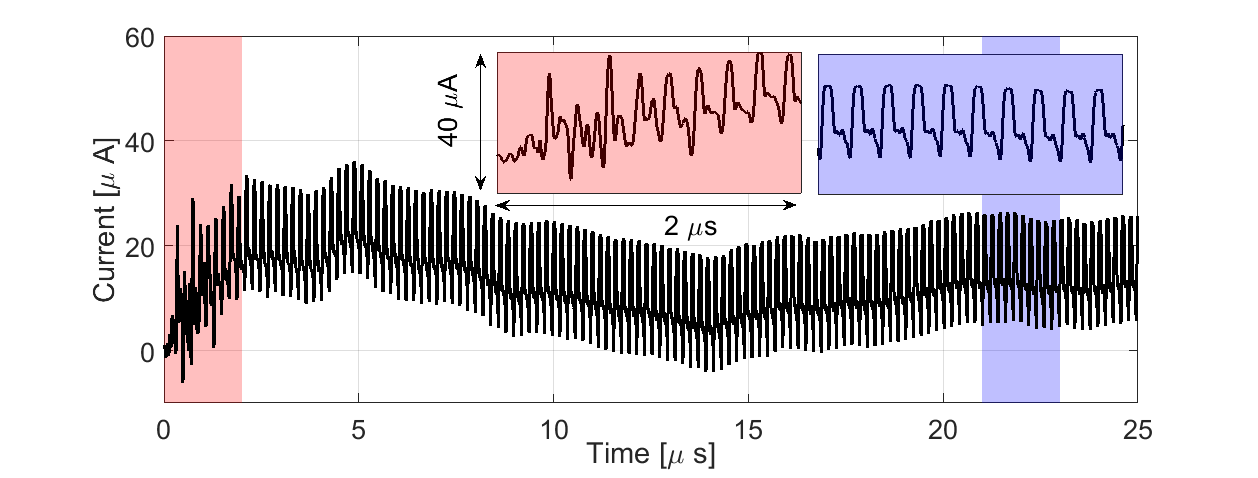}
\label{fig:microA-wcm-signal-a}
}
\subfigure[500 turns at operating point $I_F = 1.826$ A, $I_D = 1.826$ A.]{
\includegraphics[width=\textwidth]{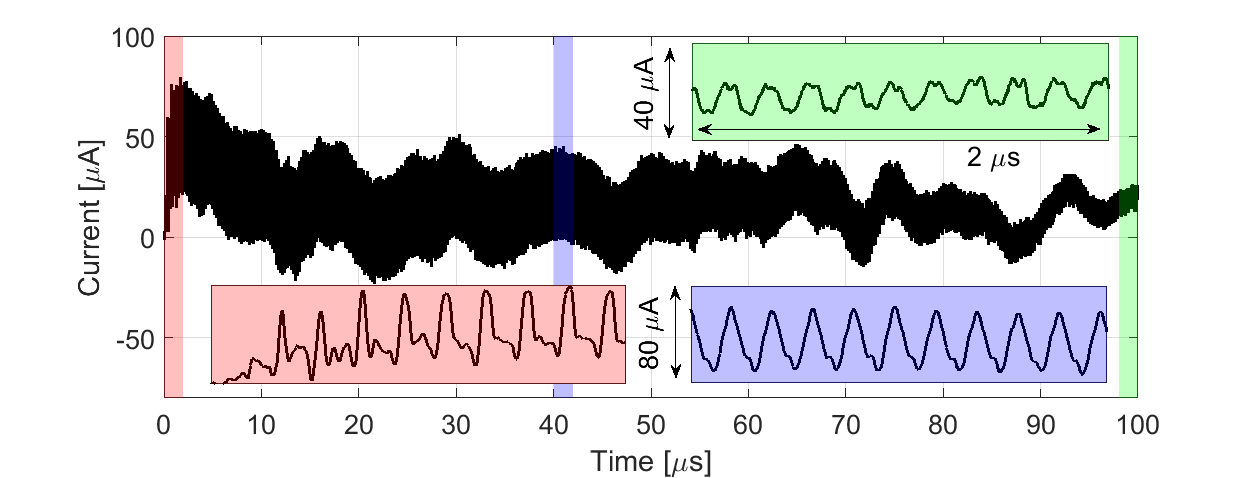}
\label{fig:microA-wcm-signal-b}
}
\caption{Typical WCM signal traces for DC beam. Longitudinal bunch shape remains fairly constant, although current is steadily lost during 100's of turns. Bias voltage is constant between figures (3.70 V) and decreased output in (a) compared to (b) is due to drift in gun output over experimental run. }
\label{fig:microA-WCM-signal}
\end{figure}

The DC beam is produced by operating the UMER electron gun in voltage amplification mode. DC refers to the fact that the bias voltage pulse is turned off and the source emits continuous current. This was done to avoid amplification of pulsed circuit ripple on the longitudinal bunch profile, and is not a fundamental quality of this beam. Longitudinal structure is formed through the pulsed magnet injection.
DC beam generation and detection of the very low-level signal were discussed in Section \ref{sec:design:lowcurrent}. 

The procedure for constructing the ``DC beam" survival plot is generally identical to the 6 mA and 0.6 mA scans described above. For the DC beam, additional care is taken to boost the signal to noise ratio, including amplifying the WCM signal and averaging over more waveforms. 16 cycles are averaged for each measurement, with statistical errors around $7\%$ as characterized in Section \ref{sec:design:lowcurrent}. 

The gun bias voltage is set to 3.70 V to produce an initially $40\ \mu$A beam.
For the DC beam, which has energy 10 keV + 3.70 eV, revolution period should be $T=196.9$ ns.     
However, the $T=197.2$ ns used above for the 6 mA and 0.6 mA beams (with energy 10 keV -30 eV) agreed better with observed revolution frequency.

Typical DC beam signals picked up by the wall current monitor (with amplification) are shown in Fig. \ref{fig:microA-WCM-signal}. There are some immediately noticeable differences when compared to the 6 mA (Fig. \ref{fig:6mA-WCM-signal}) and 0.6 mA (Fig. \ref{fig:pencil-WCM-signal}) beams. First, the AC beam signal is preserved out to 500 turns. There is no visible end erosion in Fig. \ref{fig:microA-wcm-signal-a}. In Fig. \ref{fig:microA-wcm-signal-b} there appears to be some erosion at $40\ \mu s$ (turn 200). However, by $100\ \mu$s (turn 500), the accelerated/decelerated head/tail appear to have exceeded the machine acceptance, and the pulse structure once again has a flat top. 

Although longitudinal structure is preserved, there is steady current loss per turn. This loss is plotted below in Fig. \ref{fig:loss-curves}.
A likely source of loss is that the injection match is far from optimal. Additionally, the beam current is below the threshold for BPM detection, so the steering has not been optimized in this case. There are likely scraping errors due to the large transverse beam size, which is predicted to be $\sim 0.7$ cm on average in the matched case.

\subsection{Current drift over time} \label{sec:res:drift}

\begin{figure}[!htb]
\centering
\subfigure[Tune scan for current in second turn. Color axis units are $\mu$A.]{
\includegraphics[width=0.7\textwidth]{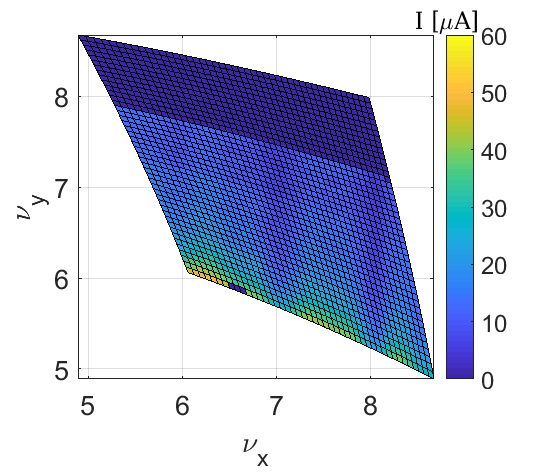}
\label{fig:microA-drift-a}}
\hspace{.5in}
\subfigure[Normalized current in second turn as a function of number of data points collected during 13 hour scan.]{
\includegraphics[width=0.7\textwidth]{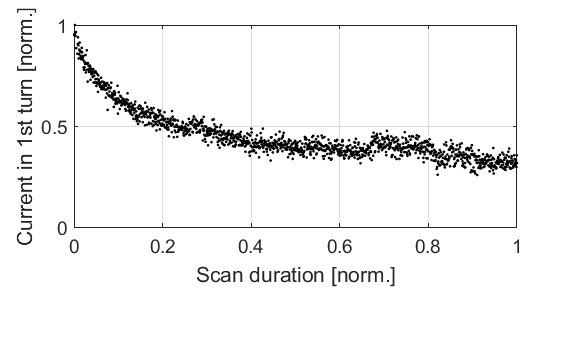}
\label{fig:microA-drift-b}}
\caption{Current drift over duration of ``DC-beam" tune scan measurements. }
\label{fig:microA-drift}
\end{figure}

During the DC beam measurements, the output current from the gun decreased significantly. Data collection took place over a period of 13 hours of continuous operation. During this time, the mid-pulse current, which was initially set to be $40\ \mu$A, decreased to $15-20\ \mu$A. The measured signal in the beam at the second turn is plotted with respect to tune in Fig. \ref{fig:microA-drift-a}. The first measured point is in the bottom left ``corner," with subsequent points collected in row major order, left to right and bottom to top. Figure \ref{fig:microA-drift-b} shows the current in the second turn as a function of scan duration. There is a clear exponential decrease in measured current from the initial $40\ \mu$A to $\sim20\ \mu$A by the end of data collection. Due to the decrease in beam current the collection was prematurely interrupted, so the collected data only spans the range $I_D = 1.65 \to 1.95$ A. Unmeasured points appear as the dark blue band in Fig. \ref{fig:microA-drift-a}. 

The cause of the ``drooping" current is not clear at this point.
One possible explanation is that the increased load on the DC circuit is draining the capacitors that hold off the $10$ keV gap voltage. However, one would expect to see this reflected in the final beam energy. No accompanying drift in the revolution frequency is observed, to a resolution of $\sim 10$ eV in beam energy.
The other possibility is temperature dependence in the current output. This intuitively makes sense with the hours-long decay time in Fig. \ref{fig:microA-drift-b}, as well as the equilibrium reached after many hours.

In either case, current drift will likely be reduced by modifying the pulse-forming circuit for operation of DC beam in a ``long-pulse" mode (this will also reduce the heat load on the cathode grid and reduce the risk of heat damage). For now, the drift is assumed to not greatly effect the dynamics. For a variation of $40\ \mu A \to 20\ \mu A$, by the simple approximation for tune shift (Equation \ref{eq:tune-spread}), we expect $\delta \nu = 0.008 \to 0.003$. This is a small enough change that we don't expect the dynamics are significantly affected.


\subsection{$40\ \mu$A DC beam tune scan results} \label{sec:res:DCbeam}

\begin{figure}[!tb]
\centering
\includegraphics[width=\textwidth]{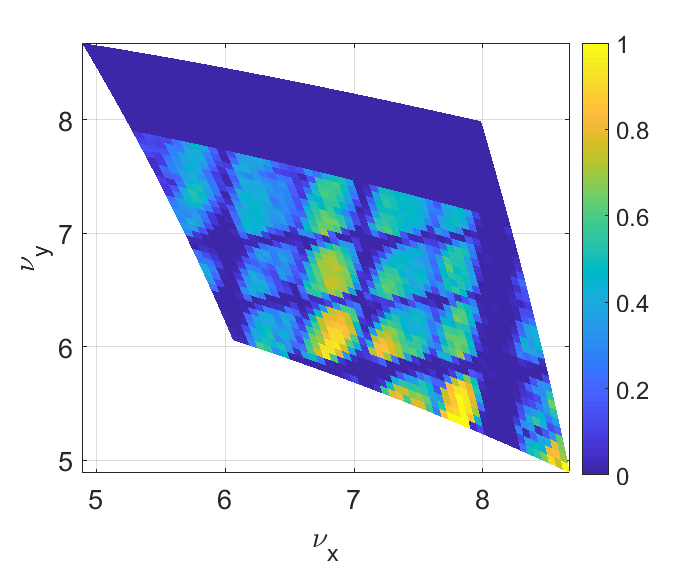}
\caption{Beam transmission for 20 - 60 $\mu$A DC-beam on turn 100 ($\sim 10\ \mu$s), plotted as fractional survival (color axis) as a function of predicted tune. }
\label{fig:microA-turn100}
\end{figure}

The resulting tune scan is shown in Fig. \ref{fig:microA-turn100}. Because the pulsed injection is used for bunch formation, on the first turn the longitudinal profile is not fully formed.\footnote{Refer to Fig. \ref{fig:microA-WCM-trace}.} Because of this, beam current is normalized with respect to the second turn. This also normalizes out the current drift. 

In tune space, all integer and half-integer stop-bands are very well-defined. Additionally, there appears to be a third order stop-band at $\nu_x = 19/3 \approx 6.3$ that did not appear at higher currents. Compared to the 0.6 mA results, the sum resonances $\nu_x + \nu_y = p$ are less apparent.

\subsection{Rate of beam current loss}

One of the proposed experiments for the quasi-integrable lattice is to measure the dependence of dynamic aperture on octupole strength. Additionally, as quasi-integrability is meant guarantee long-term stability, the octupole lattice should demonstrate stable beam transport over as long a path length as possible. For this reason, it is important to be able to accurately measure loss rates and demonstrate low-loss transport over many turns. 

\begin{figure}[]
\centering
\subfigure[Beam loss for 6 mA and 0.6 mA beams, normalized to 1st turn.]{
\includegraphics[width=0.45\textwidth]{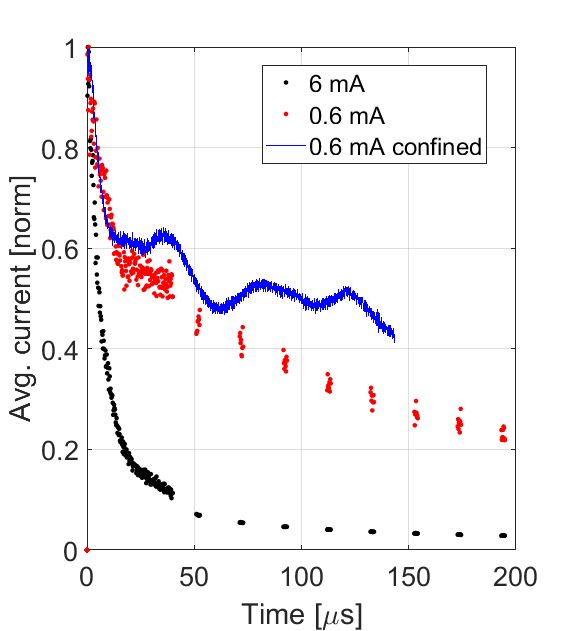}
\label{fig:loss-curves-a}}
\subfigure[Beam loss for DC beam, normalized to 10th turn.]{
\includegraphics[width=0.45\textwidth]{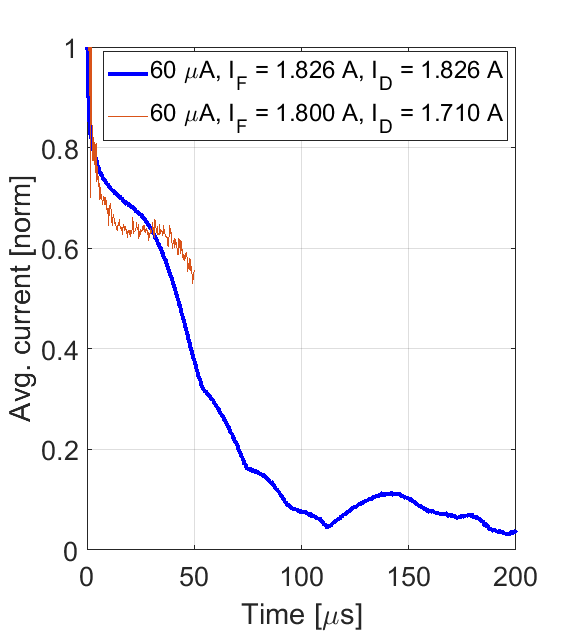}
\label{fig:loss-curves-b}}
\caption{Average current per turn for three UMER beams. }
\label{fig:loss-curves}
\end{figure}

Figure \ref{fig:loss-curves} shows measured loss rates for the three UMER beams described here. Fig. \ref{fig:loss-curves-a} shows measured loss rates for the 6 mA and 0.6 mA beams (black and red dots, respectively). As the AC signal is lost, current per turn is measured by restoring AC signal using the knock-out technique (described in \cite{Koeth2010} and \cite{Stem2012}). Here a BPM pick-up electrode is converted to a fast-pulsed kicker, which knocks out the bunch and creates an AC time signature that can be detected on the WCM.\footnote{See Section \ref{sec:apparatus:knock-out}.} As the beam ends erode until the ring is uniformly filled, the average current per turn is measured (so that no loss would result in a constant value). Here the current is normalized to the first turn average (3 mA and 0.3 mA, respectively, for the 100 ns bunch which has $50\%$ fill-factor). This measurement is done at the standard operating point $I_F = I_D =  1.826$ A.

The third (blue) curve in Fig. \ref{fig:loss-curves-a} shows the current per turn in the 0.6 mA beam when end erosion is prevented using longitudinal confining fields from an induction cell at the RC4 location. Here pulsed longitudinal electric fields are synchronized with the passing of the bunch head/tail to create a barrier against beam expansion. This technique is described in more detail in \cite{BeaudoinAAC2010,Beaudoin2011}. The use of barrier fields launches density waves across the bunch and creates addition RF ripple on the beam signal. This results in the oscillatory artifact in Fig. \ref{fig:loss-curves-a}; the beam is not actually gaining current at any point.

From Fig. \ref{fig:loss-curves-a} it is apparent for both beams that there is rapid beam loss within the first 100 turns, with a steady but decreased loss rate afterwards. There is some improvement with longitudinal confinement, which presumably removes beam loss due to the accelerated/decelerated beam head/tail exceeding the ring acceptance. However, most of the loss appears to be transverse (ie, due to scraping and envelope oscillations). 
Of most interest is the observed loss rate for the DC beam, plotted in Fig. \ref{fig:loss-curves-b}. There is an initial rapid loss, which slows at $\sim 70\%$ loss, but strangely increases again around turn 200. This behavior appears consistent between operating points. Some loss may be mitigated, as mentioned above, by improving the injection match and reducing orbit distortion. 
Improving DC beam transmission beyond the initial characterization made here will be vital for the success of the quasi-integrable experiments. This requires further studies to identify the loss mechanism and, if possible, apply correction.

\clearpage
\section{Excitation of resonances with an octupole element} \label{sec:res:octu}

\begin{figure}[tb]
   \centering
    \includegraphics[width=\textwidth]{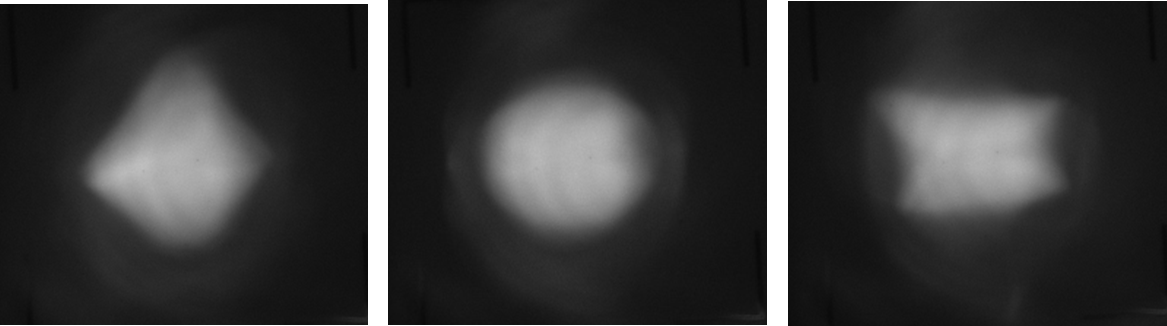}
 	\caption{6 mA beam profile measured downstream of octupole, imaged using phosphor screen. From left to right: $I_{oct} >0$, $I_{oct} =0$, $I_{oct} <0$. }
   \label{fig:beam-pass-through-octu}
\end{figure}

The true success for octupole lattice experiments will be a demonstration of resonance suppression and reduced losses compared to the linear case. One approach is to move the lattice tune towards a resonant condition, as is done in the above tune scans. Another method is to intentionally drive nearby resonances. This section describes a test of the driven case, where a single octupole element is used to increase nonlinearity in the FODO lattice.

Incoherent tune resonances are excited by driving forces of the appropriate order, typically in the form of magnetic field errors. All orders are driven in the FODO lattice, due to dipole and quad errors as well as unwanted harmonics in the PCB magnet fields. These are generally fixed, a property of the accuracy of magnet design and machine control. In order to probe beam response to the magnitude of driving term, a single octupole magnet is installed at the location of ring quadrupole QR26 (they are co-housed in a single mount). I repeat the tune scan measurement with the added octupole fields. As described in Section \ref{sec:design:octu}, the octupole contains quadrupole as well as higher order terms.

\begin{figure}[]
\centering
\subfigure[Beam survival plot at turn 20 with octupole current 1 A. The white line indicates the location of the slice plotted below.]{
\includegraphics[width=0.6\textwidth]{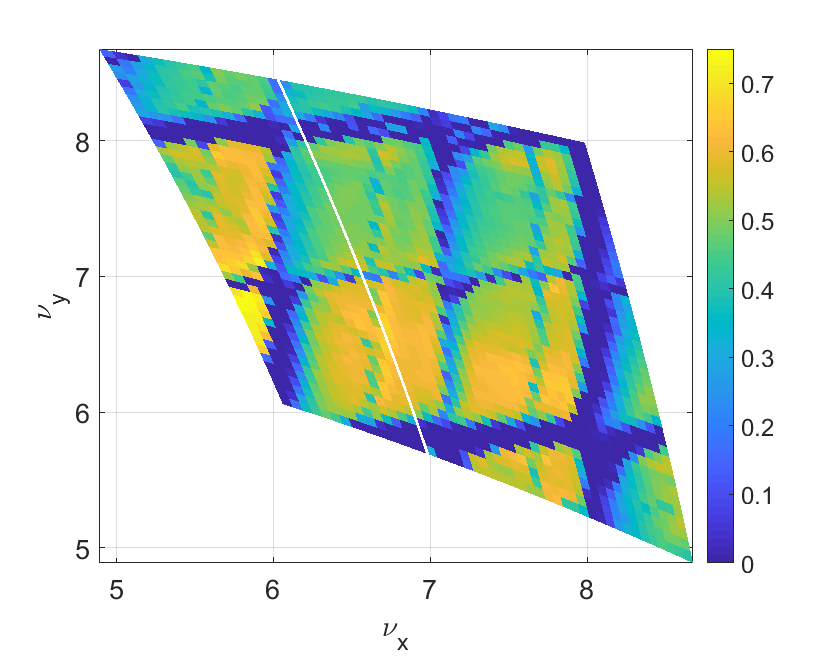}
\label{fig:6mA-with-octu-a}}
\hspace{.5in}
\subfigure[Line-out from beam survival plot comparing transmission with/without octupole fields. Vertical lines indicate resonant tune values. Solid black is first order, dashed red is second order and dotted blue is third order.]{
\includegraphics[width=\textwidth]{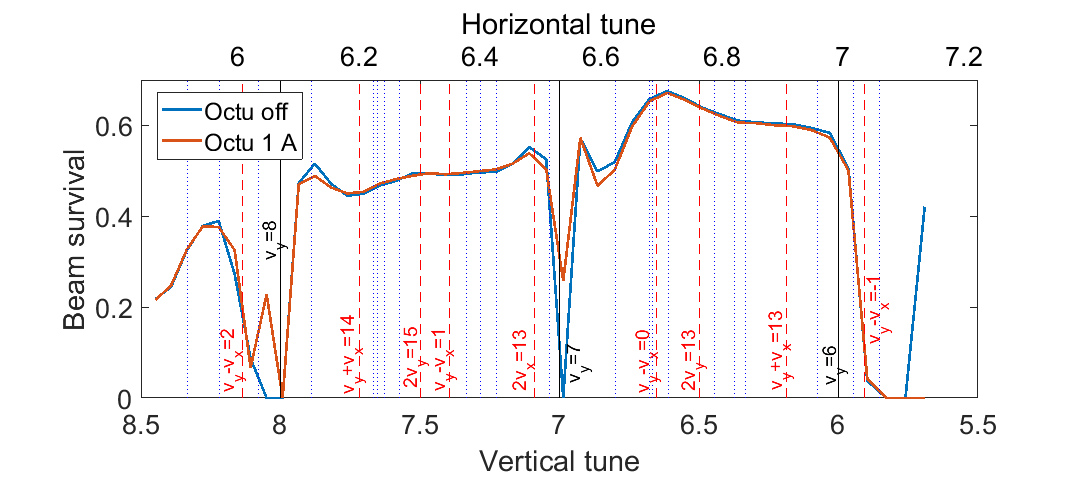}
\label{fig:6mA-with-octu-b}}
\caption{Results of tune scan with octupoles for 6 mA beam. }
\label{fig:6mA-with-octu}
\end{figure}

Results for the 6 mA tune scan with octupole excitation are shown in Fig. \ref{fig:6mA-with-octu}. While the beam survival plot (Fig. \ref{fig:6mA-with-octu-a}) appears identical to the ``linear" case shown in Fig. \ref{fig:6mA-turn20}, comparing the transmission rate along the line $I_F = 1.800$ A (Fig. \ref{fig:6mA-with-octu-b}) highlights the differences. The only significant difference is in the depth of the resonant structures. With the octupole excitation, the half-integer stop-band $\nu_x = 6.5$ actually suffers \textit{less} loss, while the integer band at $\nu_y = 8$ gains some unexpected structure. Overall, the effect of the octupole fields is almost unnoticeable, which is not surprising given the short path length over which beam transmission can reasonably be measured (20 turns) and the magnitude of the space charge force (which causes tune shift $\sim2.4$ for the 6 mA beam).

\begin{figure}[]
\centering
\subfigure[Beam survival plot at turn 50 with octupole current 3 A. The white line indicates the location of the slice plotted below.]{
\includegraphics[width=0.6\textwidth]{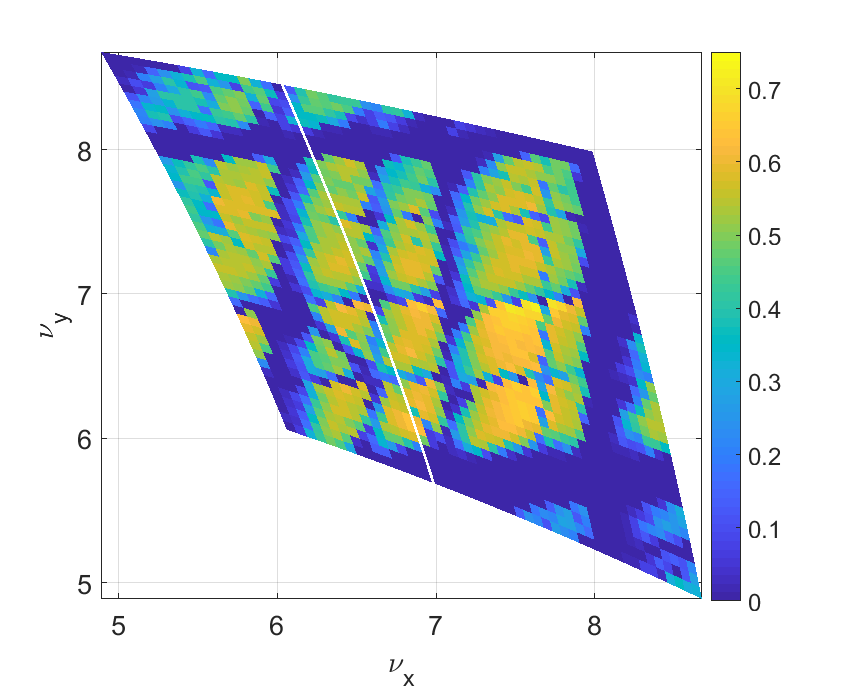}
\label{fig:pencil-with-octu-a}}
\hspace{.5in}
\subfigure[Line-out from beam survival plot comparing transmission with/without octupole fields. Vertical lines indicate resonant tune values. Solid black is first order, dashed red is second order and dotted blue is third order.]{
\includegraphics[width=\textwidth]{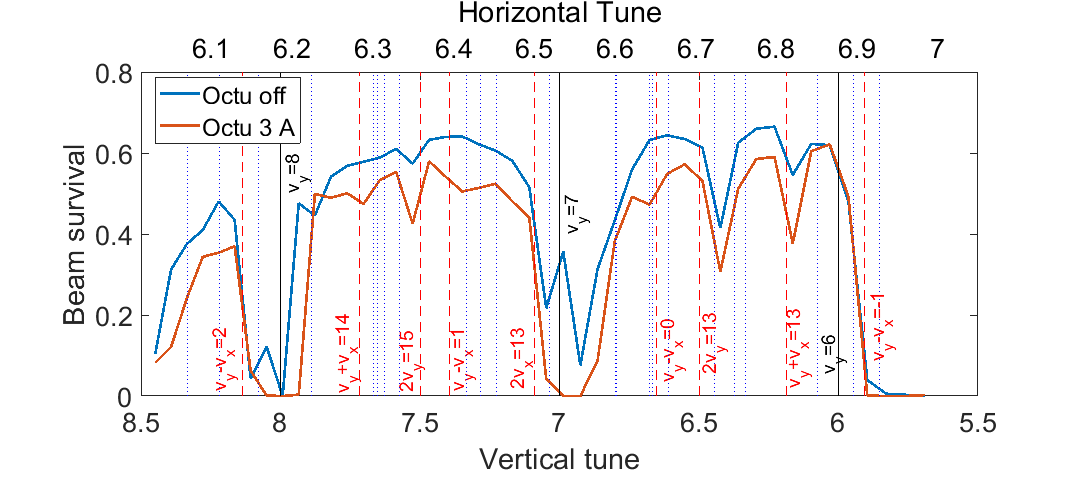}
\label{fig:pencil-with-octu-b}}
\caption{Results of tune scan with octupoles for 0.6 mA pencil beam. }
\label{fig:pencil-with-octu}
\end{figure}

Results with the 0.6 mA beam are shown in Fig. \ref{fig:pencil-with-octu}. In this case, octupole excitation was increased to 3 A. Additional resonant features appear here in comparison to the un-driven case, Fig. \ref{fig:pencil-turn50}. In Fig. \ref{fig:pencil-with-octu-a} additional resonance lines appear above and below the $\nu_y=6.5$ band, which are likely driven third order resonances $\nu_y = 22/3 \approx 7.3$ and $\nu_y = 23/3 \approx 7.7$. These features are also visible in Fig. \ref{fig:pencil-with-octu-b}. The other significant difference, in comparison to the driven 6 mA results, is that first and second order stop-bands get wider and deeper. The effect of the added driving term is to increase loss near these resonances. 

\begin{figure}[]
\centering
\subfigure[Beam survival plot at turn 100 with octupole current 3 A. The white line indicates the location of the slice plotted below.]{
\includegraphics[width=0.6\textwidth]{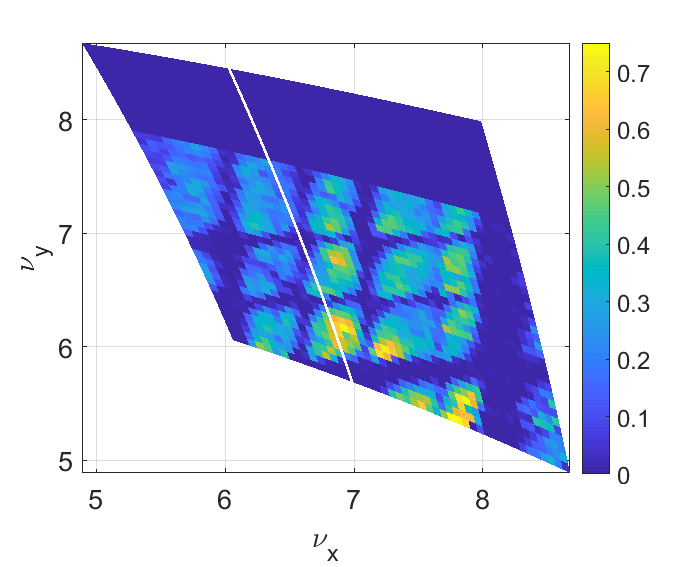}
\label{fig:microA-with-octu-a}}
\hspace{.5in}
\subfigure[Line-out from beam survival plot comparing transmission with/without octupole fields. Vertical lines indicate resonant tune values. Solid black is first order, dashed red is second order and dotted blue is third order.]{
\includegraphics[width=\textwidth]{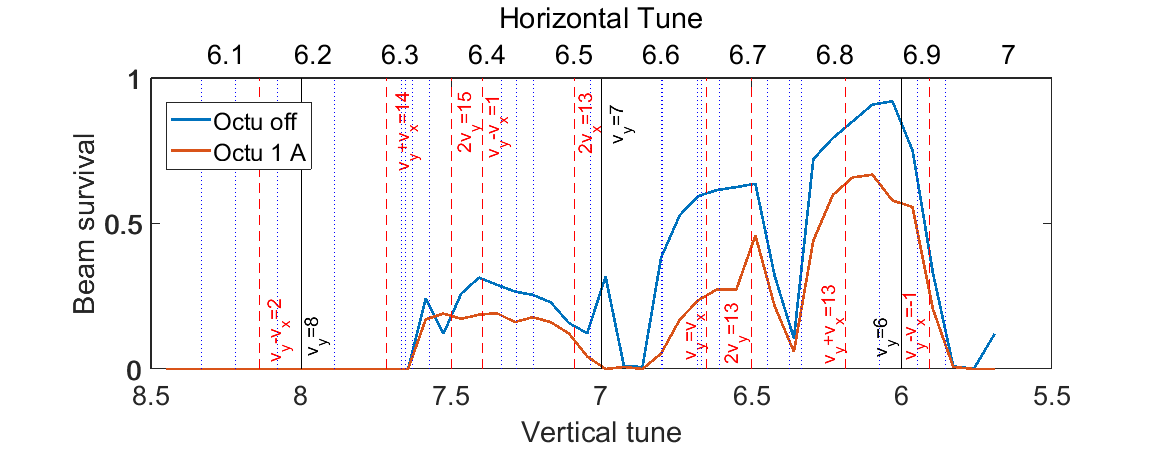}
\label{fig:microA-with-octu-b}}
\caption{Results of tune scan with octupoles for $\mu$A level DC beam. }
\label{fig:microA-with-octu}
\end{figure}

Finally, DC beam results are plotted in Fig. \ref{fig:pencil-with-octu}. While the results are less clear due to the non-uniformity of transmission across the scanned range, in general the same observations hold true: integer and half-integer stop-bands increase in both depth and width. 
Compared to the un-driven case (Fig. \ref{fig:microA-turn100}) there appears to be much more feature in the ``good transmission" regions that may be due to tune resonances, but much of the fine structure is indiscernible at this resolution. 

\begin{figure}[tb]
\centering
\includegraphics[width=\textwidth]{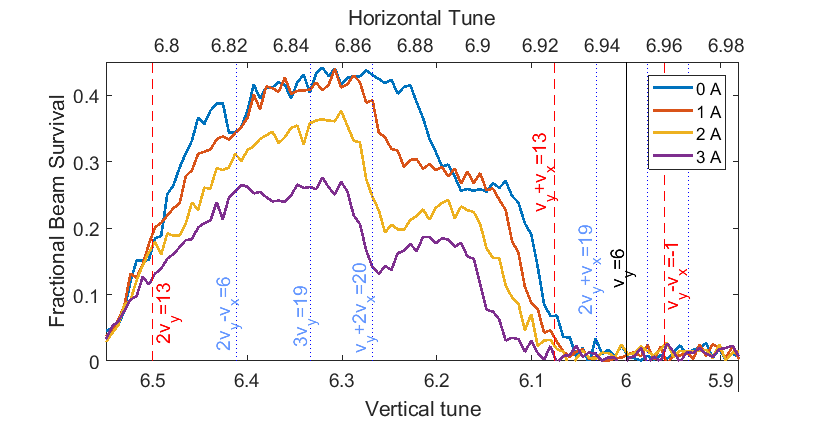}
\caption{Results from high-resolution 1D tune scan with varying octupole excitation, measuring beam transmission at turn 200. Vertical lines indicate resonant tune relationships. Solid black is first order, dashed red is second order and dotted blue is third order. }
\label{fig:microA-1D-scan}
\end{figure}

Results from a higher resolution follow-up study are plotted in Fig. \ref{fig:microA-1D-scan}. The resolution is magnified by 10, with stepsize $\Delta I = 0.001$ A in quad current. A correction factor of +0.2 from the smooth-focusing estimate is applied to the vertical tune.\footnote{See Section \ref{sec:theory:smooth} for description of the smooth-focusing approximation.} The smooth-focusing prediction for vertical tune is known to be inaccurate as dipole edge-focusing effects are not included. Here the $\nu_y = \nu_{y,smooth} + 0.2$ transformation is chosen to line up the resonance lines beam loss at $\nu_y = 6.5$ and $\nu_y = 6$.

Beam transmission curves in this figure show the emergence of resonant structure not present in the linear FODO lattice. A candidate for this resonance is the nearby line $\nu_y + 2 \nu_y = 20$. The identification could be verified by measuring machine tune at this operating point ($I_F = 1.800$ A, $I_D = 1.706$ A) with one of the higher current beams.

\section{Chapter summary}

This chapter covered a characterization of the resonance landscape of three UMER beams of varying space charge density. This includes measurement of DC beam transmission, which is a new mode of gun operation that will be used in experiments in the quasi-integrable octupole lattice. As expected, operating in the DC mode allows for many turns ($>500$) with minimal space-charge driven end erosion and mostly ``frozen-in" bunch structure. There appears to be no need for longitudinal confinement for the number of turns observed. At certain operating points there is very little beam loss over the first 100 turns. However, after 200 turns there appears to be rapid beam loss at all operating points. Transmission may be improved through tuning the injection match and steering solution. 

The FODO lattice provides a baseline for expected loss rates in the proposed experiment. However, the linear optics discussed in Chapter \ref{ch:lattice} differ greatly in focusing strength and lattice periodicity. In general, since the FODO lattice is the most efficient transport line (in terms of minimizing transverse beam size), we expect loss rates in the single-channel octupole experiment to be equal to or greater than the observations above. At this time, no effort is made to match the $40\ \mu$ A beam at injection and there are very large mismatch oscillations. It is likely that the loss rate will decrease after matching is optimized.

The other aim of this work was to investigate space-charge dependent resonant structure.
Observed loss patterns are sensitive to the strength of the space charge force, as predicted. At high space charge density (as in the 6 mA beam), fewer resonance lines are seen (only 1st order and little 2nd order structure). While this could be attributed to the fast loss of AC signal in the 6 mA beam, comparison to the same turn for lower current beams shows more pronounced first and second order resonant structures. This suggests the space charge tune spread itself may act to decohere resonant particle motion and mitigate resonant losses. 

The ability to selectively drive resonant losses has potential for the quasi-integrable experiments.
Introduction of additional driving terms in the form of a single octupole PCB magnet in the UMER lattice is observed to drive otherwise absent third order resonant structure in the 0.6 mA pencil and $40\ \mu$A DC beams. In the DC beam in particular, there is evidence for a $\nu_x + 2 \nu_y$ sum resonance that gradually emerges when the octupole current increases. Follow-up tune measurements can verify the identification of this resonance and observe other resonant effects, such as increase in beam size.  The true proof of principle demonstration for the QIO experiments will be to show that resonant losses in the linear case are mitigated when octupole fields are included. Here we have shown with an added driving term, a third order resonance has  detectable levels of beam current loss.

%% file: 12.Chapter.tex
\renewcommand{\thechapter}{12}

\chapter{Summary and Future Work}
\label{ch:future}

\section{Summary of dissertation}

This dissertation describes design of a quasi-integrable octupole (QIO) lattice that follows from the theory of nonlinear integrable optics (NLIO) proposed by Danilov and Nagaitsev.\cite{Danilov2010} 
The goal of the UMER nonlinear optics program is to experimentally demonstrate a strong nonlinear lattice for stable beam transport with resonance suppression using octupole elements.
In this dissertation, a design for experiments at low space charge concentration was proposed and numerical studies for error tolerances and lattice performance were completed (Chapters \ref{ch:qio}, \ref{ch:design} and \ref{ch:lattice}). 
This dissertation also discussed steering and resonance studies utilizing  the linear UMER lattice (Chapters \ref{ch:steering} and \ref{ch:res}, respectively). 

\subsection{Design of quasi-integrable experiments}

Chapter \ref{ch:qio} examined particle dynamics in a reduced model of the quasi-integrable octupole lattice, assuming a long octupole channel embedded in a thin-lens FOFO lattice. I verified invariant conservation and applied the technique of frequency map analysis to predict dynamic aperture and octupole-induced tune spread. The single particle Hamiltonian for the octupole lattices possesses unstable fixed points with coordinates that depend on octupole field strength. This should be straightforward to verify experimentally. 
Assuming a channel length $L=64$ cm and transverse envelope $\beta_* = 0.3$ m at the center of the octupole insert, a peak octupole gradient $G_3 = 50/ T/m^3$ should limit dynamic aperture near the edge of a $100 \mu$ m, $40 \mu$A beam and support octupole-induced tune spreads up to $\delta \nu = 0.25$ (RMS spread $\sim 0.03$). Chapter \ref{ch:qio} also demonstrates octupole-driven decoherence of mismatch oscillations and beam halo suppression. Results show that space charge acts against octupole-driven damping. For this reason, in initial octupole lattice experiments we plan to operate UMER at a current density lower than the machine design. Generation of a low-current ``DC-beam" ($\epsilon \sim 100 \mu$ m, $I \sim40 \mu$A) is discussed in Chapter \ref{ch:apparatus}. 

Chapter \ref{ch:apparatus} introduces UMER capabilities and available diagnostics, including design of PCB octupole magnets. The unwanted integrated multipole content is below the $<1\%$ threshold assumed for the UMER quadrupole and dipole designs. The octupole circuits are in-hand and a long octupole channel is currently being assembled. Chapter \ref{ch:design} presents two possibilities for the nonlinear UMER experiments: a distributed and a single-channel design. Using the reduced model from Chapter \ref{ch:qio}, we examine lattice performance (quantified by dynamic aperture and tune spread) on linear focusing errors, orbit distortions and octupole field. Based on this work, we require tolerances in the octupole element of orbit distortion $<0.2$ mm and average background field $<100$ mG. Errors in linear focusing were considered as deviations of the ring tune advance from the quasi-integrable condition. The octupole lattice is more sensitive to tune splitting $\nu_x - \nu_y$ than tune errors of the form $\Delta \nu_x = \Delta \nu_y$, but in general there is not a strong dependence and tune errors $\Delta \nu_x, \Delta \nu_y < 0.1$ should be sufficient for experiment. Including realistic octupole fields based on the PCB design does not significantly affect dynamics and a configuration for the multi-PCB long channel is proposed.

Lattice design for the single-channel octupole experiment is discussed in Chapter \ref{ch:lattice}. Here I describe a method for finding a linear lattice solution that provides transverse focusing equivalent to the thin-lens FOFO lattice and propose a lattice solution for the low-current ``DC-beam" experiments. Initial results from full ring simulations with the WARP code are shown. Due to the high-emittance beam, particle orbits sample large amplitudes where quadrupole and dipole magnets have significant nonlinearity. This shows up as externally-induced tune spreads in the linear lattice ($\delta \nu \sim 0.05$) that are comparable to the octupole-induced spread observed in WARP and the best-case estimate from the reduced model ($\delta \nu \sim 0.04$ and $\delta \nu \sim 0.25$ respectively). This is a complication specific to transport of a high-emittance beam. 
The lattice as implemented in WARP has large tune errors and does not perform well when both space charge and octupole fields are included. Better performance is expected for a further optimization of the solution.

\subsection{Tuning and characterization of UMER lattice}

Chapter \ref{ch:steering} discusses improved algorithms for orbit control to meet the tolerances required by the octupole experiment. Control of orbit distortion to within $|x| < 3.7$ mm and $|y|<4.1$ mm has been demonstrated experimentally. For experiments with the high-emittance DC beam additional vertical correction is necessary to avoid large local excursion. In the ring section identified for the long octupole channel, $|x| < 0.7$ mm and $|y|<0.1$ mm. The large horizontal distortion is limited by the background Earth field. $\mu$-metal shielding of the ambient fields around the octupole element is recommended and water cooling of the dipole steering magnets will be necessary. 

Finally, Chapter \ref{ch:res} investigates resonant structure in the UMER FODO lattice, as a function of current density both with and without an octupole driving element. It is observed that the space charge force acts to decohere higher order resonances, even when the octupole driving term is included. An octupole-driven third order sum resonance is identified in the $40\ \mu$A DC beam measurements. The ability to drive and observe third and higher order resonances is crucial for the nonlinear lattice experiments, as demonstrating octupole-induced resonance suppression is a strong demonstration of the feasibility of nonlinear integrable optics. 

The $40\ \mu$A resonance studies are the first experiments conducted with the DC beam. 1000 turns are observed, with longitudinal bunch profile preserved over transport (compared to the high-current UMER beams, where space charge drives bunch expansion and inter-penetration). Despite the lack of end erosion, the DC beam still experiences large losses ($30\%$ at turn 200 and $>90\%$ loss after 1000 turns). As long confinement times are desired for the nonlinear experiments, more work needs to be done to understand and mitigate current loss. Over the course of the 12-hour experimental run, the gun output current slowly decreased ($40\to15\ \mu$A). This is likely due to increased power load during DC operation and should be mitigated before nonlinear experiments begin.

\section{Future work}

Much work remains to be done preparing UMER for octupole lattice experiments. In this section I make recommendations for ``next step" preparations and propose experiments. Here I focus on preparations for the single-channel experiment. The distributed octupole lattice has already been implemented experimentally (Chapter \ref{ch:distr}), although should be revisited with better tuning and at lower beam current. 
The experiments suggested here could also be used with the distributed octupole configuration.

\subsection{Preparation for experiments}

The design of the single-channel experiment includes a custom $20^{\circ}$ ring section for the octupole insert. A long mount has been designed for the multi-PCB octupole channel that uses existing screw holes and can be mounted without adjustment to the $20^{\circ}$ plate. The recommended location for the octupole section is at RC9, as this is where the the most precise control of the local orbit distortion has been demonstrated (see Chapter \ref{ch:steering}). Additionally, with this orientation the two unoccupied drift regions will be aligned with the imaging screens at RC3 and RC15. 
Imaging the beam at the XY-symmetric waist (rather than at more asymmetric points) will simplify tuning of the injection match and observations during octupole lattice experiments.

The first step before implementing the single-channel octupole experiment is tuning up the linear lattice solution proposed in Chapter \ref{ch:lattice}. It would be best to do this before installing the custom $20^{\circ}$ octupole ring section, as the proposed RC9 location will replace the RC9 diagnostic chamber (which includes phosphor imaging screen). 
Orbit tolerances for the $\epsilon=100\ \mu$m DC beam will need to be tighter than the steering solutions demonstrated in Chapter \ref{ch:steering}. 
The matched beam edge extends to $\sim 1.6$ cm radially, while the pipe aperture is $r_{wall}=2.5$ cm. Ideally, first turn orbit excursions should be kept within $1$ mm. Tuning for turn-to-turn oscillation amplitude of $<1$ mm has been demonstrated using the RCDS optimization method.\cite{Huang2013} This accommodates a $\sim40\%$ mismatch without scraping. 

While enhanced vertical orbit control $|y|<1.2$ mm has been demonstrated with the addition of SSV correctors (Section \ref{sec:steering:SSV}), this not the most ideal approach as there is not space to include them in all ring sections (including the octupole section). Currently a set of radial field-canceling Helmholtz coils is being installed at each ring section, which should reduce the average background field. 
There are no plans to improve horizontal orbit control except in the octupole section.
Due to the sensitivity of particle dynamics to orbit distortion in the octupole channel, I recommend either vertical-field-canceling Helmholtz coils or $\mu$-metal shielding to reduce the average vertical field from $\sim400$ mG to $<100$ mG. 

The response of beam orbit to the steering algorithm will likely differ greatly in the proposed non-FODO lattice. All orbit characterization assumes the standard UMER FODO lattice with quadrupole currents $I_F = I_D = 1.826$ A. 
The betatron wavelength is much longer in the proposed octupole lattice, which operates at $\nu \sim 3.27$ compared to the standard $\nu \sim 6.7$. Prior to implementation, the steering algorithm should be tested in VRUMER with the quadrupole values proposed in Chapter \ref{ch:lattice}. A different approach may yield better results. 

After steering is optimized in the non-FODO lattice we can consider second order beam moments. Using an envelope solver such as MENV, the injection line focusing elements can be optimized for the high-emittance DC beam. This will need to be empirically adjusted following the procedure developed in \cite{HaoThesis}. Finally, measured properties of the linear lattice should be compared to predictions. This includes beam size at the phosphor screen locations and fractional tune measurements. Tuning of the lattice and refinement of the models may be necessary to find good agreement.

There are many open questions that can be addressed via simulation while hardware is being prepared. First, the operating point investigated with WARP PIC simulations in Section \ref{sec:lattice:warp} is very near the fourth order resonance $\nu_y=1/4$, which causes de-stabilization of particles near the beam core and hollowing of the beam distribution when octupole fields are included. Reducing the fractional tune of the lattice (by choosing a slightly larger $\beta_*$ may avoid these losses, but simulations at the new operating point should be done to investigate if similar losses are seen for higher order resonances ($\nu_y=1/5$ or $\nu_y=1/6$ and so on).
Second, for now chromatic tune spreads are assumed to be low, $\sim0.004$, for energy spreads $\sim10$ eV. Energy measurements of the ``DC beam" can establish if this is a fair assumption. In the case of large energy spread, chromaticity-correcting sextupole magnets may be required. Simulations with energy spread can determine if this is necessary by predicting the effect of chromatic spreads on dynamic aperture. 
Finally, it should be a matter of priority to benchmark simulation models against ring measurements \textit{for the proposed lattice solution}, as previous efforts have concentrated on finding agreement near the nominal FODO lattice with $I_f = I_D = 1.826$ A.

\subsection{Proposed experiments}

A variety of experiments are possible once the octupole element is in place and the solution is well-tuned. The simplest proof-of-principle experiment is demonstrating stable beam transport with the addition of octupole fields. Turn-by-turn loss curves will be measured as a function of octupole strength and compared to predicted aperture dependence. 

A low-emittance probe beam can be used to measure tune, following the approach in Section \ref{sec:apparatus:meas-tune}. The amplitude-dependent tune of the octupole lattice can be sampled by varying the initial injection amplitude of the probe beam. If good agreement is found between the envelope model and the measured lattice it will be possible to measure the invariant $H_N$ and place bounds on invariant conservation. 

Damping of mismatch oscillations (predicted in Section \ref{sec:qio:halo-studies}) can be observed by injecting an initially mismatched beam and using multi-turn phosphor screen imaging to measure the beam profile. The knock-out method will need to be adjusted for the longer betatron wavelength. If halo is observed to form in the linear case, high dynamic range imaging can be used to quantify halo dependence on injection mismatch and octupole strength.\cite{HaoThesis}

Another set of experiments involve purposefully violating the quasi-integrable condition to understand the robustness/limitations of the octupole lattice. As the octupole channel is composed of multiple PCB circuits, each independently powered, the longitudinal octupole profile can be changed arbitrarily and resulting loss curve measured. In addition, Chapter \ref{ch:lattice} proposed a method for shifting the lattice tune while maintaining a matched beam. Using this method we can conduct a small-footprint ``tune scan" around the quasi-integrable condition. Another method is to vary the beam energy to scan across a range of tunes. For all beams, fine adjustment of the 10 keV nominal energy is possible.

Finally, the tune scan method introduces the possibility of operating near resonances and observing octupole-induced resonance suppression. This is the true ``proof-of-principle" experiment for the nonlinear integrable optics theory, proving that losses due to incoherent tune resonances can be suppressed with strong nonlinear fields and without loss of dynamic aperture. 

\subsection{Extensions of the planned experiments}

Observing resonant suppression as mentioned above may be difficult. The linear lattice needs to be tuned up on a non-resonant condition, but changing linear focusing to move towards a resonant operating point will violate the quasi-integrable condition on lattice tune. Chapter \ref{ch:lattice}, Section \ref{sec:lattice:solenoids} describes the use of symmetric solenoid lenses to control the beam waist in the octupole channel without changing the lattice function in the surrounding linear lattice. Incorporation of solenoids into the proposed experiment will be a powerful tool for shifting lattice tune onto resonant frequencies. The solenoids as proposed require relatively weak field strengths and should be straightforward to design.

Future plans include extending the octupole lattice experiments to non-negligible space charge densities. This takes advantage of UMER's variable beam current and investigates the applicability of nonlinear integrable optics in a space-charge-dominated regime. However, the incoherent tune shift of the 0.6 mA beam, $\delta \nu = 0.94$, is a large increase from the $40\ \mu$A DC beam, $\delta \nu = 0.005$. Currently, a method for mid-range space charge is being developed for use in UMER.\cite{BernalAAC2017} Total beam current will be reduced by double-aperturing the beam, using an second aperture plate installed downstream of the aperture wheel. Nominal operating parameters for the double-apertured beam are $I=60\ \mu$A and $\epsilon=0.13\ \mu$m. This corresponds with tune depression $\nu/\nu_0=0.95$ and incoherent tune spread $\delta \nu = 0.3$ (in the nominal UMER FODO lattice with $\nu =6.7$). The $60\ \mu$A double-apertured beam will be used for tests of the octupole lattice in a regime relevant to existing high-intensity machines.

%% file: A.Appendix.tex
\renewcommand{\thechapter}{A}
\renewcommand{\chaptername}{Appendix}

\chapter{Additional notes on linear-focusing single particle dynamics}
\label{ap:Hamiltonian}

This appendix follows the derivation for single particle dynamics in a linear focusing accelerator with greater detail than in Chapter \ref{ch:theory}, including discussion of momentum-dependent effects. Linear focusing assumes that only dipole and quadrupole fields are present. This approach is presented more thoroughly in \cite{SYLee} and \cite{uspasEM} but summarized here.

\section{Derivation of single particle Hamiltonian}

The Hamiltonian for the general case of a charged particle moving through electromagnetic fields is

\begin{equation}
\textrm{H} = \sqrt{ m^2 c^4 + c^2\left(\vec{\Pi}-e\vec{A}\right)^2} + e \Phi.
\label{eq:H-appendix-general}
\end{equation}

\noindent for electron charge $e$, vector potential $\vec{A}$ and scalar potential $\Phi$. The canonical momentum is $\vec{\Pi} = \vec{P} + e \vec{A}$ where $\vec{P}$ is the relativistic momentum $\vec{P} = \gamma m \vec{v}$. 
The equations of motion are 
\begin{align}
\dot{x} &= \frac{\partial \textrm{H}}{\partial P_x} \\ 
\dot{P_x} &= -\frac{\partial \textrm{H}}{\partial x}
\label{eq:H-appendix-eqns-appendix}
\end{align}
\noindent and similar for $y$, $z$.

Several approximations and canonical transformations are applied to tailor the general electromagnetic case to an accelerator system. First, Eq. \ref{eq:H-appendix-general} is transformed into the beam-frame, curvilinear Frenet-Serret coordinate system (see Fig. \ref{fig:frenet-serret}). Coordinate $s$ is the propagation distance along a reference orbit and variable $\rho$ is defined as the local radius of curvature along this orbit, $\frac{1}{\rho} \equiv \frac{d\hat{x}}{ds}$.

We choose appropriate canonical transformation to use $s$ as the independent variable, rather than time $t$. 
A small angle approximation is applied, assuming transverse momenta $P_x$ and $P_y$ are much smaller than total momentum $P$: 
\begin{equation}
\tilde{H} \approx -P \left(1 + \frac{x}{P} \right) + \frac{1+x/P}{2P} \left[\left(P_x-eA_x\right)^2 + \left(P_y-eA_y\right)^2 \right] - eA_s.
\end{equation}

\noindent In the ideal accelerator, the electric potential is zero and the magnetic fields have only transverse components (therefore we safely set $A_x = A_y = 0$ and $\Phi = 0$).
With some algebra, the Hamiltonian can be expressed as 

\begin{equation}
\tilde{H} = -P \left(1 + \frac{x}{\rho} \right) \left(1-\frac{P_x^2}{2P^2} - \frac{P_y^2}{2P^2}\right) - eA_s \left(1 + \frac{x}{\rho} \right). 
\end{equation}

\noindent The last transformation made is into dimensionless momentum variables $p_x = P_x/p_0$ and $p_y = P_y/p_0$, where $p_0$ is the nominal or ``design" momentum of the beam. The resulting Hamiltonian is

\begin{equation}
H = -\frac{P}{p_0} \left(1 + \frac{x}{\rho} \right) \left(1-\frac{1}{2}p_x^2 -\frac{1}{2}p_y^2\right) - \frac{e}{p_0}A_s \left(1 + \frac{x}{\rho} \right).
\label{eq:H-appendix1}
\end{equation}

Now we assume a form for $A_s$ corresponding to a linear focusing accelerator lattice. While the preceding was applicable for any applied fields that can be expressed as $\vec{A} = \left(0,0,A_s\right)$, the following only applies to lattices with linear (quadrupole) focusing. Inside an ideal quadrupole, the magnetic potential is 

\begin{equation}
A_s = \frac{G_1(s)}{2} \left(x^2 + y^2\right)
\end{equation}

\noindent where $G_1 = dB_x/dy = dB_y/dx$ is the quadrupole gradient. Ignoring terms higher than second order and assuming design energy ($P = p_0$) the Hamiltonian becomes: 
\begin{equation}
H =  \frac{1}{2} \left(p_{x}^2 + p_{y}^2\right) + \frac{1}{2} \left(K_x(s)x^2 +K_y(s)y^2 \right).
\label{eq:H-appendix}
\end{equation}

\noindent Here, focusing strength is $K_x(s) = eG_1(s)/p_0$ and  $K_y(s) = -eG_1(s)/p_0$.\footnote{If dipole fields are included, $K_x(s) = \rho(s)^{-2} + eG_1(s)/p_0$. The following analysis is still valid.} In a transport line containing only quadrupoles, $K_x(s) = -K_y(s)$. Additionally, if $K_x(s)=K_x=\textrm{constant}$, Eq. \ref{eq:H-appendix} is equivalent to two uncoupled harmonic oscillators.

\section{Single particle equations of motion}

Applying Hamilton's equations of motion (Eq. \ref{eq:H-appendix-eqns-appendix} for appropriate coordinates), we arrive at Hill's equation:
\begin{equation}
z''(s) + K(s)z = 0.
\label{eq:diff-eqn-motion-appendix}
\end{equation}

\noindent Here $z$ is used as a general variable, $z \in \{x,y\}$. In an accelerator ring, $K(s)$ is a periodic function in s. The solution to Hill's equation has the form of a Floquet transformation
\begin{equation}
z(s) = A w(s)e^{\pm i \psi(s)}
\label{eq:eqn-motion-appendix}
\end{equation}

\noindent for amplitude constant $A$, amplitude function $w(s)$ and phase function $\psi(s)$. 
Substitution of Eq. \ref{eq:eqn-motion-appendix} into Eq. \ref{eq:diff-eqn-motion-appendix} and definition of the betatron amplitude function. $\beta(s) \equiv w(s)^2$ leads to the following constraints on orbit amplitude $\sqrt{\beta(s)}$ and phase $\phi(s)$:

\begin{align}
\psi' &= \frac{1}{\beta(s)} \label{eq:psi-prime} \\
1 &= \frac{1}{2}\beta\beta''-\frac{1}{4}\beta'^2 + K(s)\beta^2  \label{eq:beta-eq}
\end{align}

\noindent As $\omega(s)$ is the envelope function for single particle oscillations, for a particle with amplitude $A$, $A\omega(s)$ defines the maximum extent of its orbit. An ensemble of particles with emittance $\epsilon$ will have maximum extent $\sqrt{\epsilon \beta}$.

\section{Off-momentum particles}
The RMS envelope equations describe the propagation of a particle distribution at the design energy (in the case of UMER, 10 keV). However, in reality the beam distribution includes particles at a range of longitudinal momenta. Energy spread is quantified as fractional momentum deviation $\delta \def \frac{p-p_0}{p_0}$ from the design momentum $p_0$. In a synchrotron, the energy spread is set by the fill factor of the RF cycle. An initially mono-energetic beam of length $\Delta t$ will oscillate in $\Delta t - \Delta E$ space, and therefore have an energy spread $\delta = \Delta E$. The picture is not quite as straightforward when considering a coasting beam, as is typical for UMER, where $\delta$ is expected to be much smaller than in a synchrotron. 

A particle with momentum deviation $\delta > 0$ has a magnetic rigidity $B\rho=\frac{p}{q}$ greater than the on-momentum particle, and therefore experiences a smaller bending angle through the dipoles and a weaker focusing gradient $K_x = \frac{\partial B_{y} / \partial x}{B\rho}$ in the quadrupoles. The converse is true for $\delta < 0$. 

Off-momentum particles follow an orbit $x(s) = x_0(s)+D(s)\delta$ where $x_0$ is the trajectory at the design energy (the solution to Eq. \ref{eq:diff-eqn-motion-appendix}). The correction term depends on momentum deviation and the dispersion function $D(s)$. $D(s)$ satisfies the equation:

\begin{equation}
D'' + K_x(s) D = \frac{1}{\rho}
\label{eq:dispersion}
\end{equation}

\noindent The effect of dispersion on an ensemble of particles is to increase the beam size in dispersive regions.

Momentum deviation also leads to a shift in particle tune due to the energy-dependence of the focusing gradient $K_{x}$. This is described by the chromaticity, defined as the derivative of tune with fractional momentum deviation, 

\begin{equation}
C_{x,y} \equiv \frac{d\nu_{x,y}}{d\delta}.
\end{equation}

\noindent To first (linear) order, the natural (uncorrected) chromaticity is given by the integral:

\begin{equation}
C_{x,y} \approx -\frac{1}{4\pi}\oint \beta_{x,y} K_{x,y} ds
\label{eq:chromaticity}
\end{equation}

\noindent and the resulting chromatic tune spread for an ensemble of particles with momentum spread $\delta$ is
\begin{equation}
\Delta \nu_{x,y} \approx C_{x,y} \delta
\end{equation}

\noindent This formulation, following that given in \cite{SYLee}, ignores nonlinear chromaticity. Higher order corrections to the chromatic tune spread arise due to non-linearities in the lattice. For example, high dispersion may lead to off-momentum particles sampling nonlinear regions at large transverse amplitude in the focusing and bending magnets.

%% file: B.Appendix.tex
\renewcommand{\thechapter}{B}
\renewcommand{\chaptername}{Appendix}

\makeatletter
\def\env@matrix{\hskip -\arraycolsep
  \let\@ifnextchar\new@ifnextchar
  \linespread{1}\selectfont
  \renewcommand{\arraystretch}{0.5}
  \array{*\c@MaxMatrixCols c}}
\makeatother

\chapter{Matrix representation of linear focusing elements}
\label{ap:matrix}

Particle evolution in a linear focusing system can be expressed in terms of a matrix formulation. To first order, an orbit with initial condition $(x,x')_1$ at location $s_1$ can be mapped as:

\begin{equation}
\begin{bmatrix} x \\ x' \end{bmatrix}_2 = M_{s_1\to s_2} \begin{bmatrix} x \\ x' \end{bmatrix}_1
\end{equation}

\noindent where $M_{s_1\to s_2}$ is the transfer function between $s_1$ and $s_2$.

\section{Matrix representation of ring}

The following can be found in many accelerator texts, but here is referenced from \cite{SYLee}. For a transport with transfer function 

\begin{equation}
M_{s_2|s_1} =
\begin{bmatrix} C & S \\ C' & S' \\ \end{bmatrix}
\end{equation}

\noindent the Courant-Snyder parameters transform as 

\begin{equation}
\begin{bmatrix} \beta \\ \alpha \\  \gamma \\ \end{bmatrix}_{s_2} =
\begin{bmatrix} C^2 & -2SC & S^2 \\ -CC' & SC'+S'C & -SS' \\ C'^2 & -S'C' & S'^2 \\ \end{bmatrix}
\begin{bmatrix} \beta \\ \alpha \\  \gamma \\ \end{bmatrix}_{s_1}.
\label{eq:CSmatrix}
\end{equation}

\noindent Conversely, if the Courant Snyder parameters are known for two locations in a beam line, the transfer matrix $M$ is

\begin{equation}
M(s_2|s_1) = \begin{bmatrix} \sqrt{\frac{\beta_2}{\beta_1}}\left(\cos{\psi} + \alpha_1 \sin{\psi}\right) & \sqrt{\beta_1 \beta_2} \sin{\psi}  \\
-\frac{1+\alpha_1 \alpha_2}{\sqrt{\beta_1 \beta_2}}\sin{\psi}+\frac{\alpha_1-\alpha_2}{\sqrt{\beta_1 \beta_2}}\cos{\psi} & \sqrt{\frac{\beta_2}{\beta_1}}\left(\cos{\psi}-\alpha_2\sin{\psi}\right) \\ \end{bmatrix}
\label{eq:M-in-alpha-beta}
\end{equation}

\section{Introducing phase errors in the FOFO lattice}

This section shows the source of Eq. \ref{eq:tune-error-matrix}, which is used to simulate the effect for external focusing errors in the quasi-integrable octupole lattice. This is done for the simple model of the octupole lattice, consisting of octupole channel and linear-focusing, thin-lens ``T-insert."
In the ideal case, the T-insert transformation applied at $s_0$ is

\begin{equation}
T(s_0|s_0)=
\begin{bmatrix} 
1 & 0 & 0 & 0 \\ 
-k & 1 & 0 & 0 \\ 
0 & 0 & 1 & 0 \\
0 & 0 & -k & 1 \\ 
\end{bmatrix}.
\label{eq:thinlens}
\end{equation}

\noindent This transfer matrix can be derived from the general expression in Eq. \ref{eq:M-in-alpha-beta}.
For a thin focusing impulse in a periodic FOFO lattice, $\beta_1 = \beta_2$ and $\alpha_1 = - \alpha_2$. The above matrix is the special case where $\psi = 2 n \pi$. Without assuming an integer phase advance in the linear lattice, $T$ becomes

\begin{equation}
T_z = \begin{bmatrix} \cos{\psi_z} - \alpha_z \sin{\psi_z} & \beta_z \sin{\psi_z}  \\
-\frac{1-\alpha_z^2}{\beta_z}\sin{\psi_z}-\frac{2\alpha_z}{\beta_z}\cos{\psi_z} & \cos{\psi_z}-\alpha_z\sin{\psi_z} \\ \end{bmatrix}
\end{equation}

\noindent where the full matrix is simply

\begin{equation}
T(s_0|s_0) = 
\begin{bmatrix} 
\begin{bmatrix} T_x \end{bmatrix} & \begin{matrix} 0&0\\0&0 \end{matrix} \\
\begin{matrix} 0&0\\0&0 \end{matrix} & \begin{bmatrix} T_y \end{bmatrix} \\
\end{bmatrix}.
\end{equation}

\noindent The linear lattice phase advance is $\psi_{x,y} = 2\pi\nu_{x,y} + 2\pi \Delta\nu_{x,y}$ for lattice design tune $\nu_{x,y}$ and tune error $\Delta\nu_{x,y}$. As the lattice design tune is constrained to be an integer value, $\mod \psi_{x,y} = 2\pi \Delta\nu_{x,y}$, and the applied focusing is:

\begin{equation}
T_z = 
\begin{bmatrix} 
\cos{2\pi\Delta\nu_z} - \alpha_z \sin{2\pi\Delta\nu_z} & \beta_z \sin{2\pi\Delta\nu_z} \\
-\frac{1-\alpha_z^2}{\beta_z}\sin{2\pi\Delta\nu_z}-\frac{2\alpha_z}{\beta_z}\cos{2\pi\Delta\nu_z} & \cos{2\pi\Delta\nu_z}-\alpha_z\sin{2\pi\Delta\nu_z}\\
\end{bmatrix}
\end{equation}

%% file: C.Appendix.tex
\makeatletter
\def\env@matrix{\hskip -\arraycolsep
  \let\@ifnextchar\new@ifnextchar
  \linespread{1}\selectfont
  \renewcommand{\arraystretch}{0.5}
  \array{*\c@MaxMatrixCols c}}
\makeatother

\renewcommand{\thechapter}{C}
\renewcommand{\chaptername}{Appendix}

\chapter{Analytic description of a symmetric focusing (FOFO) lattice}
\label{ap:fofo}

The implementation of the quasi-integrable lattice requires the beam to come to a ``round" (XY symmetric) waist in the octupole element. This appendix shows an derivation for envelope evolution in a drift space and the matched envelope solution in a FOFO lattice with symmetry $X(s) = Y(s)$. The FOFO lattice consists of a drift (field-free) region and periodic XY-symmetric thin lens focusing elements.

\section{Symmetric beam waist and free expansion in a drift}

The evolution of the betatron function in the drift space can be reconstructed from the transfer matrices. Here I follow the approach laid out in \cite{Holzer2013}.
The transfer matrix in a drift space is simply 

\begin{equation}
M_{s_0 \to s} =
\begin{bmatrix} 1 & s-s_0 \\ 0 & 1 \\ \end{bmatrix} 
\label{eq:M-drift}
\end{equation}

\noindent Applying the transformation on the Courant-Snyder parameters (shown in Eq. \ref{eq:CSmatrix})

\begin{subequations}
\begin{equation}\beta(s) = \beta_0 - 2\alpha(s) \cdot (s-s_0) + \gamma(s) \cdot (s-s_0)^2 \end{equation}
\begin{equation}\alpha(s) = \alpha_0 - \gamma(s) \cdot (s-s_0) \end{equation}
\begin{equation}\gamma(s) = \gamma_0 \end{equation}
\end{subequations}

\noindent for $\alpha_0 = \alpha(s_0)$ and $\gamma_0 = \gamma(s_0)$. As $\alpha(s) = -\frac{1}{2} \frac{d\beta(s)}{ds}$, the beam waist occurs where $\alpha = 0$, at a longitudinal location

\begin{equation}
s_* \equiv s-s_0 = \frac{\alpha_0}{\gamma_0}
\end{equation}

\noindent and the beta-function amplitude at the waist is

\begin{equation}
\beta_* \equiv \beta(s_*) = \beta_0 - \frac{\alpha_0^2}{\gamma_0}.
\end{equation}

\noindent Through the identity between $\beta$, $\alpha$ and $\gamma$ in Eq. \ref{eq:abg}, we see that

\begin{equation}
\beta_* = \frac{\beta_0}{1+\alpha_0^2} = \frac{1}{\gamma_0}.
\label{eq:beta-star}
\end{equation}

\noindent Eq. \ref{eq:beta-star} is an important scaling law for low-beta insertions. Requiring $\beta*$ to be very small requires large $\alpha_0$. As $\alpha$ is the gradient of the $\beta$-function, this corresponds to large $\beta$ excursions outside the insertion region.

The beam expansion about the waist is described by 

\begin{equation}
\beta(s) = \beta_* + \gamma_* (s-s_*)^2.
\end{equation}

\noindent Substituting from Eq. \ref{eq:beta-star}, and keeping in mind that $\gamma(s) = \gamma_0 = \textrm{constant}$ in a drift, the expression for the betatron function near a waist is 

\begin{equation}
\beta(s) = \beta_* + \frac{(s-s_*)^2}{\beta_*}.
\label{eq:drift}
\end{equation}

\noindent As this expression is derived from the single-particle transfer matrix $M$ in Eq. \ref{eq:M-drift}, this formalism does not include space charge effects. 

\section{Matched envelope in FOFO lattice}
\label{sec:theory:fofo}

\begin{figure}[!tb]
   \centering
   \includegraphics*[width=.9\textwidth]{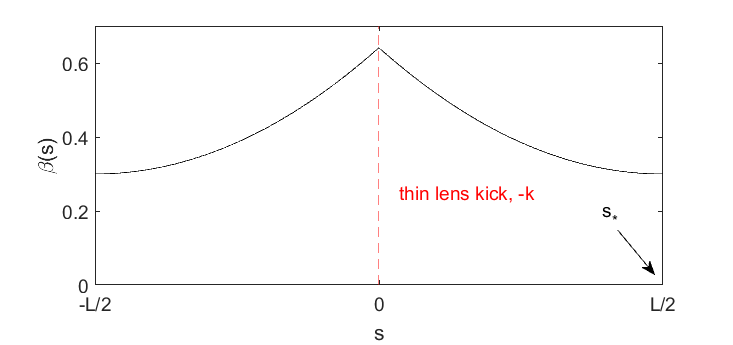}
   \caption{Matched envelope solution for FOFO lattice. A single cell is shown.}
   \label{fig:FOFOcell}
\end{figure}

The framework for an integrable lattice requires symmetric beam waist, over which a nonlinear insertion will be placed. External focusing is provided by the ``T-matrix" insert, visualized in Fig. \ref{fig:toy-model}. A periodic lattice consisting of alternating drift spaces and T-inserts can be considered a symmetric FOFO lattice. This section derives properties of a matched beam in a FOFO lattice with length $L$ and focusing strength $k$/

One period of a FOFO lattice with cell length $L$ can be constructed as a drift of length $L/2$, a thin lens with focusing strength $-k$, followed by a second drift of length $L/2$. For a thin lens transformation, focsing strength $k$ is inversely related to the focal length, $k=\frac{1}{f}$.
An example of the matched beta function is shown in Figure \ref{fig:FOFOcell} for a single FOFO cell.

As the FOFO lattice is XY symmetric, the subscripts $x$ and $y$ will be dropped and the following analysis is valid for both planes. The transfer matrix for one cell is

\begin{align}
M_{s_* \to s_*+L} &=
\begin{bmatrix} 1 & L/2 \\ 0 & 1 \\ \end{bmatrix} \cdot
\begin{bmatrix} 1 & 0 \\ -k & 1 \\ \end{bmatrix} \cdot
\begin{bmatrix} 1 & L/2 \\ 0 & 1 \\ \end{bmatrix} 
\nonumber\\
&=\begin{bmatrix} 1 -\frac{kL}{2} & L\left(1-\frac{kL}{4}\right) \\ -k & 1 -\frac{kL}{2}\\ \end{bmatrix}.
\end{align}

\noindent Applying the transfer matrix to the evolution of $[\beta,\alpha,\gamma]_{s_*}$ according to Eq. \ref{eq:CSmatrix} and enforcing the matching condition $\beta(s_*) = \beta(s_*+L)$ leads to the definition of $\beta_*$ in terms of lattice parameters $k$ and $L$:

\begin{equation}
\beta_* = \frac{L\left(1-\frac{kL}{4}\right)}{\sqrt{1-\left(1-\frac{kL}{2}\right)^2}}
\label{eq:betastar2}
\end{equation}

As it is more natural to use $\beta_*$ as the free parameter when designing an insertion region, the inverted form is more useful. Here, $k$ represents a constraint on the T-insert to give a desired waist size in the nonlinear insertion:

\begin{equation}
k = \frac{L}{L^2/4 + \beta_*^2}.
\label{eq:k}
\end{equation}

\noindent Eq. \ref{eq:betastar2} can be substituted into the expression for $\beta$-function in a drift, Eq. \ref{eq:drift}, to find an expression for $\beta(s)$ in terms of the FOFO lattice parameters. For a matched beam in a FOFO lattice cell of length $L$, $s_*=L/2$ and $\beta(s)$ is

\begin{equation}
\beta(s) = \frac{L-sk(L-s)}{\sqrt{1-\left(1-\frac{Lk}{2}\right)^2}}.
\label{eq:betas}
\end{equation} 
 
In the context of nonlinear quasi-integrable optics, the achievable tune spread scales with the phase advance through the nonlinear insertion. A large phase advance in the insertion is desired, with phase advance defined as above, Eq. \label{eq:phase-adv}. The phase advance can be calculated by integrating the inverse of Eq. \ref{eq:betas}. This has the analytic solution:

\begin{align}
\psi_{drift} &= \sqrt{1-\left(1-\frac{Lk}{2}\right)^2} \cdot \int_0^L \frac{ds}{L-sk(L-s)} \nonumber \\
&= \sqrt{1-\left(1-\frac{Lk}{2}\right)^2} \cdot \frac{2 \tanh^{-1}\sqrt{\frac{k}{L}\frac{L-2s}{\sqrt{kL-4}}}}{\sqrt{kL(kL-4)}}.
\label{eq:phi-analytic}
\end{align}

%% file: D.Appendix.tex
\renewcommand{\thechapter}{D}
\renewcommand{\chaptername}{Appendix}

\chapter{Numerical Analysis of Fundamental Frequency}
\label{ap:naff}

This appendix presents the framework for NAFF implementation in Python that was used for simulated frequency map analysis in this dissertation.

\begin{verbatim}
def naff(signal,sampr):
	import numpy as np

    # -- FFT parameters
    dens = 10**6; # Density of pts for FFT
    freq = np.fft.fftfreq(dens,sampr)

    # -- apply hanning window
    hwin = hanning(signal.size)
    windowed_signal = signal*hwin

    # -- take FFT and extract strongest frequency
    ampl = abs(np.fft.fft(signal,dens))
    f0 = freq[where(ampl==max(ampl))[0][0]]

    # -- initial guess for amplitude is rms of signal
    a0 = sqrt(2)*rms(signal)

    # -- NAFF step 
    # -- (minimize convolution of sine wave with signal)
    result = opt.minimize(overlap_integral,f0,args = \\
	(a0,signal,weight))
    ffund = result.x[0]
return [ffund]
\end{verbatim}

\begin{verbatim}
def overlap_integral(f,a,signal,weight):
    x = arange(0,size(signal)) # time-like variable (step number)
    z = a*exp(-1j*2*pi*f*x) # signal with frequency f
    overlap = -abs(sum(z*signal*weight)) # score is convolution \\
	of signal, weight and pure wave of freq. f
return overlap
\end{verbatim}

%% file: E.Appendix.tex
\makeatletter
\def\env@matrix{\hskip -\arraycolsep
  \let\@ifnextchar\new@ifnextchar
  \linespread{1}\selectfont
  \renewcommand{\arraystretch}{0.5}
  \array{*\c@MaxMatrixCols c}}
\makeatother

\renewcommand{\thechapter}{E}
\renewcommand{\chaptername}{Appendix}

\chapter{Simulation parameters for simple model of quasi-integrable octupole lattice}
\label{ap:simple-model}

This appendix describes the elements used for octupole channel and thin-lens ``T-insert" transformation in the simple model simulation used in Chapter \ref{ch:qio} and Chapter \ref{ch:design}.

\section{Elegant model}

In Elegant, the focusing impulse is applied as a 6D matrix element (\verb|MATR|), defined as

\begin{equation}
\verb|MATR| = \begin{bmatrix} 1 & 0 & 0 & 0 & 0 & 0 \\ -k & 1 & 0 & 0 & 0 & 0 \\ 
0 & 0 & 1 & 0 & 0 & 0  \\ 0 & 0 & -k & 1 & 0 & 0  \\ 
0 & 0 & 0 & 0 & 1 & 0  \\ 0 & 0 & 0 & 0 & 0 &1   \\ 
\end{bmatrix}
\end{equation}

\noindent for specified focusing strength $k$. 
The octupole fields are generated with the multipole element \verb|MULT| of order 3. The \verb|MULT| element delivers canonical kicks based on fourth-order symplectic integration. This was chosen over the alternative option in Elegant, \verb|OCT|, which uses a third-order matrix transformation and is not symplectic.\cite{elegantmanual} As \verb|MULT| amplitude is defined as an integrated strength, the ``smoothness" of the longitudinal profile depends on the number of discrete, flat-top elements used. Results in this thesis use 16 adjacent \verb|MULT| elements, each of length 4 cm, with integrated strength set according to the target strength (Eq. \ref{eq:K3s}) at the element center.

\section{Warp model}
In the WARP PIC code, I use the 2D transverse slice package, which assumes zero coupling between transverse and longitudinal motion. The focusing kick is implemented at the python level as a velocity impulse every $N$ steps, where $N = 64$ cm / stepsize. This simple transformation is shown below, where \verb|i| is iteration number, \verb|v_s| is total forward beam velocity and \verb|gamma| is the relativistic factor.

\begin{verbatim}
def thinlens():
    if mod(i,int(L/stepsize))==0:
        v_x += -k* v_s * gamma * x
        v_y += -k* v_s * gamma * y 
\end{verbatim}

\noindent In WARP, the octupole fields are generated using the magnetic multipole element \verb|mmlt|, which is a pure-harmonic multipole field with arbitrary longitudinal profile. The longitudinal profile is defined with 100 discrete steps in $s$. Therefore the octupole profile used in WARP is a closer approximation of the desired $1/\beta^3(s)$ scaling than the profile defined in the Elegant model. 

%% file: F.Appendix.tex
\makeatletter
\def\env@matrix{\hskip -\arraycolsep
  \let\@ifnextchar\new@ifnextchar
  \linespread{1}\selectfont
  \renewcommand{\arraystretch}{0.5}
  \array{*\c@MaxMatrixCols c}}
\makeatother

\renewcommand{\thechapter}{F}
\renewcommand{\chaptername}{Appendix}

\chapter{Method for testing closed orbit distortion tolerance in reduced model}
\label{ap:simple-steering}

This appendix reviews how centroid orbit corrections are made for orbit distortion tolerance calculations described in Section \ref{sec:design:steering}. 
These simulations use WARP to model a 64 cm octupole channel with an ideal linear FOFO thin-lens transformation. To simulate the effect of a closed orbit distortion, a coherent orbit correction is made for each particle at the location of the thin lens focusing transformation. The two cases examined in Section \ref{sec:design:steering} are:

\begin{enumerate}
\item Orbit distortion in otherwise shielded 64 cm section (centroid has straight trajectory between steering elements.)
\item Curved orbit distortion due to constant background field
\end{enumerate}

\section{Case 1: Straight/shielded orbit distortion}

\begin{figure}[!tb]
\centering
\includegraphics[width=\textwidth]{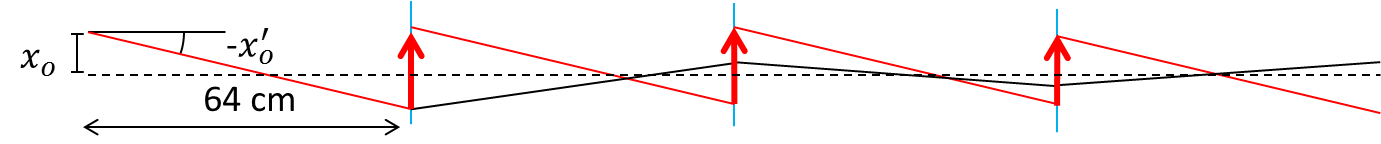}
\caption{Schematic of shielded closed orbit distortion through octupole channel for error analysis calculations. Red line is centroid motion of beam with initial offset $x_0$, $x'_0$. Thick red lines indicate thin lens centroid transformation. Black line is centroid motion without centroid transformation. }
\label{fig:straightorbitdistortion-appendix}
\end{figure}

Fig. \ref{fig:straightorbitdistortion-appendix} shows distortion of the closed orbit in the case where particle trajectories are straight between magnetic elements. The distortion is defined by initial conditions $x_0$ and $x'_0$, which represent the initial offset in beam centroid. At the same location as the FOFO thin lens transformation (that represents focusing in the linear part of the ring), a centroid transformation is also made. For a single plane, the following correction is made for each particle:

\begin{equation}
\begin{bmatrix} x \\ x' \end{bmatrix}_f = \begin{bmatrix} x \\ x' \end{bmatrix}_i - \begin{bmatrix} L x_0 \\ -k(x_0 + Lx_0') \end{bmatrix}.
\label{eq:straightorbitdistortion}
\end{equation}

\noindent The derivation follows. Consider single particle matrix equations for propagation through a focusing element (strength $k$) and a drift of length $L$. 

\begin{equation}
\begin{split}
\begin{bmatrix} x \\ x' \end{bmatrix}_f 
& = \begin{bmatrix} 1&0 \\ -k&1 \end{bmatrix}_i \ast \begin{bmatrix} 1&L \\ 0&1 \end{bmatrix}_i \ast \begin{bmatrix} x \\ x' \end{bmatrix}_i \\
& = \begin{bmatrix} x_i + Lx'_i \\ -kx_i + (1-kL)x'_i \end{bmatrix}.
\end{split}
\end{equation}

\noindent Now divide motion into single particle $x_p$ and centroid $x_c$ components, where $x_i = x_{c,i} + x_{p,i}$ and $x'_i = x'_{c,i} + x'_{p,i}$. The matrix equation for a single pass becomes:

\begin{equation}
\begin{split}
\begin{bmatrix} x \\ x' \end{bmatrix}_f 
&= \begin{bmatrix} x_{c,i} + Lx'_{c,i} \\ -kx_{c,i} + (1-kL)x'_{c,i} \end{bmatrix} + \begin{bmatrix} x_{p,i} + Lx'_{p,i} \\ -kx_{p,i} + (1-kL)x'_{p,i} \end{bmatrix} \\
&=\begin{bmatrix} x_p \\ x'_p \end{bmatrix}_f + \begin{bmatrix} x_c \\ x'_c \end{bmatrix}_i + \begin{bmatrix} L x_{c,i} \\ -k(x_{c,i} + Lx'_{c,i}) \end{bmatrix}.
\end{split}
\end{equation}

\noindent Considering $\begin{bmatrix} x \\ x' \end{bmatrix}_f = \begin{bmatrix} x_c \\ x'_c \end{bmatrix}_f + \begin{bmatrix} x_p \\ x'_p \end{bmatrix}_f$, one can examine just the centroid motion $\begin{bmatrix} x_c \\ x'_c \end{bmatrix}_f = \begin{bmatrix} x_c \\ x'_c \end{bmatrix}_i + \begin{bmatrix} L x_{c,i} \\ k(x_{c,i} + Lx'_{c,i}) \end{bmatrix}$. Let $x_{c,i} = x_0$ and $x'_{c,i} = x'_0$. The centroid will follow the same distorted path through the channel from turn to turn if $\begin{bmatrix} L x_0 \\ -k(x_0 + Lx_0') \end{bmatrix}$ is subtracted at the end of each pass, as in Eq. \ref{eq:straightorbitdistortion}.

It should be noted that in Case 1, I ignore the effect of the bending dipoles (there are two in a 64 cm channel at 32 cm intervals). This relies on two assumptions: the bending dipole field is ``flat" (bend angle does not depend on displacement in dipole) and, in a shielded environment, there is a setting for the dipoles that allows the beam to propagate centered through all elements. 

While $x_0$, $x'_0$ is a two-dimensional space of possible orbit distortion, I simplified the problem by requiring that the distortion be symmetric in a 64 cm drift, with $x_f = -x_0$ and $<x_c(s)> = 0$. In this case, $x'_0 \approx \sin{x'_0} = L/2x_0$ and the distortion is parametrized in terms of $x_0$ only. The maximum acceptable closed orbit distortion stated in Chapter \ref{ch:design} refers to this value $x_0$.

\section{Case 2: Curved orbit distortion due to background field}

\begin{figure}
\centering
\includegraphics[width=\textwidth]{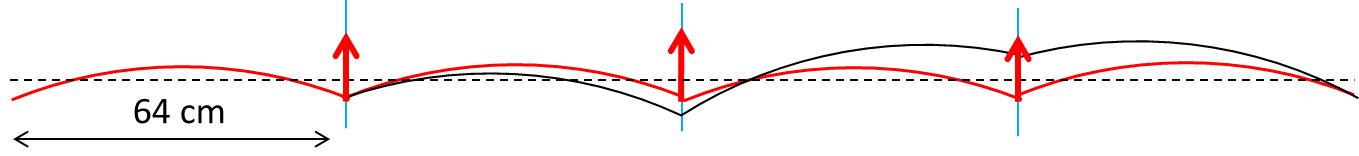}
\caption{Schematic of unshielded closed orbit distortion through octupole channel for error analysis calculations in vertical plane. Red line is centroid motion of beam with thin lens centroid transformation at thick red arrows. Black line is centroid motion without centroid transformation, just periodic thin lens focusing element. }
\label{fig:vertcurvedorbitdistortion-appendix}
\end{figure}

To model the effect of orbit  due to immersion in ambient background fields, I applied a similar thin-lens centroid transformation. Fig. \ref{fig:vertcurvedorbitdistortion-appendix} shows the case where steering corrections are made every 64 cm, which represents vertical steering with RSV steerers. The orbit is assumed to be ``as centered as possible." For a given background field, initial conditions $y_0$ and $y'_0$ are chosen so that $y_i=y_f$ across a 64 cm drift, and $\max{y}=\min{y}$ in the drift. In this case, the centroid correction is simply:

\begin{equation}
\begin{bmatrix} y \\ y' \end{bmatrix}_f = \begin{bmatrix} y \\ y' \end{bmatrix}_i + \begin{bmatrix} 0 \\ ky_0 + \theta \end{bmatrix}
\label{eq:curvedorbitdistortion-appendix}
\end{equation}

\noindent where $\theta$ is the bending angle due to the background field. Assuming a constant background field $B_x$, $\theta = \frac{LB_x}{B \rho}$ depends only on length of straight section $L$ and particle rigidity $B \rho$. To meet the condition $y_i=y_f$, $y'_0 = \theta/2$. To satisfy $\max{y}=\min{y}$, $y_0=\rho/2 ( 1-\cos{(\theta/2)} )$.    

\begin{figure}
\centering
\includegraphics[width=\textwidth]{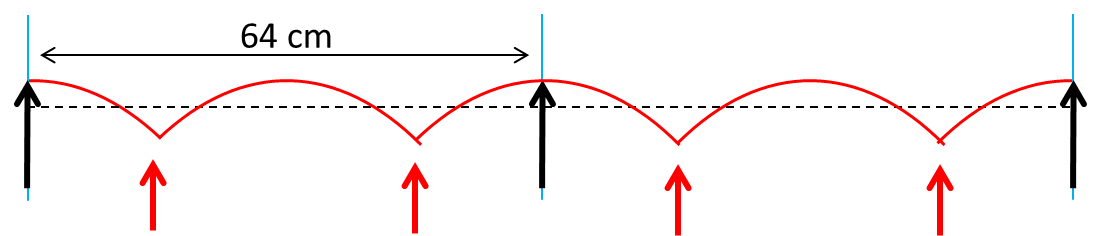}
\caption{Schematic of unshielded closed orbit distortion through octupole channel for error analysis calculations in horizontal plane. Red line is centroid motion of beam with thin lens centroid transformation at thick red arrows. Thick black arrows indicate thin-lens focusing element. }
\label{fig:horzcurvedorbitdistortion-appendix}
\end{figure}

As horizontal steerers are located every 32 cm and will be co-housed in the long 64-cm octupole channel, the horizontal plane requires a different treatment. The steering correction of Eq. \ref{eq:curvedorbitdistortion-appendix} is split and applied at different locations: 

\begin{subequations}
\begin{equation}
\begin{bmatrix} x \\ x' \end{bmatrix}_f = \begin{bmatrix} x \\ x' \end{bmatrix}_i + \begin{bmatrix} 0 \\ kx_0 \end{bmatrix}
\label{eq:horzcurvedorbitdistortion1}
\end{equation}
\begin{equation}
\begin{bmatrix} x \\ x' \end{bmatrix}_f = \begin{bmatrix} x \\ x' \end{bmatrix}_i + \begin{bmatrix} 0 \\ \theta \end{bmatrix}.
\label{eq:horzcurvedorbitdistortion2}
\end{equation}
\end{subequations}

\begin{figure}
\centering
\includegraphics[width=\textwidth]{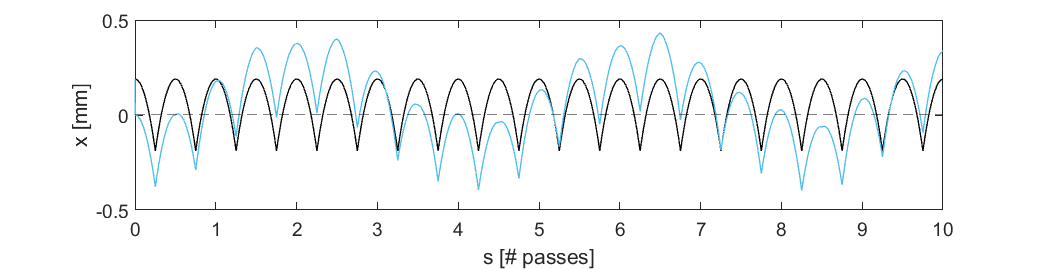}
\caption{Single particle orbits in case of unshielded orbit distortion. A particle is launched on the closed orbit (black) and with initial offset (blue) for octupole strength $G_{3,max}=0$. }
\label{fig:horsptraces-appendix}
\end{figure}

\noindent Eq. \ref{eq:horzcurvedorbitdistortion1} is applied at the ends of the 64 cm channel, same location as the thin-lens focusing kick, to cancel the kick from the thin focusing lens on the displaced centroid. Eq. \ref{eq:horzcurvedorbitdistortion2} is applied every 32 cm at dipole locations (in 64 cm channel, at $s=16,48$ cm). This is pictured in Fig. \ref{fig:horzcurvedorbitdistortion-appendix}. Fig. \ref{fig:horsptraces-appendix} shows resulting orbits over 10 passes through a 64 cm drift. No additional steering correction is made when octupole fields are included.

%% file: G.Appendix.tex
\renewcommand{\thechapter}{G}
\renewcommand{\chaptername}{Appendix}

\chapter{Models of UMER magnets in hard-edged approximation}
\label{ap:model}

Accelerator magnets are often included in simulation models using a hard-edged approximation. Under this approximation, the element is defined solely by an effective length and effective field/gradient. A hard-edged approximation of UMER elements is used in all models in this dissertation except in the WARP PIC model, where gridded elements based on Biot-Savart solutions of the PCB magnets are used instead. 
A description of the hard-edged model for UMER quadrupoles can be found in \cite{Bernal2006,Bernal1999}. The most recent calculations for the hard-edged quadrupole model in \cite{Kishek2010} are used here.

\section{Dipole edge-focusing} \label{sec:lattice:edges}

Edge focusing is the focusing impulse a particle experiences when moving from a region of high to low vertical field (or vice versa). 
The magnitude of the force is proportional to the edge angle $\zeta$ of the magnet-edge normal vector and the beam velocity vector. For $\zeta=0$ there is no focusing or defocusing dipole effect. In UMER, $<\zeta> \sim \pm 5^o$ for leading/trailing edges. This results in a focusing impulse at each edge in the vertical plane, and defocusing impulses in the horizontal plane. 
In the ring dipoles, there is also a horizontal geometric focusing due to path length dependence on transverse position. 
For an arbitrary dipole field profile, the geometric focusing and horizontal edge defocusing exactly cancel. \cite{Kishek2010-2} 
A more detailed description of dipole edge effects can be found in \cite{Brown1982} or chapter four of \cite{Livinghood1961}.
Edge-focusing is included as a quadrupole gradient inside the dipole elements. Specifics of the implementation in each code are discussed in the following sections.

\section{VRUMER model}\label{sec:model:vrumer}
\begin{table}[!t]
\centering
\begin{threeparttable}
\caption{UMER steering magnet strengths, from \cite{umerweb}. }
\label{tab:UMERsteererstrength}
\begin{tabularx}{\textwidth}{X R{1.5cm} R{1cm} R{1.5cm} R{2.7cm} R{2.5cm} R{1cm}}
Name & length & $l_{eff} $ & radius & Int. Field  & strength  & count\\
     & [cm]   & [cm]       & [cm]   & [G-cm/A]    & [$^o$/A]  & [\#]\\     
\hline\hline
BD          & 4.44 & 3.76 & 2.87 & 19.917& 3.37 & 36\\
PD          & 4.40 & 5.18 & 4.40 & 1.913 & 0.32 & 1 \\
RSV         & 3.80 & 5.67 & 5.75 & 3.886 & 0.66 & 18\\
SD $\ast$& 2.37 & 5.18 & 4.73 & 3.317 & 0.56 & 14\\
SSV         & 1.54 & 3.06 & 2.79 & 3.627 & 0.61 & 11\\
\hline
\end{tabularx}
\begin{tablenotes}
\footnotesize
\item $\ast$ Not all SD's have these parameters. SD5 is identical to the BD circuit, SD4 is identical to RSV circuit and SD6 is identical to SSV circuits.
\end{tablenotes}
\end{threeparttable}
\end{table}

Table \ref{tab:UMERsteererstrength} lists parameters of all the UMER steering/dipole magnets. Steerers of length $l_{eff}$ with appropriate strength were used in the VRUMER model.
UMER ring quads were also modeled as hard-edged, with a gradient of $G=3.609$ G/cm/A, hard-edged factor $f=0.8354$ and length $4.475$ cm. For QR1, $G=1.010$ G/cm/A, $f=0.8964$. For YQ, $G=1.110$ G/cm/A and $f=0.8965$. Although the effective length of QR1 and YQ are much longer than the ring quads (5.999 cm), in VRUMER they are modeled as length $4.475$ cm elements. An additional scaling factor is applied to account for this discrepancy. The YQ tilt and displacements were not included in the VRUMER model used here.\footnote{VRUMER has since been updated to include a transverse YQ displacement of +1.389 cm.}

At this time, the VRUMER model does not include edge-focusing or higher order terms such as magnetic nonlinearities. Additionally, it does not include the steering kick from the off-centered YQ magnet, or coupling due to skew terms from magnet rotations. These effects could be implemented if desired.

\section{Elegant model}

In Elegant, dipoles are modeled with the \verb|SBEND| sector dipole element. Dipole forces are applied as second-order transformation matrices. Edge focusing is calculated assuming $5^{\circ}$ dipole edge angles and specified edge-field integral. 
The quadrupole magnets use the \verb|QUAD| implementation with third-order matrix transformations. Both elements are not symplectic. Parameters used in the definition of Elegant parameters are given in Table \ref{tab:warp-env-sim}.

\begin{table}
\centering
\caption{Simulation parameters for Elegant model of single-channel lattice. }
\label{tab:warp-env-sim}
\vspace{10pt}
\begin{tabularx}{0.7\textwidth}{X L{4cm}}
\hline
Parameter & Value \\
\hline
Quadrupole gradient & 3.608 G/cm/A\\
Quadrupole length   & 5.164 cm \\
Dipole bend angle   & $10^{\circ}$\\
Dipole edge angle   & $\pm5^{\circ}$\\
Dipole length       & 3.760 cm \\ 
Half-gap 			& 2.87 cm \\
Fringe field integral \cite{Kishek2010-2} & 1.02 G \\
\hline
\end{tabularx}
\end{table}

\clearpage
\section{MENV model}
In MENV, edge focusing is applied as an effective hard-edged quadrupole that is the same length as the dipole (listed in Table \ref{tab:menv-sim}). I found that geometric strengths $K_x = 1.978\ m^{-2}$ and $K_y = -14.695\ m^{-2}$ for the ``edge-focusing" quads gave good agreement with WARP results over a half-cell (1.92 meters). A comparison is given in \ref{fig:benchmark}.

\begin{table}
\centering
\caption{Simulation parameters for MENV integration of $60 \mu$A ``DC-beam". }
\label{tab:menv-sim}
\vspace{10pt}
\begin{tabularx}{0.7\textwidth}{X L{4cm}}
\hline
Parameter & Value \\
\hline
Perveance          & $4.49 \times 10^{-7}$ \\
Quadrupole gradient& 3.446 G/cm/A\\
Quadrupole length  & 3.7384 cm \\
Dipole bend angle  & $8^{\circ}$\\
Dipole length      & 3.850 cm \\ 
Dipole integrated gradient  & X, 0.258 G \\
                   & Y, 1.917 G \\
\hline
\end{tabularx}
\end{table}

\clearpage
\section{WARP model}

Parameters for the WARP hard-edged model are listed in Table \ref{tab:warp-env-sim}.
The WARP hard-edged dipole does not include a built-in edge focusing term. For this work, edge-focusing was approximated by short, hard-edged quadrupole elements of length $1$ cm at the leading and trailing edge of the dipole.

Some parameters for the gridded-field models used in WARP are listed in Table \ref{tab:warp-env-sim}.
In the WARP ring model, the gridded fields length is much longer than the magnet effective length. 
The dipole fields are solved in a bent coordinate frame as described in \cite{Kishek2010-2} and are placed in a \verb|BEND| element.
\cite{Kishek2010} describes the quadrupole element.

\begin{table}
\centering
\caption{Simulation parameters for WARP envelope integrator. }
\label{tab:warp-env-sim}
\vspace{10pt}
\begin{tabularx}{0.7\textwidth}{X L{4cm}}
\hline
Parameter & Value \\
\hline
Quadrupole gradient & 3.608 G/cm/A\\
Quadrupole length   &  3.740 cm \\
Dipole bend angle   & $10^{\circ}$\\
Dipole length       & 3.757 cm \\ 
Dipole edge length  &  1.000 cm \\ 
Dipole int. grad.   & X, 0.240 G\\
                    & Y, 1.928 G\\
\hline
\end{tabularx}
\end{table}

\begin{table}
\centering
\caption{Simulation parameters for WARP PIC model. }
\label{tab:warp-pic-sim}
\vspace{10pt}
\begin{tabularx}{0.7\textwidth}{X L{4cm}}
\hline
Parameter & Value \\
\hline
Peak quadrupole gradient     & 3.608 G/cm/A\\
Quadrupole length (trunc.) & 8.2 cm \\
Dipole bend angle      & $10^{\circ}$\\
Dipole length (trunc.)	& 16.2 cm \\
Dipole bend length     & 4.4057 cm\\
\hline
\end{tabularx}
\end{table}

\clearpage
\section{Benchmarking MENV against WARP} \label{sec:model:benchmark}

\begin{figure}[htb]
\centering
\includegraphics[width=\textwidth]{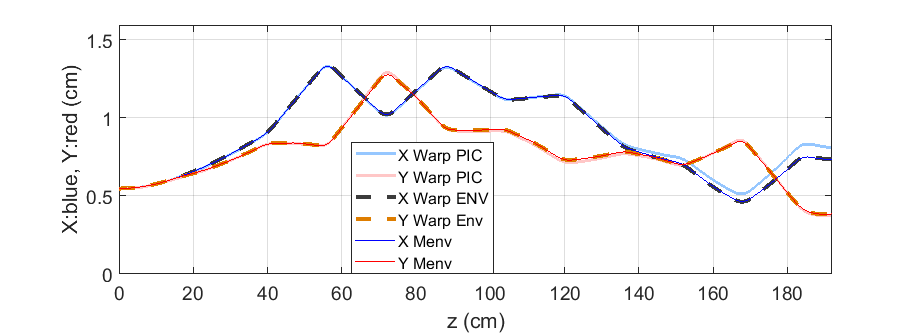}
\caption{Comparison of beam edge (2*RMS) evolution in half-cell between MENV integrator, WARP envelope integrator and WARP PIC code. }
\label{fig:benchmark}
\end{figure}

Assuming the gridded field elements are the most accurate models of the UMER magnets, I compared envelope integrator results (both MENV and WARP solver) with WARP PIC predictions of envelope evolution for the quadrupole current values specified in the proposed lattice solution for nonlinear experiments (described in Ch. \ref{ch:lattice}. I applied scalar corrections to the hard-edged field gradients to find agreement with the gridded fields. 

\subsection{Hard-edged dipole model}

I describe the "edge-focusing" model used in MENV above in Section \ref{sec:lattice:edges}. In MENV, the integrated quadrupole gradient over the length of a dipole are $\int k_x ds = 0.258 $ G and $\int k_y ds = 1.92 $ G.
Comparing the WARP envelope model to PIC predictions, I found that quad gradients $dB_y/dx = -0.0012$ T/m and $dB_x/dy = 0.00964$ T/m for the "edge-focusing" quads gave good agreement with WARP results over a half-cell (1.92 meters). The integrated quadrupole gradient for a single dipole (both "edge-focusing" quads) is $\int k_x ds = 0.24 $ G and $\int k_y ds = 1.93 $ G, very close to the integrated values used in the MENV calculation.

\subsection{Hard-edged quadrupole model}

I also applied correction factors to the quadrupole magnet gradients.
In WARP simulations, the hard-edged quad strengths were reduced to $97.5\%$ in the vertical plane (from the value in Table \ref{tab:warp-env-sim}) for best agreement between PIC and envelope predictions. No additional correction was applied in the horizontal plane. 
In MENV, quad excitation values were reduced $95.5\%$ in both planes (this is reflected by the gradient value in Table \ref{tab:menv-sim}). I further reduced the vertical focusing gradients by $97.9\%$ to obtain good agreement.

Figure \ref{fig:benchmark} compares beam evolution over a half-cell between the WARP and MENV envelope solvers. This is compared to the edge radius for a KV beam in the WARP PIC code. There is excellent agreement between envelope solvers, while the PIC solution only varies significantly in the horizontal plane. 
This may be due to the geometric focusing effect in the dipole, which is not included in envelope integration.

%% file: H.Appendix.tex
\renewcommand{\thechapter}{H}
\renewcommand{\chaptername}{Appendix}

\chapter{Parameterizing tune in the single-channel linear focusing lattice}
\label{ap:param}

This appendix reviews a technique mentioned in Chapter \ref{ch:lattice}, in which matched lattice solutions local to the solution identified in Section \ref{sec:lattice:solution} are linearized in terms of tune operating point. In this way the tune of the lattice can be shifted while maintaining a matched, round beam through the octupole insert/drift regions.

\section{Other MENV solutions}

Two nearby MENV solutions are identified in addition to the Chapter \ref{ch:lattice} solution shown in Fig. \ref{fig:p3-at7f-full} and Table \ref{tab:p3-at7f-quad-vals}. Quad current values for all three solutions are listed in Table \ref{tab:menv-quad-vals-appendix}.

\begin{figure}[tb]
\centering
\includegraphics[width=\textwidth]{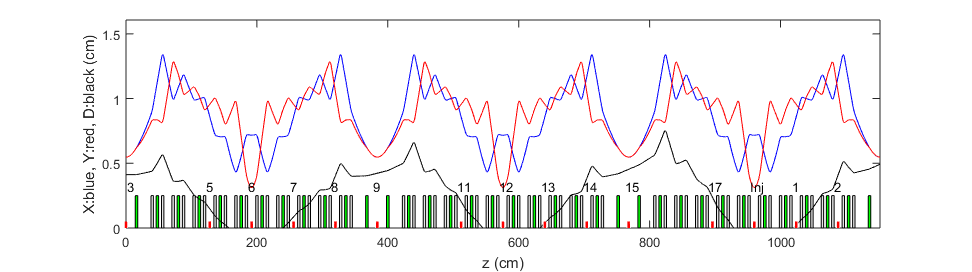}
\label{fig:p3-at5a-full:b}
\caption{MENV ``solution A" for $100$ micron, $60 \mu$A beam. Numbers indicate RC position. Dipoles are green, quads are gray. }
\label{fig:p3-at5a-full}
\end{figure}

\begin{figure}[tb]
\centering
\includegraphics[width=\textwidth]{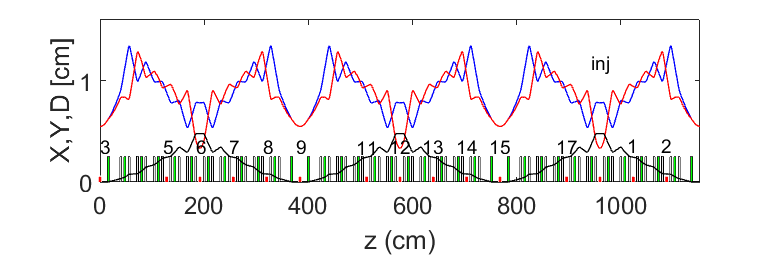}
\caption{MENV ``solution B" for $100$ micron, $60 \mu$A beam. Dipoles are green, quads are gray. Y-axis shows RMS beam size in x (blue) and y (red) and dispersion (black). Numbers indicate RC position. In this case, the solution is optimized for matched dispersion as well. }
\label{fig:p3-at7-full}
\end{figure}

Solution A of Table \ref{tab:menv-quad-vals-appendix} is the result of lattice optimization without dispersion matching, and is plotted in Fig. \ref{fig:p3-at5a-full}. 
In Solution B (Fig. \ref{fig:p3-at7-full}) the minimization function included a condition for matched dispersion $D(s)$ with zero dispersion in the drift/insert regions. However, by requiring matched dispersion the solution moves farther away from the desired tune, as seen in Table \ref{tab:lat-tunes-appendix}. Additionally, at this time UMER does not include a dispersion matching section in the injection line. 
Finally, Solution C is equivalent to the ``best solution" shown in Chapter \ref{ch:lattice} and is the result of applying iterative tune corrections to Solution B. These corrections take the form of parameterizing local matched solutions in terms of lattice tune.

\begin{table}
\centering
\caption{Quadrupole currents (in Amps) for MENV $N=3$ lattice solution. Solution A with unmatched dispersion is shown in Fig. \ref{fig:p3-at5a-full}, B in Fig. \ref{fig:p3-at7-full} has matched dispersion. C (Fig. \ref{fig:p3-at7f-full}) is nearest quasi-integrable operating point and is proposed for single-channel experiment. }
\label{tab:menv-quad-vals-appendix}
\vspace{10pt}
\begin{tabular}{c r r r r r r r r r r}
\hline
&Q1&Q2&Q3&Q4&Q5&Q6&Q7&Q8&Q9&Q10\\
\hline
A & -0.415& 1.065& -0.972& 0.616& -0.366& 0.579& -0.533& 0.902& -1.437& 0.927 \\
B & -0.415& 1.065& -0.972& 0.616& -0.332& 0.414& -0.428& 0.782& -1.313& 0.696 \\
C & -0.590& 1.247& -0.972& 0.616& -0.332& 0.414& -0.510& 0.738& -1.166& 0.762 \\
\hline
\end{tabular}
\end{table}

\begin{table}
\centering
\caption{Quantities describing matched lattice solution shown in Figures \ref{fig:p3-at5a-full}, \ref{fig:p3-at7-full} and \ref{fig:p3-at7f-full}. }
\label{tab:lat-tunes-appendix}
\vspace{10pt}
\begin{tabular}{l r r r}
\hline
Parameter & Solution A & Solution B & Solution C \\
\hline
Full ring tune 		 & $\nu_x = 3.271 $ & $\nu_x = 2.860 $  & $\nu_x = 3.270 $ \\
					 & $\nu_y = 3.283 $ & $\nu_y = 3.298 $ & $\nu_y = 3.267 $\\
Drift/insertion tune & $\nu_x = 0.269 $ & $\nu_x = 0.269 $ & $\nu_x = 0.265 $\\
					 & $\nu_y = 0.270 $ & $\nu_y = 0.270 $ & $\nu_y = 0.272 $\\
Linear lattice tune  & $\nu_x = 3.003 $ & $\nu_x = 2.592 $ & $\nu_x = 3.004 $\\
					 & $\nu_y = 3.013 $ & $\nu_y = 3.028 $ & $\nu_y = 2.995 $\\
Half-cell tune		 & $\nu_x = 0.545 $ & $\nu_x = 0.477 $ & $\nu_x = 0.545$\\
					 & $\nu_y = 0.547 $ & $\nu_y = 0.550 $ & $\nu_y = 0.544$\\
Chromaticity         & $C_x = -3.737  $ & $C_x = -3.331  $ & $C_x = -4.270$ \\
					 & $C_y = -4.028  $ & $C_y = -3.640  $ & $C_y = -3.277$ \\
\hline
\end{tabular}
\end{table}

\section{Tuning of MENV solution through parameterization of lattice tune}

\begin{figure}[!tb]
\centering
\includegraphics[width=\textwidth]{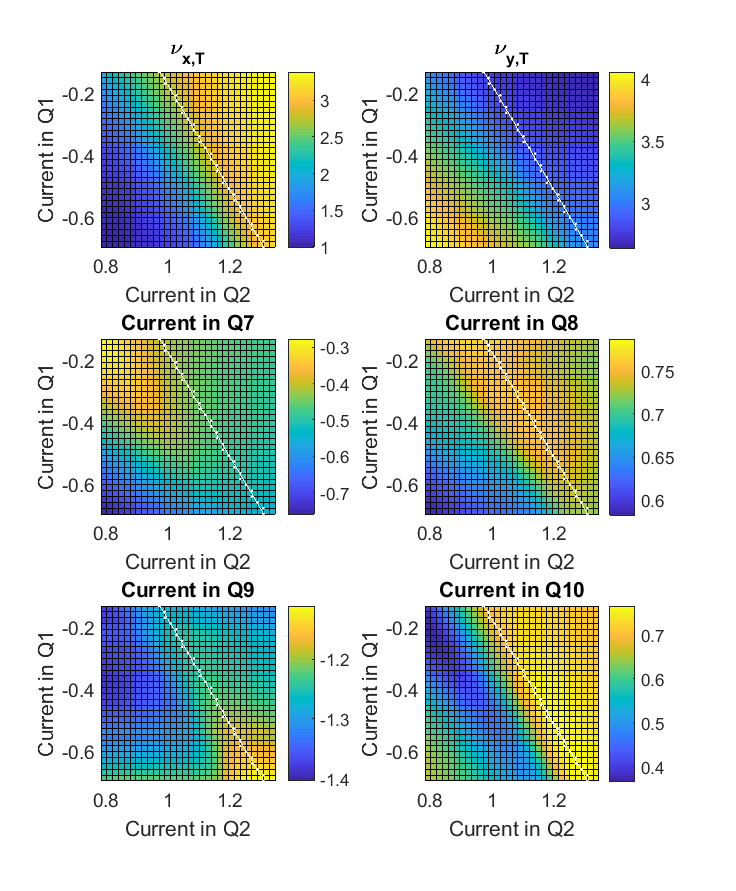}
\caption{Raster scan results starting from Solution B in Table \ref{tab:menv-quad-vals-appendix}. Color axis corresponds with quantity in figure title. All units are in Amps. Solutions with $\nu_x\approx\nu_y$ are shown as white scatter points. Best fit for $\nu_{x,T}=\nu_{y,T}$ is shown as white line. }
\label{fig:raster-matching}
\end{figure}

\begin{figure}[tb]
\centering
\includegraphics[width=0.5\textwidth]{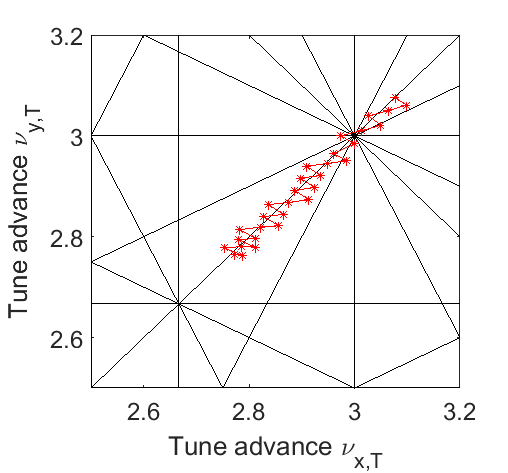}
\caption{Operating points $\nu_x\approx\nu_y$ found in 2D raster scan. }
\label{fig:tunespace-param-appendix}
\end{figure}

One concern for single channel experiments is that the lattice models do not perfectly agree with the physical magnet transfer functions. The performance of the quasi-integrable lattice depends strongly on controlling the lattice tunes $\nu_{x,T},\nu_{y,T}$ to a desired tolerance of $\delta \nu <0.1$. The lattice solutions found in MENV will not be perfectly matched when implemented in WARP simulations (using gridded field elements) or when implemented in the lab. This section discusses an approach to parameterizing the tune as a function of quadrupole strength to be used for iterative corrections of the matched solution.

Solution B in Table \ref{tab:menv-quad-vals-appendix} is used as a starting point. In MENV, two quad strengths (selected from the 10 uniquely-valued quads that composed the $N=3$ lattice) are raster-scanned. This example uses the two nearest the insert/drift region, Q1 and Q2. Strengths are scanned in a range $\pm$ the nominal value listed in Table \ref{tab:menv-quad-vals-appendix}. For each point on the 2D grid, the remaining quad strengths are varied to minimize the target $\vec{T} = \left[ X'(L/2), Y'(L/2) \right]$ at the midpoint between drift/insertion regions. While using two quads as free parameters should be sufficient to meet the two-valued target, best results were found using at least four quads. This case uses Q7-10 to optimize for a matched solution.

Fig. \ref{fig:raster-matching} shows lattice tunes and ``matching quad" currents for Q7-10 plotted versus Q1 and Q2 currents. For the set of found solutions, the sub-set of solutions nearest $\nu_{x,T} = \nu_{y,T}$ are identified and indicated in Fig. \ref{fig:raster-matching} as white points. These points are plotted in tune space in Fig. \ref{fig:tunespace-param-appendix}. As apparent in Fig. \ref{fig:tunespace-param-appendix}, among the found solutions there was one very near to the optimal working point at $\nu_{x,T},\nu_{y,T}=3$.

\begin{figure}[tb]
\centering
\subfigure[Linear fit to quad currents as a function of $i_{Q1}$.]{
\includegraphics[width=0.45\textwidth]{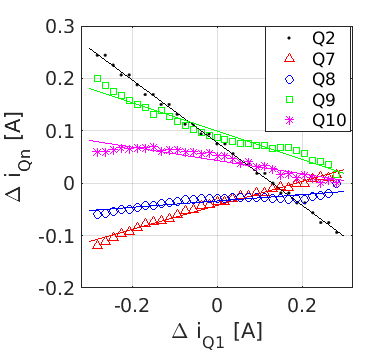}
}
\subfigure[Linear fit to tune as a function of $i_{Q1}$.]{
\includegraphics[width=0.45\textwidth]{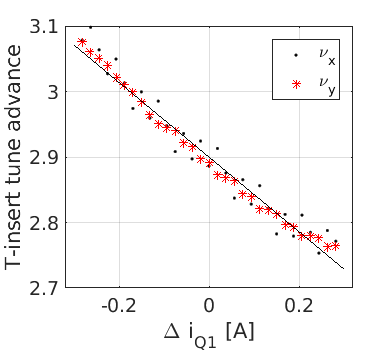}
}
\caption{Linear fits that parameterize quad current and tune with respect to $i_{Q1}$ while maintaining a matched beam in the drift/insertion region. }
\label{fig:parameterization-appendix}
\end{figure}

As the raster scan is resolution limited, linear fits are used to interpolate between grid points. For example, the dependence of Q1 current on lattice tune $\nu_{x,T}$ is given by

\begin{equation}
i_{Q1} = (-1.650 \pm 0.14) \nu_{x,T} + (4.377\pm 0.405)
\end{equation}

\noindent with error-bars indicating $95\%$ confidence. Additional linear fits for Q2, Q7-10 are given in Table \ref{tab:parameterization}. The set of points found for $\nu_{x,T} \approx \nu_{y,T}$ have the best-fit $\nu_{x,T} = (0.944 \pm 0.079) \nu_{y,T}+ (0.157 \pm 0.229)$. The desired relationship $\nu_{x,T} = \nu_{y,T}$ is within a $90\%$ confidence.

\begin{table}[tb]
\centering
\caption{Linear fits to periodic solutions in the vicinity of Solution B for tune correction of MENV model. (Table \ref{tab:menv-quad-vals-appendix} and Figure \ref{fig:p3-at7-full}). Best-fit line is expressed at $i_{Q\#} = p_1 \cdot  i_{Q1} + p_2$. }
\label{tab:parameterization}
\vspace{10pt}
\begin{tabularx}{0.5\textwidth}{X L{2cm} L{2cm}}
\hline
Quad \# & $p_1$ & $p_2$ \\
\hline
Q2 & -0.596 & 0.895 \\
Q7 & 0.230 & -0.374 \\
Q8 & 0.063 & 0.775 \\
Q9 & -0.271 & -1.326 \\
Q10& -0.127 & 0.687 \\
\hline
\end{tabularx}
\end{table}

The linearization approach was used to move the MENV solution closer to the desired operating point. I am able to find a solution within $\Delta \nu < 0.01$ by interpolating between found solutions along the $\nu_{x,T} \approx \nu_{y,T}$ line.
The solution presented in Chapter \ref{ch:lattice} (Figure \ref{fig:p3-at7f-full}) shows the most optimal working point on the line $\nu_{x,T} = 0.944\ \nu_{y,T} + 0.157$. Quadrupole currents for this new solution, calculated using the fits in Table \ref{tab:parameterization}, is also listed in Table \ref{tab:menv-quad-vals-appendix}, and tune values in Table \ref{tab:lat-tunes-appendix}.

This technique can be applied to tune up a given lattice. If the tune is known, the behavior can be parameterized around the measured/known working point and the currents adjusted accordingly. Additionally, knowing the nearby tune landscape allows one to perform a tune scan in experiment (measuring beam loss as a function of tune). By knowing the tune landscape, we can sample a desired line in tune space (such as $\nu_x = \nu_y$ or varying $\nu_x$ independent of $\nu_y$) while compensating to keep the ring lattice solution matched.

\section{Tuning of WARP model} \label{sec:param:warp}

Moving between simulation codes (MENV to WARP) provides an analogy to implementing simulation-based solutions in the lab. The lattice solution based on MENV calculations must be realizable in the WARP PIC model with gridded field elements (\verb|BGRD|) based on PCB circuit configuration. The hard-edged field models were benchmarked against the WARP gridded field elements (discussed more in Appendix \ref{ap:model}), but the agreement was not perfect. 

\begin{table}
\centering
\caption{Quadrupole currents for WARP implementation of $N=3$ lattice. }
\label{tab:warp-quad-vals-appendix}
\vspace{10pt}
\begin{tabular}{c r r r r r r r r r r}
\hline
Solution &Q1&Q2&Q3&Q4&Q5&Q6&Q7&Q8&Q9&Q10\\
\hline
D & -0.415&1.065&-0.973&0.616&-0.434&0.410&-0.370&1.055&-1.312&0.853\\
E & -0.538&1.161&-0.973&0.616&-0.434&0.410&-0.458&0.925&-1.109&0.650\\
\hline
\end{tabular}
\end{table}

I implemented solution B from Table \ref{tab:menv-quad-vals-appendix} in the WARP model. It was necessary to re-optimize in WARP to find a matched, periodic solution. Starting with the given solution, I optimized Q5-10 strengths to minimize the loss function RMS$\vec{T}$ for $\vec{T} = \vec{w}^\top \cdot \left[ X', Y', \Delta \nu_x, \Delta \nu_y \right]$ with weights $\vec{w} = \left[1,1,1.5,1.5 \right]$. The optimization was done in PIC mode, rather than the (faster) envelope solver, in order to use the gridded field models for the UMER magnets. Matching was done assuming $60\ \mu$A beam current in a $100\ \mu m$ emittance KV distribution. The quadrupole values after matching optimization are given in Table \ref{tab:warp-quad-vals-appendix} under Solution D. 

The ring tunes, as measured by integrating over the 1st turn betatron function in the linear lattice, are $\nu_{x,ring} = 3.166$ and $\nu_{y,ring}=3.150$. As the 64 cm insertion region tune is $\nu_{x,drift} = 0.263$ and $\nu_{y,drift}=0.270$, the effective T-insert tune advances are $\nu_{x,T} = 2.903$ and $\nu_{y,T}=2.880$ for tune errors of $\Delta \nu_x = -0.097 $, $\Delta \nu_y = -0.12$ and $\Delta \nu_x-\nu_y = 0.023$. 
Alternatively, by sampling lattice tune from individual particle orbits, for low-amplitude particles I measure full ring tunes $\nu_{x,ring} = 3.426$ and $\nu_{y,ring} = 3.185$. From this measurement, our tune errors are $\Delta \nu_x = 0.163 $, $\Delta \nu_y = -0.085$ and $\Delta \nu_x-\nu_y = 0.240$. The tune errors, particularly the difference between planes, is much larger than the desired tolerance and this lattice is not expected to perform very well. 
This is outside of the acceptable tune error from the quasi-integrable condition based on reduced model simulations of Chapter \ref{ch:qio}. 

\begin{figure}[htb]
\centering
\includegraphics[width=.7\textwidth]{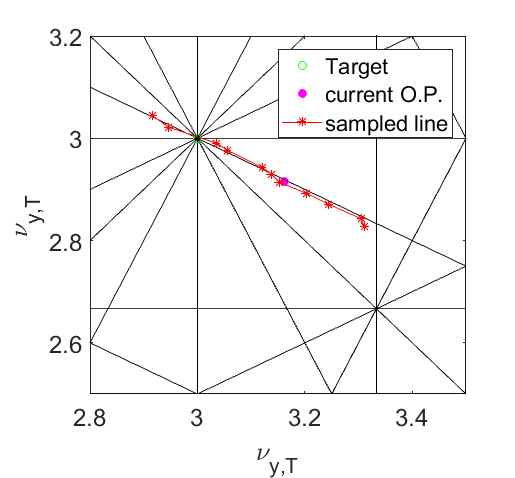}
\caption{Family of matched solutions connecting $\nu_{x,T}=3.156$, $\nu_{y,T}=2.915$ to $\nu_{x,T}=3.000$, $\nu_{y,T}=3.000$. }
\label{fig:warp-adj-tunediag}
\end{figure}

\begin{figure}[]
\centering
\subfigure[Linear fit to tune as a function of $\Delta i_{Q1}$.]{
\includegraphics[width=.45\textwidth]{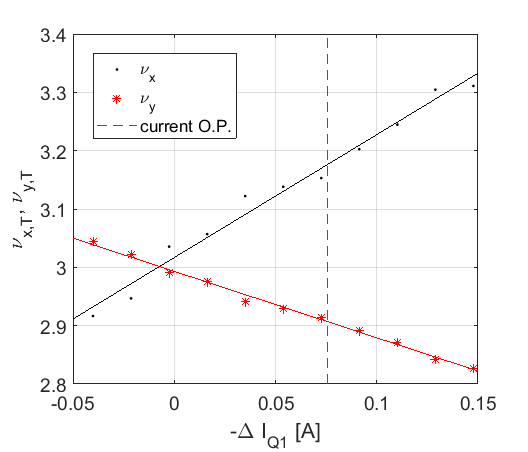}
\label{fig:warp-adj-iq-a}}
\subfigure[Linear fit to quad currents vs. $\Delta i_{Q1}$.]{
\includegraphics[width=.45\textwidth]{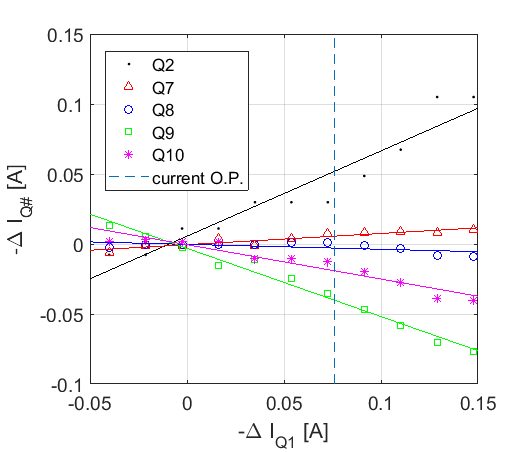}
\label{fig:warp-adj-iq-b}}
\caption{Linear fits for tune parameterization for WARP lattice correction. }
\label{fig:warp-adj-iq}
\end{figure}

The WARP solution was tuned to be closer to fractional tunes $\nu_x=\nu_y = 0.27$ using the parameterization method described above. The same quads were used to explore the local matched solutions (Q1,Q2,Q7,Q8,Q9,Q10). The desired change is $\Delta \nu_{x,T} =-0.163$ and $\Delta \nu_{y,T}  =0.085$. Therefore the local behavior was parameterized along the line $\nu_y = -0.522 \nu_x + 4.564$ as shown in Fig. \ref{fig:warp-adj-tunediag}. The dependence of tune on the current in Q1 along this line is $\nu_{x,T} = 2.107 i_{Q1} + 4.243$ and $\nu_{y,T} = -1.139 i_{Q2} + 2.330$ (plotted in Fig. \ref{fig:warp-adj-iq-a}). The remaining fits are given in Table \ref{tab:parameterization-warp} and plotted in Fig. \ref{fig:warp-adj-iq-b}. The resulting solution (E in Table \ref{tab:warp-quad-vals-appendix}) has fractional tunes $\nu_x = 0.240$ and $\nu_y = 0.259$ and is used for simulations presented in Section \ref{sec:lattice:warp}.

\begin{table}[tb]
\centering
\caption{Linear fits to periodic solutions in the vicinity of Solution B for tune correction of WARP model. Best-fit line is expressed at $i_{Q\#} = p_1 \cdot  i_{Q1} + p_2$. }
\label{tab:parameterization-warp}
\vspace{10pt}
\begin{tabularx}{0.5\textwidth}{X L{2cm} L{2cm}}
\hline
Quad \# & $p_1$ & $p_2$ \\
\hline
Q2 & 0.691 & 1.602 \\
Q7 & 0.080 & -0.462 \\
Q8 & 0.035 & 0.718 \\
Q9 & -0.488 & -1.455 \\
Q10& -0.2453 & 0.618 \\
\hline
\end{tabularx}
\end{table}

The operating point found near ideal tune is listed under solution E in Table \ref{tab:warp-quad-vals-appendix}. 
This lattice has measured tune $\nu_{x} = 0.240$ and $\nu_{y} = 0.259$. Better results are possible through iteration of this method. This lattice is used in Chapter \ref{ch:lattice} for PIC simulation of the proposed experiment.

%% file: I.Appendix.tex
\renewcommand{\thechapter}{I}
\renewcommand{\chaptername}{Appendix}

\chapter{WARP simulations of single-channel lattice at a different operating point}
\label{ap:warp-nonopt}

This appendix investigates lattice performance far from the quasi-integrable condition on tune. These simulations use the operating point listed as solution D in Table \ref{tab:warp-quad-vals-appendix}, which has fractional bare tunes $\nu_{x} = 0.426$ and $\nu_{y} = 0.185$. These results can be compared to those discussed in Chapter \ref{ch:design} at an operating point with much lower tune error.

\section{Linear WARP lattice}

\begin{figure}[]
\centering
\subfigure[Dynamic aperture over 512 turns. ]{
\includegraphics[width=0.6\textwidth]{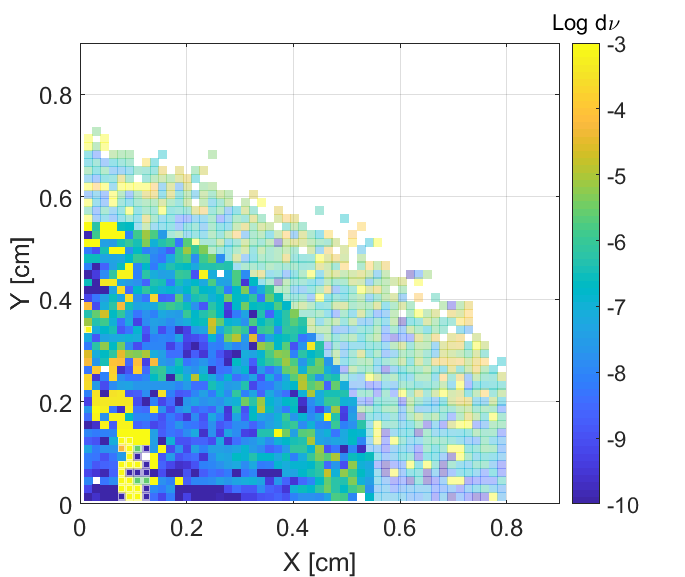}
\label{fig:warp-lin-fma-appendix-xy-appendix}
}
\hspace{.5in}
\subfigure[Tune footprint, with up to 3rd order resonance lines. The ideal quasi-integrable operating point is indicated by black dot.]{
\includegraphics[width=0.6\textwidth]{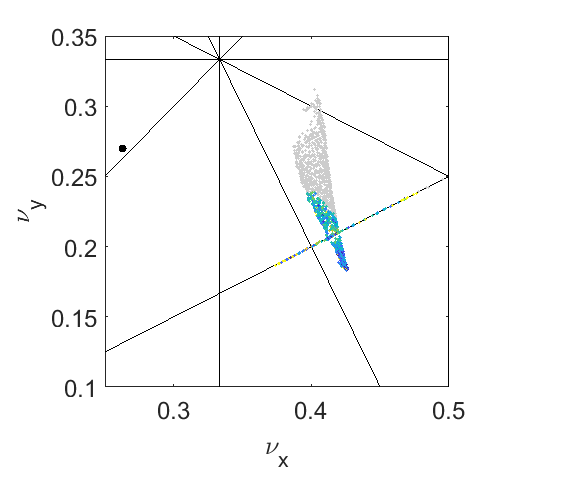}
\label{fig:warp-lin-fma-appendix-tune-appendix}
}
\caption{Frequency map analysis of full ring linear lattice in WARP for lattice solution D in Table \ref{tab:warp-quad-vals-appendix}.  The full interrogated space is shown, but a cut at $r=0.55$ cm is indicated by pixel saturation. }
\label{fig:warp-lin-fma-appendix}
\end{figure}

\begin{figure}[]
\centering
\subfigure[Dynamic aperture for 128 turns. Note the color axis is shifted when compared to most other plots.]{
\includegraphics[width=0.6\textwidth]{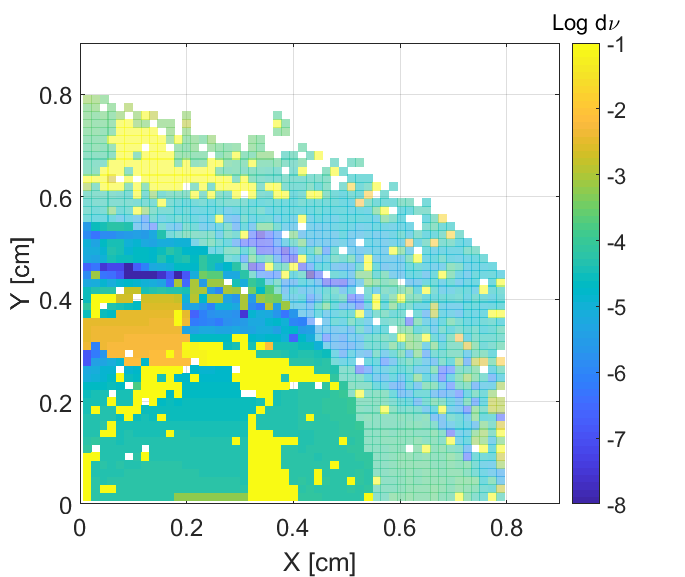}
\label{fig:warp-lin-sc-fma-appendix-xy}
}
\hspace{.5in}
\subfigure[Tune footprint, with up to 3rd order resonance lines. Footprint for zero-charge case (Fig. \ref{fig:warp-lin-fma-appendix-tune-appendix}) is shown in black.]{
\includegraphics[width=0.6\textwidth]{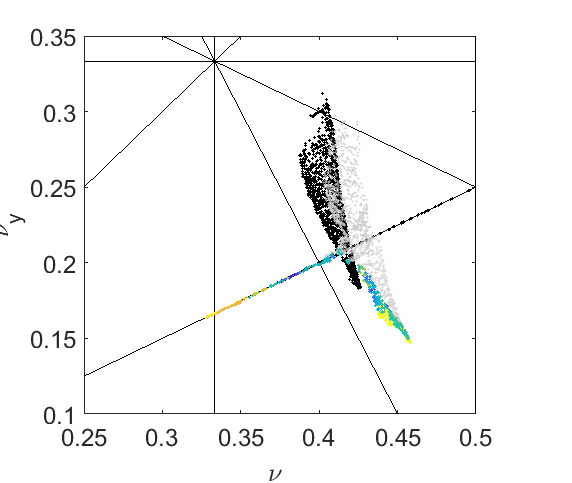}
\label{fig:warp-lin-sc-fma-appendix-tune}
}
\caption{Frequency map analysis of full ring linear lattice in WARP with $60\ \mu$A beam. A radial cut is made at $r=0.55$ cm.}
\label{fig:warp-lin-sc-fma-appendix}
\end{figure}

In the WARP model of the linear lattice ($G_{3,max} = 0$) without space charge, particles within $r=0.55$ cm inhabit a spread of tunes $\max \delta \nu_x = 0.052$ and $\max \delta \nu_y=0.054$ (RMS $\delta \nu_x=0.016$ and RMS $\delta \nu_y=0.026$). These values are comparable to the ``best-case" values in the reduced model ($\max \delta \nu = 0.113$ and RMS $\delta \nu = 0.034$), while the amplitude-dependence of the tune shift resembles that of the octupole lattice (high amplitude particles at larger tune shifts). This spread is entirely due to the nonlinearity of the UMER magnets included in the gridded field models.

When space charge is included, the tune footprint of the particle distribution is shifted from the zero-current bare tune $\nu_0$. We expect that the tune is depressed with space charge. As seen in Fig. \ref{fig:warp-lin-sc-fma-appendix}, the vertical tune is depressed but the horizontal tune experiences a small positive shift. The partial tunes for a low-amplitude particle are $ \nu_{x,ring} = 0.451$ and $\nu_{y,ring}=0.158$ (compare to $\nu_{x,ring} = 0.426$ and $\nu_{y,ring} = 0.185$ for the zero-charge case). 

Comparing the configuration space to the no-charge case (Fig. \ref{fig:warp-lin-sc-fma-appendix} with Fig. \ref{fig:warp-lin-fma-appendix}), we see stronger nonlinear behavior with the inclusion of space charge. The third order resonance seems to be driven more strongly, as particles near this line have a larger $\Delta \nu$. With stronger driving terms, we might expect to lose these particles trapped near the third order resonance. Finally, due to the tune shift, the third order resonance appears at a higher amplitude. There is no significant difference in particle tune spreads beyond the spread already present from magnet nonlinearities.

\section{Nonlinear WARP lattice}

\begin{figure}[tb]
\centering
\subfigure[Dynamic aperture for 512 turns. ]{
\includegraphics[width=0.6\textwidth]{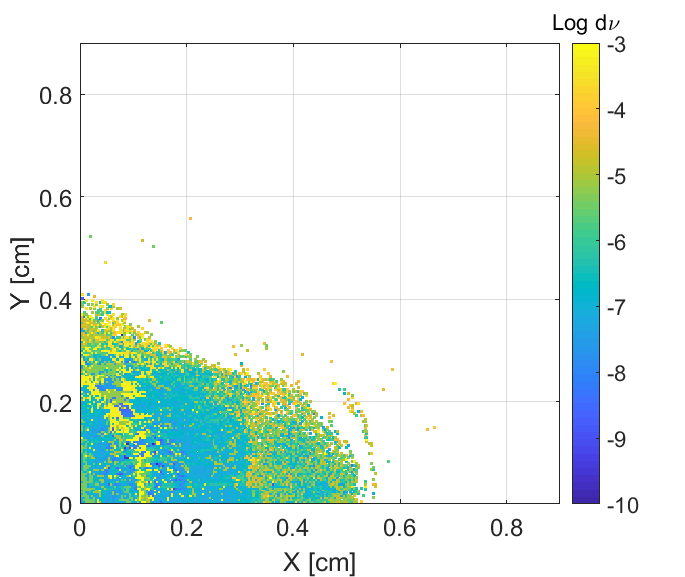}
\label{fig:warp-50-fma-appendix-xy}
}
\hspace{.5in}
\subfigure[Tune footprint with up to 3rd order resonance lines shown.]{
\includegraphics[width=0.6\textwidth]{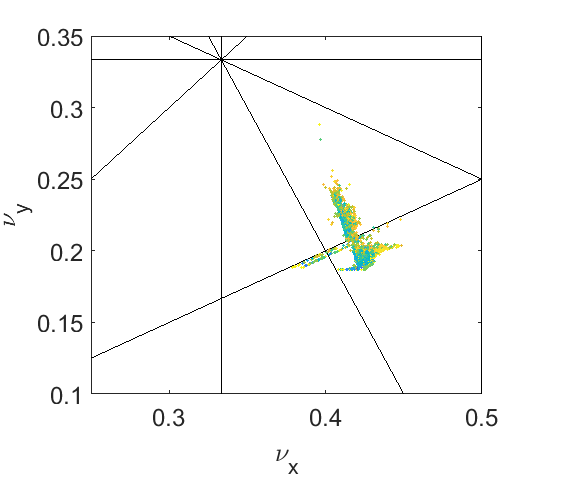}
\label{fig:warp-50-fma-appendix-tune}
}
\caption{Frequency map analysis of full ring octupole lattice at $G_{3,max}=50\ T/m^3$ in WARP with zero current. }
\label{fig:warp-50-fma-appendix}
\end{figure}

Figure \ref{fig:warp-50-fma-appendix} shows the frequency map when the octupole insertion is powered at $G_{3,max}=50\ T/m^3$ ($\sim 1$ A peak). The dynamic aperture is decreased from the linear case to $r=0.32$ cm. 
For the distribution of stable particles in Fig. \ref{fig:warp-50-fma-appendix}, $\max \delta \nu_x = 0.041$ and $\max \delta \nu_y=0.096$ while RMS $\delta \nu_x=0.010$ and RMS $\delta \nu_y=0.016$. 
Excluding outliers, the tune spreads are actually smaller than in the ``linear" lattice. This can be attributed to a reduced dynamic aperture. The asymmetry in the $XY$ spreads seems to be a property of the quad/dipole nonlinearities. Additionally, as seen in Chapter \ref{ch:design}, asymmetric tune spreads are expected for operating points \clearpage \noindent where $\nu_{x,T} \neq \nu_{y,T}$. In the example shown here, the octupoles reduce both the dynamic aperture and the tune spread in the stable particle distribution.

\begin{figure}[tb]
\centering
\subfigure[Dynamic aperture for 128 turns. Note the color axis is shifted when compared to most other plots. ]{
\includegraphics[width=0.6\textwidth]{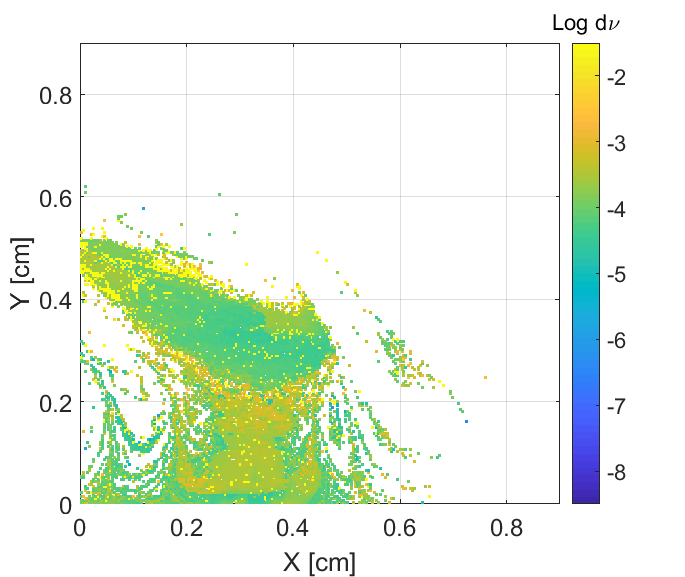}
\label{fig:warp-sc-50-fma-appendix-xy}
}
\hspace{.5in}
\subfigure[Tune footprint, with up to 3rd order resonance lines. Results at zero charge are also plotted (in black) for comparison. ]{
\includegraphics[width=0.6\textwidth]{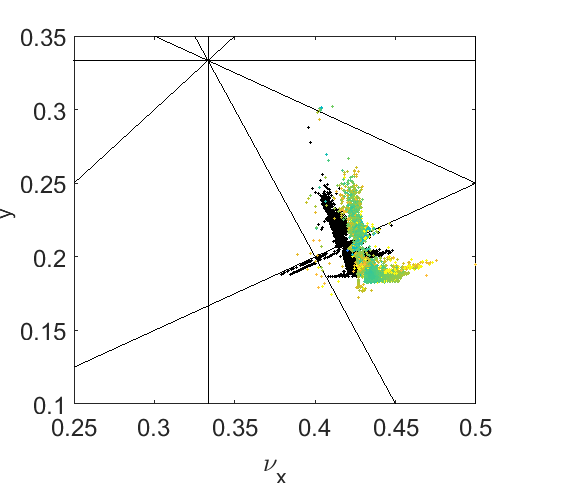}
\label{fig:warp-sc-50-fma-appendix-tune}
}
\caption{Frequency map analysis of full ring octupole lattice at $G_{3,max}=50\ T/m^3$ in WARP with $60\ \mu$A, $100\ \mu$m beam. }
\label{fig:warp-sc-50-fma-appendix}
\end{figure}

\begin{figure}[!tb]
\centering
\includegraphics[width=0.7\textwidth]{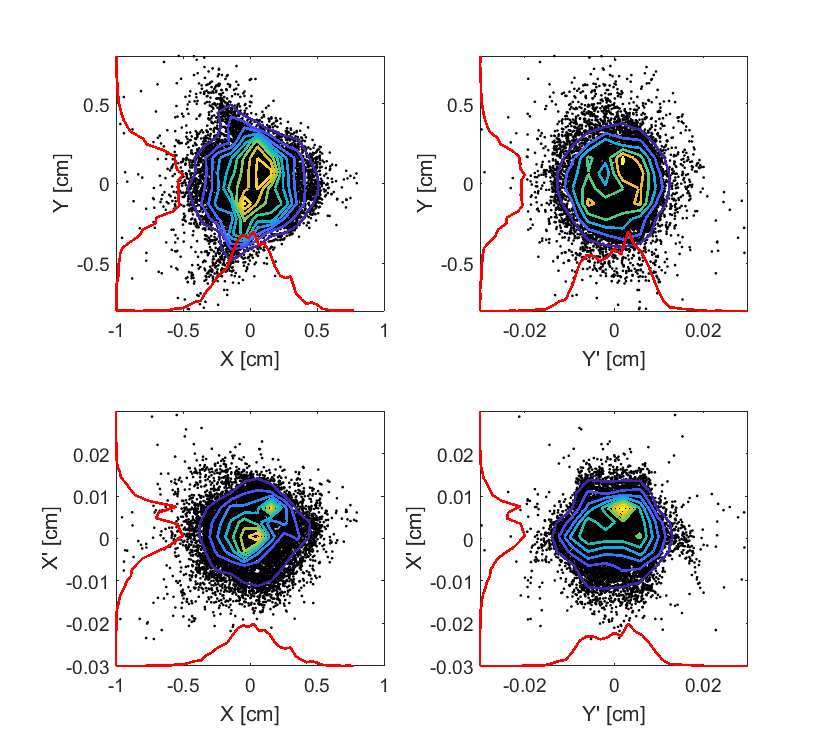}
\caption{Projections of particle distribution after 128 turns in the WARP octupole lattice at $G_{3,max}=50\ T/m^3$ for $60\ \mu$A beam. }
\label{fig:warp-sc-50-pdist-appendix}
\end{figure}

With space charge introduced as described above, the tune spreads increase. The frequency map is shown in Fig. \ref{fig:warp-sc-50-fma-appendix} for $G_{3,max}=50 T/m^3$.
For particles stable particles, $\max \delta \nu_x = 0.077$ and $\max \delta \nu_y=0.084$ while RMS $\delta \nu_x=0.012$ and RMS $\delta \nu_y=0.026$. Both the tune spread increases and the central tune shifts, as seen in Fig. \ref{fig:warp-sc-50-fma-appendix-tune}. 
However, the stable phase space area is very small due to particle losses that appear to primarily be along the third order resonance. The beam distribution after 128 turns is shown in Fig. \ref{fig:warp-sc-50-pdist-appendix}. Compared to the linear case, the loss of dynamic aperture is apparent. The apparent bunch hollowing seems to be due to low-amplitude losses driven by the third order resonance. Additionally, the transverse beam shape starts to reflect the shape of the octupole fields: the $XY$ projection gains ``wings," most noticeably in the vertical plane. The $X'\ Y'$ distribution also has this shape.

The WARP results predict poor performance when implementing this lattice solution. For the chosen operating point, space charge appears to strongly drive particle losses along the third order resonance when octupole fields are included. Additionally, the octupole-induced tune spread is barely noticeable over the tune spread in the ``linear" lattice. However, the lattice solution is known to be far from the optimal fractional tune $\nu_x = \nu_y =0.27$.

%% file: J.Appendix.tex
\renewcommand{\thechapter}{J}
\renewcommand{\chaptername}{Appendix}

\chapter{Measuring beam position with the quad-as-BPM method} \label{ap:quadasbpm}

This appendix describes the approach used to convert quadrupole response data into beam position data when using the UMER quadrupoles as ``virtual BPMs." Section \ref{sec:quadasbpm:analytic} explains the background theory and approximations made. Section \ref{ap:quadasbpm:errors} derives error propagation for the quad-as-BPM measurement. Section \ref{sec:quadasbpm:implement} goes into detail about implementation in the UMER control system.

\section{Analytic description of quadrupole response} \label{sec:quadasbpm:analytic}

In a lattice with low phase advance per cell (such as UMER, with $66.4 ^o$ and $67.5 ^o$ in the x,y planes), the particle motion is approximately sinusoidal, described by: 

\begin{equation} x(s) = A_1\cos{\frac{\sigma}{L}s} + A_2\sin{\frac{\sigma}{L}s} + x_{co}(s).\end{equation}

\noindent Here $x_{co}(s)$ is the equilibrium orbit. Generally, the equilibrium orbit is not equivalent to the design/reference orbit (the orbit that goes through the center of every quad), therefore $x_{co}(2)\neq0$. The oscillation with amplitudes $A_1$ and $A_2$ represent the betatron oscillation component of the beam motion. The derivative of this motion is

\begin{equation} x'(s) = \frac{dX}{ds} = -A_1\frac{\sigma}{L}\sin{\frac{\sigma}{L}s} + A_2\frac{\sigma}{L}\cos{\frac{\sigma}{L}s} + x'_{co}(s).\end{equation}

\noindent Consider an orbit that has been perturbed due to changing the strength of a single quad. We treat the quadrupole as a thin lens, an appropriate simplification when phase advance is low. The angular kick imparted by a quadrupole is $\Delta x' \approx tan \Delta x' = \frac{x_Q}{f}$ where $x_Q$ is the particle/centroid offset in the quad and $f$ is the focal length. For a thin quadrupole, $\frac{1}{f} = \frac{G}{B\rho}$ for integrated gradient $G$ and magnetic rigidity $B\rho$.\footnote{For UMER ring quadrupoles, $G=13.50$ [Gauss/A] and for 10 keV electrons, $B\rho = 338.85$ G-cm.} 

Consider an orbit perturbed by a quadrupole error $\Delta I$ at $s=0$: $\tilde{x}(s) = x(s) + \delta x(s)$. The initial conditions are $\tilde{x}(0) = x(0)$ and $\tilde{x'}(0) = x'(0) + \Delta x'_Q$ where $\Delta x'_Q = x_Q(\frac{G \Delta I}{B\rho})$ is the change in angle due to $\Delta I$ perturbation on the quad. Letting $\delta x(s) = B_1\cos{\sigma s / L} + B_2\sin{\sigma s / L}$ and applying these initial conditions, we find $B_1 = 0$, $B_2 = x_Q L/\sigma \times G \Delta I / B\rho$ and the perturbed orbit is:

\begin{equation} \tilde{x}(s) = x(s) + x_Q \frac{L}{\sigma} \frac{G \Delta I}{B\rho} \sin{\frac{\sigma}{L}s}. \end{equation}

\noindent With variation of the quadrupole strength, we find the dependence of the centroid position in a downstream BPM is linear in $x_Q$: 

\begin{equation} \frac{d \tilde{x}(s_{BPM})}{d\Delta I} = x_Q \frac{L}{\sigma} \frac{G}{B\rho} \sin{\frac{\sigma}{L}s_{BPM}}. \label{eq:quad-response} \end{equation}

In the quad-as-BPM approach described here, we recover the first-turn position in the quadrupoles by measuring the slope $\frac{d \tilde{x}(s_{BPM})}{d\Delta I}$. A model of the ring using the VRUMER\cite{vrumer} beam tracking code is used to calculate the constant $\frac{L}{\sigma} \frac{G}{B\rho} \sin{\frac{\sigma}{L}s_{BPM}}$ and the result is a value for $x_Q$ based on the measured beam response.
This is an essentially identical method to that described by Kamal Poor Rezaei \cite{KPRnote:2012}, with the main difference being the use of the VRUMER tracking code to calibrate position, rather than a transfer matrix calculation. This has the benefit of greater flexibility and integration with the UMER controls and data collection software. 

At an operating point of 1.826 A, the UMER 6 mA beam has a measured tune $\nu_x = 6.636$, $\nu_y=6.752$ \cite{RKnote:2010}. This corresponds to betatron wavelengths $\lambda_x = 1.736$ m, $\lambda_y=1.706$ m.
For the VRUMER parameters used in this thesis (see Appendix \ref{ap:model}), horizontal and vertical tunes are equal (no edge focusing), $\nu_x=\nu_y=6.293$. Equivalentl, Betatron wavelength $\lambda = 1.83$ m.
In this case, quadrupole strength parameters were set according to standard hard-edged approximation for the UMER quadrupoles: length$=4.475$ cm, peak strength $G=3.609$ G/cm, hard-edge factor $f=0.8354$.

\section{Error propagation for quad-as-BPM calibration} \label{ap:quadasbpm:errors}

This section discusses the systematic error introduced by differences between reality and the model used for calibration (in this case, VRUMER). Most notably, model and measured tune differ by $\Delta \nu_x = 0.340$ horizontally, $\Delta \nu_y = 0.459$ vertically. Additionally, there are measurement errors included in the measurement of position with BPM that should be propagated through to the $x_Q$ measurement as well.

We restate the formula for $x_Q$ from Eq. \ref{eq:quad-response}:

\begin{equation} x_Q = \frac{\sigma}{L} \frac{d \tilde{x}_{BPM}}{d \Delta I} \sin^{-1} \frac{\sigma}{L}s_{BPM}. \end{equation}

\noindent Applying error analysis, we find the dependence of position uncertainty on errors in the measured response slope $m$ and model phase advance $\sigma$: 

\begin{equation} \frac{\sigma_{x_Q}}{x_Q} = \bigg[1-\frac{s_{BPM}}{L}\cot{\frac{\sigma}{L}s_{BPM}} \bigg]\frac{\sigma_\sigma}{\sigma} + \frac{\sigma_m}{m}. 
\label{eq:quad-as-bpm-error}
\end{equation}

\noindent In this case, response error $\sigma_m$ should include the contributions of statistical noise and systematic errors in the collection and processing of BPM position data. For all the data shown here, $\sigma_m$ was taken to be the $95\%$ confidence bounds of the slope for the linear fit to the measured BPM position versus quadrupole strength curve. This encompasses shot-to-shot jitter and nonlinearities introduced by beam scraping. 
$\sigma_\sigma$ is an error introduced by using the VRUMER model to calibrate quadrupole response data, and can be reduced by conditioning the model for better agreement with measured tune values.

\begin{figure}
\begin{center}
\includegraphics[width=\textwidth]{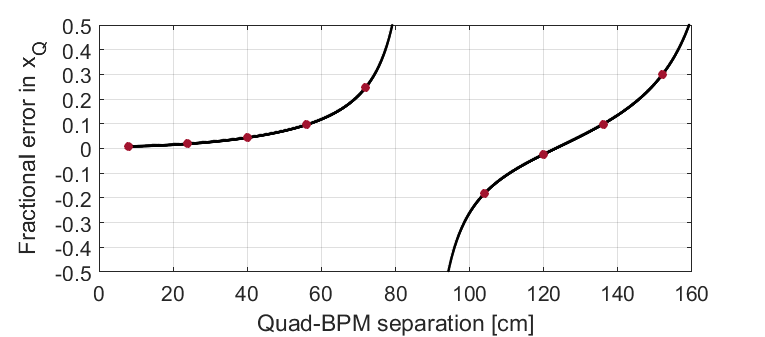}
\end{center}
\renewcommand{\baselinestretch}{1}
\small\normalsize
\begin{quote}
\caption{Fractional error in quad-as-BPM position due to phase error, versus quad and BPM separation. }
\label{fig:quad-as-bpm-error}
\end{quote}
\end{figure}

Fig. \ref{fig:quad-as-bpm-error} shows the dependence of the fractional error of measured position in the quadrupole, $\frac{\sigma_{x_Q}}{x_Q}$, as a function of separation between quadrupole and BPM used to measure quadrupole response. Here, $\sigma_\sigma$ is the different in phase advance between the VRUMER model used here and UMER measured values, $\sigma_\sigma = \mid \sigma_{sim} - \sigma_{exp} \mid = 0.06$.

\renewcommand{\baselinestretch}{2}
\small\normalsize

As seen from Eq. \ref{eq:quad-as-bpm-error}, the error value is not defined at $\frac{\sigma}{L}s_{BPM} = n\pi$. This is the point where the betatron wavelength is equal to the quad-BPM separation, a null point of the perturbed orbit $\delta x(s) = x_Q \frac{L}{\sigma} \frac{G \Delta I}{B\rho} \sin{\frac{\sigma}{L}s}$. Near this null, the returned quad-as-BPM measurement will have large errors, and in general be very sensitive to error in the BPM measurement as well as differences in the model and experiment.

\begin{table}
\centering
\caption{Quad-BPM pairs with separations near half of a betatron wavelength. Using these pairings to measure quadrupole response curves results in large errors in the quad-as-BPM method \textit{at the nominal 1.826 A operating point}. }
\vspace{10pt}
\label{tab:nulls}
\begin{tabular}{cc}
\hline
Quad \# & Nearest BPM  \\
\hline
QR14 & RC5  \\
QR38 & RC11  \\
QR62 & RC17  \\
QR70 & RC1 (turn 2)  \\
\hline
\end{tabular}
\end{table}

In standard UMER operation, the BPM spacing is 64 cm. For $\sigma_\sigma=0.06$, we expect errors $<10\%$. However, four BPM's are omitted for injection, longitudinal focusing and the wall current monitor, permitting larger errors. 
Table \ref{tab:nulls} shows quadrupole-BPM pairs with fractional error $>100\%$. The spacing for these four pairs is 88 cm, close to half the betatron wavelength, $\frac{\lambda}{2} \approx 86\ \mathrm{cm}$. 
In all data shown in this thesis, the four quads identified in Table \ref{tab:nulls} use the response measured in the next downstream BPM in order to avoid artificial blow-up or suppression of measured $x_Q$ and $y_Q$.

\section{Implementation in UMER controls} \label{sec:quadasbpm:implement}

The process for collecting quadrupole response data and converting to quadrupole position has been integrated into the UMER control system. 
For a given quadrupole, the procedure is:

\begin{enumerate}
\item Retrieve a list of functioning BPM's and choose a downstream BPM for measurement of quad response (excluding pairs listed in Table \ref{tab:nulls}).
\item Vary quadrupole over a range of $\pm 0.09$ A around nominal set-point, using 5 data points total to measure response.
\item Calculate response slopes $\frac{\Delta X_{BPM}}{\Delta I_{quad}}$ and $\frac{\Delta Y_{BPM}}{\Delta I_{quad}}$ by applying linear least squares fit to response data.
\item Run VRUMER simulation to calculate simulated response slope for given quad-BPM pair. Divide simulated position in quad $x_q$, $y_q$ by response slope to determine slope-to-position calibration factors $c_x$, $c_y$. 
\item Apply calibration factor to measured response slope to return measured position in quad. 
\item Calculate $x_Q$ position error, including $95\%$ bounds on reponse slope as well as phase advance error as a function of quad-BPM separation according to Eq. \ref{eq:quad-as-bpm-error}. 
\end{enumerate}

Arguably, the slope-to-position calibration factor could be generated using a matrix-based tracking technique (as outlined in \cite{KPRnote:2012}) and saved as a look-up table. For a given quad-BPM separation (for 72 quadrupoles there are only 8 possible separations), this number is not expected to change much and this approach would require fewer computations. However, this comes with an added loss of flexibility. The look-up table would have to be recalculated for UMER operating points with different quad focusing strengths or non-FODO orientations. For example, the experiment described in Chapter \ref{ch:distr} uses an alternative lattice configuration with half the quads turned off. 
Additionally, the time savings for using a look-up table are small. Running VRUMER takes $\sim 0.06$ 
seconds, so even for the an entire quad scan (9 points) VRUMER costs $\sim 0.5$ second per quad. This is negligible compared to the time required to measure BPM response for multiple quad settings.

There is a caveat to interpreting phase-advance contribution to error bar calculation, $\sigma_\sigma \equiv | \sigma_{exp} - \sigma_{sim}|$. Simulated phase advance $\sigma_{sim}$ is estimated from VRUMER results using the NAFF algorithm to extract fundamental frequency. Experimental phase advance $\sigma_{exp}$ is hard-coded to be measured values $\nu_x = 6.636$, $\nu_y = 6.752$ for the standard $1.826$ A operating point.  This will result in misleading error bars for different operating points and should be modified in the future to be more general.

The function \verb|calibrate_quad_as_BPM_vrumer| takes a quad-BPM pair and uses VRUMER to calculate the response slope (with units of m/A, position in BPM over current in quad). It also returns the error due to tune difference between the VRUMER model and the experiment. The code snippet below, from \verb|calibrate_quad_as_BPM_vrumer.m| (lines 56-78) , shows how error bars associated with the model-experiment tune difference are calculated:

\begin{verbatim}
% -- measured tunes at 1.826 operating point (Kamal 2010)
nuxu = 6.636;
nuyu = 6.752;
sigxu = 2*pi*nuxu/36;
sigyu = 2*pi*nuyu/36;

% -- geometry factors
% Cell length [m]
L = 0.32; 
% quad-BPM separation [m]
dels = mod(s(iB(BPMindex==BPM))-s(iQ(Qindex==Q)),11.52); 

% -- estimate vrumer tune (note equal x,y tunes)
[freq,amp,Xr]=naff(x(iQ(1:36))); 
nuvr = freq*72; sigvr = 2*pi*nuvr/36;

% -- delta between vrumer and exp. tunes
deltasigmax = abs(sigxu-sigvr);
deltasigmay = abs(sigyu-sigvr);

% -- returned error values
errx=(1-dels/L.*cot(sigxu/L*dels))*deltasigmax/sigxu;
erry=(1-dels/L.*cot(sigyu/L*dels))*deltasigmay/sigyu;
\end{verbatim}

\noindent Note that the experimental tune is hard-coded in the first two function lines, as mentioned above. The error bars due to phase difference will be calculated incorrectly for different operating points or non-FODO lattices. 

The tune error calculation can easily be generalized to any lattice function by including a tune measurement during the quad-as-BPM data acquisition. This can be added with only slight change to the structure of the quad-as-BPM code. An automated, robust tune measurement is relatively straightforward for the FODO lattice (for example, by applying the four-turn tune formula to BPM signals) but may require a more careful approach for non-FODO lattices.

One level above \verb|calibrate_quad_as_BPM_vrumer|, the function \verb|where_is_the_beam| calculates the total error-bars, combining the tune error from the VRUMER model with uncertainty in measured slope. Below is a code snippet from \verb|where_is_the_beam.m| (lines 25-29) showing calculation of error-bars based on tune error (\verb|errx|,\verb|erry|) and uncertainty \verb|herr|, \verb|verr| in measured response slopes \verb|hslope|, \verb|vslope|:

\begin{verbatim}
[cx,cy,errx,erry] = calibrate_quad_as_BPM_vrumer(Q,BPM);
xsim = cx*hslope; 
ysim = cy*vslope;
xerrsim = abs(xsim)*(abs(errx) + abs((herr-hslope)/hslope));
yerrsim = abs(ysim)*(abs(erry) + abs((verr-vslope)/vslope));
\end{verbatim}

\begin{figure}[]
\centering
\subfigure[Response of QR1, measured at RC1.]{
\includegraphics[width=0.45\textwidth]{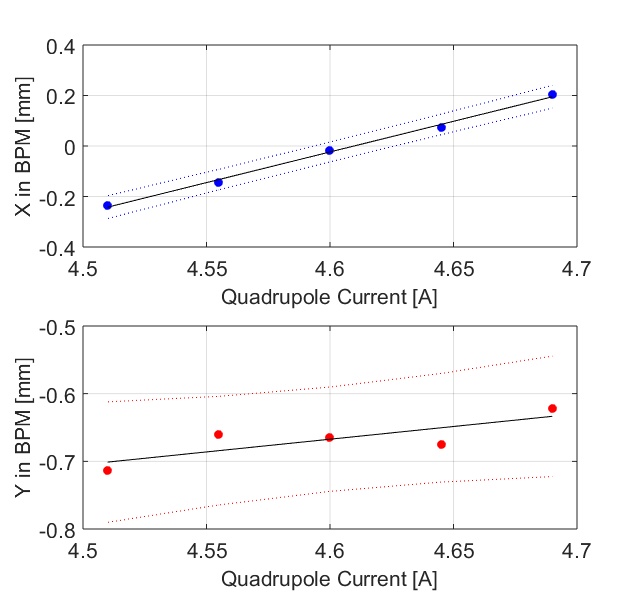}
\label{fig:QR1scan}}
\subfigure[Response of QR33, measured at RC9.]{
\includegraphics[width=0.45\textwidth]{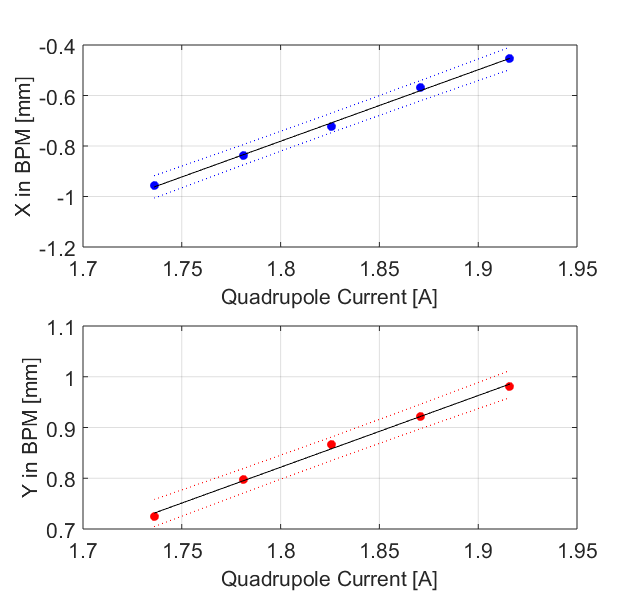}
\label{fig:QR33scan}}
\caption{Example quad response data with linear fits and 95\% confidence intervals. }
\end{figure}

\noindent An example of measured, calibrated quad response curves are shown in Fig. \ref{fig:QR1scan}.

%% file: K.Appendix.tex
\renewcommand{\thechapter}{K}
\renewcommand{\chaptername}{Appendix}

\chapter{Implementing the quad-centering method for multi-turn orbit correction} \label{ap:steering}

This appendix describes now the quad-centering steering algorithm discussed in Chapter \ref{ch:steering} is implemented in the lab. Section \ref{sec:apsteering:1} provides an overview of the method, including approach to optimizing injection and recirculation. Section \ref{sec:apsteering:2} gives a more detailed description of the injection line algorithm. Section \ref{sec:apsteering:3} discusses implementation of the quad-centering algorithm to set ring steerers, including a criterion for rejection of points while fitting the response curve. Section \ref{sec:apsteering:4} describes the approach to ``closing the orbit," which is equivalent to finding an equilibrium orbit close to the quad-centered first-turn orbit. 

\section{Overview of steering procedure} \label{sec:apsteering:1}

This steering procedure attempts to steer the beam as close as possible to the center of the quads in the first turn as well as minimize turn-to-turn oscillations about the closed orbit. The general philosophy is to first determine a good injection condition by setting 2 injection line steerers to minimize position in first few ring quads. Then, each ring steerer is set to center on a downstream quadrupole, using the algorithms outlined in Chapter \ref{ch:steering}. Finally, two steerers at the end of the first turn are used to ``close the orbit," by minimizing turn-to-turn oscillation amplitude. 
The procedure for horizontal steering is as follows:

\begin{enumerate}
\item Steer injection line by setting steerers SD1H, SD2H, SD3H, and SD4H to minimize response in injection quadrupoles.
\item Set 2 injection dipoles (typically SD5H, SD6H) by scanning currents and identifying smallest RMS offset in first few RQ's after injection. 
\item Steer through RQ3 (first turn) by setting current in D1; Repeat injection scan if change was significant.
\item Steer through quads in first turn according to centering algorithm described in Chapter \ref{ch:steering}, setting dipoles D2-34 in order and using quad-as-BPM method to measure position in quads. 
\item Close orbit by scanning D34 and D35 currents. 
\item Verify orbit quality by running quad scan for 1st turn quad-as-BPM data, and look at multi-turn BPM data to estimate orbit excursions from closed orbit.
\end{enumerate}

This procedure is nearly identical for vertical steering using RSV and SSV correctors. In this case, RSV17 and RSV18 are used to `close the orbit." In general, best behavior is seen by correcting vertical orbit first, then horizontal second. This is due to the fact that (a) the vertical orbit tends to have larger excursions, especially when uncorrected, leading to scraping and nonlinear response curves, and (b) the RSV magnets tend to have large rotational errors, leading to larger coupling between vertical steerers and horizontal orbit than horizontal steerers and vertical orbit.

\section{Injection line} \label{sec:apsteering:2}

For injection line steering, each steerer is set to minimize centroid position in the secound downstream quadrupole. Horizontal and vertical are done in tandem (ie, SD1H is set, then SD1V, then proceed to SD2H, SD2V).

\begin{enumerate}
\item Set all injection steerers to 0.
\item Measure quadrupole response in Q2 for a range of SD1H settings, set SD1H based on zero-crossing of linear fit.
\item Repeat for SD1V.
\item Continue to SD2H, and so on.
\end{enumerate}

\noindent Table \ref{tab:inj-steerers} shows the SD-quad pairs used for injection line steering. SD5H,V and SD6H,V are used for setting the injection condition.
There is no clear procedure for choosing the injection condition. In general, good results have been achieved by setting SD5H,V and SD6H,V for minimum position in QR2 and QR3. Occasionally this constraint has to be relaxed in order to achieve good recirculation (which suggests the necessity of an orbit-bump to compensate for the tilted YQ kick).

\begin{table}
\centering
\caption{Injection steerer - injection quad pairings. }
\vspace{10pt}
\begin{tabular}{cc}
\hline
Steerer name & Quad target \\
\hline
SD1H,V & Q2 \\
SD2H,V & Q3 \\
SD3H,V & Q4 \\
SD4H,V & Q6 \\
SD5H,V & -- \\
SD6H,V & -- \\
\hline
\end{tabular}
\label{tab:inj-steerers}
\end{table}

\section{Setting Ring Steerers} \label{sec:apsteering:3}
 
\begin{figure}
\centering
\subfigure[Setting RSV8 using response at QR32.]{
\includegraphics[width=0.75\textwidth]{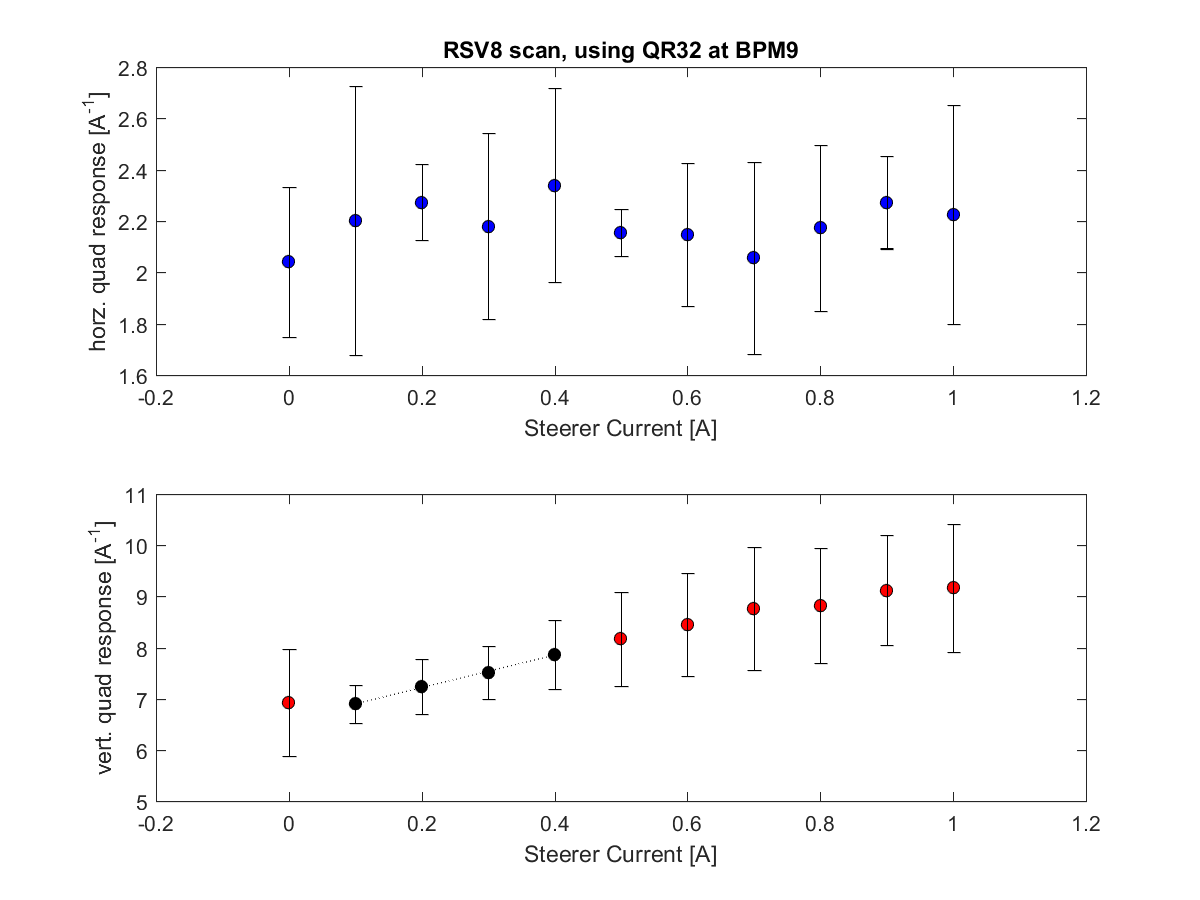}
\label{fig:RSVset-a}}
\subfigure[Setting RSV12 using response at QR48.]{
\includegraphics[width=0.75\textwidth]{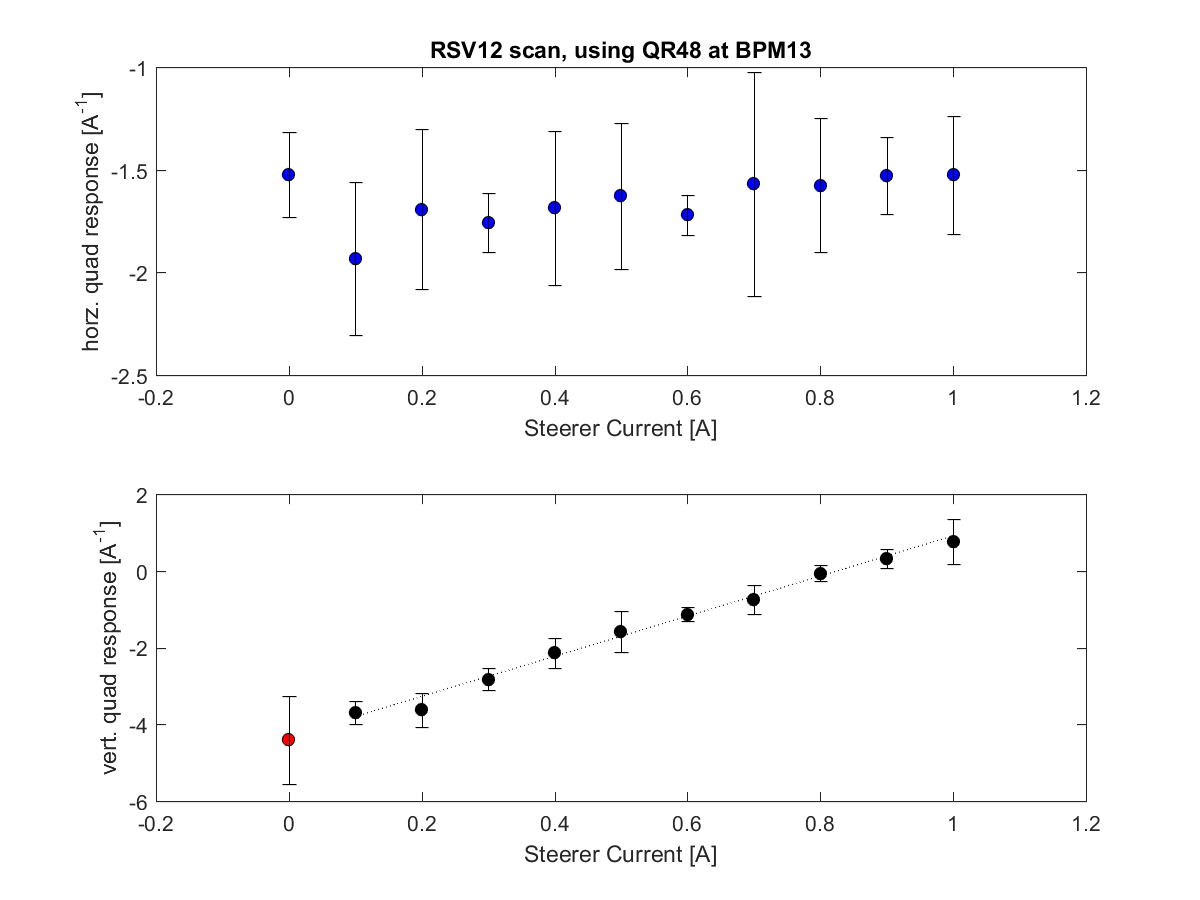}
\label{fig:RSVset-b}}
\caption{Example of measured response curves for vertical RSV steerers. For the vertical response, a linear fit is made to the black points, while red points exceed the error-bar threshold. Both plots are courtesy of L. Dovlatyan.}
\label{fig:RSVset}
\end{figure}

An example of a measured response curves for vertical steerers is shown in Fig. \ref{fig:RSVset}. In this case, RSV currents are set to minimize beam offset in the third downstream quadrupole. Vertical response is plotted in the bottom axis. Black, filled-in points meet error-bar criterion (discussed below) and are used for fitting. Red points have error-bars which exceed the tolerance and are rejected. In Fig. \ref{fig:RSVset-a}, the set-point is outside the scanned range in the opposite (negative) polarity. In Fig. \ref{fig:RSVset-b}, the optimal RSV12 set-point is approximately 0.8 A. Error-bars are calculated as outlined in Appendix \ref{ap:quadasbpm}. The top axis shows horizontal response to vertical steerer. In this case, there is clear coupling (likely due to an XY rotation of the RSV circuit). In principle, the rotation angle can be calculated from the horizontal response slope.

\subsection{Points Rejection}
Occasionally the steerer response data will not fall on a straight line. This is most likely due to scraping between the dipole and BPM, but could also be due to magnet nonlinearities. In the ring steering scripts, I have implemented a points rejection critera that throws away quad position data with error-bar $> 1$ mm. VRUMER is used to convert error-bar from response units $[A^{-1}]$ to meters. However, the steering algorithm attempts to minimize response, not position calibrated with VRUMER or other models (and therefore should be robust to model errors).

As the error is proportional to the uncertainty in the fitted slope of the quad response, large non-linearities in the quad response will manifest as large error bars. This is typical of large centroid excursion in either the quad or the BPM. Generally poor algorithm performance (resulting in large local distortions) can be attributed to nonlinearity in the quad-as-BPM method. 

Good results from quad-centering applied as described is shown in Figures \ref{fig:horz-result} and \ref{fig:vert_result}. 
The large excursion in the measurement around QR19 $\to$ QR26 is driven by poor fitting to the target function caused by a large vertical excursion at RC4. Vertical scraping skews the measurement of horizontal position versus dipole setting, resulting in nonlinear response curves. In practice, measuring the response curve for the horizontal dipoles in a range of $2.0 \to 2.8$ A results in well-behaved steering, but nonlinearities appear when the orbit distortion is large in the BPM or scraping occurs between the quad and a BPM. 
For a well-tuned solution the range may be reduced, as the needed adjustments are usually small. 
Similarly, the horizontal solution has large distortion near the back end of the ring (QR67-QR71). 
The quad-as-BPM measurement is less reliable in this region, as the VRUMER model of the Y-section is not very accurate and there appears to be a large orbit bump in the vicinity of YQ (making scraping more likely).

\section{Closing the orbit} \label{sec:apsteering:4}

\begin{figure}[]
\centering
\includegraphics[width=\textwidth,trim={0 2.7in 1in 2.7in},clip]{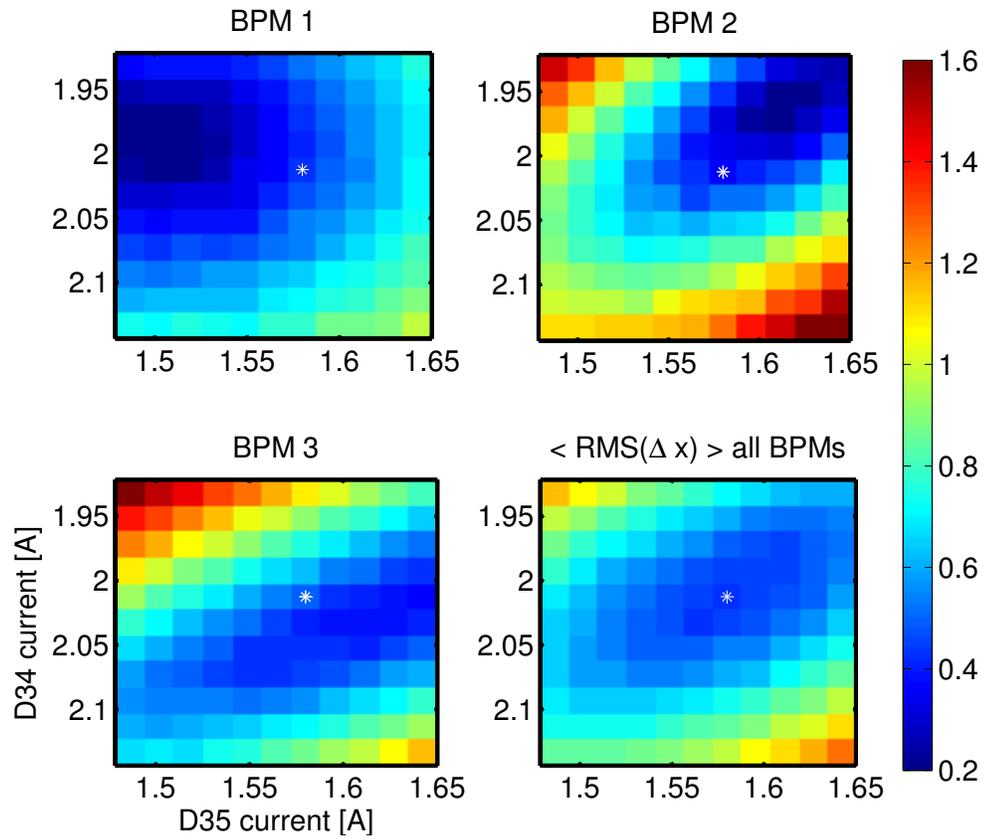}
\caption{Scan results for first 3 BPMs for iteration 3 in Figure \ref{fig:close_BPMs}. Color scale is RMS value of $\Delta x$ over first 4 turns [mm]. White asterisk indicates optimal setting (D34=2.0129 A, D35=1.5802 A). }
\label{fig:close_scan}
\end{figure}

\begin{figure}[]
\centering
\includegraphics[width=\textwidth,trim={.5in 2.7in .5in 2.7in},clip]{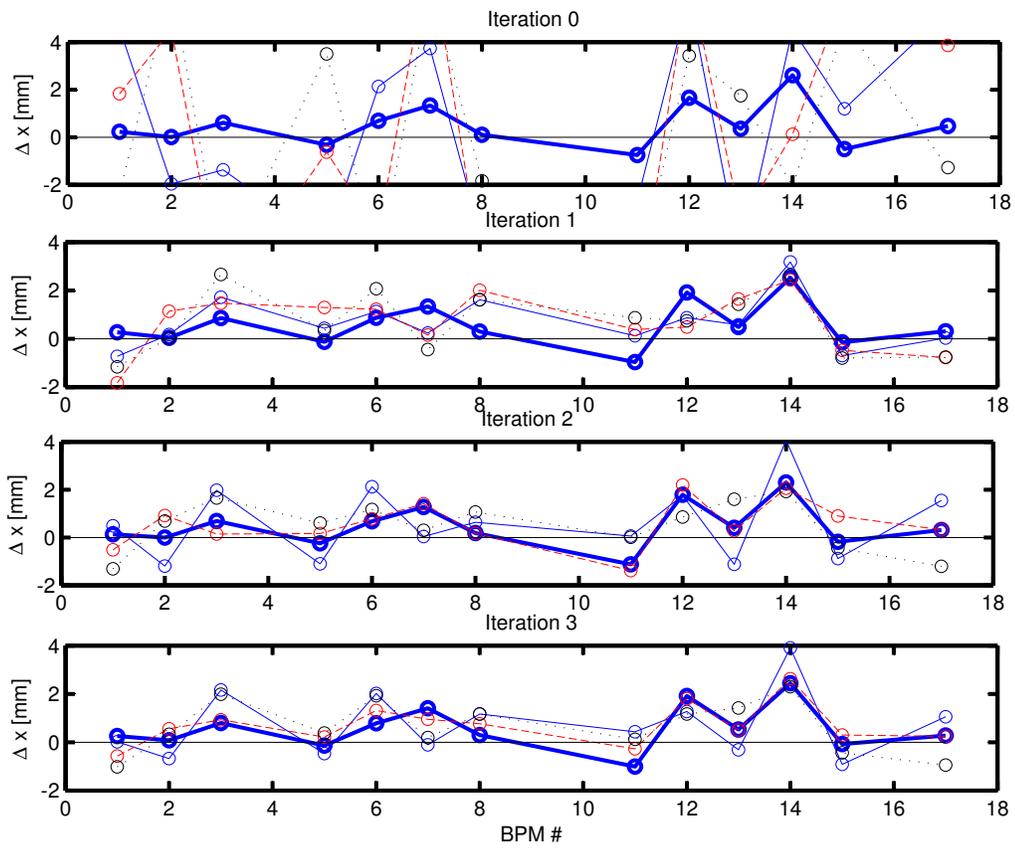}
\caption{BPM response for first 4 turns during correction of D34, D35 currents. 1st turn: heavy blue trace. 2nd turn: solid blue. 3rd turn: long dash red. 4th turn: short dash black. }
\label{fig:close_BPMs}
\end{figure}

The final step is to set the current in the dipoles near the end of the ring so that the closed orbit is close to the optimized first-turn orbit.  Two dipoles are required for control of x and x', in this thesis I used D34 and D35.  
The currents in D34 and D35 are raster scanned and position data recorded for the first four turns in the first 3 BPMs. I try to minimize the RMS change in position between turn 1 and turns 2-4 in the first three BPMs. For each BPM 2-4, I define an RMS quantity $\sqrt{\frac{1}{3}\left[ \left(x_2-x_1\right)^2+\left(x_3-x_1\right)^2+\left(x_4-x_1\right)^2\right]}$. In order to find a good closed orbit, I ran 3 scans, increasing the resolution and/or shifting the scan range for each successive scan. Each scan takes $\sim13$ minutes to read 3 BPMs for an $11\times11$ current range. 
Figure \ref{fig:close_BPMs} shows the four-turn BPM measurements for the beam after successive iterations of the D34/D35 raster scan. 
Figure \ref{fig:close_scan} shows dependence of BPM position at RC1-3 and the RMS quantity described above on the rastered dipole currents. 

Since this approach was developed, faster convergence and better results for for low turn-to-turn oscillation amplitude has been found through application of the RCDS algorithm. \cite{Huang2013,LevonRCDS} It is recommended that the raster scan approach described here be replaced with the RCDS method.

\phantom{
\cite{RuisardUndergrad}, 
\cite{RuisardIPAC2012}, 
\cite{RuisardIPAC2013} 
\cite{psf}, 
}

%% file: L.Appendix.tex
\renewcommand{\thechapter}{L}
\renewcommand{\chaptername}{Appendix}

\newpage



\section*{Supplementary material}

pipe. However, as the tilted quadrupole (YQ) in the injection section introduces problematic focusing-steering coupling, the proposed design avoids any off-axis elements in the extraction (all elements are centered on the recirculating beam pipe). Another concern with the extraction section is that, while the injected beam is a clean, well behaved beam that longitudinally fills approximately 50% of the ring, the extracted beam will be of a largely unpredictable distribution, suffering from beam halo, head-tail spreading and emittance growth. Our extraction design must have sufficient acceptance of off-axis beams consistent with observed equilibrium orbit distortions (discussed in Chapter 10). Finally, the injection kicker is a pulsed dipole, for which the rise time is severely limited by induced wall currents. To avoid the mechanical difficulties of including a glass gap in the extraction pipe, and for faster attainable rise times, we choose to employ an electrostatic "kicker" electrode (BKE).

### L.1. *Conceptual design with linear optics*

The most fundamental question concerning extraction design is placement of the kicker. Due to the dense lattice structure, there is not sufficient space between elements for a moderately kicked beam to exceed the pipe radius before encountering the next downstream steering/focusing element. The kicked beam must pass through several magnetic elements in the ring before extraction. A basic conceptual design study was conducted to identify the most optimal location for the extraction kicker with respect to the quadrupole focusing elements (assuming the typical FODO arrangement at the 1.826 Amp operating point).

Simulations of the linear focusing optics were done using thin lens transfer matrices (implemented in Matlab). The UMER quadrupole and dipole elements were

each modelled as a series of distributed thin lenses, with the electrostatic kicker approximated as a discrete, instantaneous velocity kick (which correlates to integrated field for unspecified kicker length and voltage). Infinite aperture was assumed for all magnets, anticipating that the mechanical design will include an enlarged pipe/magnets.

Initially on-axis single particles were propagated in the matrix model for a range of potential kick locations. "Ideal" is defined as maximum extraction angle versus applied kick voltage. Figure L.2 illustrates the ideal kick location identified with single-particle tracking. The hardware is visualized in Figure L.3.

Results were verified against beam tracking code PBOlab, in which all ring magnets were modelled as ideal, infinite aperture hard-edged fields. The conceptual study also considered dual-kicker configurations, but did not find any significant advantage in terms of beam angle per applied voltage that justified the additional complexity. Further details are given in [78],[79].

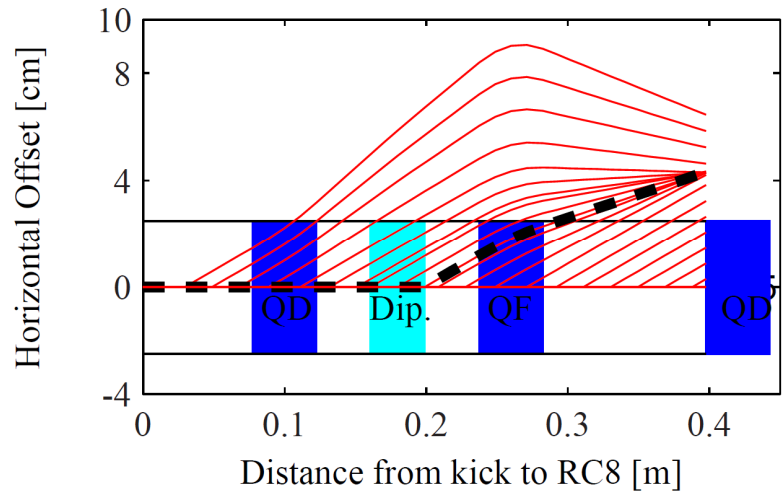

Figure L.2 Single particle trajectories (red traces) for various kick locations along s, corresponding to kick with integrated field 6.2 kV. The optimal trajectory is marked by a heavy dashed line.

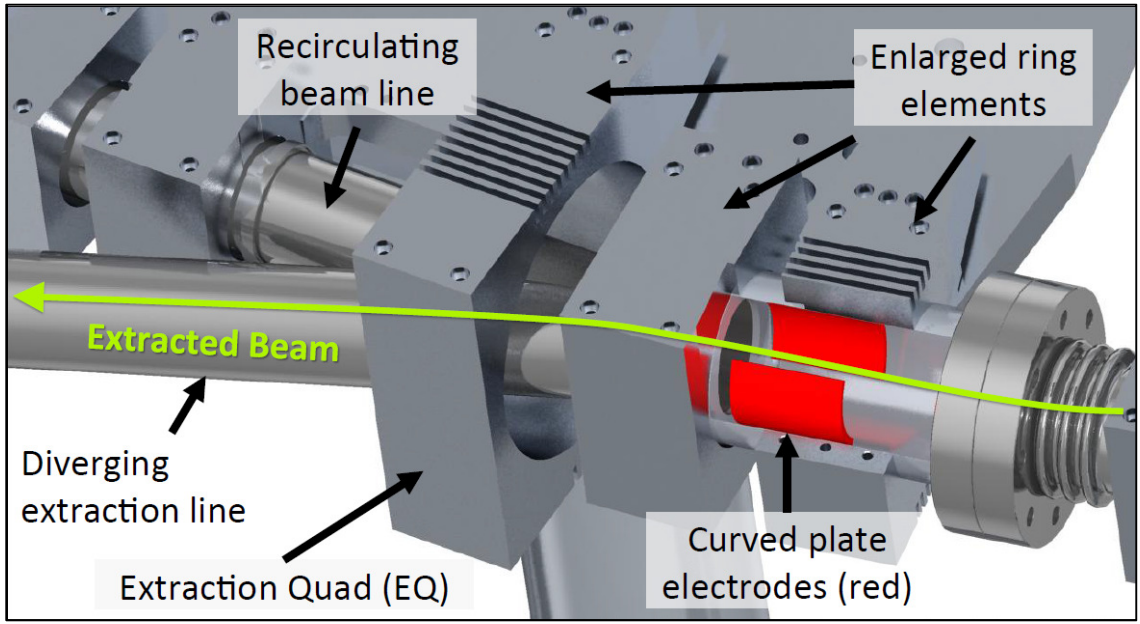

Figure L.3 CAD of conceptual design based on linear optics model.

## L.2. Refined model with WARP simulations

Higher order effects such as space charge and finite transverse beam size were examined with the WARP PIC code. The WARP transverse slice packaged is used, which probes 2D transverse dynamics assuming negligible transverse-longitudinal coupling. Fringe fields are included.

Quadrupole magnet fields were modelled using MAGLI [28]. The coil geometry is specified in an input file, and MAGLI uses a Biot-Savart solver to solve for the field contributions of each wire segment. The BKE electrostatic kicker was modeled using WARP's built-in geometry definitions[1] and field solver. The field profiles for the curved plate geometry are shown in the next section (Figure L.8). The Earth's magnetic field was included in all simulations, approximated as a constant vertical field of 0.4 T, the average measured field along the ring circumference. The horizontal Earth field is not included.

---

[1] Module defined in generateconductors.py

### L.2.1      *Implementation of the "warped" frame*

WARP allows bent coordinate systems, in order for fields to be specified in the (approximate) beam-frame rather than the lab-frame. WARP allows user-specified fields on a grid, but specifying field profiles that overlap with WARP "bend" elements is difficult due to the bent coordinate frame. To address this problem, Rami Kishek developed a method of field-solving on a bent coordinate system. First, the desired coordinate system is specified in Python, and a series of grid points are supplied to the MAGLI solver. The resulting field data file can be directly imported into WARP, as long as the coordinate systems are aligned and follow the same bending radius. This method was used for all fields in the WARP simulations described below. More description of this method can be found in [73].

### L.2.2      *WARP tracking of extracted beam*

Design orbit tracking was carried out in WARP with the 0 mA (single particle), 23 mA and 80 mA beams, using a beam matched to recirculating conditions.[2] Kicker voltage was adjusting to find the optimal extraction trajectory. WARP simulations based on the single-particle linear optics model shown in Figure L.3 and Figure L.4 (which we will refer to from now on as V1) indicated the potential for scraping losses of the extracted beam path. When considering the finite transverse profile, the beam required additional clearance through the enlarged, horizontally focusing quadrupole (which would have required increasing the magnet bore significantly). A second model, V2, was developed, with the kicker and diverging pipe upstream by approximately 16 cm. This solution had been overlooked in the initial conceptual design but has a much better prognosis in terms

---

[2] Using matched beam conditions defined in /humer/shared/Universal/Ubeams.py

of beam clearance. All subsequent explanations are relevant to the second model V2. Both configurations are shown in Figure L.4. The form of the extraction section vacuum pipe was re-engineered to accommodate the new extraction trajectory. More details are given in [80].

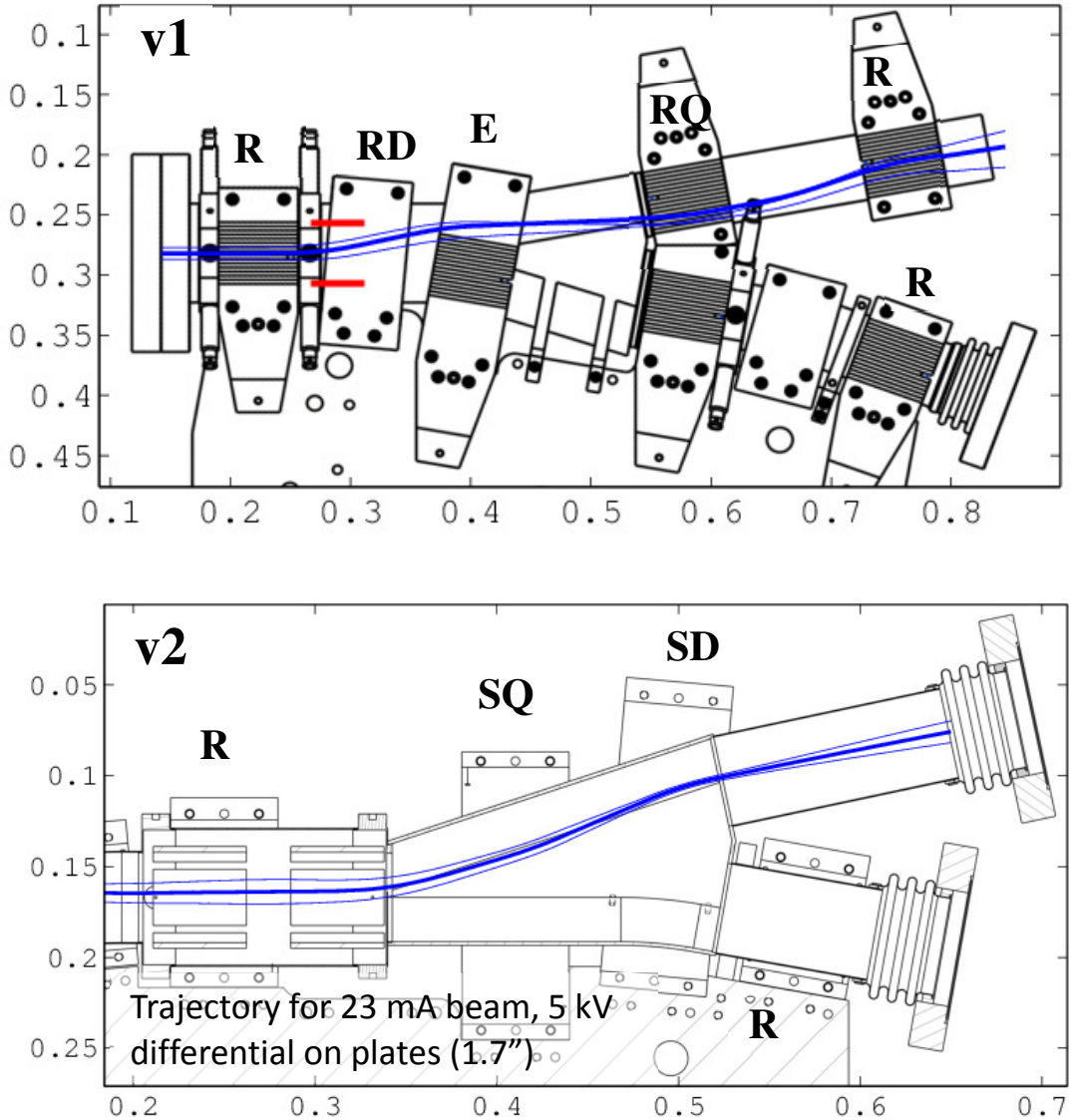

Figure L.4 Orbit tracking for 23 mA beam, for early design (v1) and refined design (v2), in which the kicker plates and Y-section have been relocated ~16 cm downstream. Both simulations show centroid (heavy line) and transverse rms beam edges (thin lines).

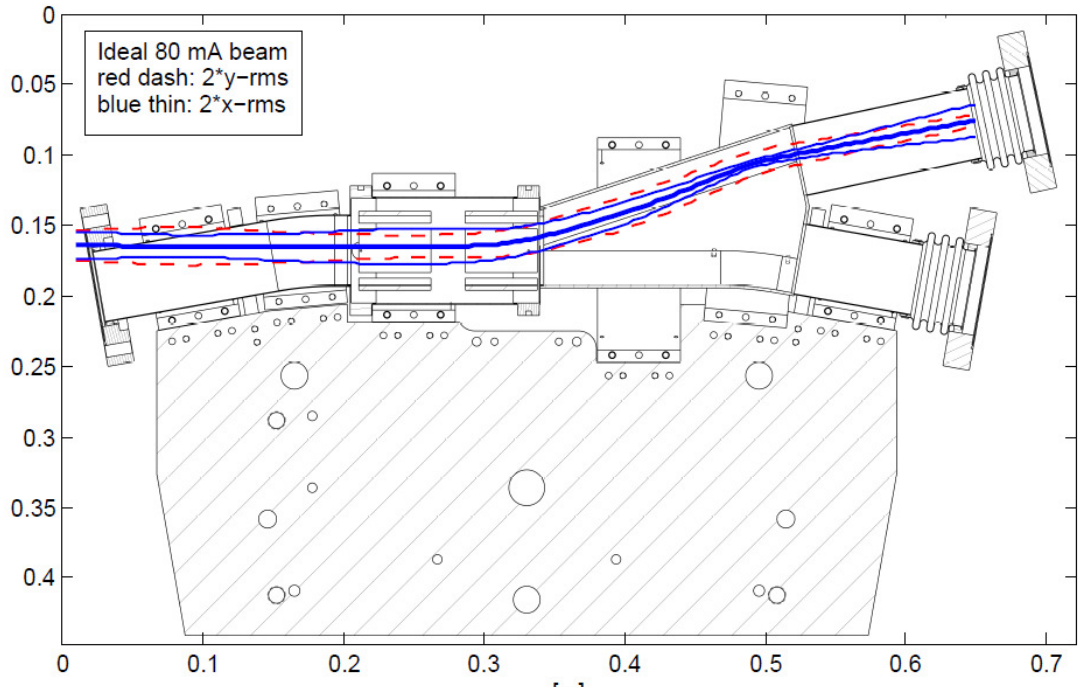

Figure L.5 Centroid and envelope tracking for matched 80 mA beam in extraction. Misalignment of beam centroid with pipe axis in beginning 16 cm is because the simulation follows the beam frame, using WARP bend elements.

### L.3.  *Hardware design*

The following section describes the mechanical design for the extraction section, including the vacuum pipe, enlarged-bore magnets and kicker electrodes. The bulk of the mechanical design work was done by Jacob Butcher and James Wissman.

### L.2.1  *Vacuum pipe design*

The extraction section requires a custom vacuum pipe with a diverging extraction pipe and enlarged pipe radius to accommodate the extraction kicker. The pipe is designed as a monolithic structure, free of demountable joints so as to accommodate the ring's dense lattice. The extraction Y-section was slightly modified from the injection Y-section, which contains a sharp edge that is troublesome due to suspected beam scraping. To avoid sharp edges and to simplify machining, we chose a 5-piece design for the custom

pipe section, which is connected through a series of welds. The entire section subtends 20° of the ring and will be a modular replacement for a standard ring section. The enlarged section of pipe houses two sets of 4-electrode plates that serve as a Beam Position Monitor (BPM) and the Beam Kicker Electrodes (BKE).

The BPM plates are coaxially located in thick-walled "collars" on which are directly welded commercially available SHV electrical feedthroughs. The BPM plates are supported solely by the center conductor of the SHV connectors, which attaches .25" from the longitudinal end of the plate for an asymmetrical mount. After assembling the BPM plates, the beam pipes are also then welded to the collar. At this point, the BPM/BKE plates will no longer be accessible. The outer (atmosphere) profile of the collars is octal. An SHV feedthrough is mounted to every other flat of the octagon. The remaining flats are blind-tapped to provide the mounting locations of the vacuum section to its support stand.

Starting at the up-stream flange, the extraction section begins with a 4.5" CF flange, followed by a short length of standard two-inch beam pipe with 10° bend, allow the first quadrupole and dipole to remain at normal UMER dimensions. Coupled to a zero-length adapter, the pipe transitions to a 3" diameter pipe. The enlarged beam pipe houses the two BPM/BKE assemblies before terminating at the Y-section. The Y-section is a diverging vacuum box constructed of flat and rolled plates welded together. The complication of rounded sides is necessary to minimize the bore of the SuperQuad SQ magnet. The Y-section provides the transverse clearance for both the recirculating beam and growing horizontal displacement of the kicked beam. At the longitudinal location

where the transverse displacement of the recirculating and extracted orbits differ by approximately a two-inch beam pipe diameter, the Y-section breaks into two discrete two-inch beam pipes: the recirculating beam pipe and the extraction line. From that point forward, the recirculating line follows the standard UMER beam pipe run. The extraction section nominally angle off at 12°, but a welded bellows (not shown in drawing) permits slight adjustment. Both down-stream beam pipes terminate in 4.5CF flanges.

To reduce undesired longitudinal fields, the BPM plates are located in recessed pockets such that the beam-facing contour lies on the same surfaces as the up- and down-stream grounded beam pipe walls. In their section of beam line, the pipe diameter is 3". Therefore, to create the effect of re-entrant pockets, rings with ID of 2.5 cm are machined and welded into the enlarged pipe. A simulation model showing the geometry and field-guarding properties of the guard rings can be seen in Figure L.9.

Due to concerns about the welds on the downstream wall of the box, a finite element analysis was performed to assure structural stability under UMER vacuum. We sought advice from the RU machine shop, who assured us that the welds should not present a structural problem. They suggested that the downstream face of the box should be a single piece of steel with two holes for the diverging pipes, rather than a 2-piece welded face as shown in the drawing. The Rutgers machine shop quoted $15k for the entire flange-to-flange pipe and BKE/BPM assembly.

### L.2.2    *Support design*

Limited work was done on designing the support structure for the extraction section. The support plate must be heavily modified to accommodate the enlarged magnet elements, as well as the electrical feedthroughs for the BPM/BKE collars. Consideration must be

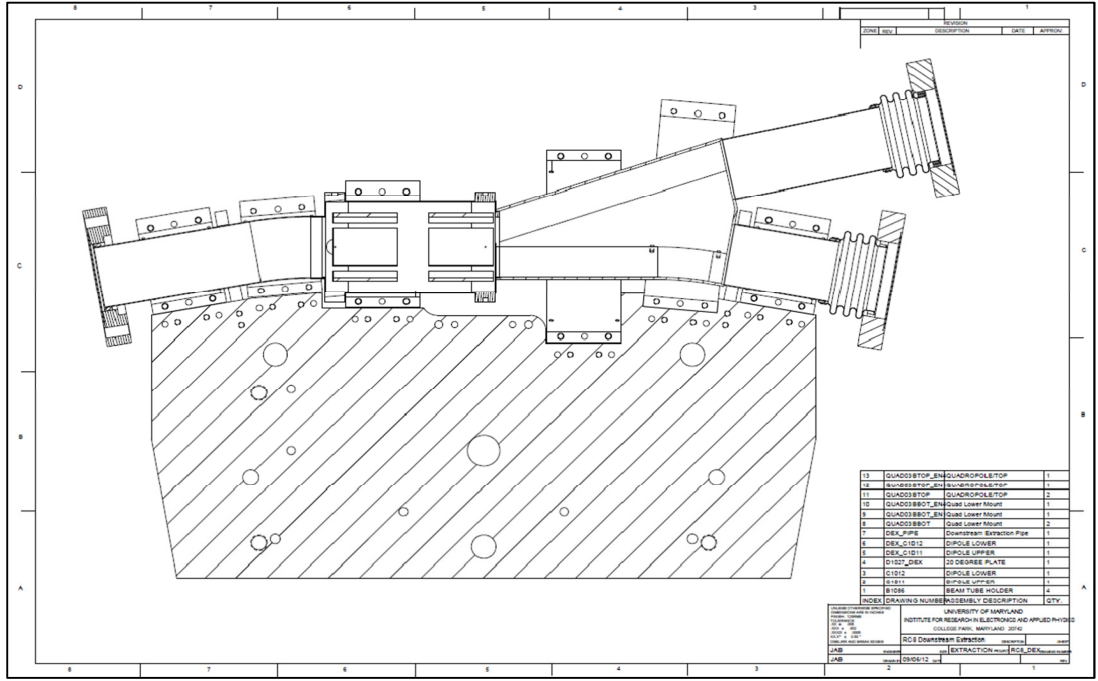

Figure L.7 Drawing of vacuum chamber, showing rough sketch of modified support plate.

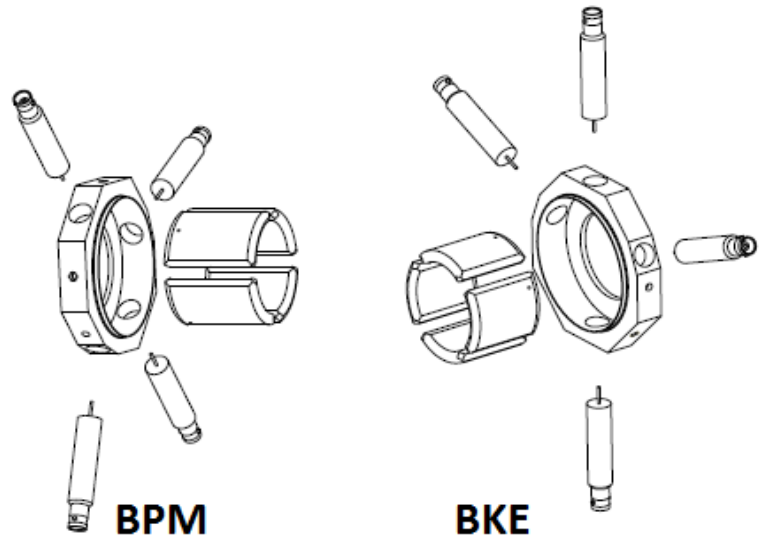

Figure L.6. BPM and BKE assemblies, from Vacuum Chamber drawings package.

made for proper alignment of the BPM/BKE assemblies. The working idea for support structure is to incorporate a cradle to directly support the octagonal BPM/kicker collars. Preliminary drawings of this cradle were made.

### L.2.3    Magnet design

The modified extraction pipe will replace the UMER lattice in a single 20° section. Enlarged magnets must be designed to accommodate the new pipe geometry. As shown in Figure 1, the extraction requires several enlarged magnet: two enlarged quadrupoles ("Extraction Quad," EQ and "Super Quad," SQ) and an enlarged dipole ("Super Dipole", SD). Extensive work on the magnet designs was carried out by Santiago Bernal and his students Ellen Wetzel and Tim Petro.

Table L.1 Magnet parameters for comparison between ring elements and custom, enlarged extraction elements

| Magnet | Dimension | Operating Current | Effective Length | Peak on-axis gradient/field | Creator |
|---|---|---|---|---|---|
| 3rd Gen. Ring Quad (RQ) | r=29.5 mm | 1.8 A | 5.164 cm | 3.61 G/cm·A | Rami/ Santiago |
| Super Quad (SQ) | r=71.06 mm* | ~10 A | 9.81 cm | 0.3826 G/cm·A | Ellen |
| Enlarged Quad (EQ) | r=42.2 mm | 3.7 A | | 1.44 G/cm·A | Tim Petro |
| Ring Dipole (RD) | r=29.5 mm | 2.97 A | 3.810 cm | 5.216 G/A | Rami/ Santiago |
| Super Dipole 5 (SD5) | $w_x$=279 mm $w_y$=40 mm $w_z$=58.93 mm | 136 A·turn | N/A | N/A | Tim Petro/ Santiago |

* SQ radius has since been updated to 80.969 mm, but the SQ calculations all reflect the original radius and simulations use smaller radius (meaning, superquad.spc has smaller radius). Change was to allow pipe clearance of downstream magnet housings.

The enlarged quadrupoles are printed circuit magnets, and their field profiles are modelled using the in-house Biot-Savart solver MAGLI [28] and included in the WARP simulations described in this appendix.

The Super Dipole design is still incomplete (therefore realistic fields have not been included in the WARP solution). Unlike the quads, we are not limited to a symmetric aperture, and the most cost-effective design can have a rectangular aperture. We hope to hand-wind the dipole coils, rather than using printed circuits. This is due to the set-up cost required for PC manufacturing (it is only cost-effective to print a large number PC's, as we lack in-house ability to DIY), the fact that we are not limited to a circular aperture, and increased flexibility (small adjustments can be made to the coils without purchasing a new set). The only requirement on the super dipole is that the field profile on the recirculating beam path must be relatively flat. This requirement is relaxed for the extracted beam path, as the beam only makes one pass through the extraction line.

The simplest SD design is a pair of rectangular coils, where the peak dipole field is centered on the recirculating beam pipe. This requires minimum horizontal aperture of ~10.9" to accommodate the pipe. Santiago considered several SD coil configurations to try to minimize negative field "wings" along the longitudinal field profile.

The reverse-field dipole (RFD), as suggested by Dave Sutter, was an attempt to provide flat field profiles in both the extraction and recirculating beamlines, inspired by wiggler geometries. The stacking of multiple coils is meant to sharpen the transition between dipole polarities. In this case, the dipole field centered over the extraction pipe will actually assist the kicker in bending the beam away from the ring (instead of counteracting the kicker). Implementation of this design would require restructuring of

the beam pipe at the dipole to take into account an increased extraction angle. Simulations by Tim Petro found that, for the proposed RFD design, the flatness of the dipole profile was much worse for recirculation. For now, it seems best to stick with a large rectangular, bore single dipole.

### L.2.4 Kicker design

The electrostatic extraction kicker BKE is based on the success of the knock-out diagnostic.[65] This is a multi-turn transverse imaging diagnostic that uses a BPM as a pulsed kicker. The BPM is rewired such that 3 plates are grounded, and the top plate is pulsed with a peak voltage of 1 kV. The pulse perturbs the beam, which experiences centroid Betatron oscillations, and terminates downstream on an off-axis phosphor screen. The extraction kicker will be a similar arrangement of 4 curved plates, with a slightly modified geometry to fit inside the extraction section pipe. The kicker design with equipotential lines is pictured in Figure L.8. The extraction kicker will fit inside the

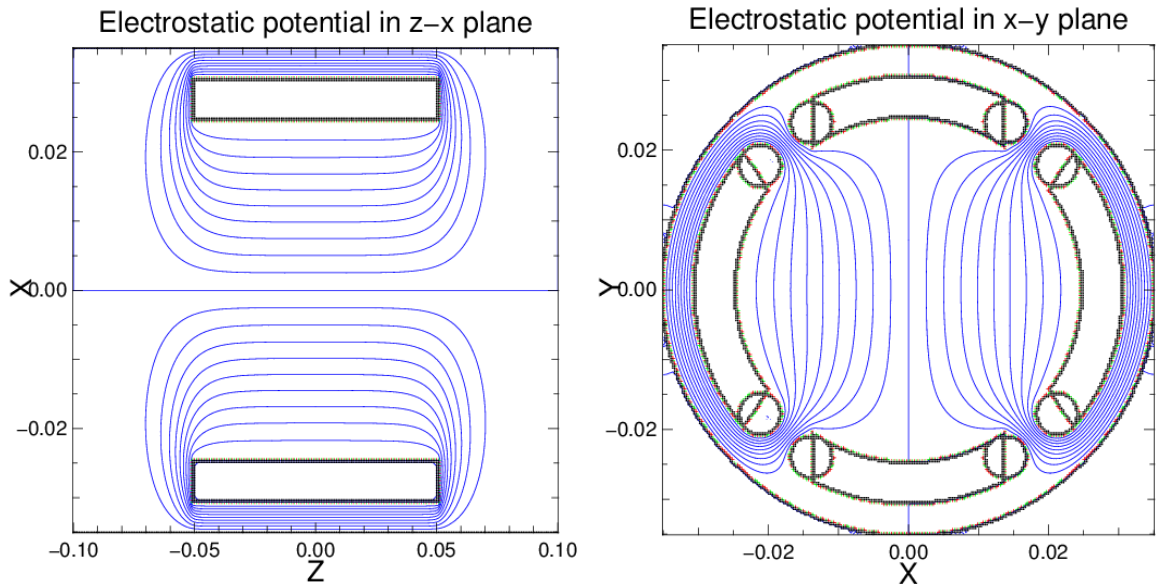

Figure L.8 Equipotential lines for kicker electrodes (BKE), generated using the WARP geometry module and 3D field solver, setup in file kplates.py

3.5" radius can, immediately downstream of an identical electrode for BPM output. A "guard" ring will be included between the 2 sets of electrodes to minimize interference.

The effect of the kicker electrodes on beam profile and kick angle was simulated to identify an optimal longitudinal length. A minimum exists around 5.5 cm, above which the beam starts to scrape on the kicker electrodes. Beyond this regime, emittance growth seems negligible. However, this study is flawed due to failure to transform the emittance measurement from the lab to the beam frame. Ultimately, kicker length was dictated by design to be 1.7" (4.318 cm), in order to allow room for the diverging pipe and neighboring BPM, which is identical to the kicker.

### L.3.4.1 Guard ring study with Poisson Superfish

Guard rings were proposed to reduce pick-up between BKE and BPM electrodes. A 2D field solution was found used Poisson Superfish [81]. The resulting field profiles suggest that the installation of guard rings will decrease interference of the BKE pulse with the BPM signal, although this will still have a large effect due to the proximity of the plates. Inner diameter 2.5 cm was chosen to not limit pipe aperture beyond standard pipe diameter. Simulation results are plotted in Figure L.10 and Figure L.9 without and with rings.

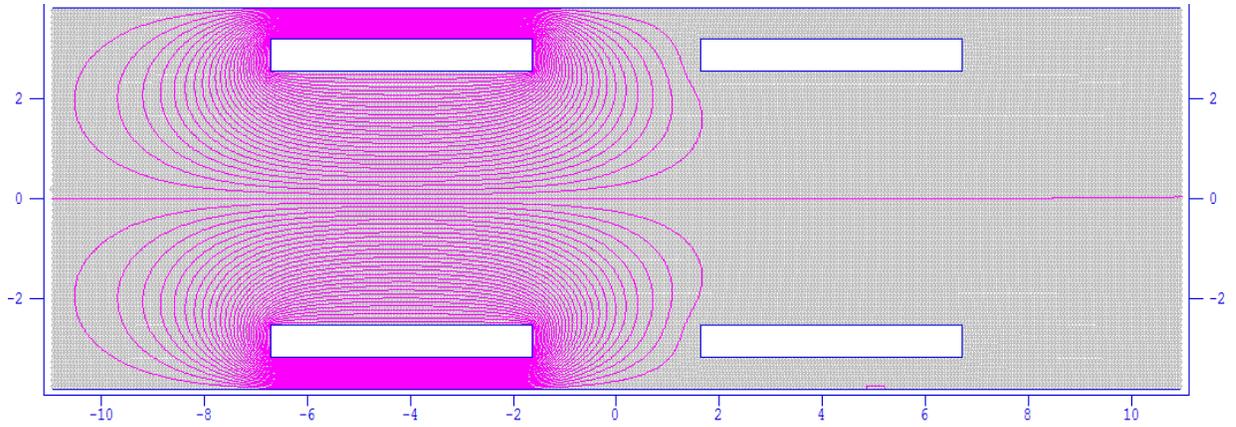

Figure L.10 Poisson Superfish simulation of biased BKE (left) and neighboring BPM without guard ring. This is a horizontal slice view of the BPM/BKE in the enlarged pipe. Units are in cm.

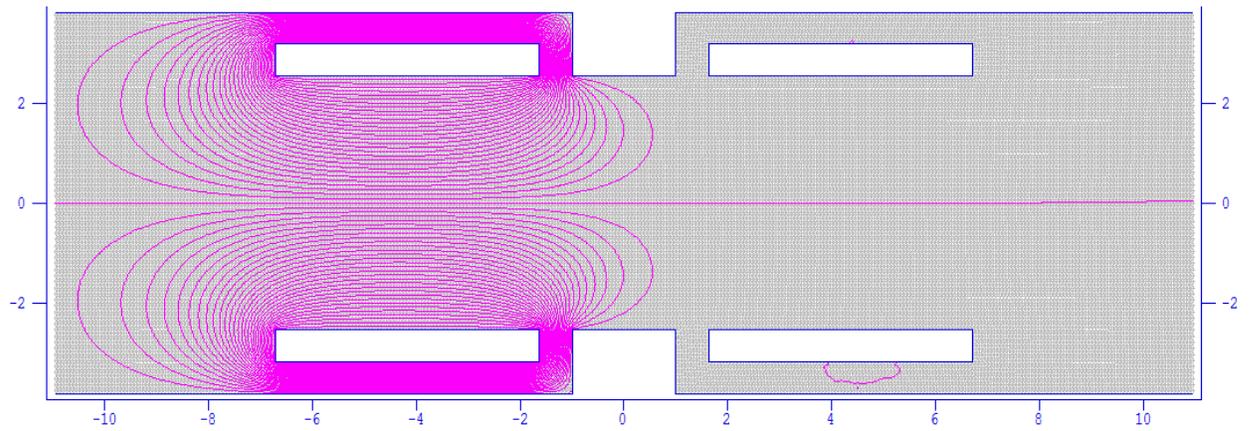

Figure L.9 Poisson Superfish simulation of biased BKE (left) and neighboring BPM with guard ring. Field lines are clearly more contained, and we can expect less interference on BPM due to pulsed BKE.

## L.4. Acceptance studies

After the mechanical design was sufficiently evolved, acceptance studies of the extracted beam were performed. The focus was on acceptance of off-axis beams, given that in general the UMER beam is not well-aligned to the pipe axis. In this study, the extraction acceptance is compared to acceptance in recirculation, for the geometries shown in Figure L.11 and Figure L.12. The recirculation case uses a "standard" 20° ring section with the typical QR quadrupole magnet, while the extracted case uses the enlarged magnets described above.

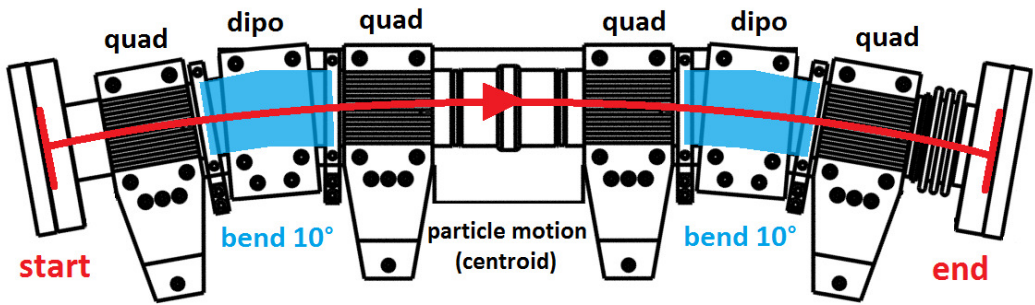

Figure L.11 Geometry of recirculation simulation (0 kV kick) in WARP. Model bends and imported bgrd fields are indicated. Simulation runs "flange to flange" for 0.64 m, or 2 periods.

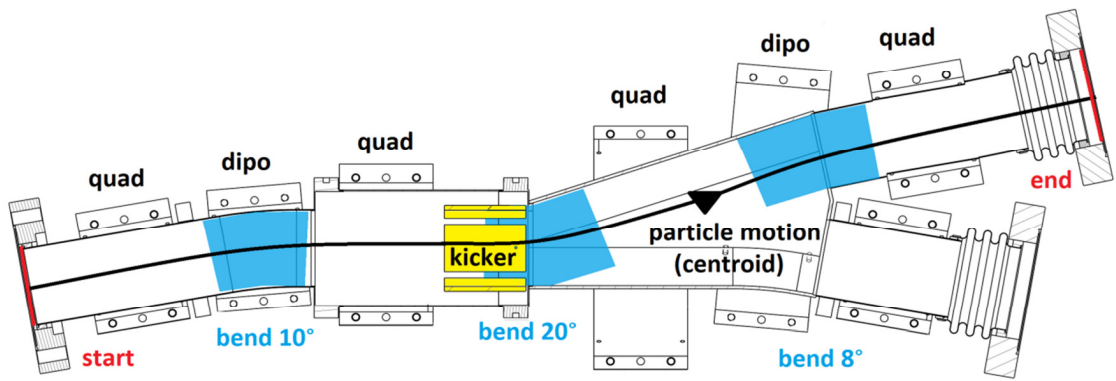

Figure L.12 Geometry of extraction simulation in WARP. Simulation runs "flange to flange" for 0.64 m.

*L.4.1      Parallel-beam acceptance studies*

For each case, a matched particle distribution is initiated at the upstream flange of the extraction section. Initial centroid position, centroid angle and kicker voltage are varied and the percentage of current to make it to the first downstream flange is recorded. In Fig. L.13, the survival percentages are plotted as a color-map versus XY centroid position (for parallel beams). Several voltage values are compared (0 kV indicates a recirculation simulation). The ideal voltage is identified as having the largest accepted (loss-free) range of initial conditions. This analysis was carried out for the 23 mA and 80 mA beams, which appear to have optimal BKE voltages of 3.8 and 3.2 kV respectively.

An interesting artifact was discovered in which, although the extraction line has smaller "parallel-beam" acceptance than the ring, the extracted parameter space has

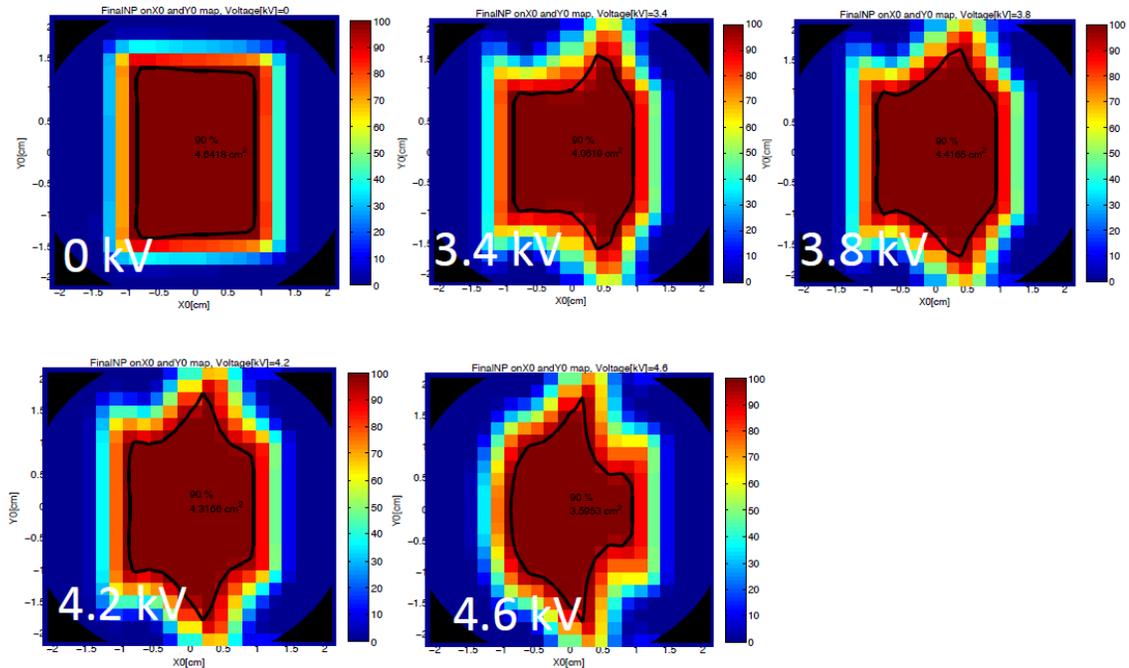

Figure L.13 Results of WARP parallel-beam acceptance studies for 23 mA beam. Each pixel is an initial beam centroid position. Color axis corresponds to percent loss of beam particles for an initially parallel (X'=Y'=0) beam at the initial XY position.

"wings" of acceptance that do not exist in the ring. Upon further investigation, this seems to arise from the fact that the SQ element has fields that extend further than the standard ring quadrupoles, which can prevent scraping in some circumstances. This design appears to be fairly tolerance to large vertical orbit errors.

### L.4.2 Single-particle acceptance

An alternative approach was used, in which the acceptance is measured by flooding the simulation with a distribution of non-interacting particles. For the initial distribution, particles were uniformly distributed spatially, completely filling the ring pipe, with uniformly distributed angular offsets in the range ±0.3 radians based on estimates of

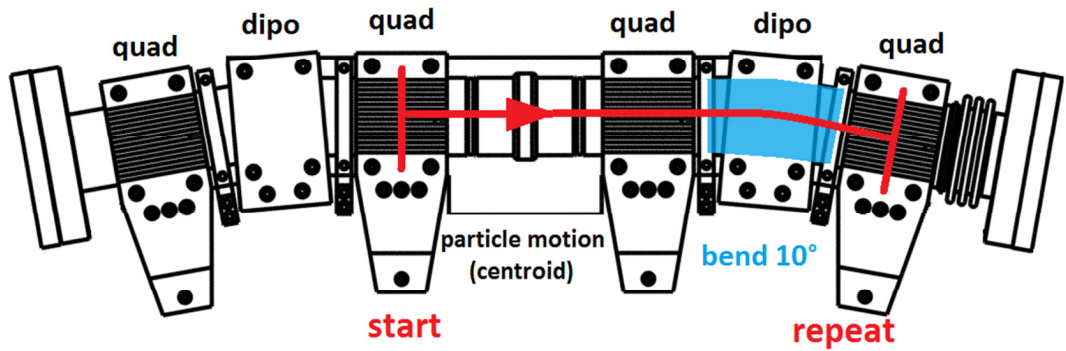

Figure L.14 Geometry of WARP simulation for recirculation acceptance. Model bends and imported bgrd fields are indicated. Simulation runs "quad to quad" and is repeated 36 times for 1 turn of ring (11.52 m).

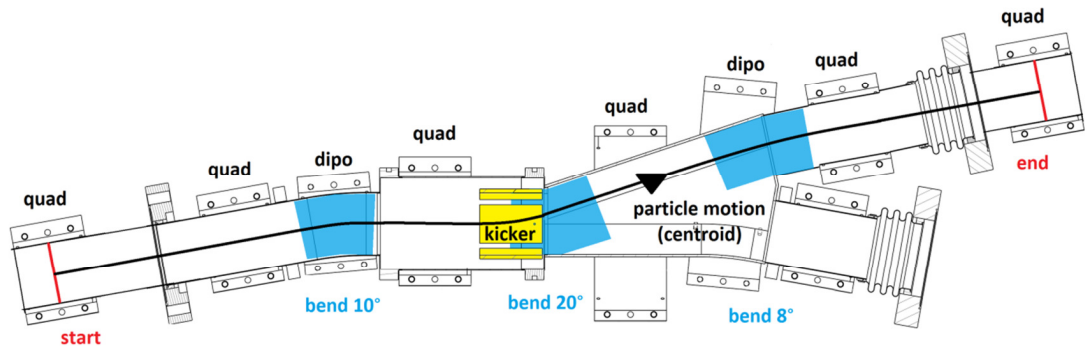

Figure L.15 Geometry of WARP simulation for extraction acceptance. Model bends and imported bgrd fields are indicated. Simulation runs "quad to quad" for 0.81 m.

maximum angular acceptance. The isotropic initial distribution was allowed to propagate through 10 periods of recirculation (3.2 meters) and through the extraction section (0.64 m), in both cases starting mid-quadrupole, as shown in Figure L.14 and Figure L.15. The preliminary results were used to estimate the maximum accepted emittance for both recirculation and extraction. The estimated acceptance value was verified by initializing an on-axis KV distribution of specified emittance and guaranteeing that there were no particle losses.

An on-axis, non-tilted, 0-current beam has acceptance $\epsilon_x = 750\ \pi$ mm-mrad and $\epsilon_y = 700\ \pi$ mm-mrad for recirculation and $\epsilon_x = 500\ \pi$ mm-mrad and $\epsilon_y = 450\ \pi$ mm-mrad through extraction. The results in the case of isotropic distribution are plotted in Figure L.16. These plots represent the uniform configuration and velocity distribution at the simulation start: the mid-plane of a horizontally focusing quadrupole. Black particles are lost during the run, magenta particles survive until the end. Yellow ellipses are added to provide a reference for comparison, enclosing areas of $\epsilon_x = 750\ \pi$ mm-mrad and $\epsilon_y = 700\ \pi$ mm-mrad.

These numbers are very promising, as they are approximately one order of magnitude above typical beam emittances. However, the complication of space charge forces was not considering in this analysis, as beam current was set negligibly small (order $10^{-11}$). Irv proposed an iterative approach to the isotropic "flooding" method, in which current is set to some non-negligible value for the initial isotropic distribution. Significant particle losses are expected after even a very small number of time steps, so every few steps the WARP beam current must be set to the desired current value. This is expected to converge on a maximum acceptance emittance for non-negligible current, but

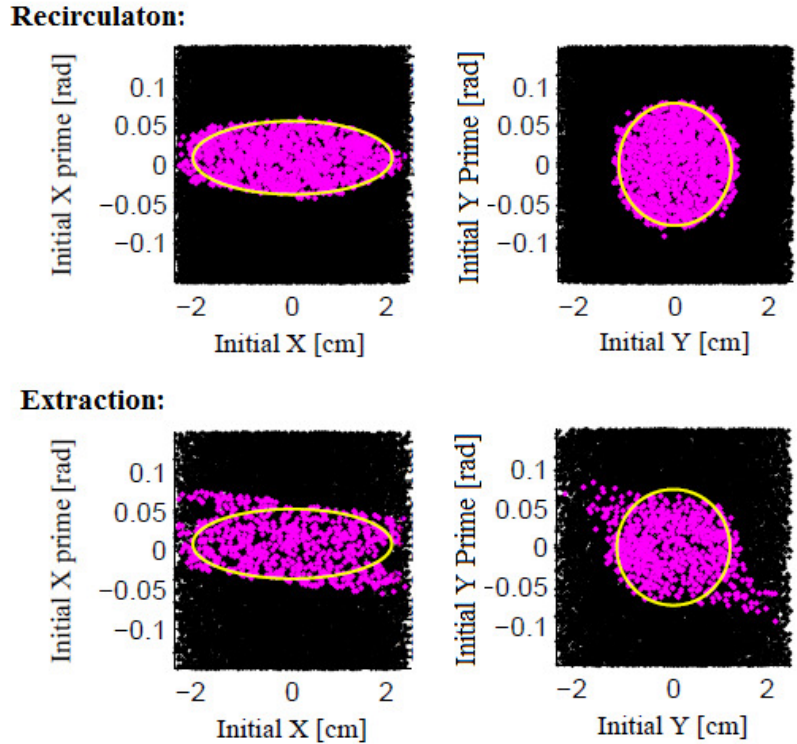

Figure L.16 0-current acceptance studies for recirculation (1 turn, 11.52 m) and extraction (.64 m, 10 kV kick voltage).

was never attempted in favor of an alternate method. As a reminder, these simulations were done with realistic magnet models except for the "Super Dipole" SD. An ideal, hard-edged super dipole was used. Simulations with the most recent super-dipole fields are shown below in Section L.7.

## L.5. Emittance growth caused by extraction section

Early on, during consideration of horizontal extraction scheme V1, recirculation simulations were run that tracked beam emittance growth due to the enlarged quadrupole elements. With the 6.5" radius quadrupole from the original design, an initially matched beam was propagated 20 turns through the ring (assumed ideal ring, no injection Y-section). The figure below shows a comparison of the beam emittance over total travel

length for two cases. Initially, the integrated gradient was scaled to be the same as the normal-sized ring quadrupoles. This caused significant emittance growth over the control case, with no enlarged elements. However, scaling the EQ gradient by 7.5% gave a much closer match to the control case over the 20 turn lifetime. This simulation was never repeated for the updated extraction scheme V2, mainly due to the need for a finalized SD design.

### L.6. 3D model of extracted beam dynamics

It was strongly suggested that I pursue 3D WARP simulations of the extraction section, as it would be a useful comparison to the 2D transverse-slice dynamics, and because the extraction section was an ideal candidate for implementing WARP 3D due to its short length. I wrote a complete 3D simulation, for a beam emitted continuously at z=0 (mid-quadrupole upstream of extraction). I included realistic pipe boundaries based on the vacuum chamber drawings. The model includes one WARP bend element used in combination with the ring dipole field (but no bend associated with the SD dipole). Quadrupoles and the SD were imported on a 3D grid, while the BKE geometry was

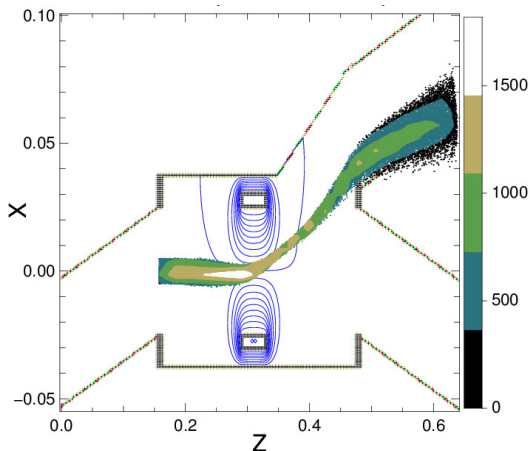

Figure L.18 Simulation results from 3D WARP model, showing particle densities and pipe geometry.

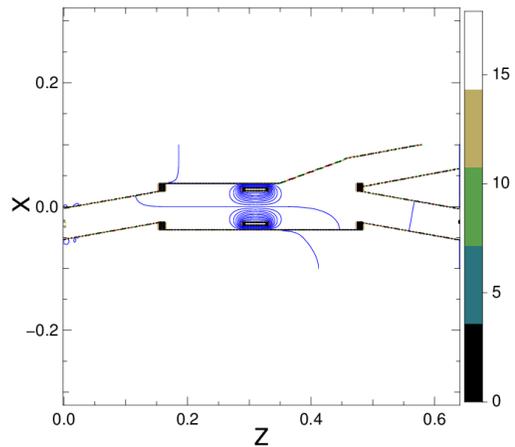

Figure L.17 To-scale 3D WARP geometry with extraction kicker.

specified along with the pipe geometry and held at a specified voltage in the WARP field solver. Even few-particle simulations (10 emitted per time step) took many hours to complete, and tuning the simulation for appropriate kicker strength was difficult. No enlightening results have been obtained using the 3D WARP model.

*L.7. Acceptance calculations with "Super Dipole" fields*

This section includes more acceptance diagrams. The 23 mA acceptance is decreased when SuperDipole SD5 fields are used (Figure L.19) compared to the case with ideal dipole fields shown in Figure L.15. Acceptable beam offsets are smaller in the case of the 80 mA beam, shown in Figure L.20 and Figure L.21.

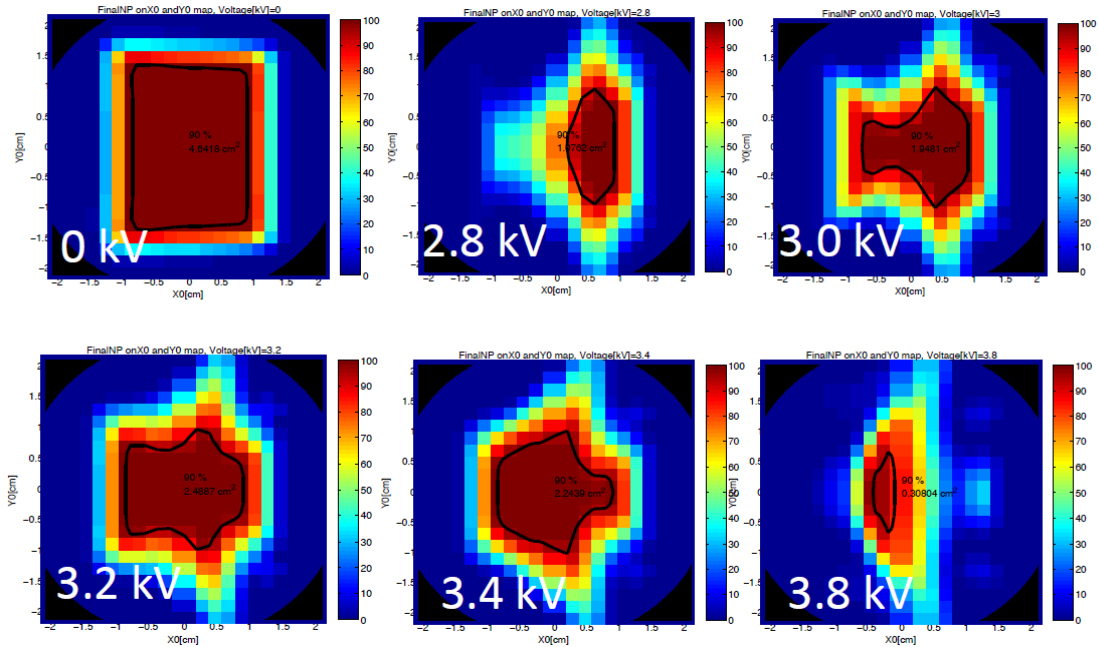

Figure L.19 Results of WARP parallel-beam acceptance studies for 23 mA beam with SuperDipole fields.

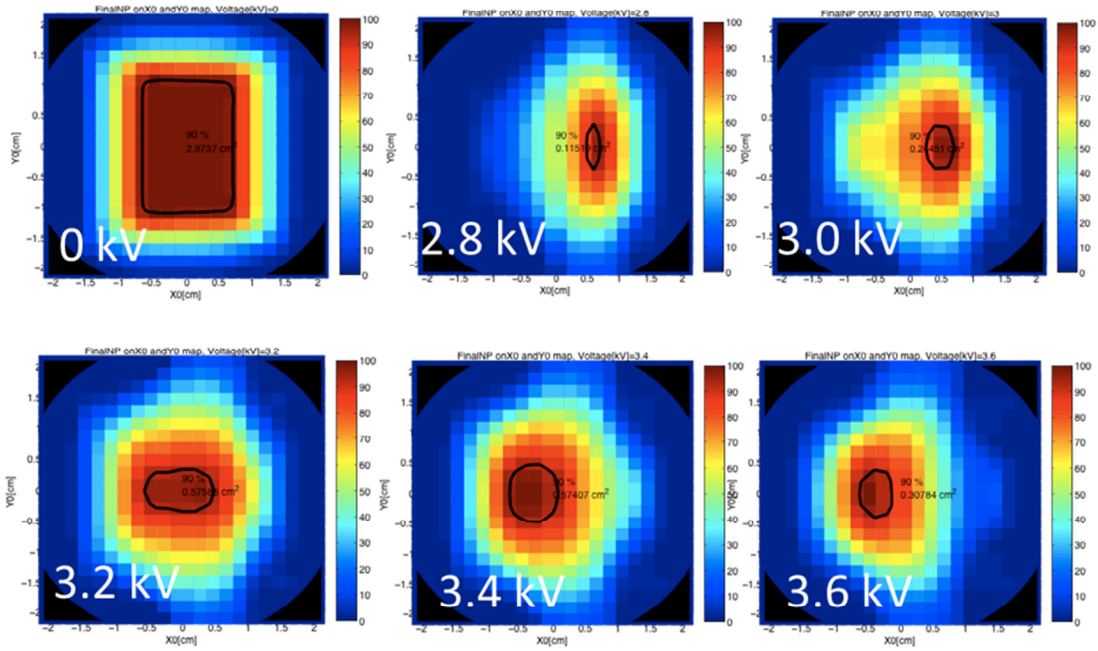

Figure L.20 Results of WARP parallel-beam acceptance studies for 80 mA beam with SuperDipole fields.

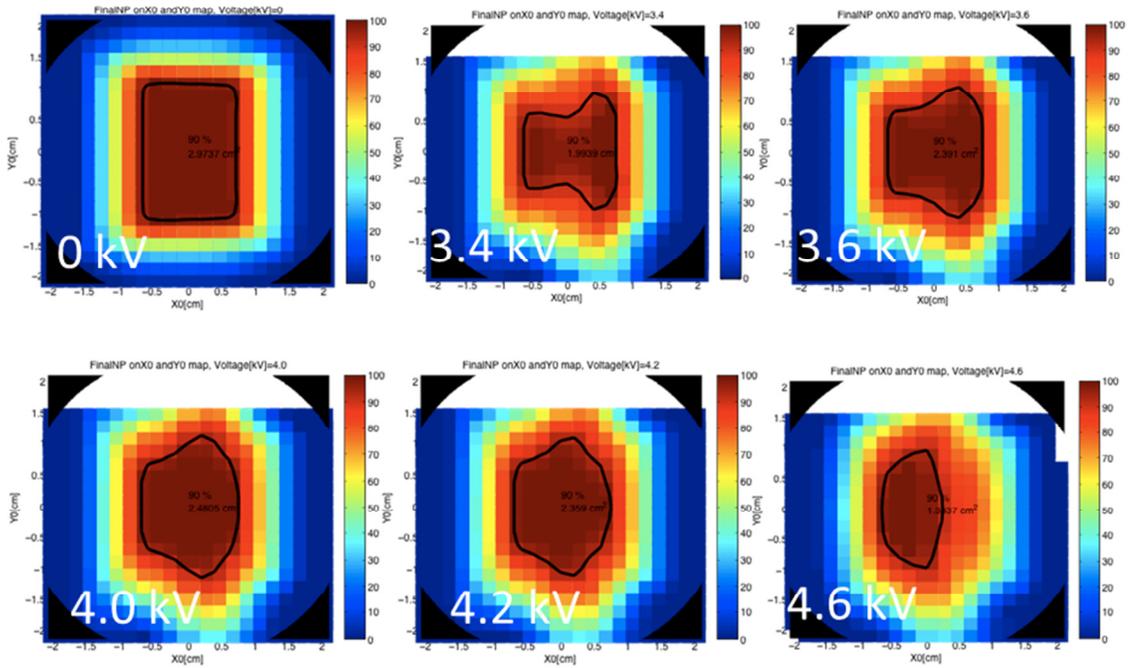

Figure L.21 Results of WARP parallel-beam acceptance studies for 80 mA beam with ideal (flat) large-aperture SD field

343

## L.8. Vertical extraction scheme

Vertical extraction was briefly discussed as an option, if horizontal schemes were untenable. Beam tracking simulations with WARP show that vertical extraction is much more straightforward from a linear optics perspective (repeating the conceptual study discussed in Section L.1. However, the difficulty of engineering and aligning such a set-up discouraged us from pursuing it further. The horizontal extraction scheme V2, was concluded to be the most feasible from an engineering/alignment perspective. Further details are discussed in [78].

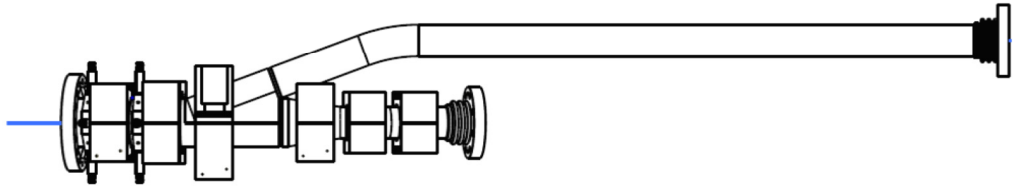

Figure L.22 Sketch of possible vertical extraction scheme. This is an identical vacuum chamber to extraction v1 that is rotated by 90°, with pipe levelling off at minimum height for clearance of magnet housings.

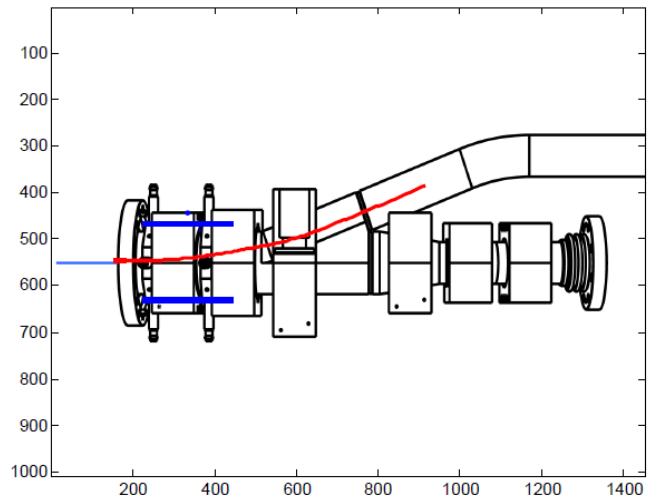

Figure L.23 Single particle trajectory using linear optics model for vertical extraction. With vertical extraction, the EQ (enlarged quad) helps increase the extraction angle. Unnormalized units. BKE plates are indicated in blue.

## L.9. Future work

Further refinement of the extraction model is needed in response to mechanical constraints. In particular, the initial angle of the diverging pipe is 20° in simulation but 18.25° in mechanical drawings. The second bend in extraction (in the SD) is 8° in sim. but 6.25° in drawings (final extraction angle is 12° in both cases). (This was done to reduce the required aperture of the enlarged magnets SD and SQ ) In addition, the BKE kicker model differs from the most recent CAD design: the biased plates sweep a greater transverse angle than the grounded plates, as seen in Figure L.8, while all 4 should be equal according to drawings. These minor mechanical details should be updated in the simulation geometry before proceeding.

A systematic study of the sensitivity of the extraction section performance to mechanical and beam errors (ie, misalignment, field errors, acceptance of mismatch) is desirable before cutting metal. This includes refining the SD model for better extraction acceptance than was calculated here.


